\documentclass[aps, prb,  nobibnotes, twocolumn, superscriptaddress]{revtex4-2}
\usepackage{graphicx}
\usepackage{amsmath,amssymb}
\usepackage{booktabs}
\usepackage{xcolor}
\usepackage{amsfonts}
\usepackage{array}
\usepackage{mathrsfs}
\usepackage{bm}
\usepackage{xspace}
\usepackage{epstopdf}
\usepackage{dcolumn}
\usepackage{longtable}
\usepackage{multirow}
\usepackage[title,titletoc]{appendix}
\usepackage{float}
\usepackage[colorlinks=true, linkcolor=blue, citecolor=blue, urlcolor=blue]{hyperref}
\usepackage{soul}
\usepackage{enumitem}
\usepackage{natbib}

\newcolumntype{C}[1]{>{\centering\arraybackslash}p{#1}}

\makeatletter

\newcommand{\Rmnum}[1]{\expandafter\@slowromancap\romannumeral #1@}
\makeatother

\newcolumntype{M}{>{$}c<{$}}
\renewcommand{\arraystretch}{2.2}

\begin{document}

\title{Directional Criticality and Higher-Order Flatness: Designing Van Hove Singularities in Three Dimensions}
\author{Hua-Yu Li}
\affiliation{Chongqing Key Laboratory of Micro $\&$ Nano Structure Optoelectronics, and School of Physical Science and Technology, Southwest University, Chongqing 400715, P. R. China}

\author{Hengxin Tan}
\email{hxtan@sjtu.edu.cn}
\affiliation{Key Laboratory of Artificial Structures and Quantum Control (Ministry of Education), School of Physics and Astronomy, Shanghai Jiao Tong University, Shanghai 200240, P. R. China}

\author{Hao-Yu Zhu}
\affiliation{Chongqing Key Laboratory of Micro $\&$ Nano Structure Optoelectronics, and School of Physical Science and Technology, Southwest University, Chongqing 400715, P. R. China}

\author{Hong-Kuan Yuan}
\affiliation{Chongqing Key Laboratory of Micro $\&$ Nano Structure Optoelectronics, and School of Physical Science and Technology, Southwest University, Chongqing 400715, P. R. China}

\author{Min-Quan Kuang}
\email{mqkuang@swu.edu.cn}
\affiliation{Chongqing Key Laboratory of Micro $\&$ Nano Structure Optoelectronics, and School of Physical Science and Technology, Southwest University, Chongqing 400715, P. R. China}

\begin{abstract}
Van Hove singularities (VHSs) play a pivotal role in driving correlated electronic phenomena. Traditional classifications focus only on critical points where the band gradient vanishes in all directions. Here we establish a unified classification of VHSs in three-dimensional systems, characterized by the number of vanishing gradient components and Hessian eigenvalues: ordinary ($M$-type), higher-order ($T_1$, $T_2$, $T_3$), noncritical ordinary ($N_0$, $N_1$, $N_2$), and noncritical higher-order ($S_1$, $S_2$) types. Noncritical VHSs exhibit directional quenching: the gradient vanishes in a two-dimensional subspace while remaining finite along the orthogonal direction, yielding finite density-of-states enhancements with distinct energy dependencies. Using an $s$-orbital tight-binding model on the pyrochlore lattice with spin-orbit coupling, we demonstrate that all singularity classes emerge at distinct high-symmetry points through controlled tuning of the hopping ratio. This work establishes directional criticality and higher-order flatness as design principles for tailoring density-of-states enhancements in three-dimensional quantum materials.
\end{abstract}
\maketitle

\section{Introduction}

The complex relationship between electronic band structure and electron-electron interactions lies at the heart of modern condensed matter physics, with Van Hove singularities (VHSs)\cite{van1953occurrence} playing a key role as non-analytic features in the density of states (DOS)\cite{yuan2020classification,patra2025high,efremov2019multicritical,yuan2019magic,classen2025high}. 
Traditional VHSs occur at stationary points in the electronic dispersion ($\nabla_{\mathbf{k}} \varepsilon(\mathbf{k}) = 0$), corresponding to topological transitions in Fermi surface configurations. When the Fermi level approaches these singularities, the significantly enhanced DOS amplifies electronic correlation effects, often triggering collective quantum phases such as magnetism\cite{hausoel2017local,sala2021van}, spin density waves\cite{makogon2011spin,liu2018chiral,isobe2018unconventional}, charge density waves\cite{wilson2024v3sb5,yin2022topological,teng2022discovery,teng2023magnetism,hu2024phonon,tan2021charge}, or superconductivity\cite{isobe2018unconventional,gonzalez2019kohn,hao2021electric,wu2021nature,xu2021tunable,liu2018chiral,stepke2017strong,tan2021charge,chichinadze20224}. Yet this paradigm of fully critical points leaves unexplored a broader landscape of singularities, including noncritical singularities where criticality is confined to a subspace, and higher-order singularities where band flattening dramatically reshapes the DOS.

Building upon this foundation, the exploration has advanced to higher-order Van Hove singularities (HOVHSs), which emerge when the electronic dispersion exhibits not only a vanishing gradient but also a zero determinant of the Hessian matrix at the stationary point\cite{classen2025high}. Such singularities, where the standard quadratic expansion becomes insufficient and cubic or higher-order terms govern the dispersion, lead to more exotic DOS behavior. For instance, a two-dimensional monkey saddle ($E \sim k_x^3 - 3 k_x k_y^2$) produces a stronger power-law divergence $g(E) \sim |E|^{-1/3}$ compared to the conventional logarithmic form\cite{shtyk2017electrons}. 
However, in three-dimensional systems, their realization generally requires extremely fine-tuning of the band structure, posing significant challenges for tunability and accessibility\cite{park2024anomalous,patra2025high,classen2025high,efremov2019multicritical,wu2024discovery,tan2024three}.

\begin{table*}[htbp]
\centering
\renewcommand{\arraystretch}{1.6} 

\caption{\textbf{Classification of Van Hove Singularities (VHSs).} 
The classification is strictly governed by the polynomial exponents $(n_\alpha, n_\beta, n_\gamma)$ of the local energy dispersion and the topological properties (eigenvalues) of the corresponding full or reduced Hessian matrices.}
\label{tab:VHS_Classification}

\begin{tabular*}{\textwidth}{@{\extracolsep{\fill}} c c c c l l @{}}
\toprule
\multicolumn{3}{c}{\textbf{Polynomial Exponents}} & 
\multirow{2}{*}{\textbf{Topological Indicator}} & 
\multirow{2}{*}{\textbf{Classification}} & 
\multirow{2}{*}{\textbf{Description}} \\
\cmidrule(lr){1-3}

\makebox[0.08\textwidth][c]{$\boldsymbol{n_\alpha}$} & 
\makebox[0.08\textwidth][c]{$\boldsymbol{n_\beta}$} & 
\makebox[0.08\textwidth][c]{$\boldsymbol{n_\gamma}$} & & & \\
\midrule

1 & 2 & 2 & 
\parbox[t]{0.26\textwidth}{\raggedright Number of negative eigenvalues \\ of the reduced $2 \times 2$ H matrix} & 
\parbox[t]{0.22\textwidth}{\raggedright $N_0$ (in-plane minimum) \\ $N_1$ (in-plane saddle) \\ $N_2$ (in-plane maximum)} & 
\parbox[t]{0.18\textwidth}{\raggedright Ordinary \\ noncritical VHSs} \\
\addlinespace[4pt] 
\midrule
1 & $\ge 2$ & $\ge 2$ & 
\parbox[t]{0.26\textwidth}{\raggedright Number of exponents \\ exceeding 2.} & 
\parbox[t]{0.22\textwidth}{\raggedright $S_1$ (one $n > 2$) \\ $S_2$ (two $n > 2$)} & 
\parbox[t]{0.18\textwidth}{\raggedright Higher-order \\ noncritical VHSs} \\
\midrule 

2 & 2 & 2 & 
\parbox[t]{0.26\textwidth}{\raggedright Morse index $\lambda$ \cite{van1953occurrence} \\ ($\equiv$ number of negative $\sigma_i$)} & 
\parbox[t]{0.22\textwidth}{\raggedright $M_0$ (minimum) \\ $M_{1,2}$ (saddles) \\ $M_3$ (maximum)} & 
\parbox[t]{0.18\textwidth}{\raggedright Ordinary VHSs} \\
\addlinespace[4pt] 
\midrule
$\ge 2$ & $\ge 2$ & $\ge 2$ & 
\parbox[t]{0.26\textwidth}{\raggedright Number of zero eigenvalues \\ of the full H matrix \\ ($\equiv$ number of $n > 2$)} & 
\parbox[t]{0.22\textwidth}{\raggedright $T_1$ (one $n > 2$) \\ $T_2$ (two $n > 2$) \\ $T_3$ (three $n > 2$)} & 
\parbox[t]{0.18\textwidth}{\raggedright Higher-order VHSs} \\

\bottomrule
\end{tabular*}
\end{table*}

Beyond these critical points, we uncover a previously underappreciated class: the noncritical singularity. Its hallmark is the directional quenching of the band gradient, where criticality is satisfied only within a two-dimensional subspace while remaining finite along the orthogonal direction---for instance, $\partial\varepsilon/\partial k_x = \partial\varepsilon/\partial k_y = 0$ while $\partial\varepsilon/\partial k_z \neq 0$. This anisotropic flattening gives rise to extended line-like critical contours in momentum space, yet yields large but finite density-of-states enhancements. The finite group velocity out of the critical plane suppresses true divergences while allowing a substantially enhanced DOS to persist over a finite energy window. This mixed-dimensionality character manifests in ordinary ($N$-type) and higher-order ($S$-type) noncritical families. We establish a unified algebraic framework that classifies all singularities into ordinary ($M$-type), higher-order ($T$-type), noncritical ordinary ($N$-type), and noncritical higher-order ($S$-type) classes, as detailed in the following.

We further demonstrate that the pyrochlore lattice serves as a natural platform realizing the entire taxonomy. Through tight-binding (TB) modeling on the pyrochlore lattice, we show that all singularity classes described above emerge at distinct high-symmetry points, with quantitative agreement between analytical predictions and numerical tight-binding calculations. Our findings establish a unified paradigm of directional criticality and higher-order flatness, transforming Van Hove singularities from serendipitous band features into designable elements of quantum materials and providing a new route to engineering correlation-driven phenomena in three dimensions.

\section{Systematic Classification of Van Hove Singularities}
\label{sec:classification}
Ordinary VHSs serve as the foundation for our extended framework. In one dimension, band edges yield a square-root divergence $\sim |E-\epsilon_0|^{-1/2}$ \cite{bohm1991material,zeng2008charge}; in two dimensions, saddle points yield a logarithmic divergence $\sim -\ln|E-\epsilon_0|$ \cite{yuan2019magic,li2010observation,seiler2022quantum}; in three dimensions, band edges give parabolic DOS edges while saddle points form finite cusps with divergent derivatives \cite{wu2021nature,tamai2008fermi}. The singularity strength diminishes from one to three dimensions, with nontrivial saddle points giving the most pronounced signatures.

To establish a rigorous classification encompassing both ordinary and higher-order VHSs, we introduce a generalized three-dimensional polynomial energy dispersion near $\mathbf{k}=0$, assumed separable in Cartesian coordinates with no cross terms:
\begin{equation}
\varepsilon(\mathbf{k}) = \varepsilon_0 + \sum_{i=x,y,z} \sigma_i c_i k_i^{n_i},
\label{eq:dispersion}
\end{equation}
where $\sigma_i = \pm 1$, $c_i>0$, and $n_i\in\mathbb{N}^+$. The parity of $n_i$ is governed by local symmetry: at time-reversal invariant momenta (TRIMs), $\varepsilon(\mathbf{k})=\varepsilon(-\mathbf{k})$ forces $n_i$ even; at generic non-TRIM points, odd exponents are permitted.

The gradient and Hessian at the origin are:
\begin{align}
\left.\frac{\partial\varepsilon}{\partial k_i}\right|_{\mathbf{k}=0} &= \sigma_i c_i \delta_{n_i,1}, \label{eq:grad}\\
\left.H_{ii}\right|_{\mathbf{k}=0} &= 2\sigma_i c_i \delta_{n_i,2}. \label{eq:Hess}
\end{align}
Eq.\eqref{eq:grad} distinguishes \textit{critical} points ($n_i\ge2$ for all $i$) from \textit{noncritical} points (exactly one $n_i=1$) (see Table.~\ref{tab:VHS_Classification}). Cases with two or three linear directions are excluded as they do not correspond to VHSs.  Eq.\eqref{eq:Hess} acts as a topological switch: $n_i=2$ gives a nonzero eigenvalue, while $n_i>2$ yields a zero eigenvalue.

Crucially, in the context of our classification, the ``topological indicator'' refers to the topology of the constant-energy contours (Fermi surfaces) in momentum space undergoing Lifshitz transitions\cite{volovik2003universe}. Eq.\eqref{eq:Hess} acts as a rigorous topological switch grounded in Morse theory\cite{milnor1963morse}: an exponent $n_i=2$ gives a nonzero eigenvalue, indicating a non-degenerate Morse critical point. Conversely, when $n_i>2$, the corresponding second derivative inherently vanishes, yielding a zero eigenvalue in the Hessian matrix. Mathematically, this causes the determinant of the Hessian to vanish, transforming the critical point into a degenerate (non-Morse) singularity\cite{efremov2019multicritical}. Furthermore, our algebraic framework inherently encompasses vanishing derivatives of arbitrary higher orders ($>2$). Whenever the polynomial index $n_i > 2$ (e.g., $n_i = 4$ or $6$), Taylor's theorem dictates that not only the first and second derivatives vanish at the singularity, but all subsequent higher-order derivatives up to the $(n_i - 1)$-th order are also strictly and simultaneously zero. Therefore, this formulation naturally covers the entire singularity hierarchy with arbitrary high-order vanishing derivatives, ensuring an extended region of higher-order flatness.

\begin{figure}[tbp]
\centering
\includegraphics[width=\linewidth]{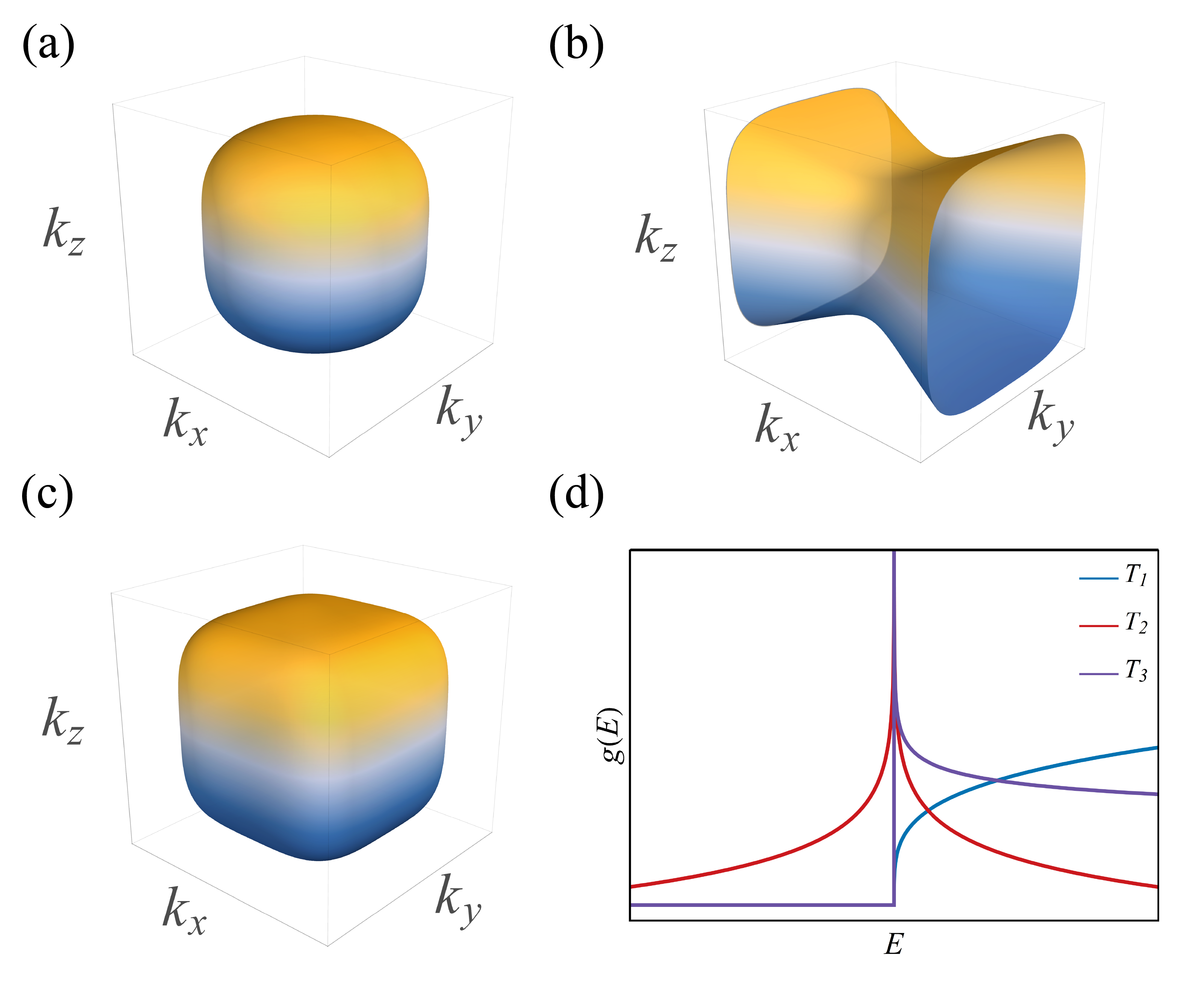}
\caption{\textbf{Representative higher-order VHSs.} 
(a)-(c) Fermi surfaces for $T_1$, $T_2$, and $T_3$ types. 
(d) Density of states $g(E)$ showing distinct scaling behaviors: finite peak with divergent derivative ($T_1$, blue), logarithmic divergence ($T_2$, red), and power-law divergence $|E-\varepsilon_0|^{-1/4}$ ($T_3$, purple).
Dispersions: $T_1$: $k_x^2+k_y^2+k_z^4$, $T_2$: $k_x^2-k_y^4-k_z^4$, $T_3$: $k_x^4+k_y^4+k_z^4$. See Supplemental Material for additional exponent combinations and parity regimes.} 
\label{figi}
\end{figure}

For critical points ($n_x,n_y,n_z\ge2$), the $3\times3$ Hessian determines the class. Ordinary VHSs ($M$-type) occur when $n_x=n_y=n_z=2$, for which the signs of the eigenvalues unambiguously determine the standard topological change of the Fermi surface (e.g., from an ellipsoid to a hyperboloid). The Morse index $\lambda$ (number of negative $\sigma_i$) is well-defined and gives $M_0$ (minimum), $M_{1,2}$ (saddles), and $M_3$ (maximum) (see Fig.S1 of the Supplemental Material). However, the rank and the presence of zero eigenvalues strictly serve to distinguish higher-order classes. Higher-order VHSs ($T$-type) arise when at least one $n_i>2$, rendering the Hessian singular (Figs.~\ref{figi}(a),~(b),~(c)). Physically, this vanishing eigenvalue fundamentally alters the universality class of the topological Lifshitz transition, forcing the Fermi surface to adopt higher-order geometrical connections (such as monkey saddles) rather than standard quadratic ones. The number of zero eigenvalues defines $T_1$ (one $n_i>2$), $T_2$ (two $n_i>2$), and $T_3$ (three $n_i>2$). 

\begin{figure}[htbp]
\centering
\includegraphics[width=\linewidth]{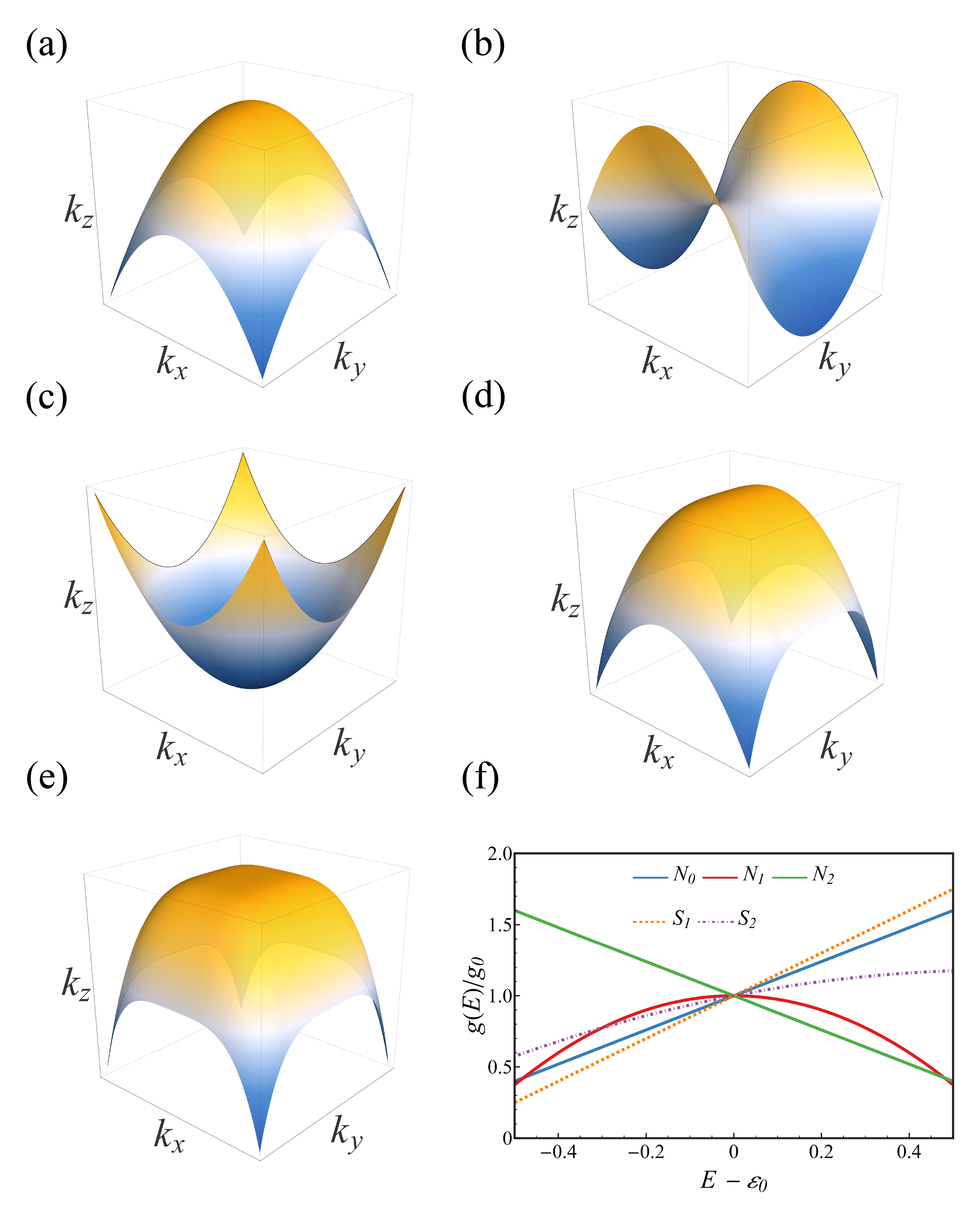}
\caption{\textbf{Representative noncritical VHSs.} 
(a)-(e) Fermi surfaces for ordinary noncritical ($N_0$, $N_1$, $N_2$) and higher-order noncritical ($S_1$, $S_2$) types. 
(f) Corresponding DOS: $N_0$ (blue, linear increase), $N_1$ (red, quadratic peak symmetric about $\varepsilon_0$), $N_2$ (green, linear decrease), $S_1$ (yellow dashed, linear increase), $S_2$ (purple dashed, nearly linear with weak quadratic correction). All curves are normalized to unity at $\varepsilon_0$ and exhibit finite values at the singularity.
Dispersions:
$N_0$: $\varepsilon = k_x^2 + k_y^2 + k_z$,
$N_1$: $\varepsilon = k_x^2 - k_y^2 + k_z$,
$N_2$: $\varepsilon = -k_x^2 - k_y^2 + k_z$,
$S_1$: $\varepsilon = k_x^2 + k_y^4 + k_z$,
$S_2$: $\varepsilon = k_x^4 + k_y^4 + k_z$.
See Supplemental Material for additional exponent combinations and parity regimes.}
\label{figii}
\end{figure}

For noncritical points, exactly one direction is linear ($n_\alpha=1$) while the remaining two satisfy $n_\beta,n_\gamma\ge2$ (where $\alpha, \beta, \gamma$ represent a generic permutation of the Cartesian axes $x, y, z$, structurally determined by the local point-group symmetry of the specific momentum point). Projecting onto the stationary $k_\beta\text{-}k_\gamma$ plane gives a reduced $2\times2$ Hessian (Figs.~\ref{figii}(a)--(e)). Ordinary noncritical VHSs ($N$-type) occur when $n_\beta=n_\gamma=2$, with subtypes $N_0$ (in-plane minimum), $N_1$ (saddle), and $N_2$ (maximum) determined by the sign combination $\sigma_\beta\sigma_\gamma$. Higher-order noncritical VHSs ($S$-type) arise when at least one transverse exponent exceeds 2: $S_1$ (one $n_i>2$) and $S_2$ (both $n_i>2$).

We now summarize the DOS for each class (Fig.~\ref{figi}(d) and Fig.~\ref{figii}(f)). For $T_1$, the DOS exhibits a finite peak near $\varepsilon_0$ with divergent derivative \cite{tan2024three}. For $T_2$ and $T_3$, power-law or logarithmic divergences emerge depending on the exponents. For $N_0$ and $N_2$, the DOS shows a constant background with linear energy correction. For $N_1$, the two-dimensional logarithmic divergence is quenched into a constant DOS with quadratic correction. For $S_1$, the DOS exhibits a constant background with linear correction. For $S_2$, the DOS shows both linear and quadratic corrections, with the linear term vanishing only in symmetry-protected cases (both transverse exponents $n_x, n_y$ odd, or isotropic hyperbolic saddles with $n_x = n_y$ and $\sigma_x\sigma_y = -1$.) The mathematical framework establishing these hallmark signatures is shown in the following section, and detailed derivations and explicit formulas are given in the Supplemental Material, Secs.I--III.

\section{Analytical Derivation of the DOS Scaling Laws}
\label{sec:Analytical_Derivations}
To systematically establish the energy-dependent DOS scaling behaviors summarized in Sec.~\ref{sec:classification}, we adopt the general definition in $d$ dimensions:
\begin{equation}
   g(E) = \int_{\Omega} \frac{\delta(E-\varepsilon(\mathbf{k}))}{(2\pi)^{d}} \, d\mathbf{k},
   \label{dos4all}
\end{equation}
where the integration is performed over the Brillouin zone $\Omega$. Before delving into the specific analytical evaluations, it is imperative to address the fundamental structural stability of these scaling laws by evaluating the impact of cross terms. We first demonstrate in Sec.~\ref{sec:Cross Terms} that the separable dispersion in Eq.~\ref{eq:dispersion} captures the universal asymptotic scaling despite the presence of cross terms. We then derive the DOS scaling laws for higher-order critical singularities (Sec.~\ref{sec:Scaling T-type}) and noncritical singularities (Sec.~\ref{sec:Scaling S-type}), respectively.

\subsection{Robustness of the Separable Dispersion and Impact of Cross Terms}
\label{sec:Cross Terms}
The separable dispersion defined in Eq.~(\ref{eq:dispersion}) serves as the canonical leading-order asymptotic representation for our classification. While cross terms are ubiquitous in real materials, they do not limit the generality of this framework. Fundamentally, any quadratic cross terms (e.g., $k_x k_y$) can be strictly eliminated through a local coordinate rotation. By diagonalizing the Hessian matrix around the stationary point, the system is transformed into its principal momentum axes. Because the eigenvalues of the Hessian---and thus the Morse index and the number of zero modes---are strictly invariant under such orthogonal transformations, the topological classification of the singularity types remains absolutely unchanged.

Regarding higher-order cross terms (e.g., $k_x^2 k_y^2$ or $k_x^2 k_z^2$), their influence is systematically governed by asymptotic scaling in the low-energy limit ($E \to \varepsilon_0$). As explicitly demonstrated via asymptotic scaling analysis in the Supplemental Material (Secs.V--VII), higher-order cross terms intrinsically possess higher scaling dimensions. For instance, if the principal axes dictate a scaling of $k_x, k_y \sim \mathcal{O}(E^{1/2})$ and $k_z \sim \mathcal{O}(E^{1/4})$, a mixed term like $k_x^2 k_z^2$ scales as $\mathcal{O}(E^{3/2})$. Such terms are sub-leading infinitesimals compared to the principal terms of $\mathcal{O}(E)$. It is important to note that this scaling truncation is rigorously justified in the low-energy limit under the standard condition that the Taylor coefficients are of natural order---a condition naturally satisfied in generic band structures. In the rare event of anomalously large cross-term coefficients, such as those occurring near a band inversion, these terms may temporarily compete with the leading ones before the extreme low-energy limit is reached, potentially giving rise to a new class of hybrid singularities beyond the present classification. However, under natural conditions, while these higher-order perturbations might slightly deform the momentum-space contours, they mathematically cannot alter the dominant fractional scaling exponents, nor do they trigger topological phase transitions of the singularity classes.

\subsection{Higher-Order Critical Singularities}
\label{sec:Scaling T-type}
For higher-order critical points ($T$-type), the band gradient vanishes in all directions, and the full 3D integral in Eq.~\eqref{dos4all} is evaluated directly. The fundamental scaling law can be analytically extracted via scaling invariance. We define a generalized topological exponent index $\nu = \sum 1/n_i$. By applying a uniform energy scaling factor $\lambda > 0$ and transforming momentum coordinates as $q_i = \lambda^{-1/n_i} k_i$, the DOS integration universally obeys:
\begin{equation}
g(E) \propto |E - \varepsilon_0|^{\nu - 1}.
\label{eq:T_scaling}
\end{equation}
The exact prefactors and piecewise boundaries depend strictly on the parity of the exponents and the transverse geometric topology (elliptic vs. hyperbolic), as exhaustively tabulated in the Supplemental Material [Sec.~I]. Nevertheless, several universal hallmarks can be identified directly from Eq.~\eqref{eq:T_scaling}. For $T_1$-type singularities, where exactly one exponent $n_z \ge 3$, the index $\nu = 1/2 + 1/2 + 1/n_z > 1$, which strictly dictates a finite DOS peak characterized by a divergent derivative ($dg/dE \to \infty$) at $\varepsilon_0$. For $T_2$-type singularities, the behavior is governed by $\nu = 1/2 + 1/n_y + 1/n_z$. Remarkably, when $n_y = n_z = 4$ in a saddle-point geometry, the scaling exponent vanishes exactly ($\nu = 1$), yielding a pure logarithmic divergence $g(E) \sim -\ln|E - \varepsilon_0|$---a direct consequence of the dimensional reduction from 3D to 2D. For $T_3$-type flat bands, where all $n_i \ge 3$, the exponent satisfies $\nu \le 11/12$, mandating a robust fractional power-law divergence $g(E) \sim |E|^{-|\nu - 1|}$. The detailed piecewise integration constants for all $T$-type sub-regimes, including the distinct behaviors on opposite sides of $\varepsilon_0$ for $T_1$ extrema and saddles, are exhaustively tabulated in the Supplemental Material.

\subsection{Noncritical Singularities}
\label{sec:Scaling S-type}
For noncritical singularities ($N$-type and $S$-type), the exact quenching of the gradient is confined to a 2D subspace, leaving a non-vanishing linear dispersion along the orthogonal direction (aligned here with $k_z$). The 3D DOS is evaluated via a dimensionality reduction method (slice integration). Defining the effective transverse energy as $\Lambda = (E - \varepsilon_0) - c_z k_z$, the 3D DOS is obtained by integrating the 2D subsystem DOS $g_{2D}(\Lambda)$ over the longitudinal Brillouin zone:
\begin{equation}
g(E) = \frac{1}{2\pi} \int_{-R}^{R} g_{2D}(E - \varepsilon_0 - c_z k_z) \, dk_z.
\label{eq:slice_integral}
\end{equation}
Because the linear term $c_z k_z$ monotonically spans across $\varepsilon_0$, the constant-energy contours traverse continuously without encountering a classical turning point along $k_z$. Consequently, the Dirac delta integration smoothly crosses $E = \varepsilon_0$, effectively bypassing the sharp Heaviside step functions characteristic of conventional band edges. The finite energy bandwidth over which the enhanced DOS persists is instead inherently bounded by the physical momentum cutoffs (e.g., $\pm R$) of the Brillouin zone.

This integration acts as a mathematical convolution that systematically quenches the singularities of the 2D subsystem. For ordinary $N_1$-type singularities, the transverse subsystem forms a hyperbolic saddle, yielding a 2D logarithmic divergence $g_{2D}(\Lambda) \propto -\ln|\Lambda|$. Upon integration along the linear out-of-plane direction, this 2D divergence is completely quenched into a large constant background accompanied by a symmetric quadratic decay $\delta g(E) \propto -(E - \varepsilon_0)^2$ (see the Supplemental Material, Sec. II.2 for the exact piecewise form and the definition of the logarithmic cutoff). For $N_0$- and $N_2$-type singularities, the transverse subsystem is elliptic (minimum or maximum), yielding a step-like $g_{2D}(\Lambda) \propto \Theta(\pm \Lambda)$; the slice integration then produces a constant background with a linear correction $\delta g(E) \propto \pm (E - \varepsilon_0)$.

For higher-order noncritical $S_1$-type singularities, where one transverse direction is quadratic and the other is higher-order ($n_y \ge 3$), the 2D subsystem follows $g_{2D}(\Lambda) \propto |\Lambda|^{1/2 + 1/n_y - 1}$. The slice integration yields a constant background with a linear correction, $\delta g(E) \propto A (E - \varepsilon_0)$, whose coefficient $A$ is determined by the phase-space asymmetry between positive and negative $\Lambda$ (see the Supplemental Material,Sec.III.1).

For $S_2$-type singularities, where both transverse directions are higher-order ($n_x, n_y \ge 3$), the 2D subsystem itself follows a fractional power law $g_{2D}(\Lambda) \propto |\Lambda|^{\nu-1}$, with $\nu = 1/n_x + 1/n_y < 1$. The slice integration generically yields a non-trivial linear and sub-leading quadratic energy correction: 
\begin{equation}
\delta g(E) \propto A (E - \varepsilon_0) + B (E - \varepsilon_0)^2,
\label{eq:S2_delta_g}
\end{equation}
where the linear coefficient $A$ is determined by the phase-space asymmetry $C_+ - C_-$ (see the Supplemental Material, Sec.III.2 for the exact definitions of $C_\pm$). Importantly, this linear term is not universally present. It vanishes identically in two symmetry-enforced scenarios: (i) isotropic hyperbolic saddles with $n_x = n_y$ and $\sigma_x \sigma_y = -1$, where discrete exchange symmetry enforces $C_+ = C_-$; and (ii) the fully non-TRIM regime where both $n_x$ and $n_y$ are odd, where coordinate inversion symmetry similarly forces $C_+ = C_-$. In these special cases, the DOS reduces to a pure quadratic correction $\delta g(E) \propto B (E - \varepsilon_0)^2$ with a negative curvature. The exact prefactors, the definitions of $C_\pm$, and the complete piecewise forms for all $S$-type sub-regimes are exhaustively tabulated in the Supplemental Material, Sec.III.

\section{Pyrochlore Lattice: Realization of the Classification}

Having established the unified classification of VHSs in Sec.~\ref{sec:classification}, we now turn to a concrete material realization. The pyrochlore lattice, with its rich band structure and tunable parameters, provides an ideal platform to demonstrate all four VHS classes---$M$-, $T$-, $N$-, and $S$-types---within a single system. Below, we construct TB models and systematically identify the predicted singularities at high-symmetry points of the Brillouin zone.

To model the 3D band structure of the pyrochlore lattice under symmetry constraints, the MagneticTB package \cite{zhang2022magnetictb} was employed to construct electronic TB Hamiltonians. The Wyckoff position $16c$ (or $16d$) exhibits the $\overline{3}m$ site symmetry associated with the $D_{3d}$ point group, corresponding to the space group $Fd\overline{3}m$ (No. 227). The primitive lattice vectors were chosen as $\mathbf{a}_{1} = (0, a/2, a/2)$, $\mathbf{a}_{2} = (a/2, 0, a/2)$, and $\mathbf{a}_{3} = (a/2, a/2, 0)$. Two classes of TB models were developed: (i) a single $s$-orbital basis including only nearest-neighbor (NN) hopping without spin-orbit coupling (SOC), and (ii) a single $s$-orbital basis including NN hoppings with SOC. The high-symmetry points evaluated in the Brillouin zone are $\Gamma$ (0,0,0), $L$ (0.5,0.5,0.5), $X$ (0.5,0,0.5), $W$ (0.5,0.25,0.75), $K$ (0.375,0.375,0.75), and $U$ (0.625,0.25,0.625).

To preserve readability while establishing a rigorous mathematical foundation, we present these Hamiltonians in a compact analytical form. For the non-SOC limit, the Hamiltonian is governed solely by the on-site energy $e_1$ and the conventional spin-conserving nearest-neighbor hopping $t_1$. In momentum space, the matrix elements between sublattices $m$ and $n$ are explicitly defined by the real-space connections:
\begin{equation}
H_{mn}^{\text{no-SOC}}(\mathbf{k}) = e_1 \delta_{mn} + t_1 \sum_{\mathbf{d}_{mn}} e^{i \mathbf{k} \cdot \mathbf{d}_{mn}} (1-\delta_{mn}),
\end{equation}
where $\mathbf{d}_{mn}$ represents the real-space nearest-neighbor vectors connecting the sublattices.(see also the Supplemental Material, Sec.IV.1)

When spin-orbit coupling is introduced, the basis naturally expands to include spin degrees of freedom, decoupling the nearest-neighbor interactions into two distinct physical channels. Utilizing the Pauli matrices $\boldsymbol{\sigma} = (\sigma_x, \sigma_y, \sigma_z)$ and the identity matrix $\sigma_0$, the SOC-inclusive Hamiltonian elements are compactly formulated as:
\begin{equation}
\begin{aligned}
H_{mn}^{\text{SOC}}(\mathbf{k}) &= e_1 \sigma_0 \delta_{mn} + \sum_{\mathbf{d}_{mn}} e^{i \mathbf{k} \cdot \mathbf{d}_{mn}} [ t_2 \sigma_0 \\
&  + i t_1 \mathbf{v}_{mn} \cdot \boldsymbol{\sigma} ] (1-\delta_{mn}).
\end{aligned}
\end{equation}
In this representation, $t_2$ strictly dictates the spin-conserving hopping channel, while $t_1$ explicitly characterizes the spin-flip hopping channel representing the effective SOC. This spin-flip process is governed by the asymmetric spin-orbit vector $\mathbf{v}_{mn}$, which is uniquely and intrinsically determined by the spatial orientation of the specific $m$-$n$ bond on the pyrochlore tetrahedra. This formulation clearly demonstrates how the spatial lattice symmetries and SOC act together to support the predicted band dispersions. (See also the Supplemental Material, Sec.IV.2).

Applying this TB framework to the pyrochlore lattice, we find that its inherent geometric frustration and tunable electronic structure provide an ideal platform to realize our unified classification scheme in a single material. By setting the on-site energy $e_1 = 0$ and utilizing the spin-flip hopping $t_1 = -1.0$ as the fundamental baseline energy scale, this SOC-inclusive model yields all four Van Hove singularity classes at distinct high-symmetry points as a function of the dimensionless ratio $t_2/t_1$. Analytical expressions for the energy dispersions were derived by exact diagonalization, directly confirming the theoretical predictions of Sec.~\ref{sec:classification}.

\begin{figure}[tbp]
\centering
\includegraphics[width=\linewidth]{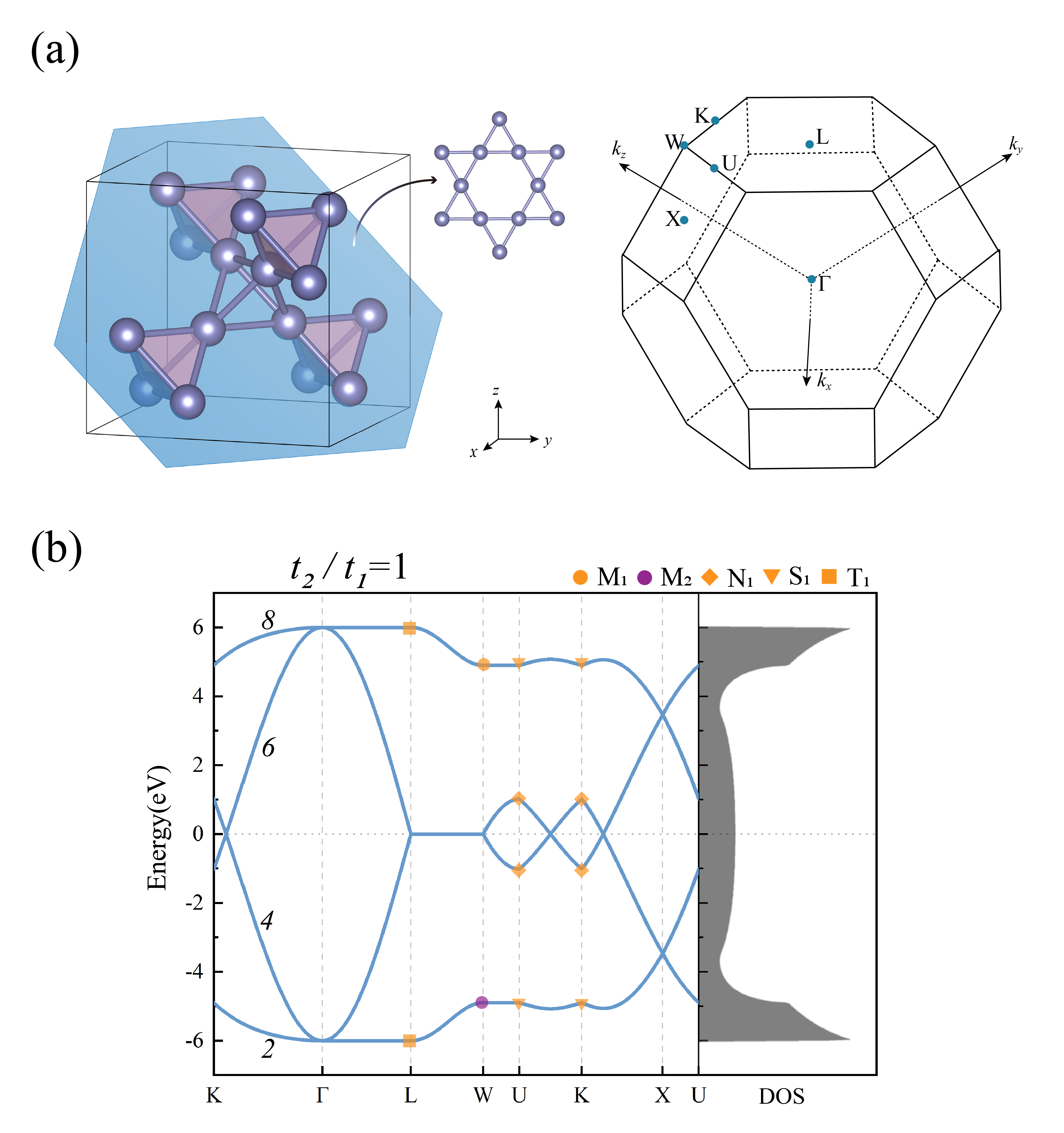}
\caption{\textbf{Lattice geometry and representative electronic structure of pyrochlore lattice.} (a) Crystal structure and Brillouin zone. (b) Band structure and DOS at $t_2=-1$, hosting $N_1$-type (band 4), $S_1$-type (band 2) at $K$, and $T_1$-type at $L$. See Supplemental Material for other $t_2$ values.}
\label{figiii}
\end{figure}

Fig.~\ref{figiii}(b) shows the band structure and DOS at $t_2/t_1 = 1$ (with $e_1 = 0$, $t_1 = t_2 = -1$). At this parameter value, three distinct singularity classes coexist: $N_1$-type (band 4) and $S_1$-type (band 2) at the $K$ point, and $T_1$-type at the $L$ point. The effective dispersion at the $L$ point is $\varepsilon \approx a(k_x^2 + k_y^2) + b k_z^4$, with coefficients obtained from numerical fitting, yielding a finite tunable peak near $\varepsilon_0$ with height $\propto R^{1/2}$, exactly the behavior predicted for $T_1$ singularities with $n_z = 4$.

At the $K$ point, band 4 exhibits an in-plane saddle with linear $k_z$ dispersion, corresponding to an $N_1$-type singularity. The logarithmic divergence of the pure two-dimensional saddle is completely quenched by the linear $k_z$ dispersion, resulting in a constant DOS followed by a quadratic correction. Band 2 at the same $K$ point hosts an $S_1$-type noncritical HOVHS, characterized by a single zero eigenvalue in the projected $2\times2$ Hessian. The effective dispersion reduces to $\varepsilon \approx \varepsilon_0 + v_z q_z - \frac{1}{2}|\lambda_1| q_1^2$, a parabolic cylinder with linear out-of-plane propagation, yielding a linear correction in the DOS. Other values of $t_2/t_1$ host additional singularity classes: $T_3$-type, $N_0$-type and $N_2$-type, and flat band features. (Detailed band structures and DOS calculations for these cases are provided in the Supplemental Material, Sec.IV.2, and Secs.V---VII).

While tuning the primary hopping ratio $t_2/t_1$ over a broad range theoretically demonstrates the capacity of the pyrochlore framework, experimental realization typically avoids drastic global chemical substitutions. In realistic physical systems, the electronic structure is governed by a high-dimensional parameter space that includes extended interactions, such as third-neighbor hoppings ($s_1, s_2$) and second-neighbor spin-orbit couplings($r_1, r_2, r_3$). When these realistic extended interactions are incorporated into our model (as explicitly detailed in the Supplemental Material, Sec.VIII, where the resulting phase diagrams are presented in Fig. S8), the isolated topological critical points expand into continuous, densely packed topological boundaries. Crucially, this reveals that experimentally accessible small physical perturbations---such as hydrostatic pressure or epitaxial strain that subtly modulate these extended hopping integrals---are completely sufficient to drive the system across the phase boundaries. This provides a highly pragmatic tuning mechanism, allowing for the simultaneous transformation of singularity types without requiring drastic changes to the primary lattice backbone. Such strain-tuning of electronic singularities has been strongly corroborated by recent experiments in prototype correlated materials\cite{hicks2014strong, lin2024uniaxial}.

\section{Discussion}

This unified classification yields several key insights. First, directional criticality---where the gradient vanishes only within a momentum subspace---provides a general mechanism for generating large but non-divergent DOS enhancements. Unlike conventional VHSs, which require fine-tuned Fermi level alignment to a diverging peak\cite{yuan2019magic,xu2021tunable,park2021tunable, wu2021nature}, noncritical singularities offer enhanced finite DOS that persists over a finite energy window. Second, the quenching in $N_1$-type singularities---where a 2D logarithmic divergence becomes a constant DOS with quadratic correction---reveals that introducing a noncritical dimension systematically reduces singularity order while retaining memory of its lower-dimensional origin. Third, $S_1$-type singularities demonstrate that in-plane quasi-1D flatness can combine with noncriticality, yielding tunable anisotropic DOS responses.

Fundamentally, a divergent or substantially enhanced DOS at or near the Fermi level alters the system's many-body instabilities by magnifying the electronic susceptibility in both the particle-particle (pairing) and particle-hole (density wave or magnetic) channels\cite{kiesel2013unconventional,classen2020competing}. In weak-to-intermediate coupling regimes, this DOS enhancement can exponentially boost the superconducting transition temperature within the BCS framework and satisfy the Stoner criterion for itinerant ferromagnetism or spin-density waves under much weaker repulsive interactions\cite{yuan2019magic,newns1992saddle}. This mechanism has been extensively demonstrated in various prototypical systems, such as cuprate superconductors\cite{markiewicz1997survey}, doped graphene\cite{nandkishore2012chiral}, and recently discovered kagome metals\cite{wu2021nature}, where unconventional superconductivity, charge density waves, and exotic magnetic orders are directly driven by ordinary or higher-order Van Hove singularities. By constructing a 3D framework that incorporates directional criticality, our work explicitly provides a tunable platform to manipulate these very DOS profiles, thereby offering a controlled route to engineer the corresponding collective phases.

Crucially, for realizing controlled instabilities in bulk systems, the robust energy plateau of the $N_1$-type and the broadened energetic width of the $T_1$-type singularities offer distinct advantages over ordinary 3D saddle points ($M$-type). The fundamental physical distinction lies in their resilience against realistic experimental factors, such as doping fluctuations ($\Delta E_F$) and impurity disorder broadening (mathematically parametrized as a finite imaginary self-energy $\Gamma$ in the standard many-body Green's function formalism\cite{bruus2004many}). Ordinary 3D $M$-type saddle points merely produce a finite DOS with a weak kink (diverging derivative). Under convolution with realistic disorder, such weak kinks are easily smeared out, rendering the correlation effects notoriously fragile. In contrast, the mathematical convolution over the extended energy plateau of the noncritical $N_1$-type merely smooths the edges of its quadratic maximum, leaving the integrated effective DOS remarkably well-preserved. Similarly, the higher-order flatness (e.g., $k^4$ dispersion) of the $T_1$-type singularity significantly broadens the energetic width and magnitude of the DOS peak, ensuring the integrated effective DOS remains highly robust without the need for stringent fine-tuning.

The theoretical robustness discussed above is not merely a formal mathematical advantage but has been increasingly corroborated by recent experimental discoveries in prominent quantum material families. They underpin emergent phenomena in Moir\'{e} superlattices---such as twisted bilayer graphene and transition metal dichalcogenides (TMDs) at magic angles---where the moir\'{e} potential essentially flattens the quadratic band bottoms into higher-order extrema \cite{yuan2019magic,pullasseri2024chern,hsu2021spin,wang2021moire,wu2023pair,lu2024fractional}. Furthermore, recent theoretical and experimental studies have confirmed their presence in kagome superconductors (e.g., CsTi$_3$Bi$_5$) \cite{hu2022rich,kang2022twofold,wang2025higher,patra2025high} and Co-based kagome metals (e.g., YbCo$_6$Ge$_6$ and MgCo$_6$Ge$_6$) \cite{tan2024three}, where sublattice interference induces the simultaneous flattening of three-dimensional saddle points. Similarly, strongly anisotropic $S=1/2$ triangular lattice antiferromagnets have also been experimentally demonstrated to host higher-order Van Hove singularities near the Brillouin zone boundaries \cite{park2024anomalous}.

The pyrochlore lattice naturally realizes this complete theoretical framework (Fig.~\ref{figiii} and Supplemental Material). At TRIMs $\Gamma$ and $L$, symmetry forces the gradient to vanish in all directions, enabling ordinary ($M$-type) and higher-order ($T$-type) critical singularities. At non-TRIM $W$, the gradient also vanishes in all directions, producing a conventional critical singularity. In contrast, non-TRIM points $K$ and $U$ exhibit noncritical singularities, with a nonvanishing gradient along a single direction that confines critical behavior to a 2D subspace. To explicitly map the generic noncritical notation to this lattice, at the high-symmetry point $K$, the linear gradient occurs along the $k_z$ direction (mapping $\alpha \to z$), while the stationary plane is spanned by $\beta, \gamma \to x, y$. Conversely, at the topologically equivalent $U$ point, the local symmetry rotates the linear direction to $k_y$ (mapping $\alpha \to y$), leaving the stationary plane as $x, z$. Within $Fd\bar{3}m$, $K$ and $U$ emerge as symmetry-enforced topological partners: they remain isoenergetic on the same band and retain their noncritical structure across the entire parameter space (See the Supplemental Material, Sec.IV.3).

Having established the spatial distribution of singularity types in the pyrochlore lattice, we now examine how this coexistence influences the superconducting pairing symmetry. The coexistence of these singularity types has profound implications for unconventional superconductivity. TRIM points $L$ and $\Gamma$ are even-parity (type-I) VHSs, favoring singlet superconductivity. Specifically, because a TRIM point maps onto itself under the time-reversal operation, the Pauli exclusion principle strictly enforces a geometric constraint on the scattering amplitudes within these momentum patches. As systematically established in the theoretical classification by Yao and Yang\cite{yao2015topological}, this inherent symmetry constraint strongly suppresses effective odd-parity triplet pairing interactions. Consequently, under weak-to-intermediate repulsive electron-electron interactions, the renormalization group flows inevitably drive the system toward even-parity singlet superconducting instabilities (such as $d$-wave or extended $s$-wave states). Non-TRIM $W$ is an odd parity (type-II) VHS, which can stabilize triplet $p$-wave pairing with ferromagnetic fluctuations\cite{yao2015topological}. Most strikingly, non-TRIM points $K$ and $U$ host noncritical singularities. Although their DOS lacks a strict divergence, they evade TRIM parity constraints. As topological partners, they may mediate pairing via interpatch scattering\cite{meng2015evidence}. Thus, the pyrochlore lattice uniquely hosts type-I, type-II, and noncritical VHSs, potentially enabling exotic superconductivity beyond canonical $d$-wave or $p+ip$ paradigms\cite{meng2015evidence,yao2015topological}.

Having established how different singularity classes influence pairing symmetries, we now turn to the question of experimental accessibility and control over these phases. The accessibility of these diverse singular phases is intimately tied to parameter tuning. As established in our pyrochlore model analysis, the inclusion of extended interactions naturally broadens isolated critical points into continuous topological manifolds within a high-dimensional parameter space. This provides a profound practical implication: one does not need to rely solely on drastic chemical substitutions to traverse the singularity hierarchy. Instead, synthesizing a material natively positioned near these dense phase boundaries allows small, continuous physical perturbations (such as strain engineering) to effectively switch the topological class of the Van Hove singularity, offering a highly controlled route to probe correlation-driven phase transitions.

In summary, this work establishes a comprehensive classification of Van Hove singularities in three dimensions that unifies ordinary critical points, higher-order extrema, and noncritical hybrid geometries within a single theoretical framework. More broadly, our classification provides a systematic language for singularities in multiband systems with anisotropic dispersions and spin-orbit coupling. The framework naturally accommodates dimensional crossovers in quasi-2D materials, thin films, and topological semimetals. This work expands the traditional focus on fully critical, quadratic band extrema and saddle points into a unified framework where partial criticality, higher-order flatness, and their noncritical hybrids are placed on equal footing. By engineering the dimensionality of criticality and the order of band flattening, one can systematically tailor the energy dependence, magnitude, and divergence character of the electronic DOS. Ultimately, we believe this systematic classification will serve as a concrete theoretical blueprint to stimulate future first-principles high-throughput screening and ARPES measurements. In forthcoming studies, constructing tight-binding models across various space groups will allow for the systematic investigation of the crystallographic and hopping prerequisites for noncritical singularities, utilizing these precise theoretical criteria to actively screen for new candidate materials within existing real-material databases.

\section{Conclusion}

We have constructed a unified taxonomic framework for Van Hove singularities in three-dimensional systems, encompassing ordinary ($M$-type), higher-order ($T$-type), noncritical ordinary ($N$-type), and noncritical higher-order ($S$-type) classes. Realized in the pyrochlore lattice, this entire hierarchy is shown to be physically attainable through continuous tuning of the parameter \(t_2/t_1\). This work establishes a new paradigm for understanding and engineering electronic singularities, transforming them from serendipitous band features into designable elements of quantum materials. By systematically controlling the dimensionality of criticality and the order of band flattening, one can now intentionally sculpt the density of states across a wide spectrum---from sharp power-law or logarithmic divergences ($T_2$,$T_3$) to finite peaks ($T_1$) and to finite enhancements with linear or quadratic energy dependencies ($N_0$, $N_1$, $N_2$, $S_1$, $S_2$)---providing a powerful route to correlation-driven emergent phenomena in three-dimensional quantum materials.

\textit{Note added}---Recent work on the pyrochlore superconductor CsBi$_2$ \cite{morita2026saddlepoints} reports ordinary VHSs at $L$ and $W$, and noncritical ordinary VHS at $U$, in agreement with the classification presented here.

\section{Acknowledgments}

M.Q. Kuang acknowledges the support from the Natural Science Foundation of Chongqing (Grant No. CSTB2024NSCQ-MSX0080) and the National Natural Science Foundation of China (NSFC, Grant No. 11704315). H.T. is supported by the NSFC with Grant No.12574270 and the Science and Technology Commission of Shanghai Municipality with Grant No. 24PJA051.

%
\clearpage
\onecolumngrid 

\vspace*{0.5cm}
\begin{center}
  \textbf{\large Supplemental Material for "Directional Criticality and Higher-Order Flatness: Designing Van Hove Singularities in Three Dimensions"}\\[0.5cm]
  
  Hua-Yu Li,$^1$ Hengxin Tan,$^{2,*}$ Hao-Yu Zhu,$^1$ Hong-Kuan Yuan,$^1$ and Min-Quan Kuang$^{1,\dagger}$\\[0.3cm]
  
  {\small \it 
  $^1$Chongqing Key Laboratory of Micro \& Nano Structure Optoelectronics, and School of Physical Science and Technology, Southwest University, Chongqing 400715, P. R. China\\[0.1cm]
  $^2$Key Laboratory of Artificial Structures and Quantum Control (Ministry of Education), School of Physics and Astronomy, Shanghai Jiao Tong University, Shanghai 200240, P. R. China
  }
\end{center}
\vspace{0.8cm}

\setcounter{equation}{0}
\setcounter{figure}{0}
\setcounter{table}{0}
\setcounter{page}{1}
\setcounter{section}{0}
\setcounter{secnumdepth}{3}

\renewcommand{\theequation}{S\arabic{equation}}
\renewcommand{\thefigure}{S\arabic{figure}}
\renewcommand{\thetable}{S\arabic{table}}
\renewcommand{\thesection}{\Roman{section}} 
\renewcommand{\figurename}{Fig.}

\section{DOS near the higher-order Van Hove singularity}
According to the number of zero eigenvalues of the 3$\times$3 Hessian matrix, higher-order Van Hove singularities (HOVHS) in 3D systems can be classified into three types, denoted as $T_1$, $T_2$, and $T_3$ (see Tables.~\ref{table_T1_DOS}, ~\ref{table_T2_DOS_Unified}, ~\ref{table_T3_DOS}). 

Before delving into the specific density of states (DOS) derivations for the $T_1$, $T_2$, and $T_3$ higher-order Van Hove singularities, we must establish two universal classification principles. These principles fundamentally govern the integration strategies and topological features of all HOVHS variations. 

As established in the main text, any quadratic cross terms can be strictly eliminated through a local orthogonal transformation (diagonalizing the Hessian matrix), which inherently preserves the topological singularity classification. Regarding higher-order cross terms, assuming that they are either precluded by local point-group symmetries or are negligibly higher-order (as governed by asymptotic scaling and natural-order Taylor coefficients), the generalized HOVHS dispersion can be analytically diagonalized into a canonical decoupled polynomial model: 
\begin{equation}
\varepsilon(\mathbf{k}) = \varepsilon_0 + \sum \sigma_i c_i k_i^{n_i}.
\end{equation}
Within this idealized framework, the scaling behavior is strictly dictated by two physical properties: the parity of the exponents $\{n_i\}$ and the relative signs of the curvature coefficients $\{\sigma_i\}$.

\vspace{8pt}
\noindent \textbf{Principle I: Parity Regimes (Governed by Local Symmetries)} \\
Crucially, the parity (even or odd) of the higher-order exponents $n_i$ is not strictly dictated by the global spatial inversion symmetry of the crystal, but rather by the local symmetry of the specific momentum vector. We universally classify the system into distinct parity regimes based on Time-Reversal Invariant Momenta (TRIMs):
\begin{itemize}
    \item \textbf{TRIM Regime (All Even Exponents):} At TRIMs (e.g., the $\Gamma$ point), local symmetry strictly enforces $E(\mathbf{k}) = E(-\mathbf{k})$, requiring all odd-powered terms in the Taylor expansion to identically vanish. Coupled with the defining HOVHS prerequisite that quadratic terms are depleted along specific principal axes, the surviving higher-order exponents must be strictly even integers ($n_i = 2m \ge 4$). The bounded nature of these even powers mathematically necessitates Heaviside step-function truncations during the DOS integration.
    \item \textbf{Mixed-Parity Regime (Even and Odd Exponents):} At specific non-TRIM symmetry axes (e.g., $U$ or $K$ valleys in certain space groups like SG 227), local inversion parity may be explicitly broken along specific directions while retaining transverse mirror symmetries. This permits a coexistence of even and odd exponents.
    \item \textbf{Fully Non-TRIM Regime (All Odd Exponents):} At generic asymmetric non-TRIM locations, all higher-order exponents can be odd integers ($n_i = 2m+1 \ge 3$). In these cases, as well as in noncritical singularities where linear terms ($n=1$) are present, the local dispersion inherently includes odd-power terms (i.e., a finite band gradient along those directions). This means the energy is not a true local extremum in all directions. As a result, the constant-energy contours traverse continuously through the singularity energy $\varepsilon_0$ without encountering a classical turning point. When integrating the Dirac delta function $\delta(E - \varepsilon(\mathbf{k}))$ over these specific momentum directions, the evaluation smoothly crosses $E = \varepsilon_0$, thereby mathematically bypassing the sharp Heaviside step functions (e.g., $\Theta(E - \varepsilon_0)$) that typically truncate the DOS at conventional band edges. Nevertheless, because this localized polynomial expansion is strictly valid only in the vicinity of the critical point, the momentum integration cannot be extended to infinity. The integration bounds are instead dictated by the physical Brillouin zone boundaries, which we parameterize as momentum cutoffs (e.g., $R$). It is these physical momentum cutoffs, rather than an artificial mathematical step function at $E = \varepsilon_0$, that ultimately determine the finite energy bandwidth over which the enhanced DOS plateau persists.
\end{itemize}

\vspace{8pt}
\noindent \textbf{Principle II: Transverse Topological Categories (Governed by Curvature Signs)} \\
When evaluating the multi-dimensional momentum integrals, we routinely group two polynomial components (e.g., $k_i$ and $k_j$) in a transverse subspace. The global geometry of this subspace is dictated by the product of their signs ($\sigma_i \sigma_j$):
\begin{itemize}
    \item \textbf{Category A ($\sigma_i \sigma_j = +1$):} Identically signed terms forming closed \textit{generalized elliptic} transverse contours (which physically reduce to strict ellipses only when $n_i = n_j$).
    \item \textbf{Category B ($\sigma_i \sigma_j = -1$):} Oppositely signed terms forming open \textit{generalized hyperbolic saddle} contours.
\end{itemize}
Guided by these two universal principles, the exhaustive theoretical permutations for $T_1$, $T_2$ and $T_3$ singularities can be systematically derived.

\begin{figure}[htbp]
	\includegraphics[width=\linewidth]{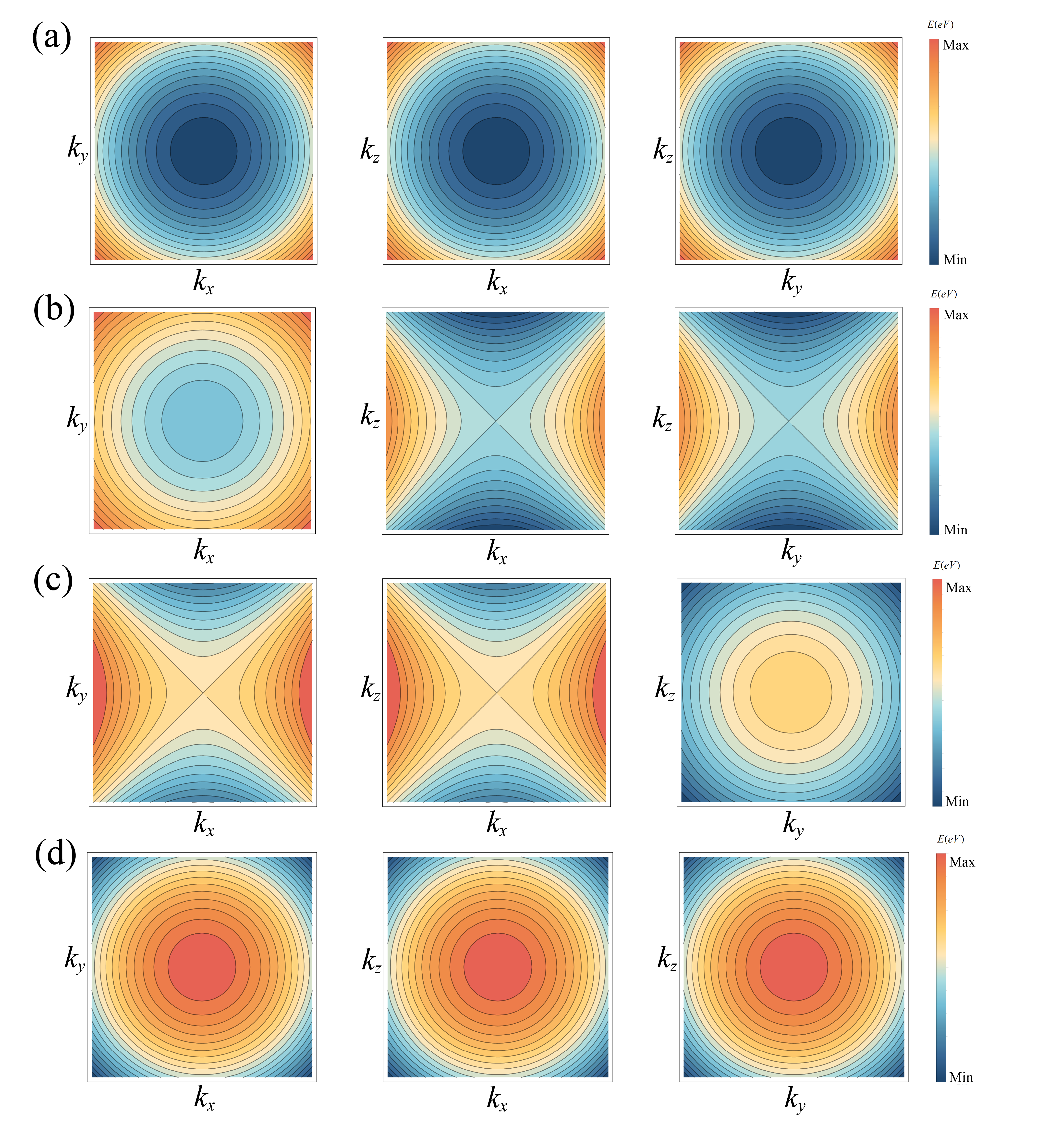}
\caption{(a)-(d) Energy contours of ordinary Van Hove singularities in a three-dimensional system. The color bar alongside each row encodes energy magnitude. Panels (a)-(d) correspond to distinct 3D VHS types: \(M_0\)-type (a), \(M_1\)-type (b), \(M_2\)-type (c), and \(M_3\)-type (d). These plots depict the hallmark momentum-space energy distributions for each VHS category.}
\label{figs1}
\end{figure}

\begin{figure}[htbp]
\centering
\includegraphics[width=\linewidth]{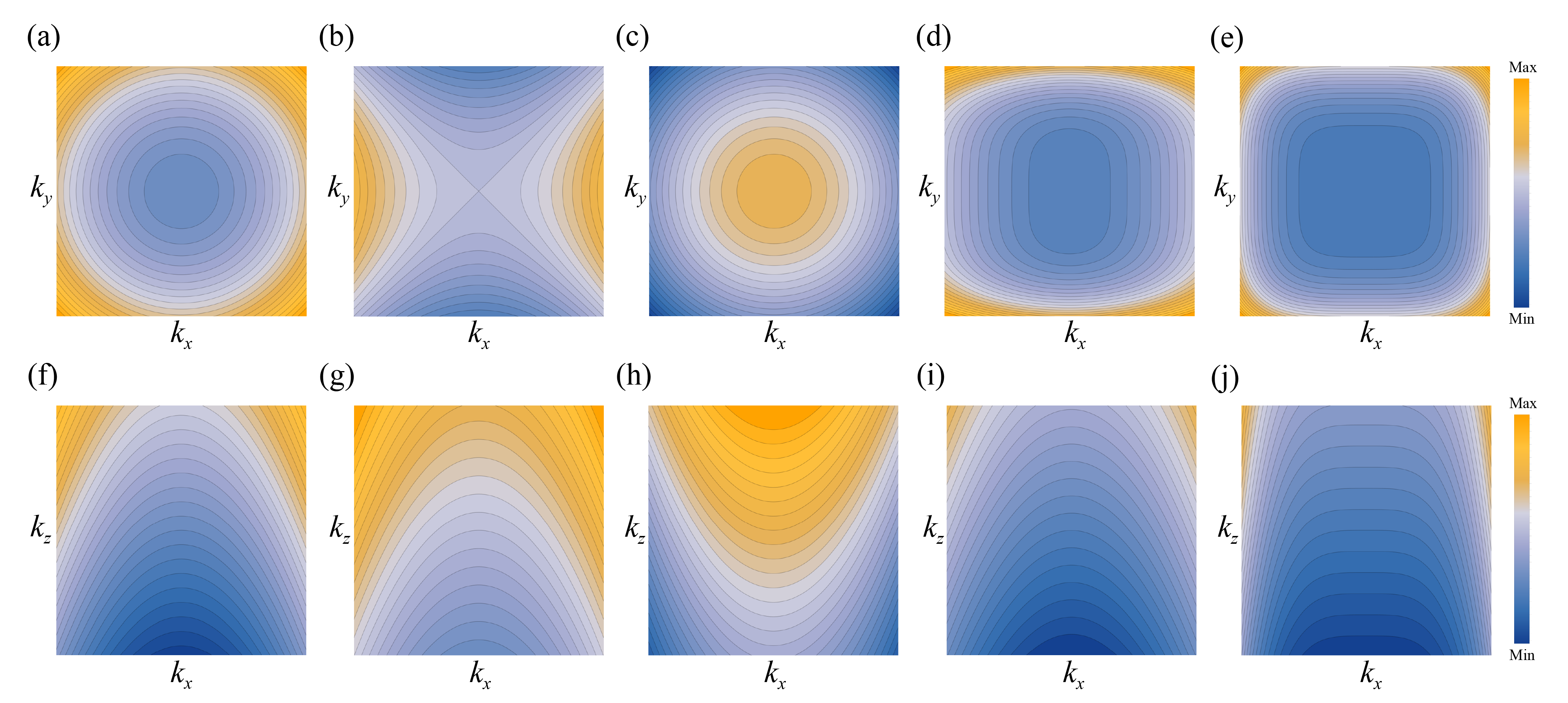}
\caption{Energy contours of representative noncritical Van Hove singularities in a three-dimensional system.
(a)-(e) Corresponding $k_x$-$k_y$ isocontours for ordinary noncritical ($N_0$, $N_1$, $N_2$) and higher-order noncritical ($S_1$, $S_2$) types, with color denoting the energy.
(f)-(j) Corresponding $k_x$-$k_z$ isocontours, where blue and orange represent the energy minimum and maximum, respectively.
The corresponding energy dispersions are defined as:
$N_0$: $\varepsilon = k_x^2 + k_y^2 + k_z$,
$N_1$: $\varepsilon = k_x^2 - k_y^2 + k_z$,
$N_2$: $\varepsilon = -k_x^2 - k_y^2 + k_z$,
$S_1$: $\varepsilon = k_x^2 + k_y^4 + k_z$,
$S_2$: $\varepsilon = k_x^4 + k_y^4 + k_z$.
}
\label{figs6}
\end{figure}

\begin{table*}[htbp]
\centering
\scriptsize
\setlength{\tabcolsep}{0pt} 
\renewcommand{\arraystretch}{2.8} 

\caption{\textbf{Analytical Density of States (DOS) for $T_1$-type Higher-Order Van Hove Singularities.} The table categorizes $T_1$ hybrid dispersions into distinct topological geometries based on the parity of the higher-order exponent $n_z$ (TRIM vs. Non-TRIM regimes). Column 2 provides exact analytical expressions in the low-energy limit.}

\begin{tabular*}{\textwidth}{@{\extracolsep{\fill}} ccc @{}}
\toprule
\parbox{0.18\textwidth}{\centering \textbf{Dispersion} \\ $\boldsymbol{\varepsilon(\mathbf{k}) - \varepsilon_0}$} & 
\parbox{0.50\textwidth}{\centering \textbf{Analytical DOS} $g(E)_{E \to \varepsilon_0}$} & 
\parbox{0.30\textwidth}{\centering \textbf{Schematic DOS}} \\
\midrule

\parbox{0.18\textwidth}{\centering $c_{x} k_{x}^{2} + c_{y} k_{y}^{2} + c_{z} k_{z}^{n_z}$ \\ \vspace{2pt} \scriptsize \textit{3D Extremum} \\ \textit{TRIM Regime} \\ (Even $n_z \ge 4$)} & 
\parbox{0.50\textwidth}{\centering $\displaystyle \frac{1}{4\pi^2\sqrt{c_x c_y} c_z^{1/n_z}} 
\begin{cases} 
0 & (E \le \varepsilon_0)\\
( E - \varepsilon_0 )^{1/n_z} & (\varepsilon_0 < E \le \varepsilon_0+ R^2)
\end{cases}$} & 
\parbox{0.30\textwidth}{\centering\includegraphics[width=0.95\linewidth]{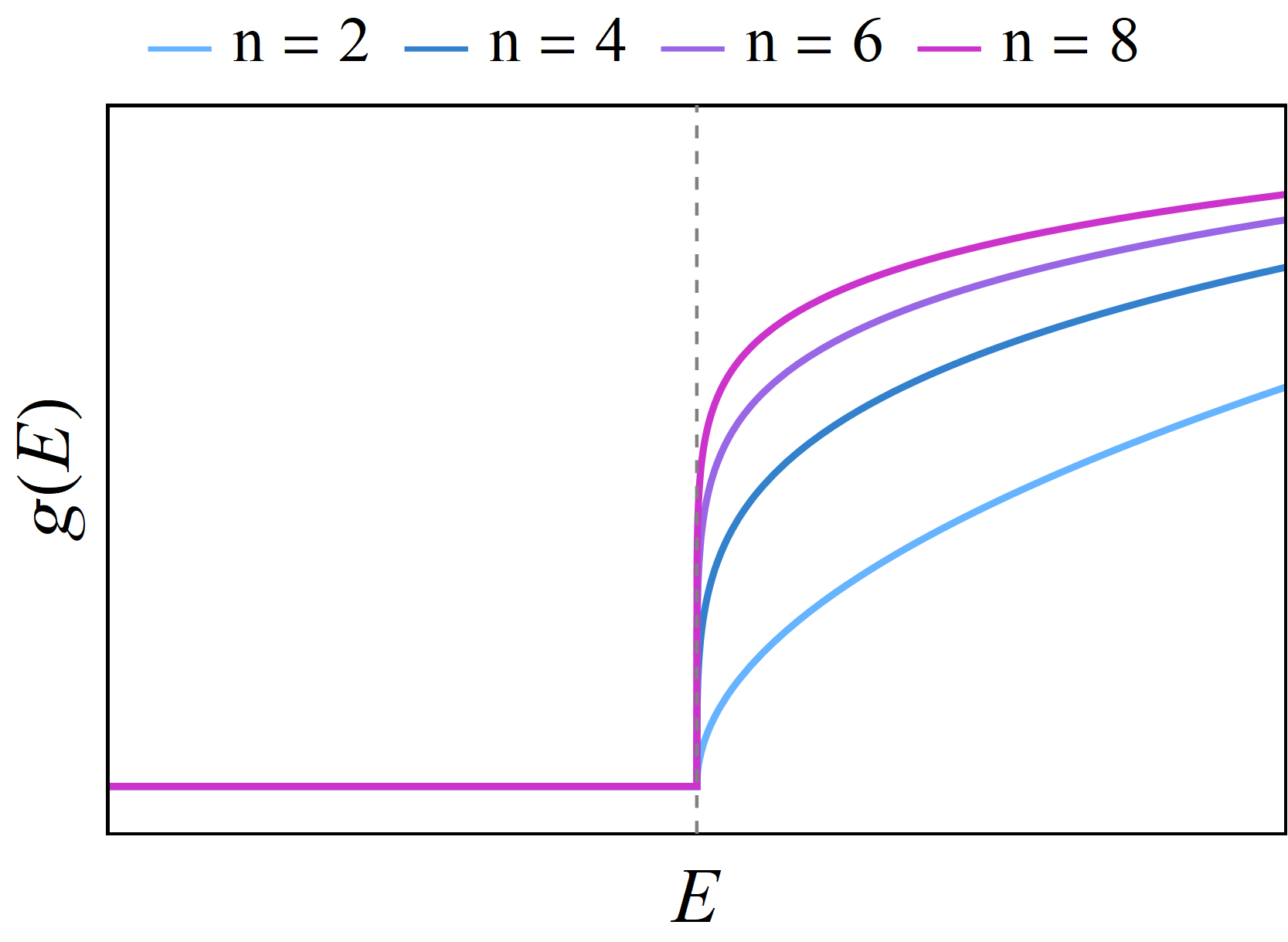}} \\ 

\parbox{0.18\textwidth}{\centering $c_{x} k_{x}^{2} + c_{y} k_{y}^{2} - c_{z} k_{z}^{n_z}$ \\ \vspace{2pt} \scriptsize \textit{3D Saddle (Elliptic)} \\ \textit{TRIM Regime} \\ (Even $n_z \ge 4$)} & 
\parbox{0.50\textwidth}{\centering $\displaystyle \frac{1}{4\pi^2 \sqrt{c_x c_y} c_z^{1/n_z}} 
\begin{cases} 
(R^2 + \varepsilon_0 - E)^{1/n_z} - (\varepsilon_0 - E)^{1/n_z} & (E \le \varepsilon_0) \\
(R^2 + \varepsilon_0 - E)^{1/n_z} & (E > \varepsilon_0) 
\end{cases}$} & 
\parbox{0.30\textwidth}{\centering\includegraphics[width=0.95\linewidth]{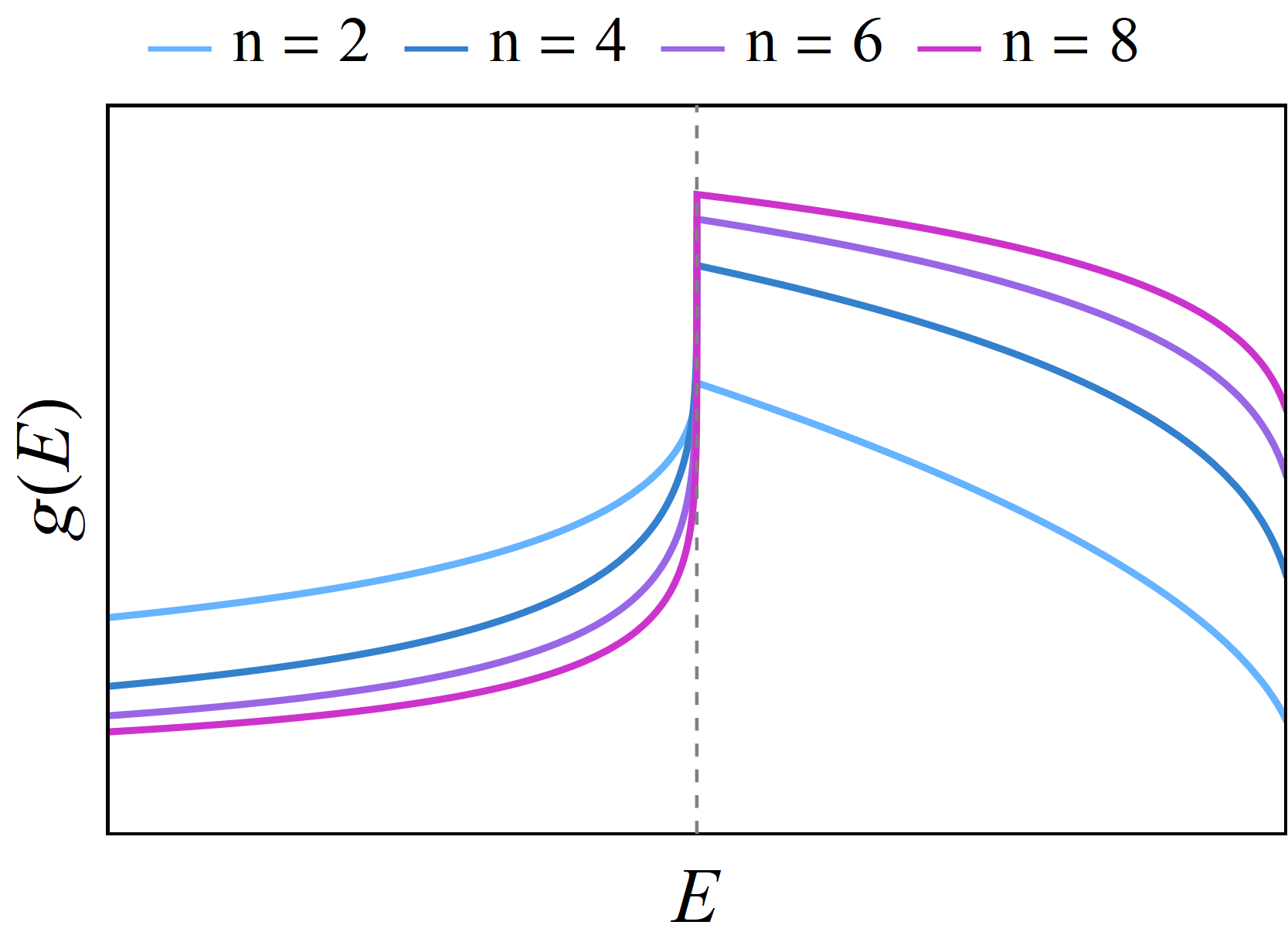}} \\

\parbox{0.18\textwidth}{\centering $c_{x} k_{x}^{2} - c_{y} k_{y}^{2} - c_{z} k_{z}^{n_z}$ \\ \vspace{2pt} \scriptsize \textit{3D Saddle (Hyperbolic)} \\ \textit{TRIM Regime} \\ (Even $n_z \ge 4$)} & 
\parbox{0.50\textwidth}{\centering $\displaystyle \frac{\sqrt{\pi} \Gamma\left(\frac{1}{n_z}\right)}{2\pi^3 n_z \sqrt{c_x c_y} c_z^{\frac{1}{n_z}} \Gamma\left(\frac{1}{n_z} +\frac{1}{2}\right)} 
\begin{cases} 
\displaystyle \int_{\sqrt{\varepsilon_0-E}}^{R} \frac{q^{2/n_z}dq}{\sqrt{E - \varepsilon_0 + q^2}}    \quad (E \le \varepsilon_0)\\
\displaystyle \int_0^{R} \frac{q^{2/n_z} dq}{\sqrt{E - \varepsilon_0 + q^2}}  \quad(E > \varepsilon_0)
\end{cases}$} & 
\parbox{0.30\textwidth}{\centering\includegraphics[width=0.95\linewidth]{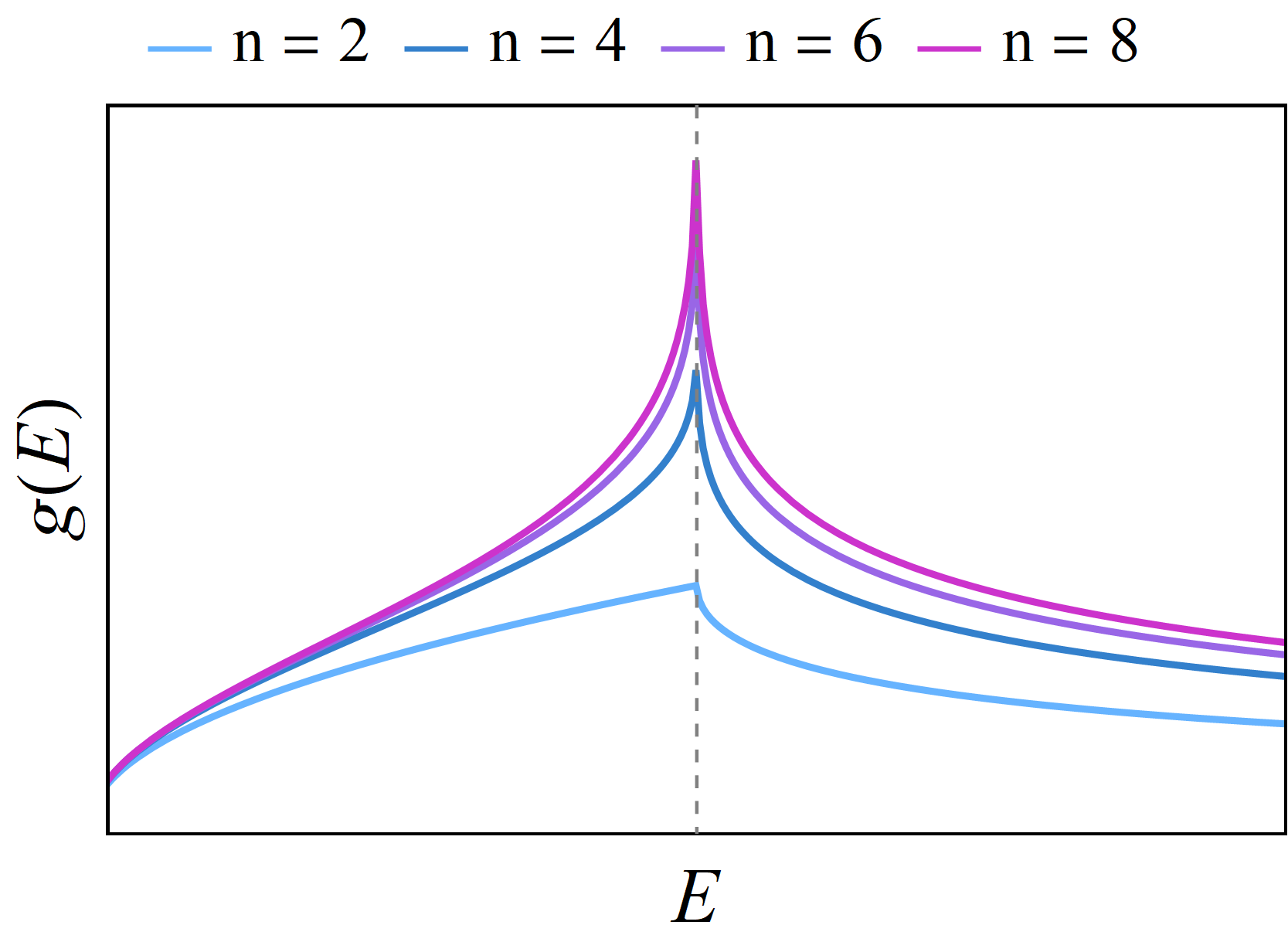}} \\

\parbox{0.18\textwidth}{\centering $c_{x} k_{x}^{2} + c_{y} k_{y}^{2} + c_{z} k_{z}^{n_z}$ \\ \vspace{2pt} \scriptsize \textit{Elliptic Topology} \\ \textit{Non-TRIM Regime} \\ (Odd $n_z \ge 3$)} & 
\parbox{0.50\textwidth}{\centering $\displaystyle \frac{1}{4\pi^2 \sqrt{c_x c_y} c_z^{1/n_z}} 
\begin{cases} 
(R^2 + |E - \varepsilon_0|)^{1/n_z} - |E - \varepsilon_0|^{1/n_z} & (E \le \varepsilon_0) \\
(R^2 - E + \varepsilon_0)^{1/n_z} + (E - \varepsilon_0)^{1/n_z} & (E > \varepsilon_0) 
\end{cases}$} & 
\parbox{0.30\textwidth}{\centering\includegraphics[width=0.95\linewidth]{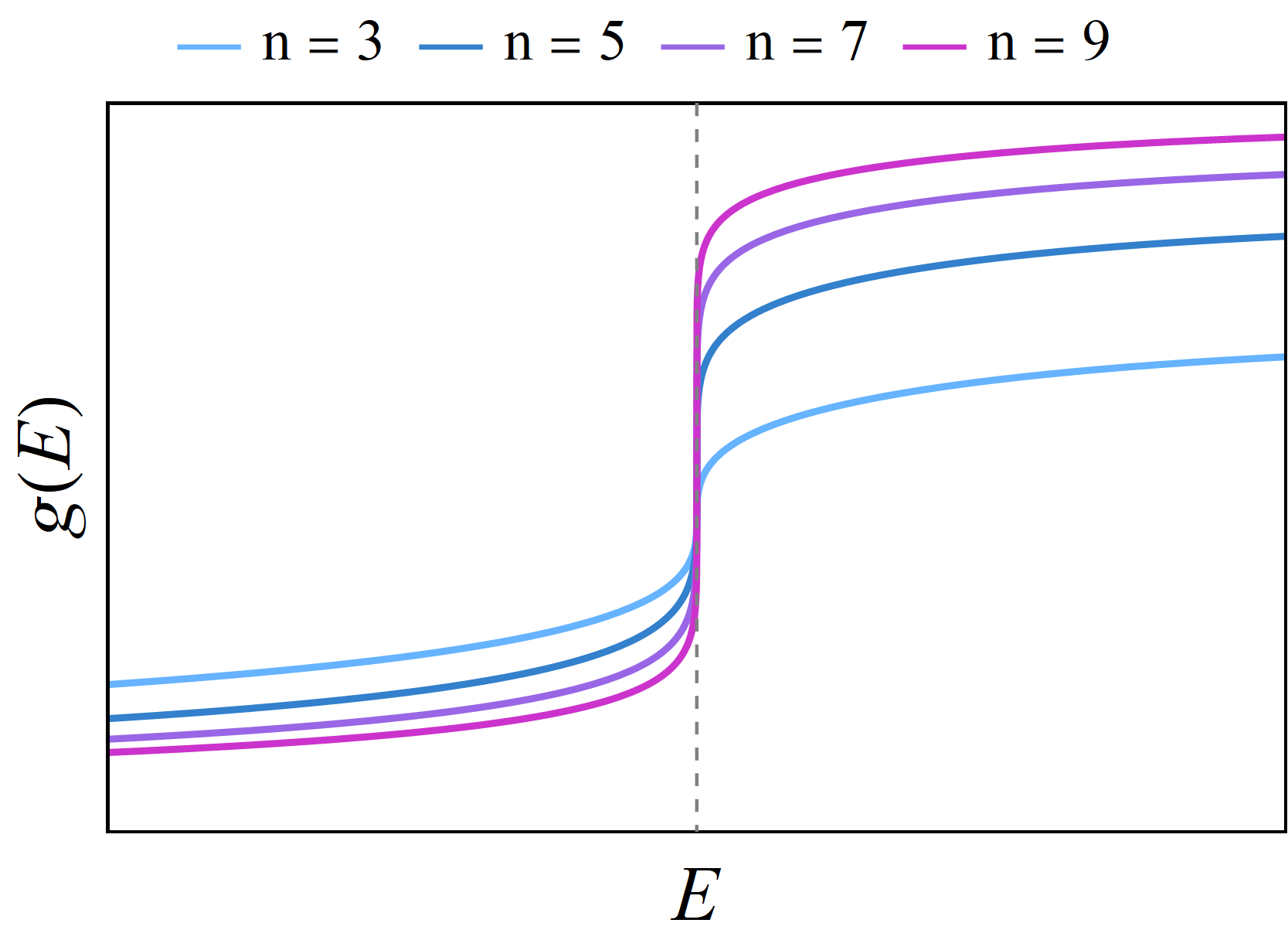}} \\

\parbox{0.18\textwidth}{\centering $c_x k_x^2 - c_y k_y^2 + c_z k_z^{n_z}$ \\ \vspace{2pt} \scriptsize \textit{Hyperbolic Topology} \\ \textit{Non-TRIM Regime} \\ (Odd $n_z \ge 3$)} & 
\parbox{0.50\textwidth}{\centering $\displaystyle g_B^{back} + \frac{|E - \varepsilon_0|^{1/n_z}}{16\pi^3 n_z \sqrt{c_x c_y} c_z^{1/n_z}} 
\begin{cases} 
\mathcal{K}_{B-}^{odd} & (E \le \varepsilon_0) \\
\mathcal{K}_{B+}^{odd} & (E > \varepsilon_0) 
\end{cases}$} & 
\parbox{0.30\textwidth}{\centering\includegraphics[width=0.95\linewidth]{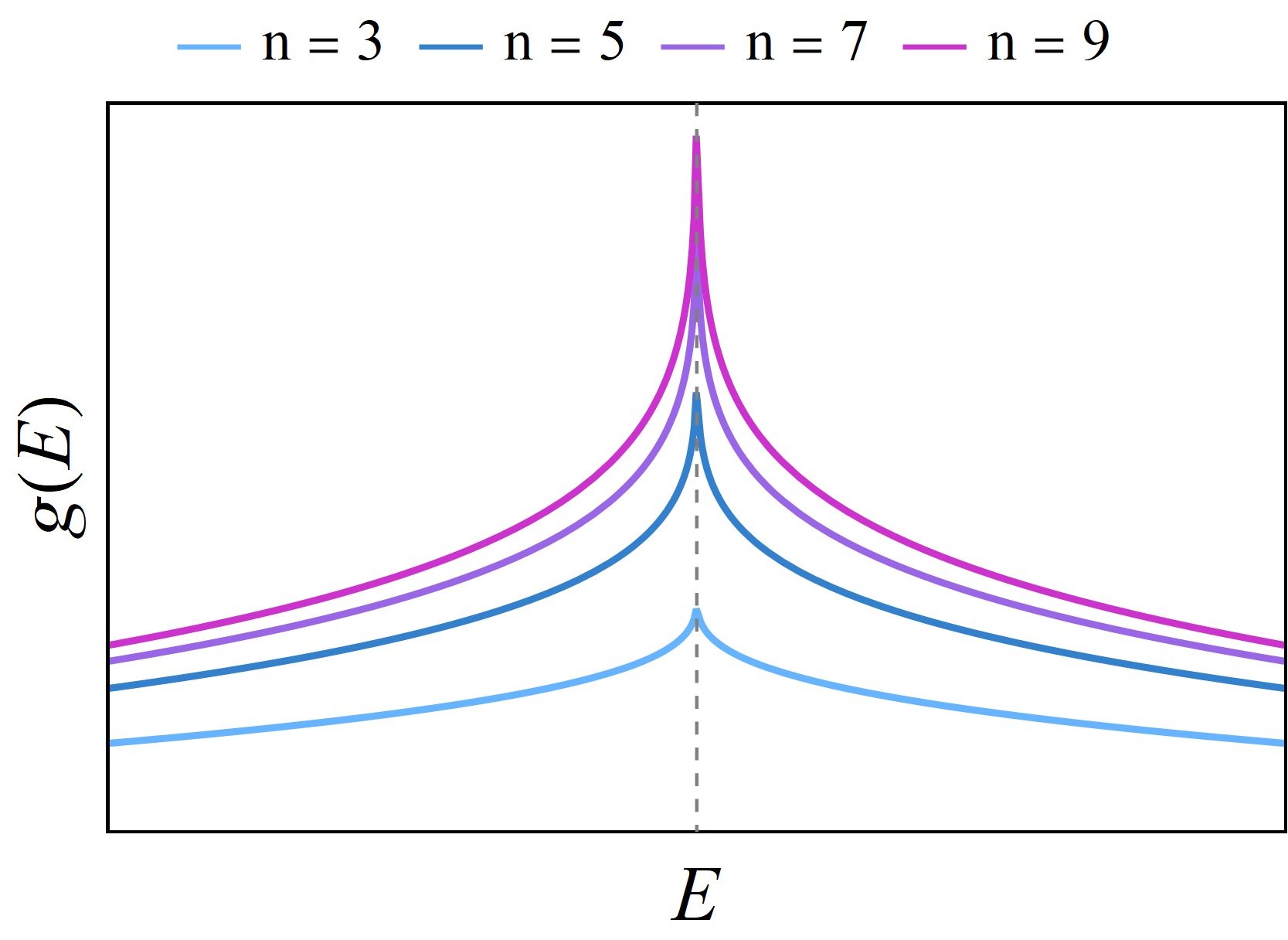}} \\

\bottomrule
\end{tabular*}

\vspace{6pt}
\begin{minipage}{\textwidth}
    \raggedright
    \scriptsize
    \renewcommand{\arraystretch}{1.2}
    \begin{tabular}{@{}ll}
    $R, \Gamma(n)$:  Cutoff wavevector and Gamma function, respectively. \\
    $g_B^{back}, \mathcal{K}_{B\pm}^{odd}$: Constant background and dimensionless integrals for the Non-TRIM saddle.(see Section~\ref{sec:iii.1} for exact piecewise definitions).
    \end{tabular}
\end{minipage}
\label{table_T1_DOS} 
\end{table*}
\begin{table*}[htbp]
\centering
\scriptsize
\setlength{\tabcolsep}{0pt}
\renewcommand{\arraystretch}{2.6}

\caption{\textbf{Comprehensive Analytical Density of States (DOS) for $T_2$-type Higher-Order Van Hove Singularities.} The table summarizes the analytical scaling forms for different topological configurations. Column 1 defines the dispersion and regime, Column 2 provides the exact low-energy expressions with explicit prefactors, and Column 3 presents the schematic DOS profiles.} 
\label{table_T2_DOS_Unified}

\begin{tabular*}{\textwidth}{@{\extracolsep{\fill}} ccc @{}}
\toprule
\parbox{0.16\textwidth}{\centering\textbf{Dispersion} \\ $\boldsymbol{\varepsilon(\mathbf{k}) - \varepsilon_0}$} & 
\parbox{0.58\textwidth}{\centering\textbf{Analytical DOS} $g(E)_{E \to \varepsilon_0}$} & 
\parbox{0.26\textwidth}{\centering\textbf{Schematic DOS}} \\
\midrule

\parbox{0.16\textwidth}{\centering $c_x k_x^2 + c_y k_y^{n_y} + c_z k_z^{n_z}$ \\ \vspace{2pt} \tiny \textit{3D Extremum} \\ \textit{TRIM Regime} \\ (Even $n_y, n_z$)} & 
\parbox{0.58\textwidth}{\centering $\displaystyle \frac{\sqrt{\pi} \Gamma\left(\frac{1}{n_y}\right) \Gamma\left(\frac{1}{n_z}\right) |E - \varepsilon_0|^{\frac{1}{n_y} + \frac{1}{n_z} - \frac{1}{2}}}{2\pi^3 n_y n_z \sqrt{c_x} c_y^{1/n_y} c_z^{1/n_z} \Gamma\left(\frac{1}{n_y} + \frac{1}{n_z} + \frac{1}{2}\right)}
\begin{cases} 
0 & (E \le \varepsilon_0)\\
1 & (E > \varepsilon_0)
\end{cases}$} & 
\parbox{0.26\textwidth}{\centering\includegraphics[width=0.95\linewidth]{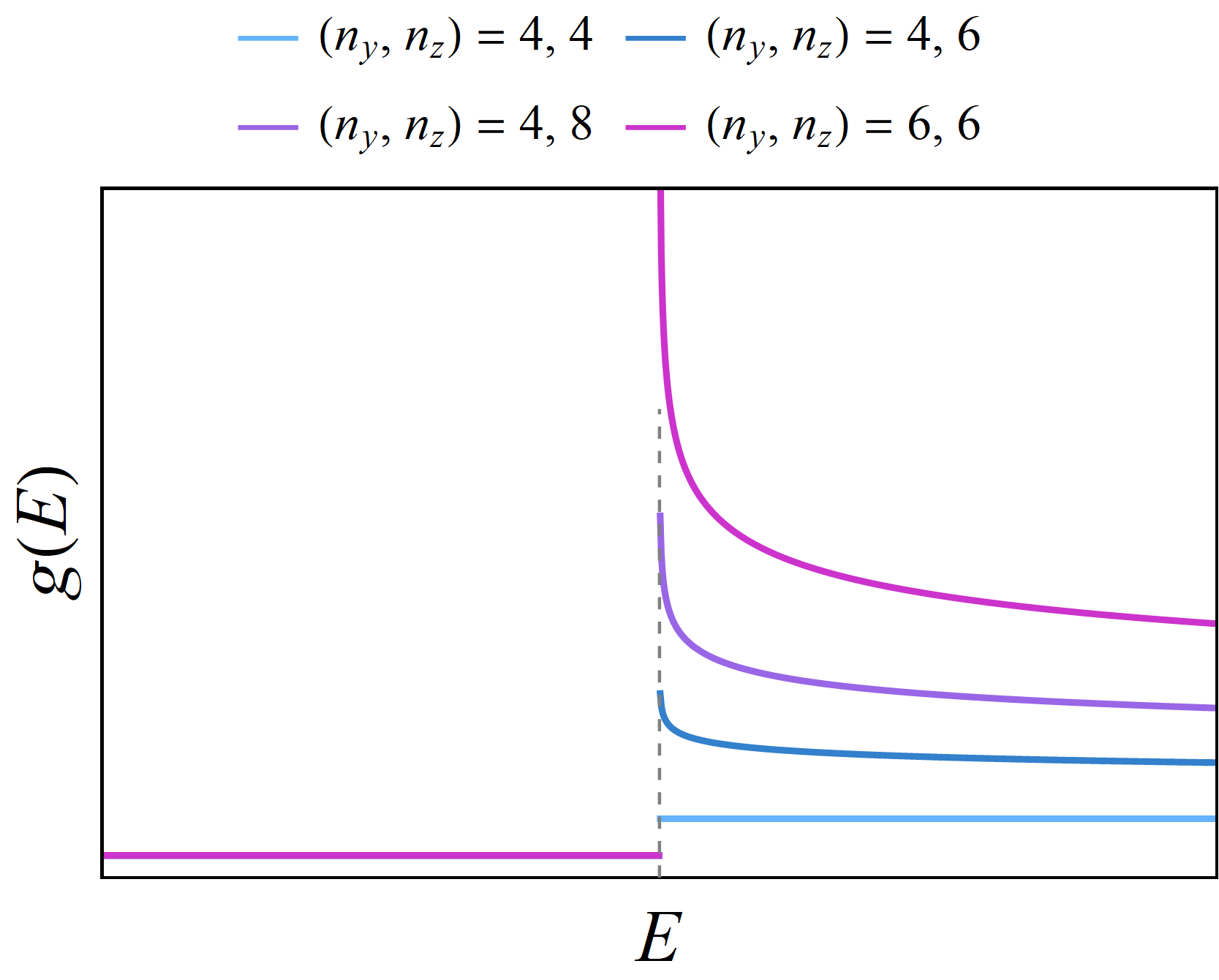}} \\ 

\parbox{0.16\textwidth}{\centering $c_x k_x^2 - c_y k_y^{n_y} - c_z k_z^{n_z}$ \\ \vspace{2pt} \tiny \textit{3D Saddle} \\ \textit{TRIM Regime} \\ (Even $n_y, n_z$)} & 
\parbox{0.58\textwidth}{\centering $\displaystyle \frac{\mathcal{A}_{yz} \cdot |E - \varepsilon_0|^{\frac{1}{n_y} + \frac{1}{n_z} - \frac{1}{2}}}{\pi^3 \sqrt{c_x} c_y^{1/n_y} c_z^{1/n_z}} 
\begin{cases} 
\mathcal{I}_{A2}^- & (E \le \varepsilon_0) \\
\mathcal{I}_{A2}^+ & (E > \varepsilon_0)
\end{cases}$} & 
\parbox{0.26\textwidth}{\centering\includegraphics[width=0.95\linewidth]{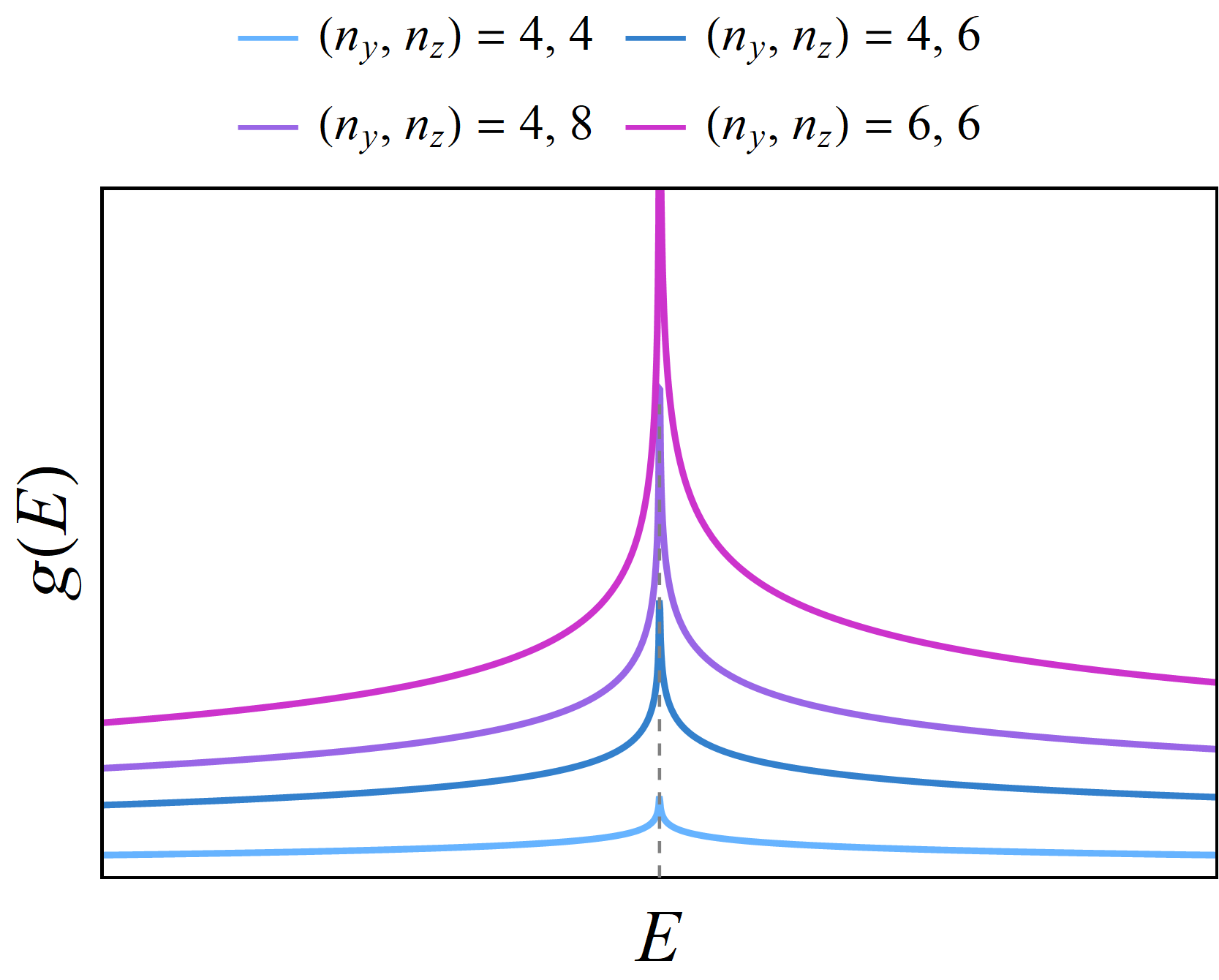}} \\

\parbox{0.16\textwidth}{\centering $c_x k_x^2 - c_y k_y^{n_y} - c_z k_z^{n_z}$ \\ \vspace{2pt} \tiny \textit{Hyperbolic Topology} \\ \textit{TRIM Regime} \\ (Even $n_y, n_z$)} & 
\parbox{0.58\textwidth}{\centering $\displaystyle \frac{\sqrt{\pi} \Gamma\left(\frac{1}{n_y}\right) \cdot |E - \varepsilon_0|^{\frac{1}{n_y} + \frac{1}{n_z} - \frac{1}{2}}}{2\pi^3 n_z \sqrt{c_x} c_y^{1/n_y} c_z^{1/n_z} \Gamma\left(\frac{1}{n_y} + \frac{1}{2}\right)}
\begin{cases} 
\mathcal{I}_{B-} & (E \le \varepsilon_0) \\
\mathcal{I}_{B+} & (E > \varepsilon_0) 
\end{cases}$} & 
\parbox{0.26\textwidth}{\centering\includegraphics[width=0.95\linewidth]{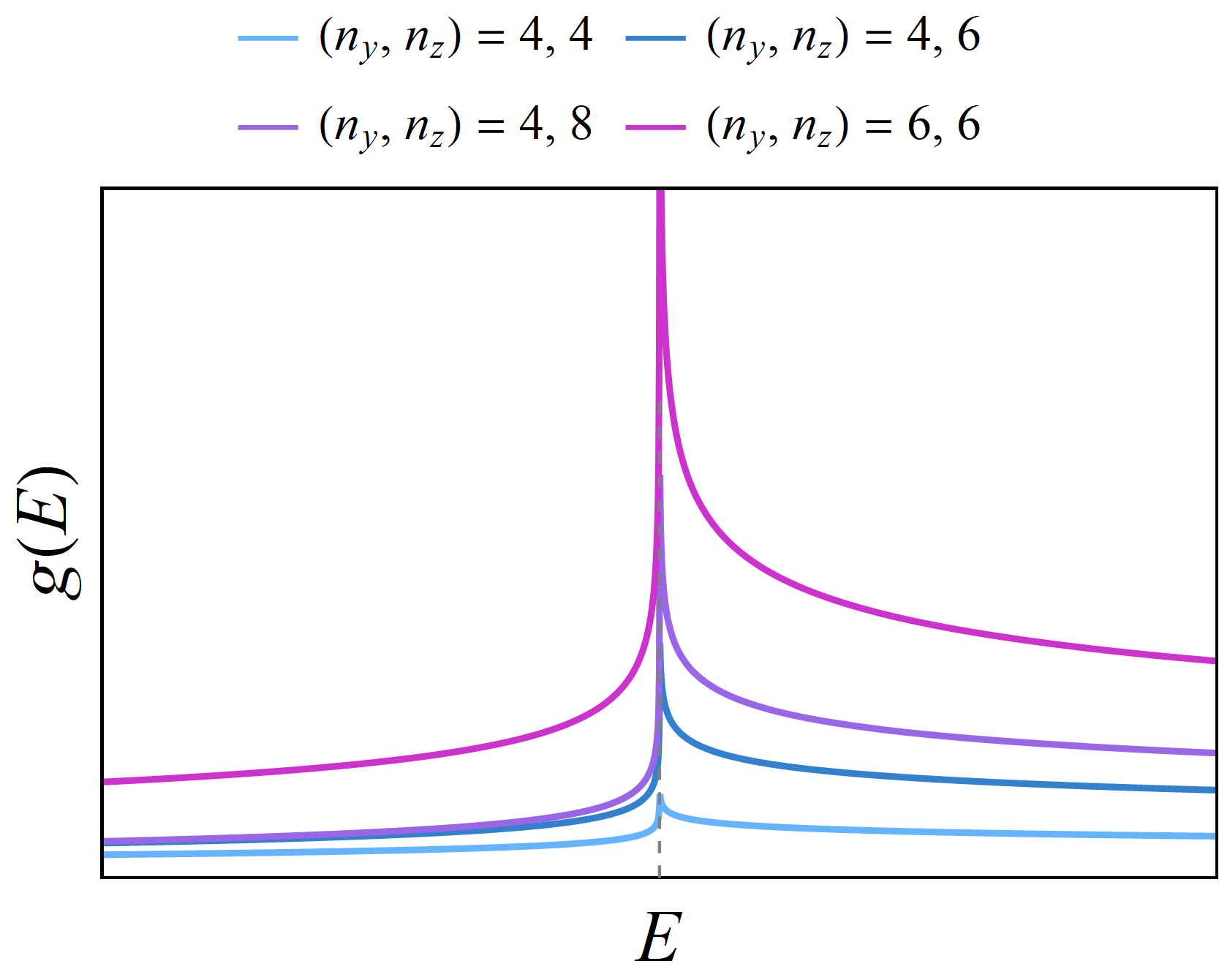}} \\

\parbox{0.16\textwidth}{\centering $c_x k_x^2 + c_y k_y^{n_y} + c_z k_z^{n_z}$ \\ \vspace{2pt} \tiny \textit{Elliptic Topology} \\ \textit{Mixed-Parity} \\ (Even $n_y$, Odd $n_z$)} & 
\parbox{0.58\textwidth}{\centering $\displaystyle \frac{\sqrt{\pi} \Gamma\left(\frac{1}{n_y}\right) \cdot |E - \varepsilon_0|^{\frac{1}{n_y} + \frac{1}{n_z} - \frac{1}{2}}}{4\pi^3 n_z \sqrt{c_x} c_y^{1/n_y} c_z^{1/n_z} \Gamma\left(\frac{1}{n_y} + \frac{1}{2}\right)}
\begin{cases} 
\mathcal{I}_{A-}^{mixed} & (E \le \varepsilon_0) \\
\mathcal{I}_{A+}^{mixed} & (E > \varepsilon_0) 
\end{cases}$} & 
\parbox{0.26\textwidth}{\centering\includegraphics[width=0.95\linewidth]{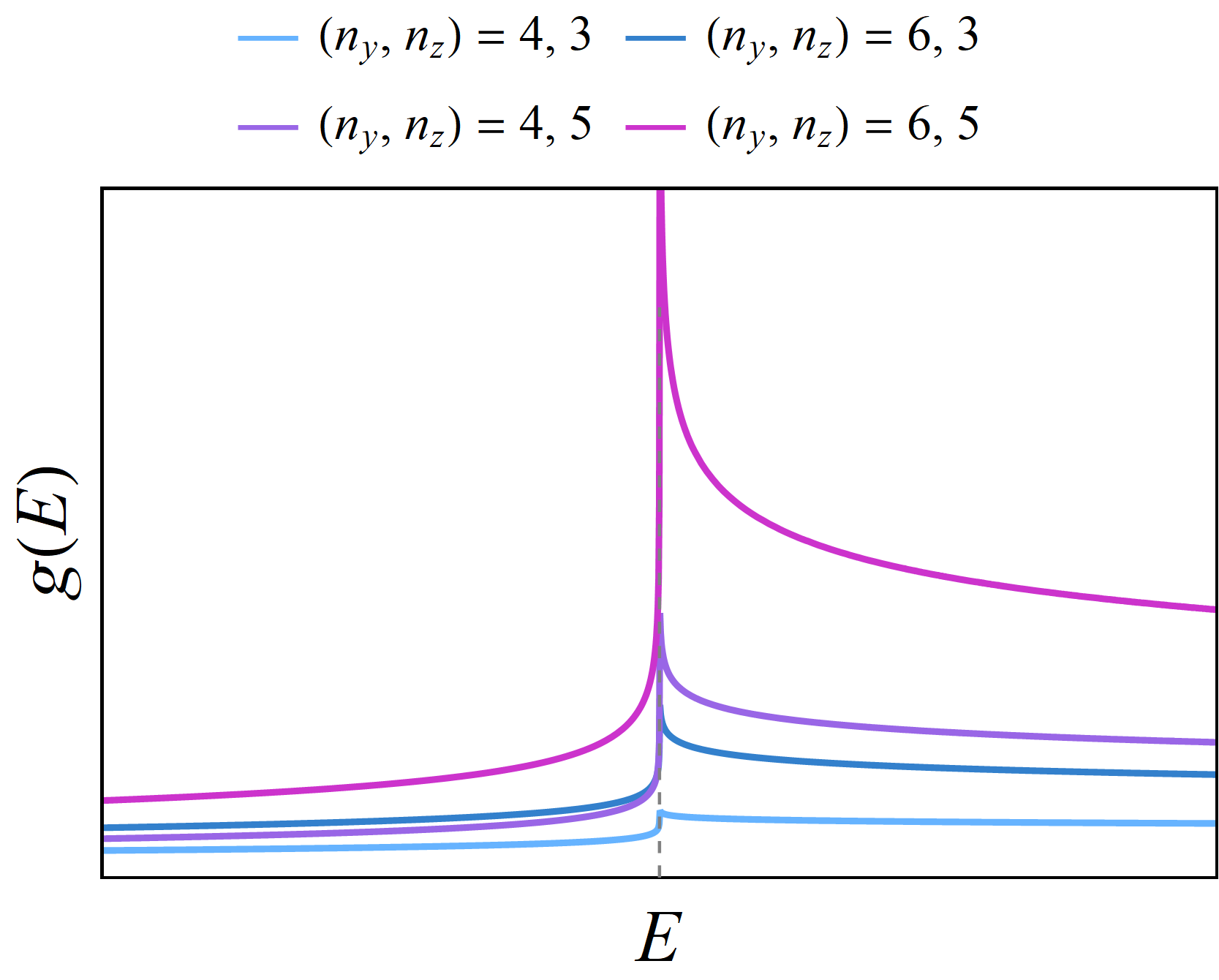}} \\

\parbox{0.16\textwidth}{\centering $c_x k_x^2 + c_y k_y^{n_y} - c_z k_z^{n_z}$ \\ \vspace{2pt} \tiny \textit{Hyperbolic Topology} \\ \textit{Mixed-Parity} \\ (Even $n_y$, Odd $n_z$)} & 
\parbox{0.58\textwidth}{\centering $\displaystyle \frac{8 \cdot |E - \varepsilon_0|^{\frac{1}{n_y} + \frac{1}{n_z} - \frac{1}{2}}}{(2\pi)^3 n_y n_z \sqrt{c_x} c_y^{1/n_y} c_z^{1/n_z}}
\begin{cases} 
\mathcal{I}_{B-}^{mixed} & (E \le \varepsilon_0) \\
\mathcal{I}_{B+}^{mixed} & (E > \varepsilon_0) 
\end{cases}$} & 
\parbox{0.26\textwidth}{\centering\includegraphics[width=0.95\linewidth]{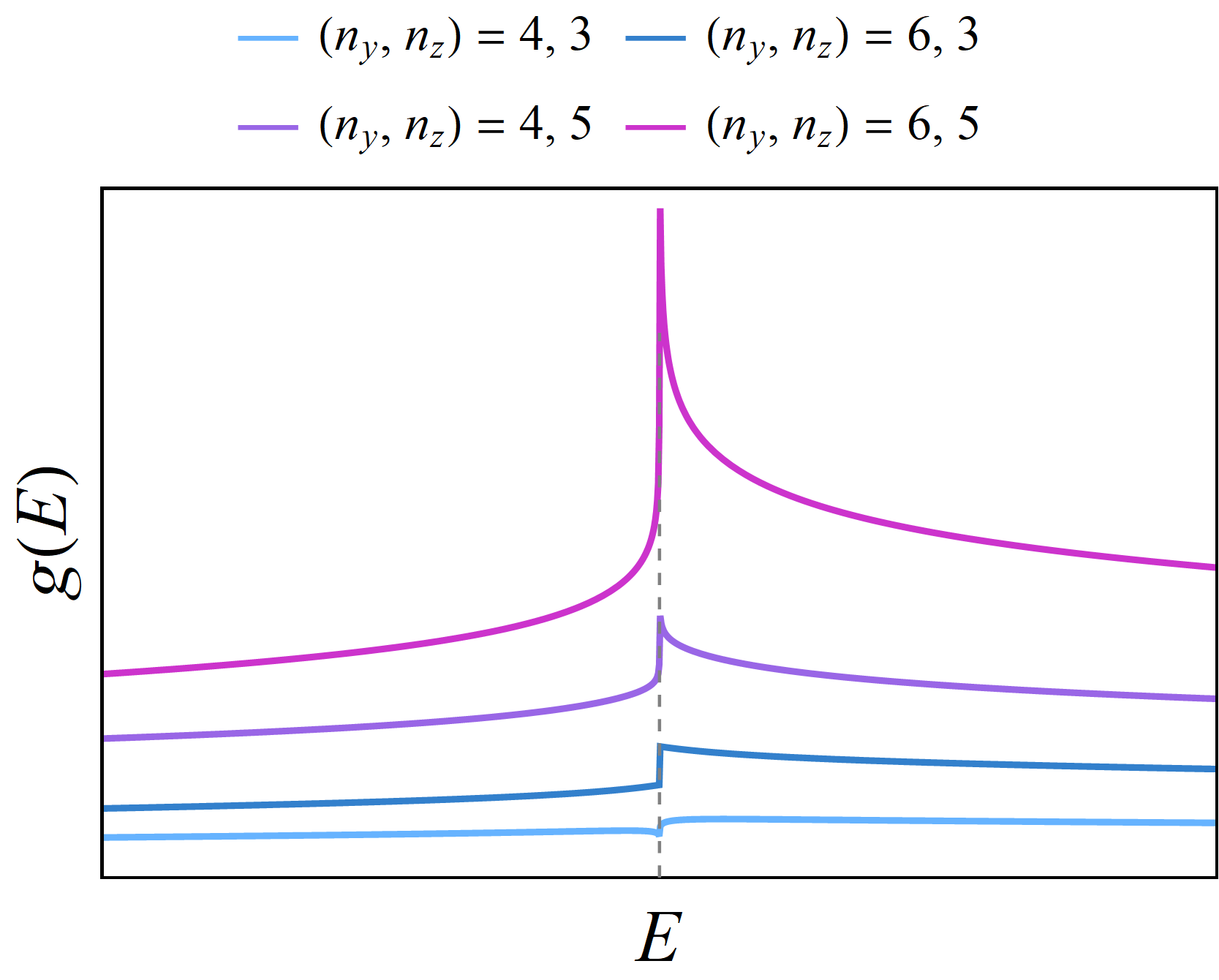}} \\

\parbox{0.16\textwidth}{\centering $c_x k_x^2 + c_y k_y^{n_y} + c_z k_z^{n_z}$ \\ \vspace{2pt} \tiny \textit{Universal Topology} \\ \textit{Fully Non-TRIM} \\ (Odd $n_y, n_z$)} & 
\parbox{0.58\textwidth}{\centering $\displaystyle \frac{|E - \varepsilon_0|^{\frac{1}{n_y} + \frac{1}{n_z} - \frac{1}{2}}}{(2\pi)^3 n_z \sqrt{c_x} c_y^{1/n_y} c_z^{1/n_z}}
\begin{cases} 
\mathcal{K}_-^{odd} & (E \le \varepsilon_0) \\
\mathcal{K}_+^{odd} & (E > \varepsilon_0) 
\end{cases}$} & 
\parbox{0.26\textwidth}{\centering\includegraphics[width=0.95\linewidth]{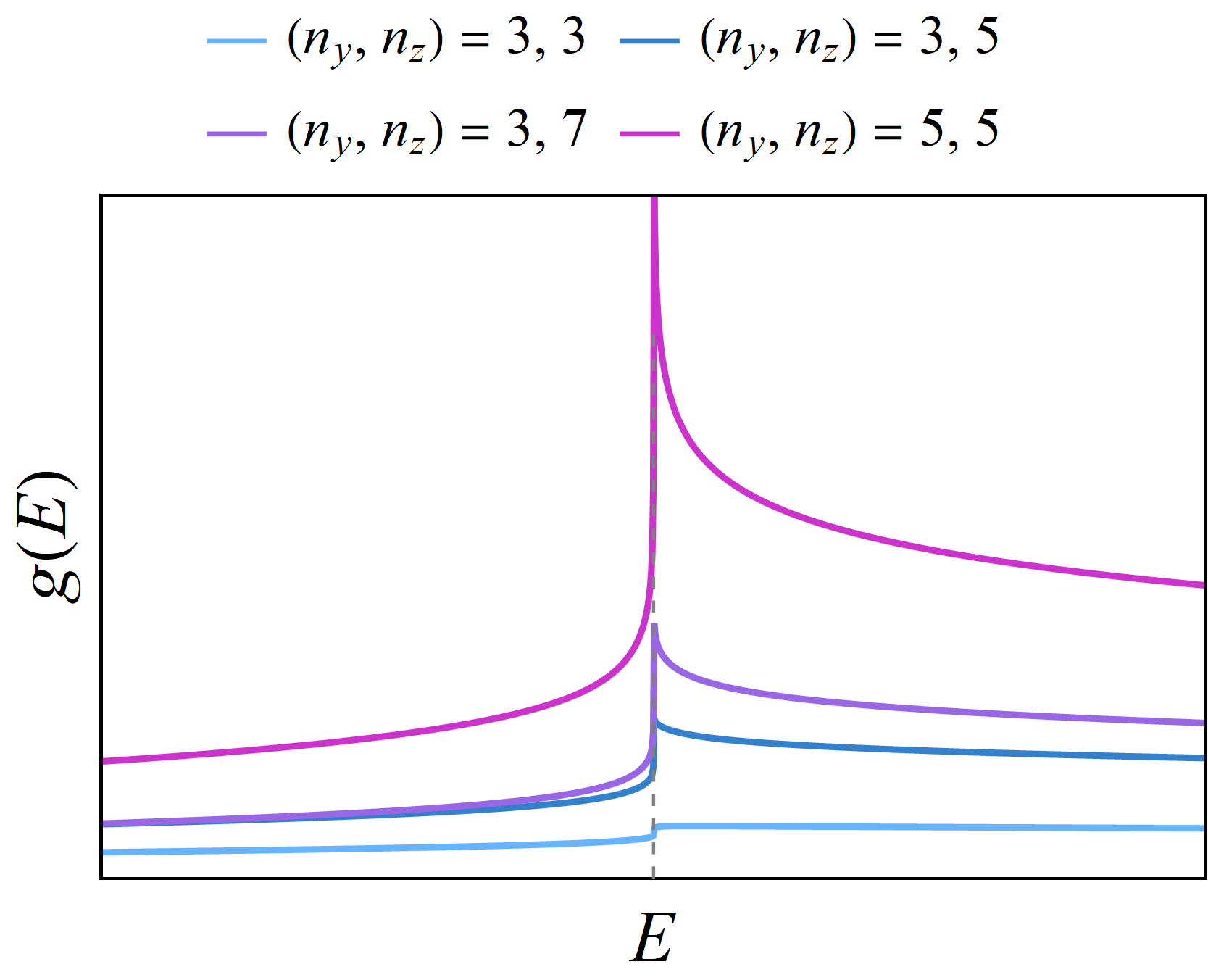}} \\

\bottomrule
\end{tabular*}

\vspace{6pt}
\begin{minipage}{\textwidth}
    \raggedright
    \tiny
    \renewcommand{\arraystretch}{1.2}
    \begin{tabular}{@{}ll@{}}
    $\mathcal{A}_{yz}$:  Dimensionless constant $\Gamma(1/n_y)\Gamma(1/n_z) / \left[ n_y n_z \Gamma(1/n_y + 1/n_z) \right]$ arising from the transverse Dirichlet integration. \\
    $\mathcal{I}, \mathcal{K}$:  Convergent dimensionless integrals defining the singular phase-space weights (see Section~\ref{sec:iii.2} for exact piecewise definitions). \\
    $R, \Gamma(n)$:  Brillouin zone cutoff and Gamma function, respectively.
    \end{tabular}
\end{minipage}
\end{table*}
\begin{table*}[htbp]
\centering
\scriptsize
\setlength{\tabcolsep}{0pt}
\renewcommand{\arraystretch}{2.6}

\caption{\textbf{Analytical Density of States (DOS) for $T_3$-type Higher-Order Van Hove Singularities.} The table summarizes the analytical scaling behaviors for systems where all three principal directions exhibit higher-order polynomial scaling ($n_x, n_y, n_z \ge 3$). Column 2 provides the exact analytical forms with explicit scaling exponents, and Column 3 presents the corresponding schematic DOS profiles.} 
\label{table_T3_DOS}

\begin{tabular*}{\textwidth}{@{\extracolsep{\fill}} ccc @{}}
\toprule
\parbox{0.16\textwidth}{\centering\textbf{Dispersion} \\ $\boldsymbol{\varepsilon(\mathbf{k}) - \varepsilon_0}$} & 
\parbox{0.58\textwidth}{\centering\textbf{Analytical DOS} $g(E)_{E \to \varepsilon_0}$} & 
\parbox{0.26\textwidth}{\centering\textbf{Schematic DOS}} \\
\midrule

\parbox{0.16\textwidth}{\centering $c_x k_x^{n_x} + c_y k_y^{n_y} + c_z k_z^{n_z}$ \\ \vspace{2pt} \tiny \textit{3D Extremum} \\ \textit{TRIM Regime} \\ (Even $n_i$)} & 
\parbox{0.58\textwidth}{\centering $\displaystyle \frac{\Gamma\left(\frac{1}{n_x}\right) \Gamma\left(\frac{1}{n_y}\right) \Gamma\left(\frac{1}{n_z}\right) \cdot |E - \varepsilon_0|^{\frac{1}{n_x} + \frac{1}{n_y} + \frac{1}{n_z} - 1}}{\pi^3 n_x n_y n_z C_{xyz} \Gamma\left(\frac{1}{n_x} + \frac{1}{n_y} + \frac{1}{n_z}\right)}
\begin{cases} 
0 & (E \le \varepsilon_0)\\
1 & (E > \varepsilon_0)
\end{cases}$} & 
\parbox{0.26\textwidth}{\centering\includegraphics[width=0.95\linewidth]{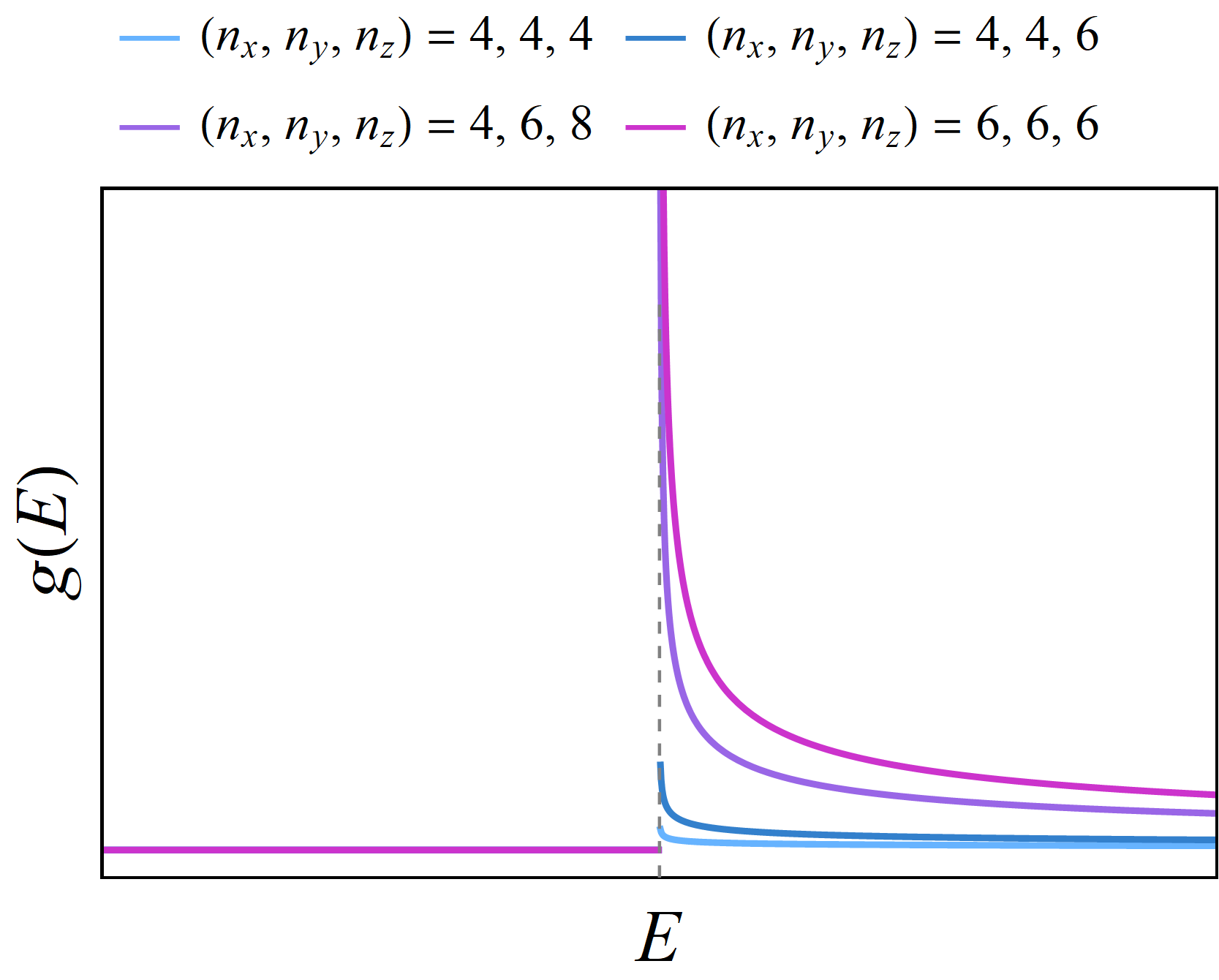}} \\ 

\parbox{0.16\textwidth}{\centering $c_x k_x^{n_x} + c_y k_y^{n_y} - c_z k_z^{n_z}$ \\ \vspace{2pt} \tiny \textit{3D Saddle} \\ \textit{TRIM Regime} \\ (Even $n_i$)} & 
\parbox{0.58\textwidth}{\centering $\displaystyle \frac{2 \Gamma\left(\frac{1}{n_x}\right) \Gamma\left(\frac{1}{n_y}\right) \cdot |E - \varepsilon_0|^{\frac{1}{n_x} + \frac{1}{n_y} + \frac{1}{n_z} - 1}}{\pi^3 n_x n_y n_z C_{xyz} \Gamma\left(\frac{1}{n_x} + \frac{1}{n_y}\right)}
\begin{cases} 
\mathcal{J}_{B-}^{even} & (E \le \varepsilon_0) \\
\mathcal{J}_{B+}^{even} & (E > \varepsilon_0) 
\end{cases}$} & 
\parbox{0.26\textwidth}{\centering\includegraphics[width=0.95\linewidth]{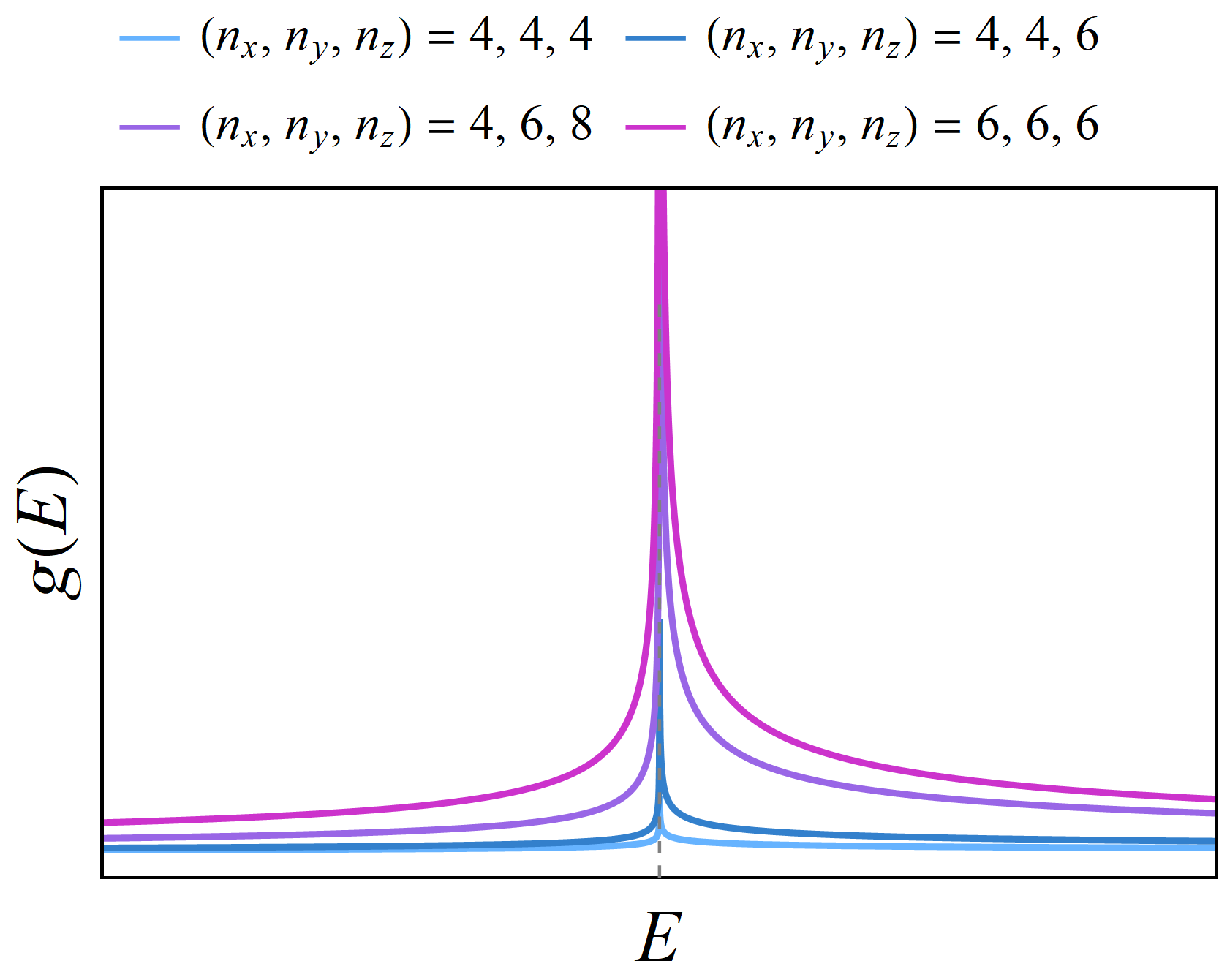}} \\

\parbox{0.16\textwidth}{\centering $c_x k_x^{n_x} + c_y k_y^{n_y} + c_z k_z^{n_z}$ \\ \vspace{2pt} \tiny \textit{Elliptic Topology} \\ \textit{Single-Odd Regime} \\ (Even $n_{x,y}$, Odd $n_z$)} & 
\parbox{0.58\textwidth}{\centering $\displaystyle \frac{\Gamma\left(\frac{1}{n_x}\right) \Gamma\left(\frac{1}{n_y}\right) \cdot |E - \varepsilon_0|^{\frac{1}{n_x} + \frac{1}{n_y} + \frac{1}{n_z} - 1}}{\pi^3 n_x n_y n_z C_{xyz} \Gamma\left(\frac{1}{n_x} + \frac{1}{n_y}\right)}
\begin{cases} 
\mathcal{I}_{A-}^{1-odd} & (E \le \varepsilon_0) \\
\mathcal{I}_{A+}^{1-odd} & (E > \varepsilon_0) 
\end{cases}$} & 
\parbox{0.26\textwidth}{\centering\includegraphics[width=0.95\linewidth]{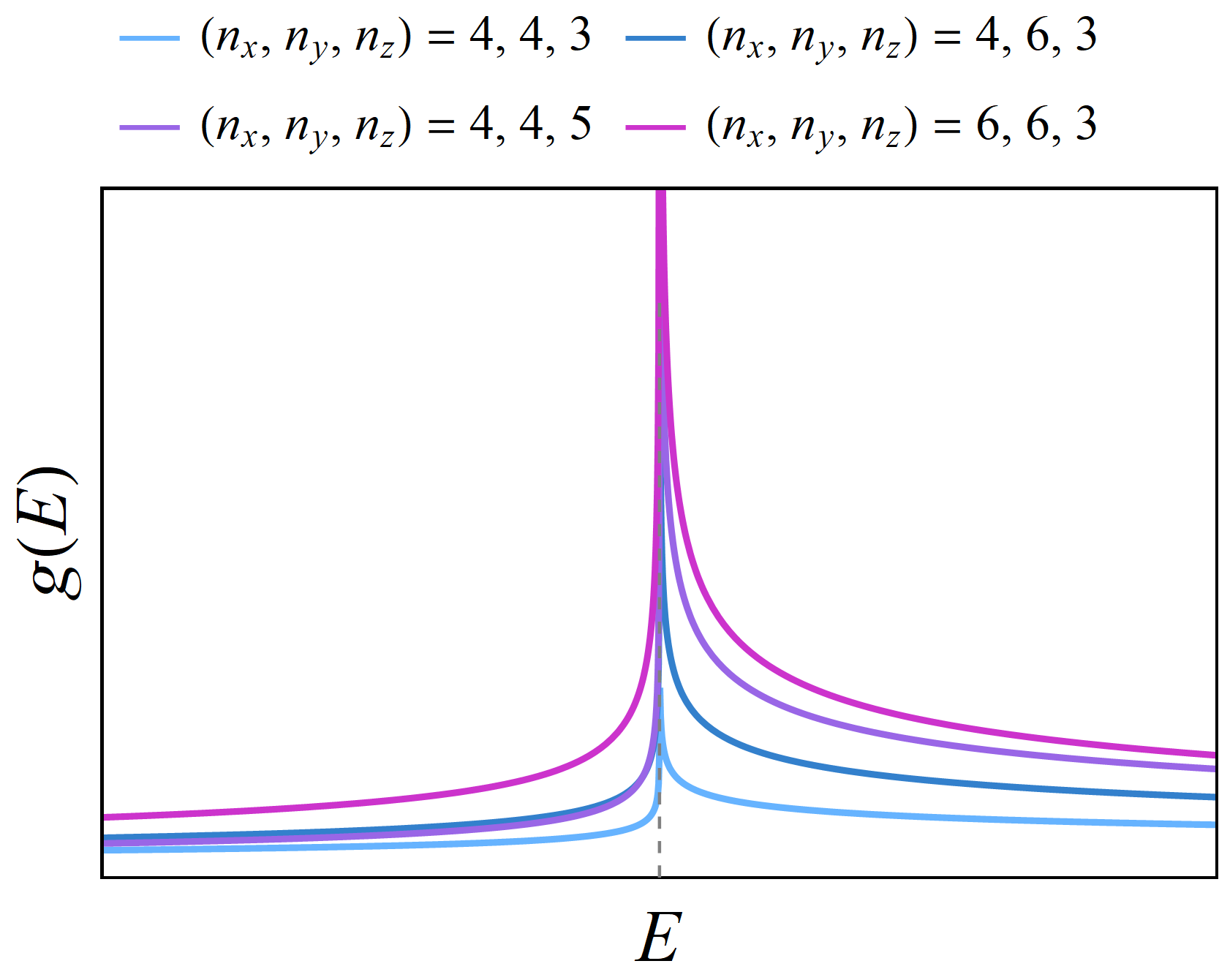}} \\

\parbox{0.16\textwidth}{\centering $c_x k_x^{n_x} - c_y k_y^{n_y} + c_z k_z^{n_z}$ \\ \vspace{2pt} \tiny \textit{Hyperbolic Topology} \\ \textit{Single-Odd Regime} \\ (Even $n_{x,y}$, Odd $n_z$)} & 
\parbox{0.58\textwidth}{\centering $\displaystyle \frac{4 \cdot |E - \varepsilon_0|^{\frac{1}{n_x} + \frac{1}{n_y} + \frac{1}{n_z} - 1}}{(2\pi)^3 n_z C_{xyz}}
\begin{cases} 
\mathcal{I}_{B-}^{1-odd} & (E \le \varepsilon_0) \\
\mathcal{I}_{B+}^{1-odd} & (E > \varepsilon_0) 
\end{cases}$} & 
\parbox{0.26\textwidth}{\centering\includegraphics[width=0.95\linewidth]{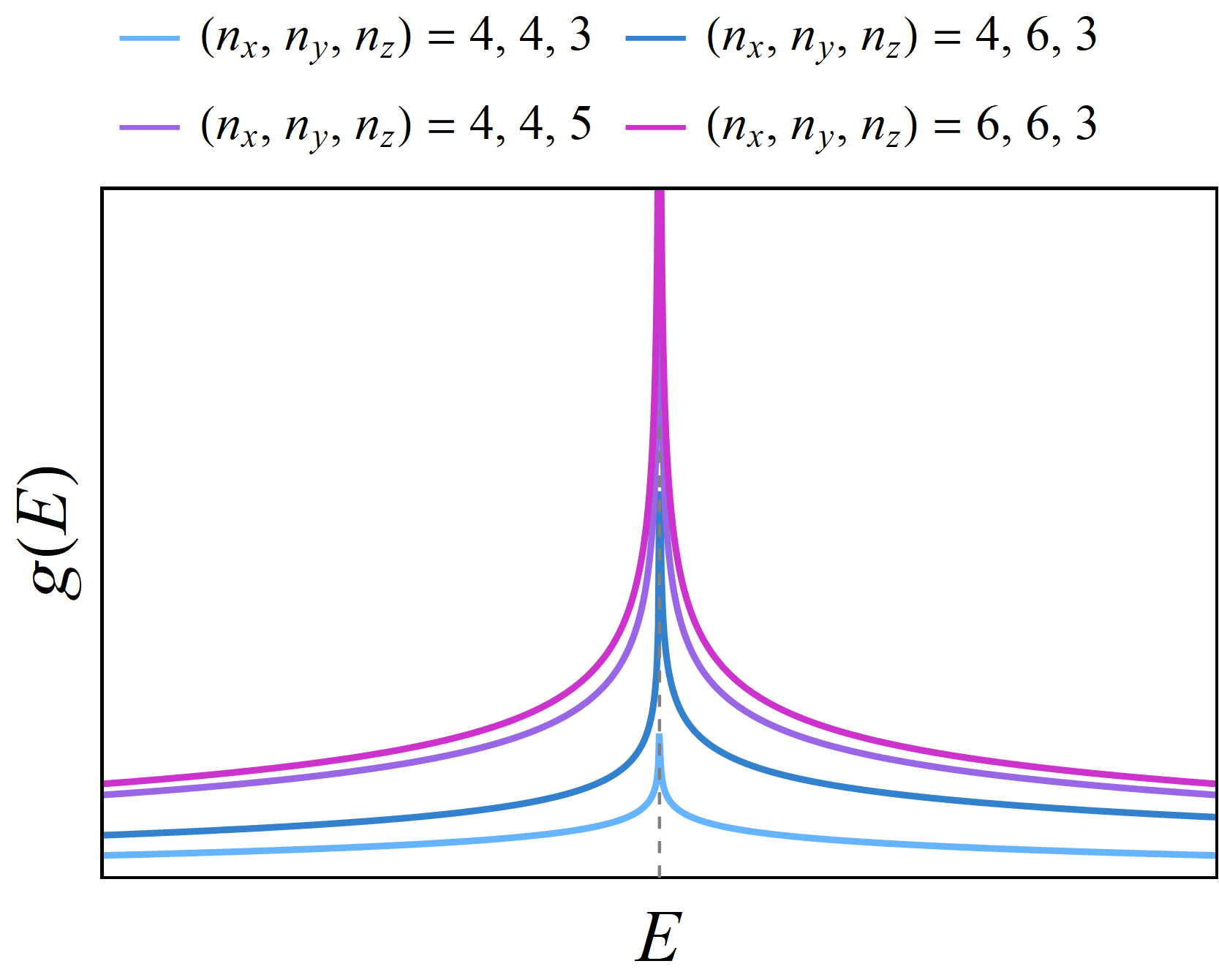}} \\

\parbox{0.16\textwidth}{\centering $c_x k_x^{n_x} + c_y k_y^{n_y} + c_z k_z^{n_z}$ \\ \vspace{2pt} \tiny \textit{Universal Topology} \\ \textit{Double-Odd Regime} \\ (Even $n_x$, Odd $n_{y,z}$)} & 
\parbox{0.58\textwidth}{\centering $\displaystyle \frac{2 \cdot |E - \varepsilon_0|^{\frac{1}{n_x} + \frac{1}{n_y} + \frac{1}{n_z} - 1}}{(2\pi)^3 n_z C_{xyz}}
\begin{cases} 
\mathcal{J}_{-}^{2-odd} & (E \le \varepsilon_0) \\
\mathcal{J}_{+}^{2-odd} & (E > \varepsilon_0) 
\end{cases}$} & 
\parbox{0.26\textwidth}{\centering\includegraphics[width=0.95\linewidth]{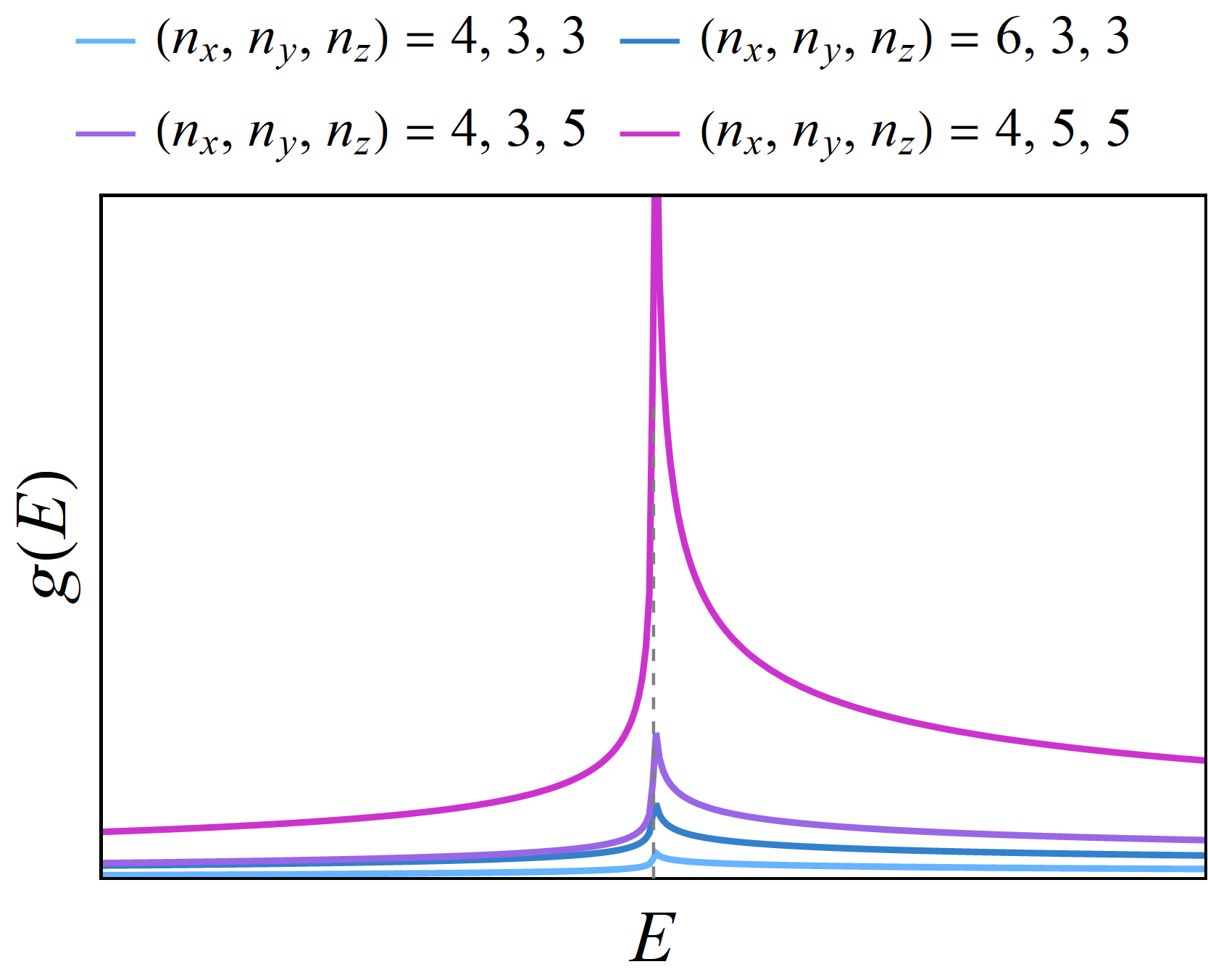}} \\

\parbox{0.16\textwidth}{\centering $c_x k_x^{n_x} + c_y k_y^{n_y} + c_z k_z^{n_z}$ \\ \vspace{2pt} \tiny \textit{Universal Topology} \\ \textit{Fully Non-TRIM} \\ (Odd $n_{x,y,z}$)} & 
\parbox{0.58\textwidth}{\centering $\displaystyle \frac{|E - \varepsilon_0|^{\frac{1}{n_x} + \frac{1}{n_y} + \frac{1}{n_z} - 1}}{(2\pi)^3 n_z C_{xyz}}
\begin{cases} 
\mathcal{K}_-^{3-odd} & (E \le \varepsilon_0) \\
\mathcal{K}_+^{3-odd} & (E > \varepsilon_0) 
\end{cases}$} & 
\parbox{0.26\textwidth}{\centering\includegraphics[width=0.95\linewidth]{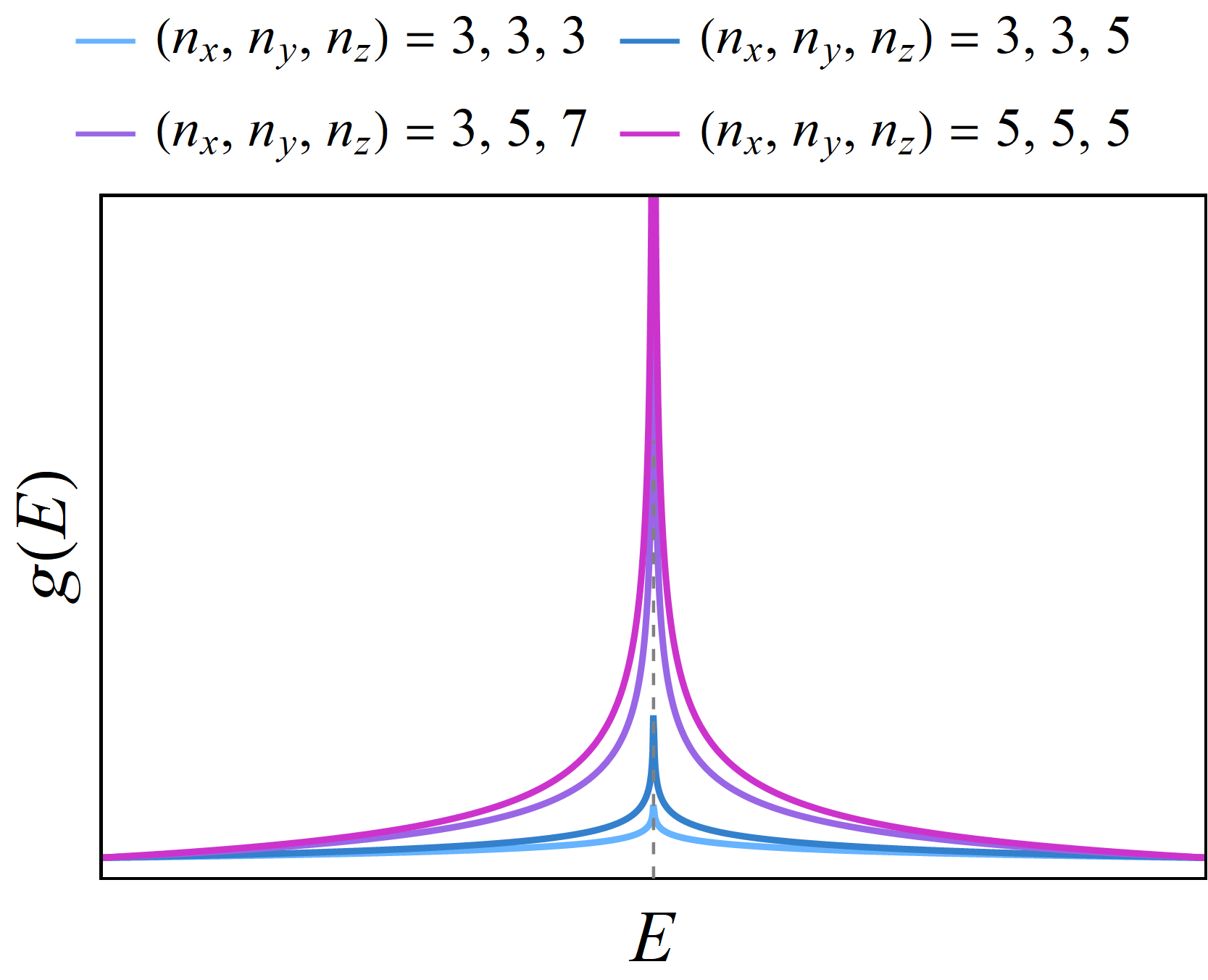}} \\

\bottomrule
\end{tabular*}

\vspace{6pt}
\begin{minipage}{\textwidth}
    \raggedright
    \tiny
    \renewcommand{\arraystretch}{1.2}
    \begin{tabular}{@{}ll@{}}
    $C_{xyz}$: & Anisotropic scaling prefactor defined as $c_x^{1/n_x} c_y^{1/n_y} c_z^{1/n_z}$. \\
    $\mathcal{J}, \mathcal{I}, \mathcal{K}$: & Convergent dimensionless integrals defining the singular phase-space weights (see Section~\ref{sec:iii.3} for exact piecewise definitions). \\
    $R, \Gamma(n)$: & Brillouin zone cutoff and Gamma function, respectively.
    \end{tabular}
\end{minipage}
\end{table*}

\begin{table*}[htbp]
\centering
\setlength{\tabcolsep}{0pt} 
\renewcommand{\arraystretch}{2.5} 

\caption{\textbf{DOS for ordinary noncritical Van Hove singularities in 3D electronic systems.} The table summarizes the analytical results for mixed-dimensional dispersions where the transverse components are strictly quadratic ($n=2$).}
\begin{tabular*}{\textwidth}{@{\extracolsep{\fill}} c c c }
\toprule
\textbf{Type} & 
\textbf{Dispersion $\boldsymbol{\varepsilon(\mathbf{k}) - \varepsilon_0}$} & 
\textbf{DOS $g(E)_{E \to \varepsilon_0}$} \\
\midrule

$N_0$ & 
$c_x k_x^2 + c_y k_y^2 + c_z k_z$ & 
\parbox{0.50\textwidth}{\centering $\displaystyle 
\frac{R}{8\pi^2\sqrt{c_x c_y}} + \frac{E - \varepsilon_0}{8\pi^2 c_z \sqrt{c_x c_y}}$} \\

$N_1$ & 
$c_x k_x^2 - c_y k_y^2 + c_z k_z$ & 
\parbox{0.50\textwidth}{\centering $\displaystyle 
\frac{R}{2\pi^3 \sqrt{c_x c_y}} \left[1 + \ln\left(\frac{W}{c_z R}\right)\right] - \frac{(E - \varepsilon_0)^2}{4\pi^3 c_z^2 R \sqrt{c_x c_y}}$} \\

$N_2$ & 
$-c_x k_x^2 - c_y k_y^2 + c_z k_z$ & 
\parbox{0.50\textwidth}{\centering $\displaystyle 
\frac{R}{8\pi^2\sqrt{c_x c_y}} - \frac{E - \varepsilon_0}{8\pi^2 c_z \sqrt{c_x c_y}}$} \\

\bottomrule
\end{tabular*}

\vspace{6pt}

\begin{minipage}{\textwidth}
    \raggedright 
    \small
    \renewcommand{\arraystretch}{1.2}
    \begin{tabular}{@{}ll} 
    $R$: Brillouin zone cutoff wavevector along the $k_z$ direction. \\
    $W$:  Effective transverse energy cutoff associated with the logarithmic singularity. (see Section~\ref{sec:iv.2} for exact piecewise definitions).
    \end{tabular}
\end{minipage}
\label{table_Ordinary_Noncritical_DOS} 
\end{table*}
\begin{table*}[htbp]
\centering
\scriptsize
\setlength{\tabcolsep}{0pt} 
\renewcommand{\arraystretch}{2.8} 

\caption{\textbf{Analytical Density of States (DOS) for $S_1$-type Higher-Order Van Hove Singularities.} The table summarizes the analytical scaling for noncritical singularities where one transverse direction is strictly quadratic ($n_x=2$) and the other is higher-order ($n_y \ge 3$).} 
\label{table_S1_DOS}

\begin{tabular*}{\textwidth}{@{\extracolsep{\fill}} ccc @{}}
\toprule
\parbox{0.18\textwidth}{\centering \textbf{Dispersion} \\ $\boldsymbol{\varepsilon(\mathbf{k}) - \varepsilon_0}$} & 
\parbox{0.47\textwidth}{\centering \textbf{Analytical DOS} $g(E)_{E \to \varepsilon_0}$} & 
\parbox{0.35\textwidth}{\centering \textbf{Schematic DOS}} \\
\midrule

\parbox{0.18\textwidth}{\centering $c_x k_x^2 + c_y k_y^{n_y} + c_z k_z$ \\ \vspace{2pt} \scriptsize \textit{Elliptic Topology} \\ \textit{TRIM Regime} \\ (Even $n_y \ge 4$)} & 
\parbox{0.47\textwidth}{\centering $\displaystyle g_0 + \frac{C_+ \Delta_z^{\mu - 1}}{2\pi c_z \sqrt{c_x} c_y^{1/n_y}} (E - \varepsilon_0)$} & 
\parbox{0.35\textwidth}{\centering\includegraphics[width=0.98\linewidth]{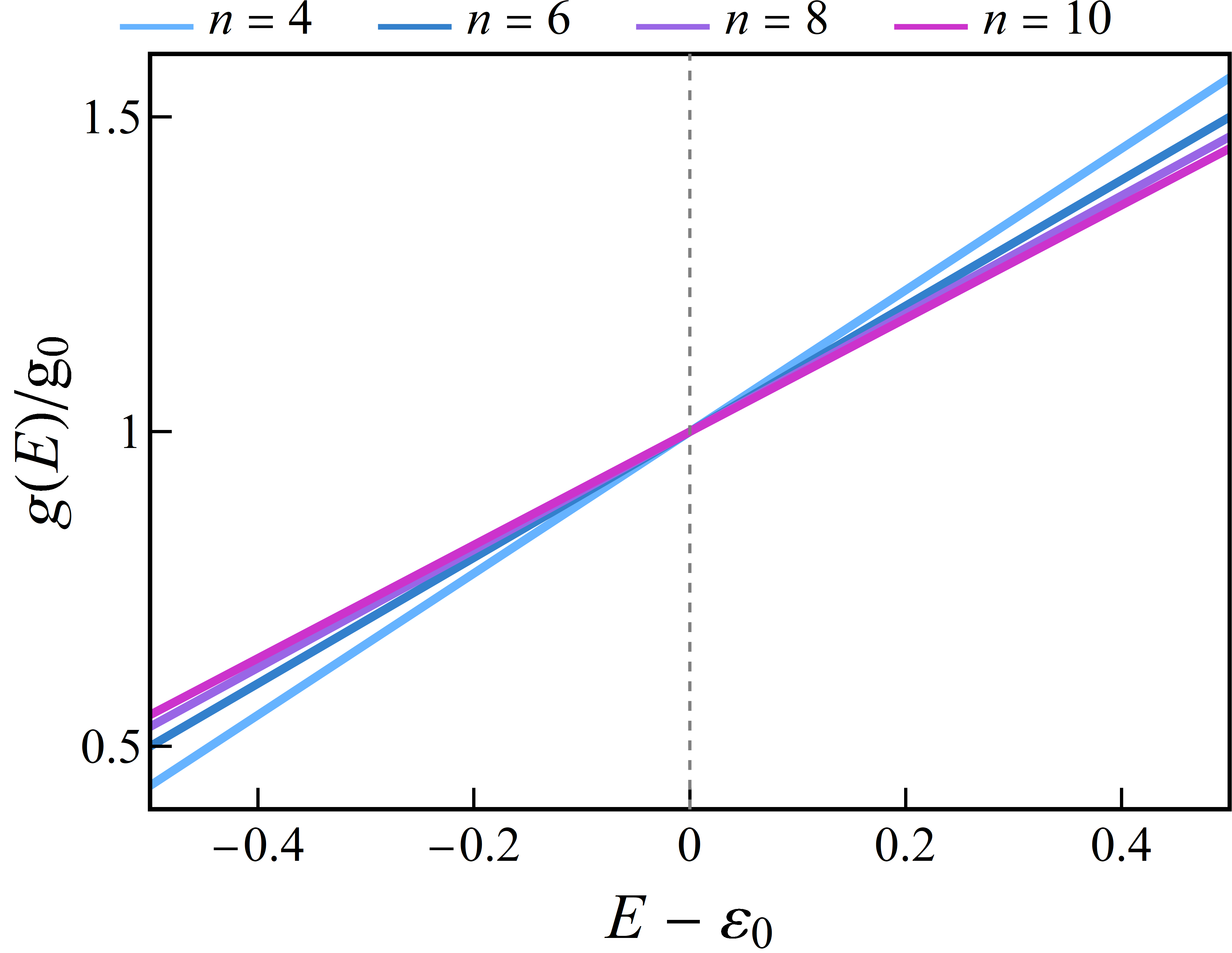}} \\ 

\parbox{0.18\textwidth}{\centering $c_x k_x^2 - c_y k_y^{n_y} + c_z k_z$ \\ \vspace{2pt} \scriptsize \textit{Hyperbolic Topology} \\ \textit{TRIM Regime} \\ (Even $n_y \ge 4$)} & 
\parbox{0.47\textwidth}{\centering $\displaystyle g_0 + \frac{(C_+ - C_-) \Delta_z^{\mu - 1}}{2\pi c_z \sqrt{c_x} c_y^{1/n_y}} (E - \varepsilon_0)$} & 
\parbox{0.35\textwidth}{\centering\includegraphics[width=0.98\linewidth]{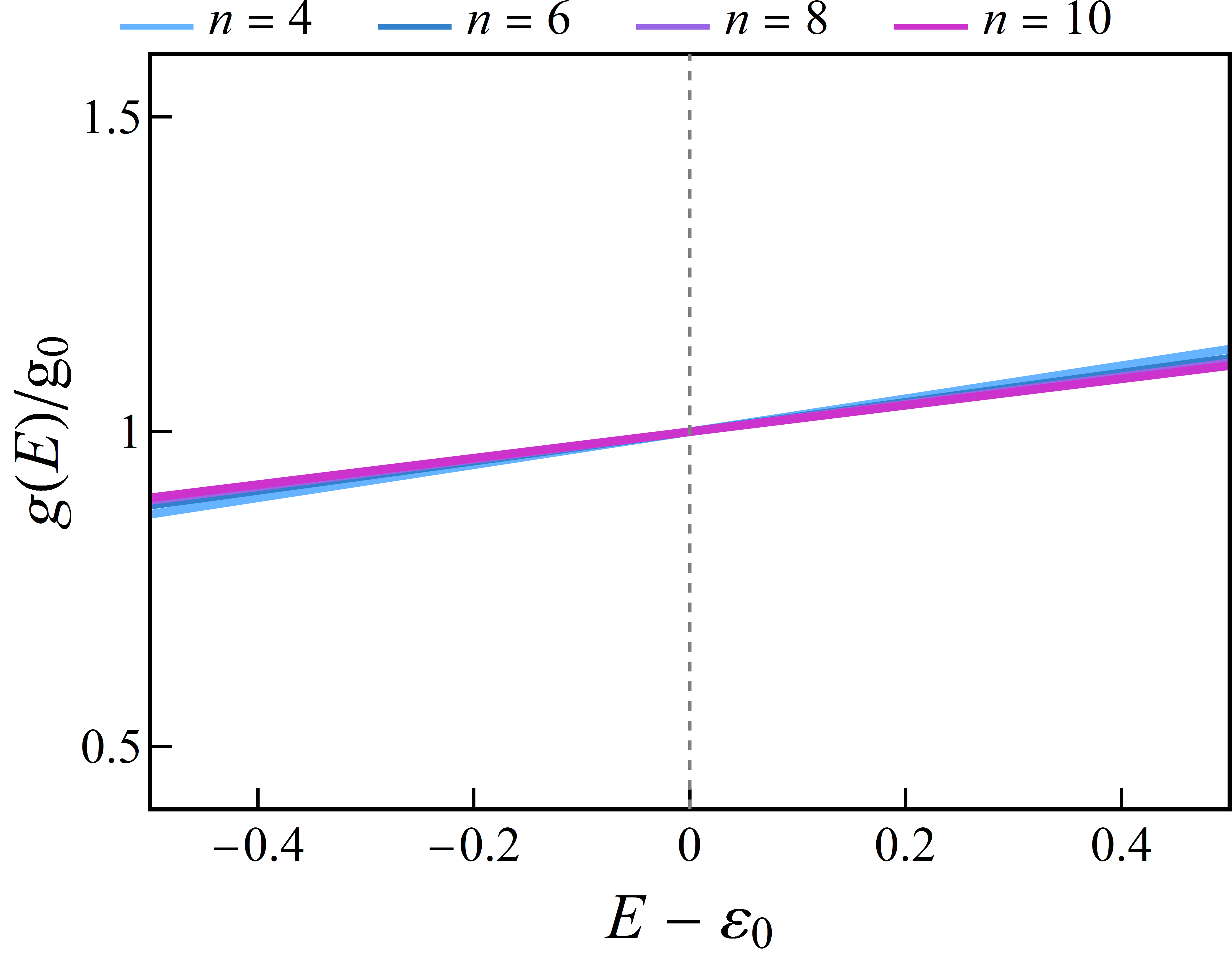}} \\

\parbox{0.18\textwidth}{\centering $c_x k_x^2 \pm c_y k_y^{n_y} + c_z k_z$ \\ \vspace{2pt} \scriptsize \textit{Universal Topology} \\ \textit{Non-TRIM Regime} \\ (Odd $n_y \ge 3$)} & 
\parbox{0.47\textwidth}{\centering $\displaystyle g_0 + \frac{(C_+ - C_-) \Delta_z^{\mu - 1}}{2\pi c_z \sqrt{c_x} c_y^{1/n_y}} (E - \varepsilon_0) \quad (C_+ \neq C_-)$} & 
\parbox{0.35\textwidth}{\centering\includegraphics[width=0.98\linewidth]{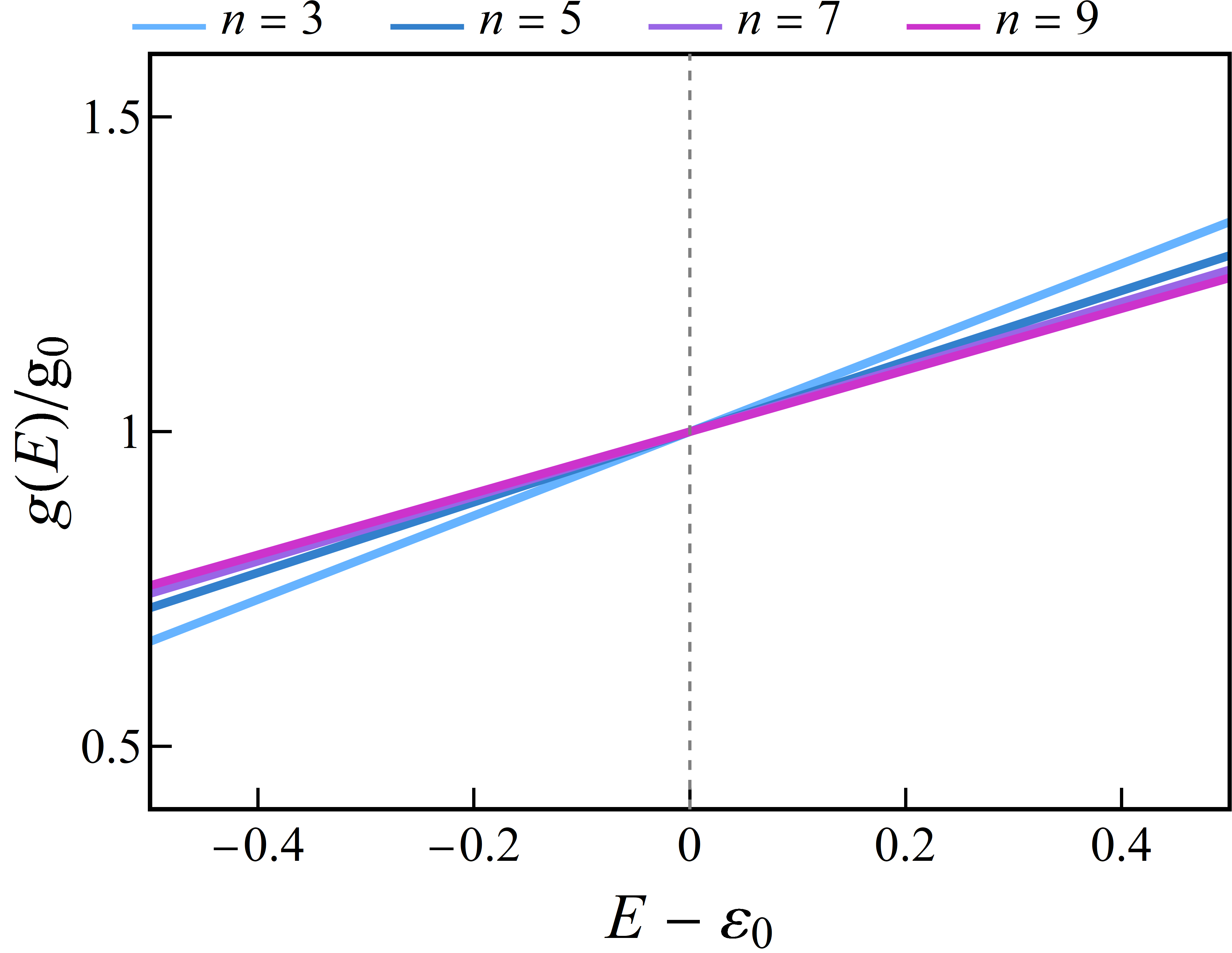}} \\

\bottomrule
\end{tabular*}

\vspace{6pt}
\begin{minipage}{\textwidth}
    \raggedright
    \scriptsize
    \renewcommand{\arraystretch}{1.2}
    \begin{tabular}{@{}ll}
    $\mu$: & Characteristic transverse exponent $1/2 + 1/n_y$. \\
    $\Delta_z$: & Longitudinal half-bandwidth $c_z R$. \\
    $g_0$: & Non-zero constant background DOS resulting from the longitudinal integration. \\
    $C_\pm$: & Dimensionless phase-space weights determined by the transverse geometric contours. (see Section~\ref{sec:v.1} for exact piecewise definitions).
    \end{tabular}
\end{minipage}
\end{table*}
\begin{table*}[htbp]
\centering
\scriptsize
\setlength{\tabcolsep}{0pt} 
\renewcommand{\arraystretch}{3.0} 

\caption{\textbf{Analytical Density of States (DOS) for $S_2$-type Higher-Order Van Hove Singularities.} The table summarizes the analytical scaling for noncritical singularities where both transverse directions exhibit higher-order polynomial scaling ($n_x, n_y \ge 3$). The universal transverse scaling exponent is defined as $\nu = 1/n_x + 1/n_y < 1$.} 
\label{table_S2_DOS}

\begin{tabular*}{\textwidth}{@{\extracolsep{\fill}} ccc @{}}
\toprule
\parbox{0.18\textwidth}{\centering \textbf{Dispersion} \\ $\boldsymbol{\varepsilon(\mathbf{k}) - \varepsilon_0}$} & 
\parbox{0.50\textwidth}{\centering \textbf{Analytical DOS} $g(E)_{E \to \varepsilon_0}$} & 
\parbox{0.30\textwidth}{\centering \textbf{Schematic DOS}} \\
\midrule

\parbox{0.18\textwidth}{\centering $c_x k_x^{n_x} + c_y k_y^{n_y} + c_z k_z$ \\ \vspace{2pt} \tiny \textit{Elliptic Topology} \\ \textit{TRIM Regime} \\ (Even $n_x, n_y$)} & 
\parbox{0.50\textwidth}{\centering $\displaystyle g_0 + \frac{C_A \Delta_z^{\nu - 1}}{2\pi c_z c_x^{1/n_x} c_y^{1/n_y}} (E - \varepsilon_0) - \delta g^{(2)}(E)$} & 
\parbox{0.30\textwidth}{\centering\includegraphics[width=0.95\linewidth]{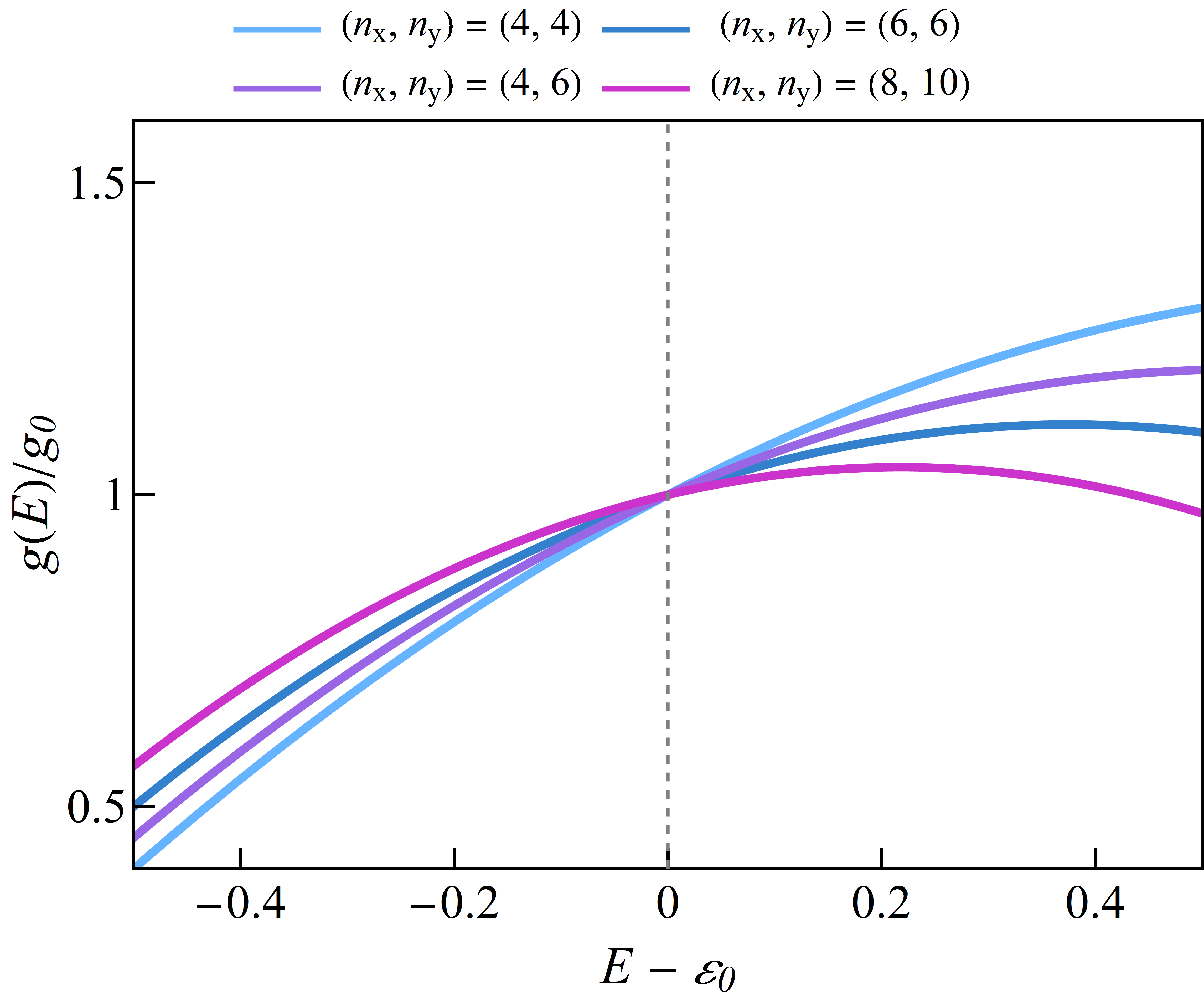}} \\ 

\parbox{0.18\textwidth}{\centering $c_x k_x^{n_x} - c_y k_y^{n_y} + c_z k_z$ \\ \vspace{2pt} \tiny \textit{Anisotropic Saddle} \\ \textit{TRIM Regime} \\ (Even $n_x \neq n_y$)} & 
\parbox{0.50\textwidth}{\centering $\displaystyle g_0 + \frac{(C_+ - C_-) \Delta_z^{\nu - 1}}{2\pi c_z c_x^{1/n_x} c_y^{1/n_y}} (E - \varepsilon_0) - \delta g^{(2)}(E)$} & 
\parbox{0.30\textwidth}{\centering\includegraphics[width=0.95\linewidth]{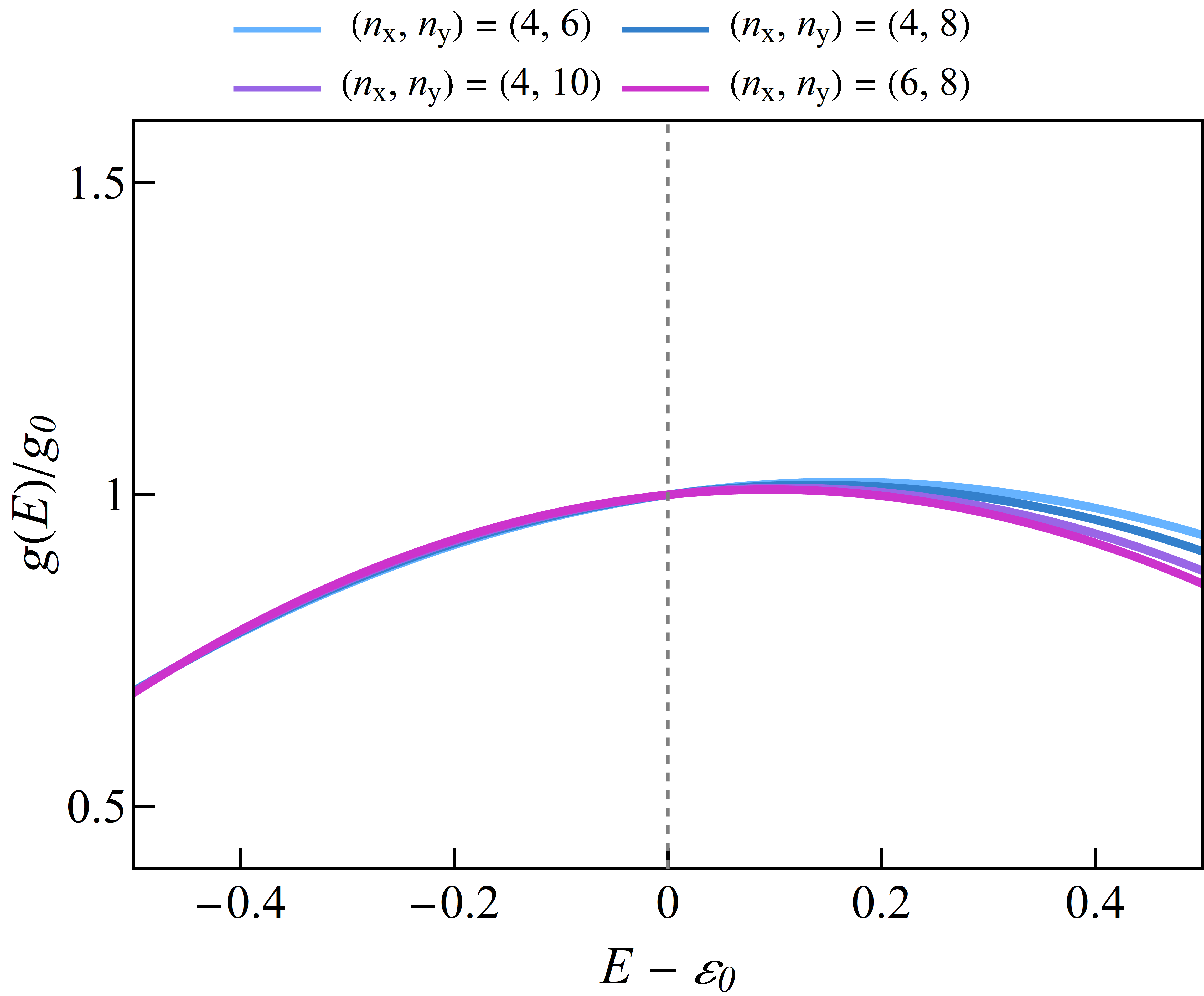}} \\

\parbox{0.18\textwidth}{\centering $c_x k_x^{n_x} - c_y k_y^{n_y} + c_z k_z$ \\ \vspace{2pt} \tiny \textit{Isotropic Saddle} \\ \textit{TRIM Regime} \\ (Even $n_x = n_y = n$)} & 
\parbox{0.50\textwidth}{\centering $\displaystyle g_0 - \frac{(1 - \nu) C_{iso} \Delta_z^{\nu - 2}}{2\pi c_z c_x^{1/n} c_y^{1/n}} (E - \varepsilon_0)^2 \quad (C_+ = C_-)$} & 
\parbox{0.30\textwidth}{\centering\includegraphics[width=0.95\linewidth]{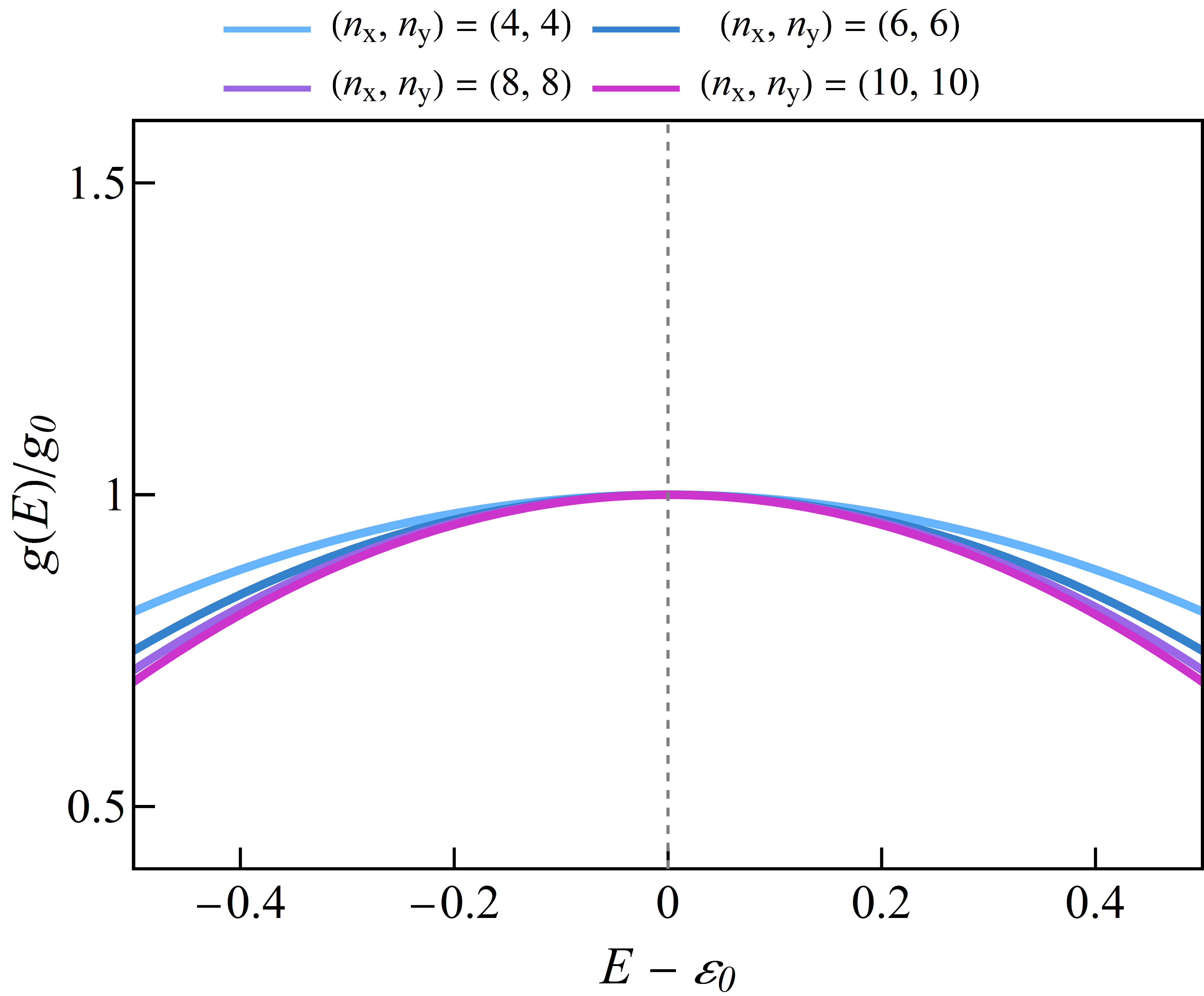}} \\

\parbox{0.18\textwidth}{\centering $c_x k_x^{n_x} \pm c_y k_y^{n_y} + c_z k_z$ \\ \vspace{2pt} \tiny \textit{Universal Topology} \\ \textit{Mixed-Parity} \\ (Even $n_x$, Odd $n_y$)} & 
\parbox{0.50\textwidth}{\centering $\displaystyle g_0 + \frac{(C_+ - C_-) \Delta_z^{\nu - 1}}{2\pi c_z c_x^{1/n_x} c_y^{1/n_y}} (E - \varepsilon_0) - \delta g^{(2)}(E)$} & 
\parbox{0.30\textwidth}{\centering\includegraphics[width=0.95\linewidth]{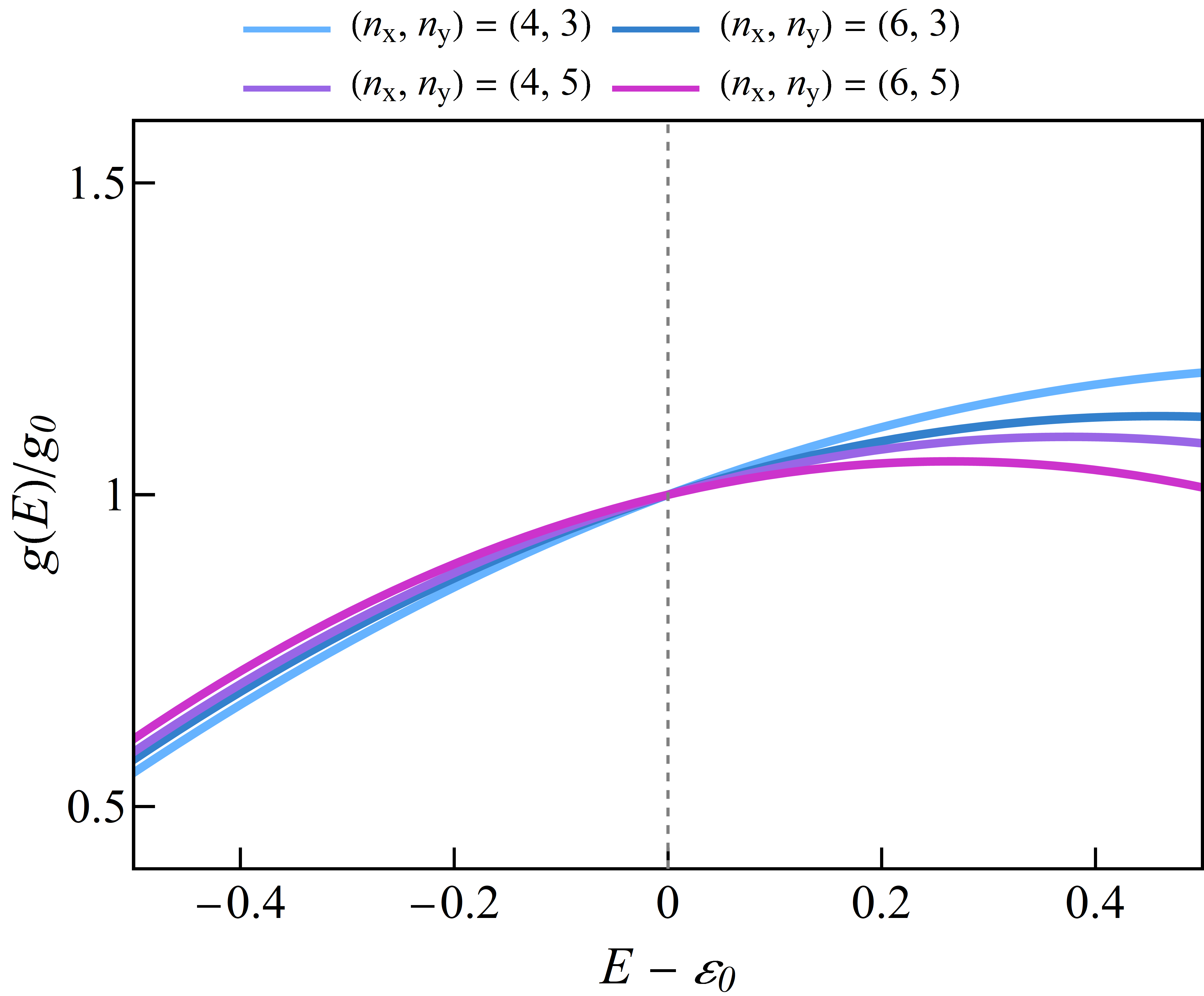}} \\

\parbox{0.18\textwidth}{\centering $c_x k_x^{n_x} + c_y k_y^{n_y} + c_z k_z$ \\ \vspace{2pt} \tiny \textit{Universal Topology} \\ \textit{Fully Non-TRIM} \\ (Odd $n_x, n_y$)} & 
\parbox{0.50\textwidth}{\centering $\displaystyle g_0 - \frac{(1 - \nu) C_{odd} \Delta_z^{\nu - 2}}{2\pi c_z c_x^{1/n_x} c_y^{1/n_y}} (E - \varepsilon_0)^2 \quad (C_+ = C_-)$} & 
\parbox{0.30\textwidth}{\centering\includegraphics[width=0.95\linewidth]{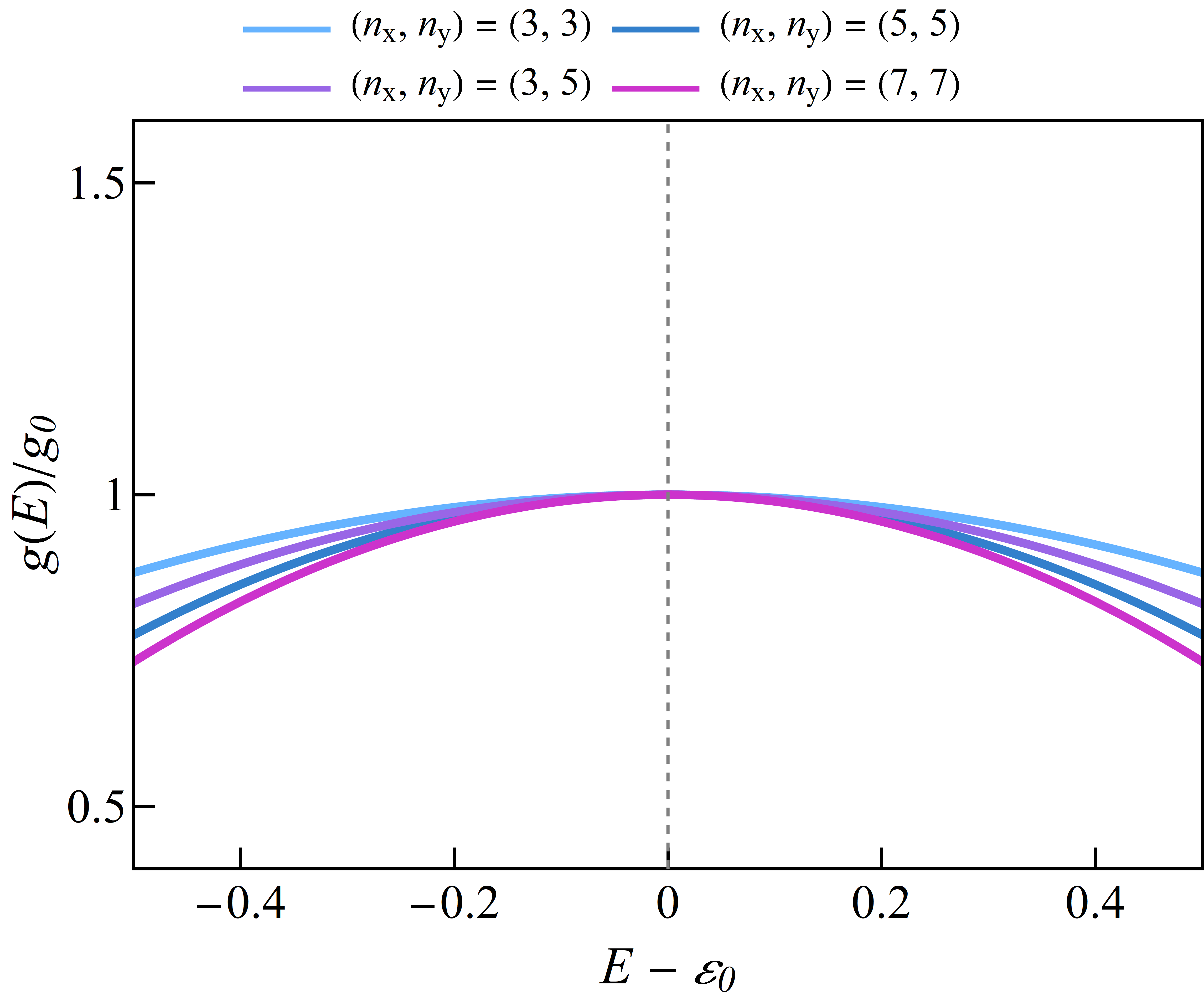}} \\

\bottomrule
\end{tabular*}
\vspace{6pt}
\begin{minipage}{\textwidth}
    \raggedright
    \tiny
    \renewcommand{\arraystretch}{1.2}
    \begin{tabular}{@{}ll@{}}
    $\Delta_z$:Longitudinal half-bandwidth $c_z R$.\\
    $\delta g^{(2)}(E)$:Sub-leading negative quadratic curvature correction.(see Section ~\ref{sec:V.2} for exact piecewise definitions). \\
    $C_\pm$:Dimensionless integration constants dictating the phase-space weight for positive and negative effective energies \\
    $C_{iso}, C_{odd}$:Unified geometric weights arising strictly from the spatial exchange symmetry of isotropic saddles or the global coordinate inversion symmetry of fully\\ \qquad  non-TRIM regimes, respectively.
 \end{tabular}
\end{minipage}
\end{table*}

\begin{figure}[htbp]
	\includegraphics[width=\linewidth]{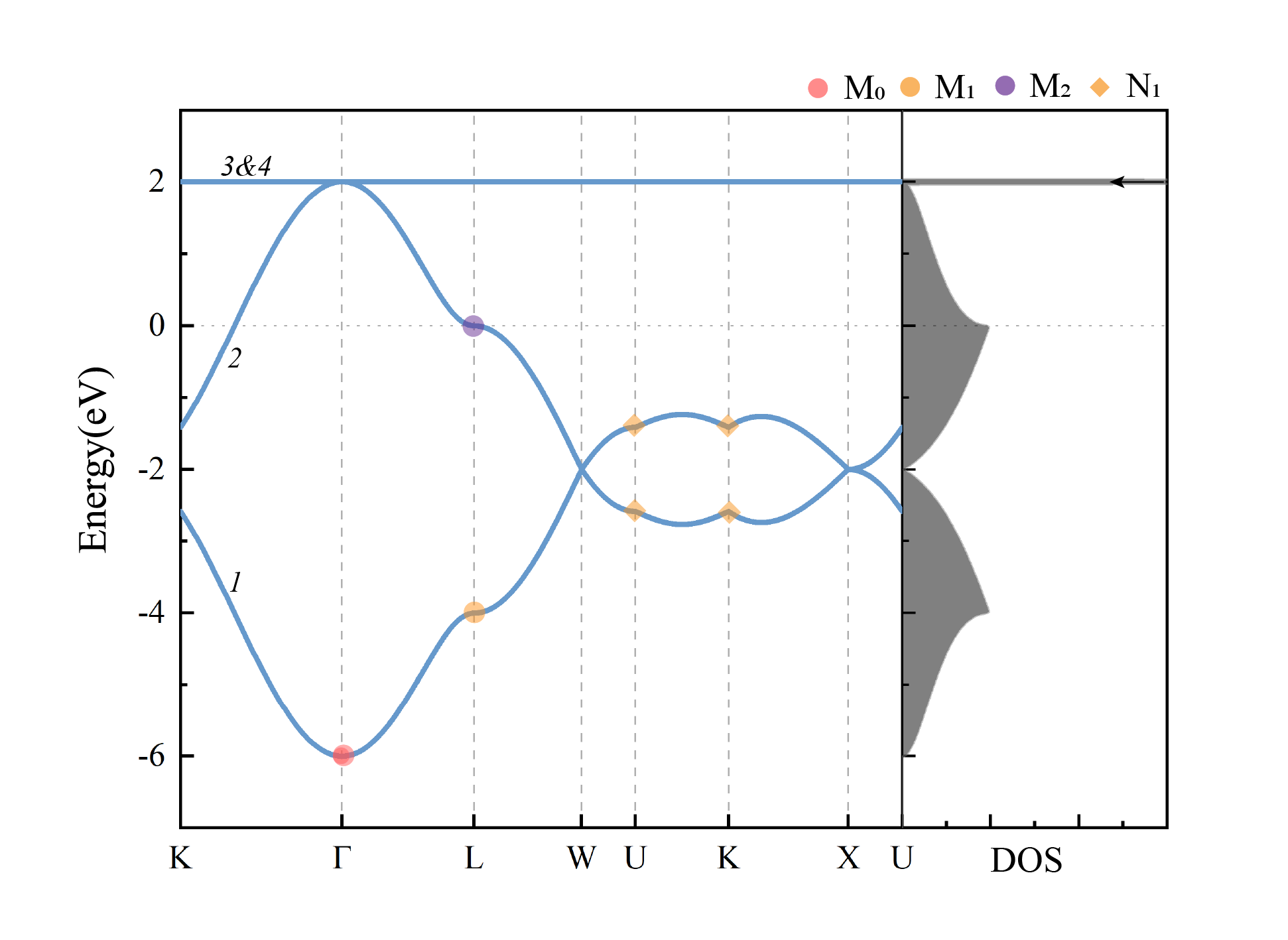}
\caption{Band structure and DOS from the \textit{s}-orbital tight-binding model of space group 227, without spin-orbit coupling. 
Model parameters: $e_{1} \to 0$, $t_{1} \to -1$. 
The blue curve represents the TB band structure along the high-symmetry path; numbers (1, 2, 3\&4) label distinct bands, and colored symbols (legend-coded) mark Van Hove singularity types. The right panel displays the corresponding DOS.}
\label{figs2}
\end{figure}

\begin{figure}[htbp]
	\includegraphics[width=\linewidth]{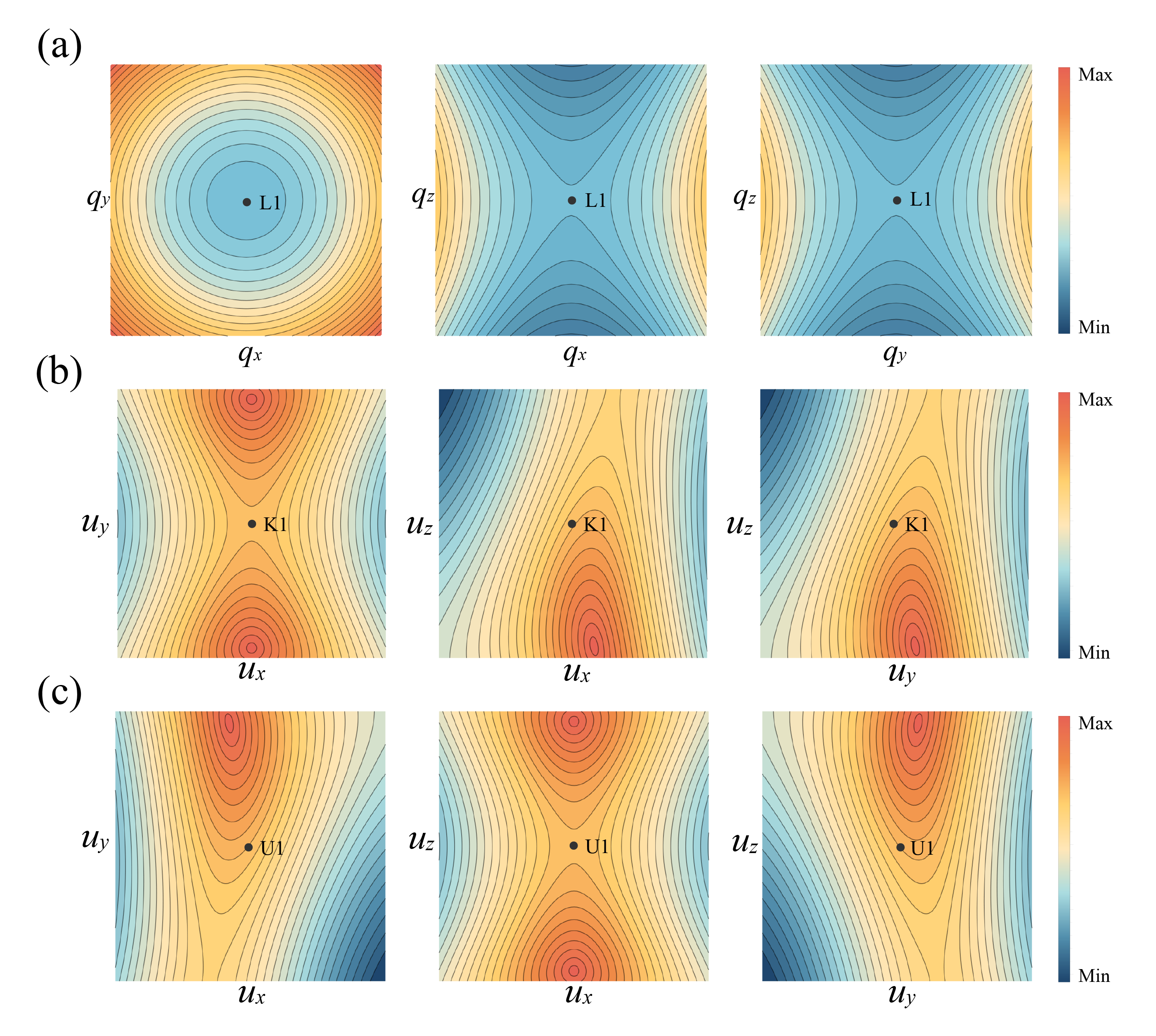}
\caption{(a)-(c) Momentum-space energy contours of distinct Van Hove singularities in the $s$-orbital tight-binding model of space group 227, without spin-orbit coupling:
- Panel (a): $L_1$ point $M_1$-type) contours;
- Panel (b): $K_1$ point ($N_1$-type) contours;
- Panel (c): $U_1$ point (also $N_1$-type) contours.
Left-to-right subfigures per panel show contours in the labeled momentum planes.}
\label{figs3}
\end{figure}
\begin{figure}[htbp]
    \centering 
    \includegraphics[
        keepaspectratio,
        width=0.95\textwidth, 
        height=0.95\textheight  
    ]{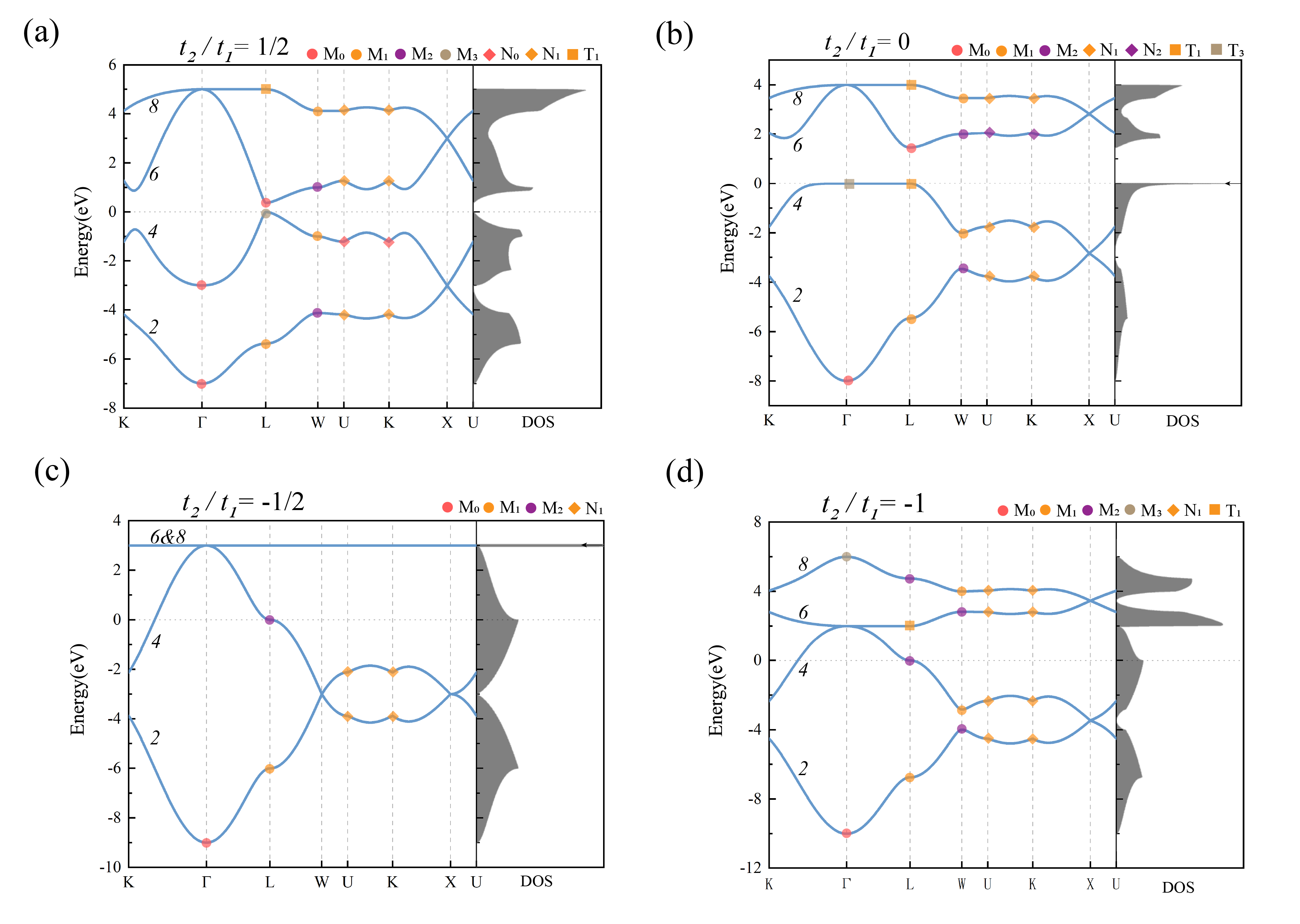}
\caption{(a)-(d) Band structure and density of states (DOS) of the pyrochlore lattice with SOC at various hopping ratios. The tight-binding parameters are defined with the on-site energy $e_1 = 0$ and the spin-flip nearest-neighbor hopping strictly locked at $t_1 = -1.0$  to serve as the baseline energy scale. The panels map the topological evolution driven by the spin-conserving hopping parameter $t_2$, corresponding to distinct dimensionless ratios $t_2/t_1$:
(a) $t_2/t_1 = 1/2$.
(b) $t_2/t_1 = 0$.
(c) $t_2/t_1 = -1/2$.
(d) $t_2/t_1 = -1$.
}
\label{figs4}
\end{figure}

\begin{figure}[htbp]
    \centering 
    \includegraphics[
        keepaspectratio,
        width=0.95\textwidth, 
        height=0.95\textheight  
    ]{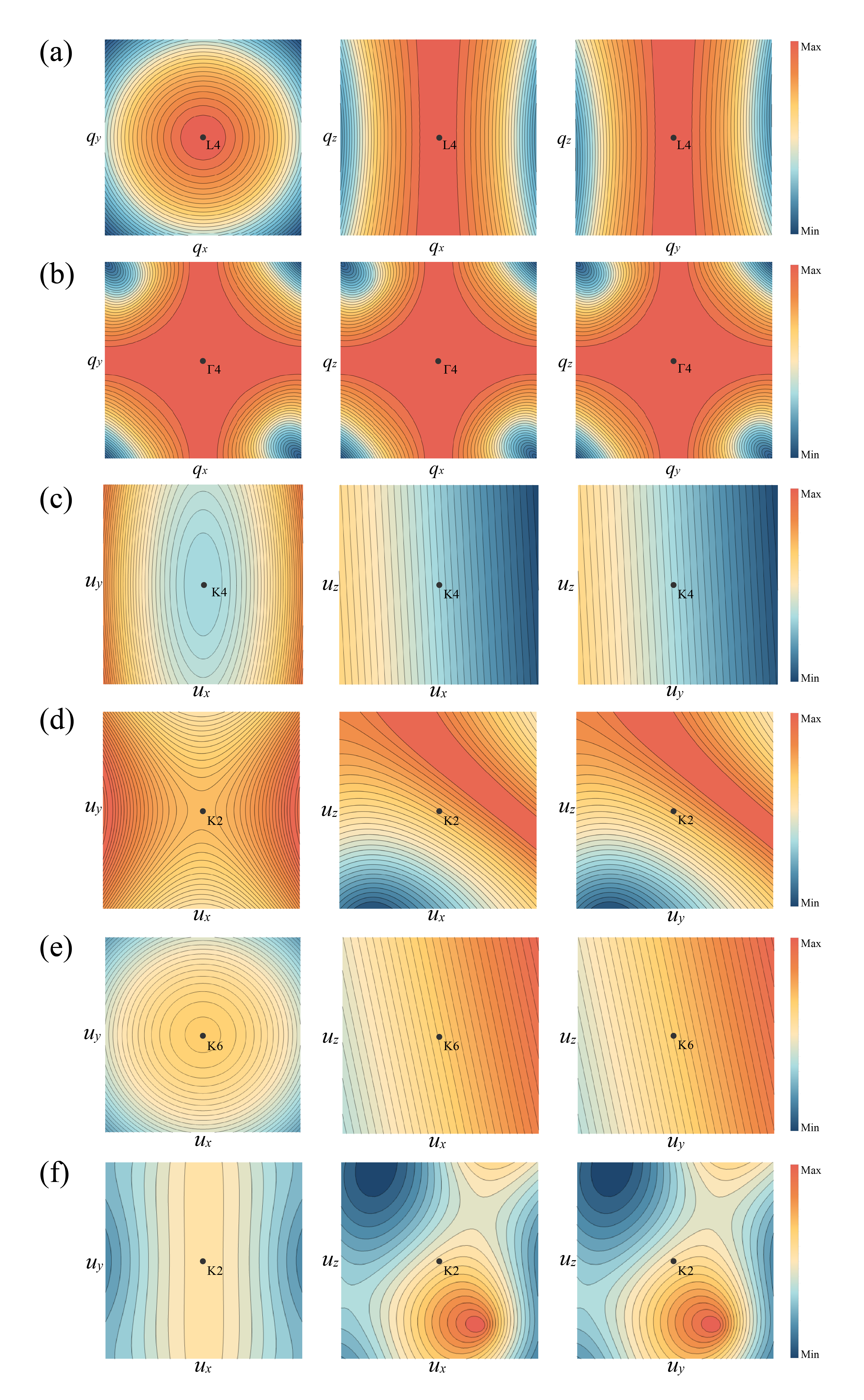}
\caption{(a)-(f) Momentum-space energy contours from the $s$-orbital tight-binding model, including spin-orbit coupling. Model parameters are fixed as $e_{1} = 0$, $t_{1} = -1$, with $t_{2} = 0, -0.5, -1$ for different panels:
 - Panels (a-b) ($t_{2} = 0$): (a) Contours for the $L_4$ point ($T_1$ type); (b) for the $\Gamma_4$ point ($T_3$ type);
- Panel (c) ($t_{2} = -0.5$): Contours for the $K_4$ point ($N_0$ type);
- Panel (d) ($t_{2} = -1$): Contours for the $K_2$ point ($N_1$ type);
- Panel (e) ($t_2 = 0$): Contours for the $K_6$ point ($N_2$ type);
- Panel (f) ($t_2 = -1$): Contours for the $K_2$ point ($S_1$ type).
}
\label{figs5}
\end{figure}

\subsection{$T_1$-type HOVHS}
\label{sec:iii.1}
For $T_1$-type HOVHS, exactly one principal direction exhibits a higher-order polynomial dependence ($n_z \ge 3$), while the remaining transverse plane retains quadratic dispersions. According to our universal principles, we evaluate the DOS for the fundamental topological categories across different parity regimes, revealing that this topological class is uniquely characterized by a continuous density of states but a strongly divergent energy derivative (see Table.~\ref{table_T1_DOS}).

\subsubsection{TRIM Regime (Even $n_z \ge 4$)}

\noindent \textbf{Category A: Elliptic Transverse Dispersions $(\sigma_x \sigma_y = +1)$}

When the transverse dispersion is elliptic, the sign of the higher-order longitudinal term ($\sigma_z$) fundamentally alters the 3D topological geometry, distinguishing between a true 3D extremum and a 3D saddle point.

\vspace{6pt}
\noindent \textit{(i) 3D Extremum Dispersions $(\sigma_x=\sigma_y=\sigma_z=+1)$}

For a generalized 3D minimum, the canonical dispersion is $\varepsilon_{A1}(\mathbf{k}') = \varepsilon_0 + (k_x')^2 + (k_y')^2 + (k_z')^{n_z}$. Introducing transverse polar coordinates ($q^2 = (k_x')^2 + (k_y')^2$) and substituting $u = (k_z')^{n_z}$ (with $dk_z' = \frac{1}{n_z} u^{\frac{1}{n_z}-1} du$), the Dirac delta function isolates the root $u = E - \varepsilon_0 - q^2$. The physical requirement of a real momentum ($u \ge 0$) intrinsically restricts the transverse momentum to $q^2 \le E - \varepsilon_0$. Changing the variable to $t = q^2$, the DOS integral is dynamically bounded by $\min(R^2, E - \varepsilon_0)$:
\begin{equation}
g_{A1}^{even}(E) = \frac{1}{4\pi^2 n_z \sqrt{c_x c_y} c_z^{1/n_z}} \int_{0}^{\min(R^2, E - \varepsilon_0)} (E - \varepsilon_0 - t)^{\frac{1}{n_z}-1} dt 
\end{equation}
Evaluating this definite integral yields the exact piecewise DOS for the higher-order extremum:
\begin{equation}
g_{A1}^{even}(E) = \frac{1}{4\pi^2 \sqrt{c_x c_y} c_z^{1/n_z}} 
\begin{cases} 
0 & (E \le \varepsilon_0)\\
( E - \varepsilon_0 )^{1/n_z} & (\varepsilon_0 < E \le \varepsilon_0+ R^2) \\
( E - \varepsilon_0 )^{1/n_z} - (E - \varepsilon_0 - R^2)^{1/n_z} & (E > \varepsilon_0 + R^2)
\end{cases}
\end{equation}
This confirms that at a true higher-order band edge, the DOS inherently scales as a pure fractional power law $\propto (E-\varepsilon_0)^{1/n_z}$ near $\varepsilon_0$.

\vspace{6pt}
\noindent \textit{(ii) 3D Saddle Dispersions $(\sigma_x=\sigma_y=+1, \sigma_z=-1)$}

For the hybrid saddle configuration $\varepsilon_{A2}(\mathbf{k}') = \varepsilon_0 + (k_x')^2 + (k_y')^2 - (k_z')^{n_z}$, applying the identical coordinate transformation shifts the physical root to $u = \varepsilon_0 + q^2 - E \ge 0$. This condition alters the lower integration bound to $\max(0, E - \varepsilon_0)$, yielding the continuous piecewise result characteristic of saddle points:
\begin{equation}
g_{A2}^{even}(E) = \frac{1}{4\pi^2 \sqrt{c_x c_y} c_z^{1/n_z}} 
\begin{cases} 
( \varepsilon_0 + R^2 - E )^{1/n_z} - (\varepsilon_0 -  E)^{1/n_z}  & (E \le \varepsilon_0)\\
( \varepsilon_0 + R^2 - E )^{1/n_z} & (\varepsilon_0 < E \le \varepsilon_0+ R^2)
\end{cases}
\end{equation}

\vspace{8pt}
\noindent \textbf{Category B: Hyperbolic Transverse Dispersions ($\sigma_x \sigma_y = -1$)}

Because the transverse plane intrinsically forms an open saddle geometry, a sign flip in the longitudinal axis merely corresponds to a global particle-hole inversion coupled with a $90^\circ$ spatial rotation. Thus, we uniquely evaluate the canonical representation $\varepsilon_B(\mathbf{k}') = \varepsilon_0 + (k_x')^2 - (k_y')^2 - (k_z')^{n_z}$.

Exploiting mirror symmetries, we group the identically signed terms and introduce weighted generalized polar coordinates: $k_y' = q \cos \theta$, $k_z'=  q^{2/n_z} \sin^{2/n_z} \theta$, with the Jacobian determinant $d k_y' d k_z' = \frac{2}{n_z} q^{2/n_z} \sin^{\frac{2}{n_z}-1} \theta \, d\theta \, dq$. The angular integral cleanly resolves into the Beta function $\frac{\sqrt{\pi} \Gamma(1/n_z)}{2\Gamma(1/n_z + 1/2)}$. Integrating over $k_x'$ via the delta function property yields an absolute derivative factor bounded by $\Theta(E - \varepsilon_0 + q^2)$. The global DOS is thereby reduced to a 1D radial representation:
\begin{equation}
g_B^{even}(E) = \frac{\sqrt{\pi} \Gamma\left(\frac{1}{n_z}\right)}{2\pi^3 n_z \sqrt{c_x c_y} c_z^{1/n_z} \Gamma\left(\frac{1}{n_z} + \frac{1}{2}\right)}
\begin{cases} 
\displaystyle \int_{\sqrt{ \varepsilon_0-E}}^{R} \frac{q^{2/n_z}}{ \sqrt{E - \varepsilon_0 + q^2 }} \, dq & (\varepsilon_0 - R^2 \le E \le \varepsilon_0 )\\
\displaystyle \int_0^R \frac{q^{2/n_z}}{\sqrt{ E - \varepsilon_0 + q^2 }} \, dq & (E > \varepsilon_0) 
\end{cases}
\end{equation}

\subsubsection{Non-TRIM Regime (Odd $n_z \ge 3$)}

Unlike the TRIM regime, an odd exponent $n_z$ renders the longitudinal dispersion term monotonically unbounded over $(-\infty, \infty)$. Consequently, a sign permutation ($\sigma_z \to -\sigma_z$) is mathematically equivalent to a mere coordinate inversion ($k_z \to -k_z$). Since the integration spans the entire symmetric momentum axis, the sign $\sigma_z$ trivially decouples from the total DOS evaluation. We evaluate the canonical positive branch ($+c_z k_z^{n_z}$) for both transverse categories.

\vspace{6pt}
\noindent \textbf{Category A: Elliptic Transverse Dispersions ($\sigma_x \sigma_y = +1$)}

For $\varepsilon_A = \varepsilon_0 + c_x k_x^2 + c_y k_y^2 + c_z k_z^{n_z}$, integrating over the unbounded $k_z'$ variable directly yields the inverse derivative factor $\frac{1}{n_z} |E - \varepsilon_0 - q^2|^{\frac{1}{n_z}-1}$. Defining the relative energy $E' = E - \varepsilon_0$ and transforming the remaining transverse variables to polar coordinates ($t = q^2$), the total DOS rigorously integrates to:
\begin{equation}
g_A^{odd}(E) = \frac{1}{4\pi^2 \sqrt{c_x c_y} c_z^{1/n_z}} 
\begin{cases} 
(R^2 + |E'|)^{1/n_z} - |E'|^{1/n_z} & (E \le \varepsilon_0) \\
(R^2 - E')^{1/n_z} + E'^{1/n_z} & (E > \varepsilon_0) 
\end{cases}
\end{equation}

\vspace{8pt}
\noindent \textbf{Category B: Hyperbolic Transverse Dispersions ($\sigma_x \sigma_y = -1$)}

For the hyperbolic saddle $\varepsilon_B(\mathbf{k}') = \varepsilon_0 + (k_x')^2 - (k_y')^2 + (k_z')^{n_z}$, integrating over the odd $k_z'$ direction yields $\frac{1}{n_z} |E' - (k_x')^2 + (k_y')^2|^{\frac{1}{n_z}-1}$. To extract the low-energy scaling, we introduce standard transverse polar coordinates and perform the dimensionless variable substitution $t = |E'|x$. This factors the energy scaling analytically out of the integral kernel, allowing the DOS to be decomposed into a finite constant background $g_B^{back}(R)$ and a relative fractional correction:
\begin{equation}
g_B^{odd}(E) = g_B^{back}(R) + \frac{|E - \varepsilon_0|^{1/n_z}}{16\pi^3 n_z \sqrt{c_x c_y} c_z^{1/n_z}} 
\begin{cases} 
\mathcal{K}_{B-}^{odd} & (E \le \varepsilon_0) \\
\mathcal{K}_{B+}^{odd} & (E > \varepsilon_0) 
\end{cases} 
\end{equation}
where the momentum-cutoff-dependent background and the convergent dimensionless geometric constants $\mathcal{K}_{B\pm}^{odd}$ are explicitly defined as:
\begin{equation}
\begin{aligned}
g_B^{back}(R) &= \frac{R^{2/n_z}}{16\pi^3 \sqrt{c_x c_y} c_z^{1/n_z}} \int_0^{2\pi} |\cos(2\theta)|^{\frac{1}{n_z}-1} d\theta \\
\mathcal{K}_{B-}^{odd} &= \int_0^{2\pi} d\theta \int_0^\infty \left( \left| -1 - x \cos(2\theta) \right|^{\frac{1}{n_z}-1} - \left| x \cos(2\theta) \right|^{\frac{1}{n_z}-1} \right) dx \\
\mathcal{K}_{B+}^{odd} &= \int_0^{2\pi} d\theta \int_0^\infty \left( \left| 1 - x \cos(2\theta) \right|^{\frac{1}{n_z}-1} - \left| x \cos(2\theta) \right|^{\frac{1}{n_z}-1} \right) dx
\end{aligned}
\end{equation}

\subsubsection{Universal Hallmarks of $T_1$-type HOVHS: The Derivative Singularity}

Synthesizing the derivations across all parity regimes, the macroscopic physical signature of the $T_1$-type HOVHS is universally dictated by the fractional scaling exponent $\alpha = 1/n_z$. Unlike conventional lower-dimensional singularities or strongly correlated $T_2/T_3$ flat bands, the DOS $g(E)$ in the $T_1$ class \textit{does not diverge} to infinity at $E = \varepsilon_0$. Instead, it converges to zero or maintains a finite, continuous background value.

The true topological signature of the $T_1$ regime is uniquely embedded in its derivative. While the DOS itself remains finite, its first derivative with respect to energy exhibits a profound fractional divergence:
\begin{equation}
\frac{dg(E)}{dE} \propto |E - \varepsilon_0|^{\frac{1}{n_z} - 1} \xrightarrow{E \to \varepsilon_0} \infty \quad (n_z \ge 3)
\end{equation}
Since the exponent $(1/n_z - 1)$ is strictly negative, it mathematically dictates an infinite slope exactly at the critical energy point. The underlying spatial parity strictly governs how this macroscopic derivative anomaly visually manifests in the energy spectrum:
\begin{itemize}
    \item \textbf{In the TRIM Regime (Even $n_z$):} The spatial inversion symmetry necessitates Heaviside step-function truncations, restricting the accessible states to specific energy half-spaces. This hard cutoff intrinsically translates the derivative singularity into a vertical launch from zero (for 3D extremum topologies) or a finite-height sharp kink (for 3D saddle topologies).
    
    \item \textbf{In the Non-TRIM Regime (Odd $n_z$):} The relaxation of spatial inversion symmetry allows the DOS to organically span across $\varepsilon_0$. Consequently, the derivative singularity seamlessly bridges the underlying topological geometries without piecewise bounds, manifesting as a continuous curve that traverses the critical point with an infinite slope (yielding an S-shaped step for generalized 3D extrema, or a sharp fractional cusp for generalized 3D saddles).
\end{itemize}

This establishes that within the $T_1$ topological hierarchy, regardless of parity breakings or transverse geometric differences, the fundamental macroscopic hallmark is invariably a topologically protected derivative divergence.

\subsection{$T_2$-type HOVHS}
\label{sec:iii.2}
For a $T_2$-type HOVHS, exactly two principal directions exhibit higher-order polynomial scaling ($n_y, n_z \ge 3$), leaving one quadratic direction. Following our universal principles, the generalized dispersion is formulated as:
\begin{equation}
\varepsilon(\mathbf{k}) = \varepsilon_0 + c_x k_x^2 + \sigma_y c_y k_y^{n_y} + \sigma_z c_z k_z^{n_z}
\end{equation}
where the scaling coefficients $c_x, c_y, c_z > 0$, and the sign parameters $\sigma_y, \sigma_z \in \{+1, -1\}$. Without loss of generality, we fix the quadratic term to be positive ($c_x > 0$), as a negative quadratic term merely corresponds to a global particle-hole inversion ($E - \varepsilon_0 \to -(E - \varepsilon_0)$). The permutations of the higher-order signs ($\sigma_y, \sigma_z = \pm 1$) demarcate the system into Category A (Elliptic, $\sigma_y \sigma_z = +1$) and Category B (Hyperbolic, $\sigma_y \sigma_z = -1$). 

Applying the universal scaling transformation $k_x' = \sqrt{c_x} k_x, k_y' = c_y^{1/n_y} k_y, k_z' = c_z^{1/n_z} k_z$, we obtain the canonical form:
\begin{equation}
\varepsilon(\mathbf{k}') = \varepsilon_0 + (k_x')^2 + \sigma_y(k_y')^{n_y} + \sigma_z(k_z')^{n_z}
\end{equation}

\subsubsection{TRIM Regime (Even $n_y$, Even $n_z$)}

\noindent \textbf{Category A: Elliptic Transverse Dispersions $(\sigma_y \sigma_z = +1)$}

When the higher-order terms share the same sign, their relative sign against the quadratic term $k_x^2$ distinguishes a 3D extremum from a 3D saddle. 

\vspace{6pt}
\noindent \textit{(i) 3D Extremum Dispersions $(\sigma_y = \sigma_z = +1)$}

For the canonical extremum $\varepsilon_{A1}(\mathbf{k}') = \varepsilon_0 + (k_x')^2 + (k_y')^{n_y} + (k_z')^{n_z}$, the strict even-parity required at TRIMs folds the integration domain into the positive octant ($k_x', k_y', k_z' \ge 0$) with a symmetry factor of $2^3 = 8$. Defining $E' = E - \varepsilon_0$, we integrate over $k_x'$ first, yielding a derivative factor restricted by the Heaviside function $\Theta(E' - (k_y')^{n_y} - (k_z')^{n_z})$. Introducing fractional substitutions $u = (k_y')^{n_y}$ and $v = (k_z')^{n_z}$, the remaining 2D integral maps onto a classic Dirichlet integral over the bounded domain $u+v \le E'$. Utilizing the multidimensional Beta function, the DOS resolves exactly to:
\begin{equation}
g_{A1}^{even}(E) = 
\begin{cases} 
0 & (E \le \varepsilon_0) \\
\frac{\sqrt{\pi} \Gamma\left(\frac{1}{n_y}\right) \Gamma\left(\frac{1}{n_z}\right)}{2\pi^3 n_y n_z \sqrt{c_x} c_y^{1/n_y} c_z^{1/n_z} \Gamma\left(\frac{1}{n_y} + \frac{1}{n_z} + \frac{1}{2}\right)} (E - \varepsilon_0)^{\frac{1}{n_y} + \frac{1}{n_z} - \frac{1}{2}} & (E > \varepsilon_0) 
\end{cases} 
\end{equation}

\vspace{6pt}
\noindent \textit{(ii) 3D Saddle Dispersions ($\sigma_y = \sigma_z = -1$)}

For the hybrid 3D saddle $\varepsilon_{A2}(\mathbf{k}') = \varepsilon_0 + (k_x')^2 - (k_y')^{n_y} - (k_z')^{n_z}$, we first integrate over the higher-order transverse plane ($k_y', k_z'$). Defining the effective energy parameter $\Lambda = (k_x')^2 - E'$, the condition for real momentum states strictly requires $(k_y')^{n_y} + (k_z')^{n_z} = \Lambda \ge 0$. Integrating the Dirac delta function over the positive quadrant via Dirichlet substitutions identically yields the 2D DOS for this subspace:
\begin{equation}
I_{yz}(\Lambda) = \frac{\Gamma\left(\frac{1}{n_y}\right)\Gamma\left(\frac{1}{n_z}\right)}{n_y n_z \Gamma\left(\frac{1}{n_y} + \frac{1}{n_z}\right)} \Lambda^{\frac{1}{n_y} + \frac{1}{n_z} - 1} \Theta(\Lambda) \equiv \mathcal{A}_{yz} \Lambda^{\frac{1}{n_y} + \frac{1}{n_z} - 1} \Theta(\Lambda)
\end{equation}

Substituting this exact 2D evaluation back into the remaining 1D integral over $k_x'$, the Heaviside step function dynamically dictates the integration domain. To analytically extract the energy dependence, we perform the exact dimensionless substitution $k_x' = \sqrt{|E'|} x$ (with $dk_x' = \sqrt{|E'|} dx$). Factoring out the energy terms, the DOS splits into two exact branches:
\begin{equation}
g_{A2}^{even}(E) = \frac{\mathcal{A}_{yz}}{\pi^3 \sqrt{c_x} c_y^{1/n_y} c_z^{1/n_z}} |E - \varepsilon_0|^{\frac{1}{n_y} + \frac{1}{n_z} - \frac{1}{2}} 
\begin{cases} 
\mathcal{I}_{A2}^- & (E \le \varepsilon_0) \\
\mathcal{I}_{A2}^+ & (E > \varepsilon_0)
\end{cases} 
\end{equation}
where the dimensionless geometric integrals defining the negative and positive energy branches are given by:
\begin{equation}
\mathcal{I}_{A2}^- = \int_0^{R/\sqrt{|E'|}} (x^2 + 1)^{\frac{1}{n_y} + \frac{1}{n_z} - 1} dx, \quad
\mathcal{I}_{A2}^+ = \int_1^{R/\sqrt{E'}} (x^2 - 1)^{\frac{1}{n_y} + \frac{1}{n_z} - 1} dx 
\end{equation}

For $\frac{1}{n_y} + \frac{1}{n_z} < \frac{1}{2}$, these dimensionless integrals converge to finite mathematical constants as $|E'| \to 0$, with the upper integration limits extending to infinity. The DOS conclusively collapses to the pure fractional power law.

\vspace{6pt}
\noindent \textbf{Logarithmic Divergence at the Topological Threshold ($\frac{1}{n_y} + \frac{1}{n_z} = \frac{1}{2}$)}

Crucially, when $n_y = n_z = 4$, the scaling exponent sum hits the strict boundary $\frac{1}{4} + \frac{1}{4} = \frac{1}{2}$. Consequently, the energy prefactor strictly vanishes ($|E'|^{\frac{1}{2} - \frac{1}{2}} = |E'|^0 = 1$), indicating a profound topological phase transition. The dimensionless integrals are now explicitly evaluated via standard inverse hyperbolic functions: $\int (x^2 \pm 1)^{-1/2} dx = \text{arsinh}(x)$ and $\text{arcosh}(x)$, respectively. 

Evaluating the upper limits as $|E'| \to 0$ (where $x \gg 1$), we utilize the asymptotic expansions $\text{arsinh}(x), \text{arcosh}(x) \approx \ln(2x)$. The integrals uniquely evaluate to $\ln(2R) - \frac{1}{2}\ln|E - \varepsilon_0|$ for both energy branches. This exact analytical resolution mathematically proves that at the specific topological threshold $\frac{1}{n_y} + \frac{1}{n_z} = \frac{1}{2}$, the fractional power law structurally transforms into a pure \textbf{logarithmic Van Hove singularity}. Substituting $n_y = n_z = 4$, the final DOS rigorously evaluates to:
\begin{equation}
g_{A2}^{even}\Big|_{n_y=n_z=4}(E) \approx \frac{\mathcal{A}_{yz}}{\pi^3 \sqrt{c_x} (c_y c_z)^{1/4}} \left[ \ln(2R) - \frac{1}{2}\ln|E - \varepsilon_0| \right]
\end{equation}

\vspace{8pt}
\noindent \textbf{Category B: Hyperbolic Transverse Dispersions ($\sigma_y \sigma_z = -1$)}

For the saddle dispersion $\varepsilon_B(\mathbf{k}') = \varepsilon_0 + (k_x')^2 + (k_y')^{n_y} - (k_z')^{n_z}$, integrating over the positive octant for $k_z'$ yields the root factor. To unify the exponents, we construct weighted generalized polar coordinates $k_x' = q^{n_y/2} \cos\theta, k_y' = q \sin^{2/n_y}\theta$, with the Jacobian $dk_x' dk_y' = q^{n_y/2} \sin^{\frac{2}{n_y}-1}\theta \, dq d\theta$. 

The angular integration elegantly evaluates to $\frac{\sqrt{\pi}\Gamma(1/n_y)}{2\Gamma(1/n_y + 1/2)}$. To analytically extract the singular energy dependence, we perform a dimensionless substitution $q = |E'|^{1/n_y} x$ in the low-energy limit ($|E'| \to 0$). The radial integral perfectly factors out the fractional power. By absorbing the resulting convergent dimensionless $x$-integrals into a constant $\mathcal{C}_B^{even}$, the DOS formally collapses to:
\begin{equation}
g_B^{even}(E) = \mathcal{C}_B^{even} |E - \varepsilon_0|^{\frac{1}{n_y} + \frac{1}{n_z} - \frac{1}{2}}
\end{equation}

\subsubsection{Mixed-Parity Regime (Even $n_y$, Odd $n_z$)}

In polar or uniaxial systems, spatial inversion is explicitly broken along the $z$-axis ($n_z$ odd, $n_y$ even). By definition of the $T_2$ class, the $x$-direction strictly retains its quadratic dispersion ($n_x=2$). The odd parity of $n_z$ renders the longitudinal term $\sigma_z(k_z')^{n_z}$ a strictly monotonic bijection over $(-\infty, \infty)$, allowing a sign permutation ($\sigma_z \to -\sigma_z$) to decouple via a trivial coordinate inversion ($k_z' \to -k_z'$). 

Thus, we fix the positive branch $+ (k_z')^{n_z}$ without loss of generality, and evaluate the canonical mixed-parity dispersion $\varepsilon_{mixed}(\mathbf{k}') = \varepsilon_0 + (k_x')^2 + \sigma_y(k_y')^{n_y} + (k_z')^{n_z}$. Integrating over the unbounded $k_z'$ variable first yields a continuous absolute value derivative factor $\frac{1}{n_z} \left| E' - (k_x')^2 - \sigma_y(k_y')^{n_y} \right|^{\frac{1}{n_z}-1}$.

Because $n_x=2$ and $n_y$ is even, the transverse $x\text{-}y$ plane strictly preserves its mirror symmetry regardless of $\sigma_y$. Folding the 2D integration into the first quadrant ($\times 4$ multiplicity) and applying the same weighted generalized polar coordinates, the dimensionless substitution $q = \sqrt{|E'|}x$ analytically extracts the energy dependence $|E - \varepsilon_0|^{\frac{1}{n_y} + \frac{1}{n_z} - \frac{1}{2}}$.

For \textbf{Elliptic Transverse Dispersions ($\sigma_y = +1$)}, the angular dependence algebraically vanishes ($\cos^2\theta + \sin^2\theta = 1$), isolating the Beta function $\frac{1}{2}B(1/n_y, 1/2)$. The DOS resolves exactly to:
\begin{equation}
g_A^{mixed}(E) = \frac{\sqrt{\pi} \Gamma\left(\frac{1}{n_y}\right)}{2\pi^3 n_y n_z \sqrt{c_x} c_y^{1/n_y} c_z^{1/n_z} \Gamma\left(\frac{1}{n_y} + \frac{1}{2}\right)}|E - \varepsilon_0|^{\frac{1}{n_y} + \frac{1}{n_z} - \frac{1}{2}}
\begin{cases} 
\mathcal{I}_{A}^- & (E \le \varepsilon_0) \\
\mathcal{I}_{A}^+ & (E > \varepsilon_0) 
\end{cases} 
\end{equation}
where $\mathcal{I}_{A}^- = \int_0^\infty x^{2/n_y} \left( 1 + x^2 \right)^{\frac{1}{n_z}-1} dx$ and $\mathcal{I}_{A}^+ = \int_0^\infty x^{2/n_y} \left| 1 - x^2 \right|^{\frac{1}{n_z}-1} dx$.

Conversely, for \textbf{Hyperbolic Transverse Dispersions ($\sigma_y = -1$)}, the angular term intrinsically evaluates to $\cos(2\theta)$, which intricately couples the angular and radial integrations:
\begin{equation}
g_B^{mixed}(E) = \frac{8 |E - \varepsilon_0|^{\frac{1}{n_y} + \frac{1}{n_z} - \frac{1}{2}}}{(2\pi)^3 n_y n_z \sqrt{c_x} c_y^{1/n_y} c_z^{1/n_z}} 
\begin{cases} 
\mathcal{I}_{B-}^{mixed} & (E \le \varepsilon_0) \\
\mathcal{I}_{B+}^{mixed} & (E > \varepsilon_0) 
\end{cases} 
\end{equation}
where the dimensionless geometric constants are explicitly defined by the convergent double integrals:
\begin{equation}
\begin{aligned}
\mathcal{I}_{B-}^{mixed} &= \int_0^{\pi/2} \sin^{\frac{2}{n_y}-1}\theta \left[ \int_0^\infty x^{2/n_y} \left| 1 + x^2 \cos(2\theta) \right|^{\frac{1}{n_z}-1} dx \right] d\theta \\
\mathcal{I}_{B+}^{mixed} &= \int_0^{\pi/2} \sin^{\frac{2}{n_y}-1}\theta \left[ \int_0^\infty x^{2/n_y} \left| 1 - x^2 \cos(2\theta) \right|^{\frac{1}{n_z}-1} dx \right] d\theta
\end{aligned}
\end{equation}

By defining these purely dimensionless integral arrays as independent macroscopic geometric constants $\mathcal{C}_{A}^{mixed}$ and $\mathcal{C}_{B}^{mixed}$ respectively, both configurations conclusively yield the exact universal scaling $g_{A/B}^{mixed}(E) = \mathcal{C}_{A/B}^{mixed} |E - \varepsilon_0|^{\frac{1}{n_y} + \frac{1}{n_z} - \frac{1}{2}}$.

\subsubsection{Fully Non-TRIM Regime (Odd $n_y$, Odd $n_z$)}

When both higher-order exponents are odd integers, the polynomial terms monotonically span $(-\infty, \infty)$. Any sign permutation of these odd terms is mathematically absorbed by a coordinate inversion over the infinite momentum domain. Consequently, the topological distinction between elliptic and hyperbolic geometries collapses, and we evaluate the canonical odd-odd dispersion $\varepsilon_{odd}(\mathbf{k}') = \varepsilon_0 + (k_x')^2 + (k_y')^{n_y} + (k_z')^{n_z}$.

Integrating over the unbounded $k_z'$ yields the absolute value factor. Splitting the odd-powered $k_y'$ integration into positive and negative halves, the total integral bifurcates. Applying the dimensionless variable substitutions $k_x' = |E'|^{1/2} x$ and $k_y' = |E'|^{1/n_y} y$, the energy variable $|E'|^{\frac{1}{n_y} + \frac{1}{n_z} - \frac{1}{2}}$ factors out from the integral core:
\begin{equation}
g^{odd}(E) = \frac{|E - \varepsilon_0|^{\frac{1}{n_y} + \frac{1}{n_z} - \frac{1}{2}}}{(2\pi)^3 n_z \sqrt{c_x} c_y^{1/n_y} c_z^{1/n_z}} 
\begin{cases} 
\mathcal{K}_-^{odd} & (E \le \varepsilon_0) \\
\mathcal{K}_+^{odd} & (E > \varepsilon_0) 
\end{cases} 
\end{equation}
where the dimensionless geometric constants for the negative and positive energy branches are:
\begin{equation}
\begin{aligned}
\mathcal{K}_-^{odd} &= 2 \iint_{x,y \ge 0} \left( \left| -1 - x^2 - y^{n_y} \right|^{\frac{1}{n_z}-1} + \left| -1 - x^2 + y^{n_y} \right|^{\frac{1}{n_z}-1} \right) dx \, dy \\
\mathcal{K}_+^{odd} &= 2 \iint_{x,y \ge 0} \left( \left| 1 - x^2 - y^{n_y} \right|^{\frac{1}{n_z}-1} + \left| 1 - x^2 + y^{n_y} \right|^{\frac{1}{n_z}-1} \right) dx \, dy 
\end{aligned}
\end{equation}

By utilizing the absolute value property $|-A| = |A|$ and performing a simple integration variable shift for the second term in $\mathcal{K}_-^{odd}$, it rigorously follows that $\mathcal{K}_-^{odd} \equiv \mathcal{K}_+^{odd}$. This mathematical invariance indicates that the phase-space volume is symmetric under energy-sign reversal in the fully non-TRIM regime, proving the topological superposition $g^{odd}(E) = \frac{1}{2} [ g_A^{mixed}(E) + g_B^{mixed}(E) ] = \mathcal{C}_{odd} |E - \varepsilon_0|^{\frac{1}{n_y} + \frac{1}{n_z} - \frac{1}{2}}$.

\vspace{6pt}
\noindent \textbf{Topological Prohibition of Logarithmic Singularities}

A profound mathematical consequence inherently emerges from this universal scaling exponent $\alpha = \frac{1}{n_y} + \frac{1}{n_z} - \frac{1}{2}$. The critical threshold for a logarithmic Van Hove singularity strictly requires the exponent to vanish ($\alpha = 0$).

This constraint yields the Diophantine equation $\frac{1}{n_y} + \frac{1}{n_z} = \frac{1}{2}$, which can be algebraically factorized as $(n_y - 2)(n_z - 2) = 4$. For integer exponents $n_y, n_z \ge 3$, the only mathematically permissible solution sets are $(4,4)$, $(3,6)$, and $(6,3)$. Crucially, \textit{none of these solutions consist of exclusively odd integers}.

This absolute algebraic constraint rigorously proves that a pure logarithmic DOS divergence is topologically forbidden in the fully non-TRIM (odd-odd) regime of a $T_2$-type HOVHS. The system is structurally forced to exhibit either a fractional power-law divergence (e.g., $n_y=n_z=5$, yielding $\alpha = -0.1 < 0$) or a continuous V-shaped convergence to zero (e.g., $n_y=n_z=3$, yielding $\alpha = 1/6 > 0$). The emergence of a perfect logarithmic singularity is thus a symmetry-protected hallmark, exclusively reserved for systems preserving at least one even-parity direction (such as the isotropic $n_y=n_z=4$ saddle).

\subsubsection{Universal Hallmarks of $T_2$-type HOVHS: The Topological Crossover}

Synthesizing the exact analytical derivations across all parity regimes, the $T_2$-type HOVHS uniquely acts as the overarching topological crossover regime between non-divergent boundary states and strongly correlated divergent flat bands. Governed by the universal scaling exponent $\alpha = \frac{1}{n_y} + \frac{1}{n_z} - \frac{1}{2}$, the macroscopic DOS accommodates three fundamentally distinct singular phases:
\begin{itemize}
    \item \textbf{The Convergent Phase ($\alpha > 0$):} When the higher-order scaling is relatively weak (e.g., $n_y = 4, n_z = 3$), the phase-space volume expansion overwhelmingly suppresses the DOS divergence. Although the DOS converges to a finite background value at the critical point, the \textit{morphology} of this convergence is topologically protected. Generalized saddle configurations enforce an angular phase-space cancellation, carving out a deep, asymmetric V-shaped valley; whereas generalized extremum configurations accumulate highly skewed phase-space volumes, manifesting as an asymmetric step-like cliff anchored by Heaviside bounds.
    
    \item \textbf{The Dimensional Reduction Phase ($\alpha = 0$):} Uniquely permitted when the exponent combination rigorously satisfies the Diophantine boundary condition $\frac{1}{n_y} + \frac{1}{n_z} = \frac{1}{2}$. The macroscopic 3D integral perfectly cancels the kinetic energy offset, driving a profound dimensional collapse. The 3D bulk states effectively degenerate into a 2D electronic gas system, precisely mimicking canonical 2D topological signatures: exhibiting a Heaviside step-function plateau for generalized extremum geometries, and a pure logarithmic Van Hove singularity for generalized saddle geometries.
    
    \item \textbf{The Power-Law Divergence Phase ($\alpha < 0$):} As the degree of band flattening intensifies (e.g., $n_y, n_z \ge 5$), the accumulation of states overcomes the 3D phase-space volume expansion. The density of states invariably diverges to infinity at the critical point, universally manifesting as a sharp, dominant fractional cusp $\delta g \propto |E-\varepsilon_0|^{-|\alpha|}$ regardless of the underlying geometric classifications. 
\end{itemize}
Ultimately, the progression across these three fundamental divergent phases is invariably determined by the universal algebraic threshold of the integer exponents.

\subsection{$T_3$-type HOVHS}
\label{sec:iii.3}
For a $T_3$-type HOVHS, the Hessian matrix becomes a null matrix with three zero eigenvalues, indicating that quadratic dispersions vanish entirely along all principal directions. Generalizing to anisotropic exponents, the dispersion is formulated as:
\begin{equation}
\varepsilon(\mathbf{k}) = \varepsilon_0 + \sum_{i=x,y,z} \sigma_i c_i k_i^{n_i} \quad (n_i \ge 3, c_i > 0, \sigma_i = \pm 1)
\end{equation}
The parity combinations of $n_i$ generate four topological equivalence classes (3 Even, 2 Even + 1 Odd, 1 Even + 2 Odd, and 3 Odd). We first derive the universal scaling law before evaluating specific regimes.

\subsubsection{Universal Scaling Law and Critical Exponent}
The fundamental energy dependence of the DOS can be analytically extracted via scaling invariance. We define a generalized topological exponent index $\nu = \sum 1/n_i$. By applying a uniform energy scaling factor $\lambda > 0$ and transforming momentum coordinates as $q_i = \lambda^{-1/n_i} k_i$, the dispersion satisfies the relation $\varepsilon(\lambda^{1/n_i} q_i) - \varepsilon_0 = \lambda [\varepsilon(\mathbf{q}) - \varepsilon_0]$. Applying this to the DOS integral ($d^3k = \lambda^\nu d^3q$) yields:
\begin{equation}
g(\varepsilon_0 + \lambda \Delta E) = \lambda^{\nu - 1} \frac{1}{(2\pi)^3} \int \delta\left(\Delta E -[\varepsilon(\mathbf{q}) - \varepsilon_0]\right) d^3q = \lambda^{\nu - 1} g(\varepsilon_0 + \Delta E)
\end{equation}
This mandates that the DOS universally obeys a fractional power-law near the critical point:
\begin{equation}
g(E) \propto |E - \varepsilon_0|^{\nu - 1}
\label{eq:T3_Scaling_Law}
\end{equation}
The specific prefactors are extracted below using the canonical form $\varepsilon(\mathbf{k}') = \varepsilon_0 + \sum \sigma_i (k_i')^{n_i}$ via the mapping $k_i' = c_i^{1/n_i} k_i$.

\subsubsection{TRIM Regime (Even $n_x, n_y, n_z$)}
At TRIMs, exponents are restricted to even integers. Sign permutations collapse into 3D Extrema ($+++$) and 3D Saddles ($++-$).

\vspace{6pt}
\noindent \textbf{Category A: 3D Extremum Dispersions ($\sigma_x = \sigma_y = \sigma_z = +1$)}:
The strict even-parity restricts the integration to the positive octant. Applying Dirichlet fractional substitutions $u_i = (k_i')^{n_i}$, the integral exactly resolves via the multidimensional Beta function:
\begin{equation}
g_A^{even}(E) = 
\begin{cases} 
0 & (E \le \varepsilon_0) \\
\frac{\Gamma\left(\frac{1}{n_x}\right) \Gamma\left(\frac{1}{n_y}\right) \Gamma\left(\frac{1}{n_z}\right)}{\pi^3 n_x n_y n_z C_{xyz} \Gamma\left(\nu\right)} (E - \varepsilon_0)^{\nu - 1} & (E > \varepsilon_0) 
\end{cases} 
\end{equation}
where $C_{xyz} = (c_x^{1/n_x} c_y^{1/n_y} c_z^{1/n_z})^{-1}$.

\vspace{6pt}
\noindent \textbf{Category B: 3D Saddle Dispersions ($\sigma_z = -1$)}:
Integrating over the longitudinal $k_z'$ yields two symmetric roots with a factor $2/n_z$. Folding the transverse plane into the first quadrant and using weighted polar coordinates ($k_x' = q^{2/n_x} \cos^{2/n_x}\theta, k_y' = q^{2/n_y} \sin^{2/n_y}\theta$), the energy dependence is extracted via $q = \sqrt{|E'|}x$, yielding:
\begin{equation}
g_B^{even}(E) = \frac{2 \Gamma\left(\frac{1}{n_x}\right) \Gamma\left(\frac{1}{n_y}\right) |E - \varepsilon_0|^{\nu - 1}}{\pi^3 n_x n_y n_z C_{xyz} \Gamma\left(\frac{1}{n_x} + \frac{1}{n_y}\right)} 
\begin{cases} 
\mathcal{J}_{B-}^{even} & (E \le \varepsilon_0) \\
\mathcal{J}_{B+}^{even} & (E > \varepsilon_0) 
\end{cases} 
\end{equation}
where $\mathcal{J}_{B-}^{even} = \int_0^\infty x^{\frac{2}{n_x} + \frac{2}{n_y} - 1} (x^2 + 1)^{\frac{1}{n_z}-1} dx$ and $\mathcal{J}_{B+}^{even} = \int_1^\infty x^{\frac{2}{n_x} + \frac{2}{n_y} - 1} (x^2 - 1)^{\frac{1}{n_z}-1} dx$.

\subsubsection{Single-Odd Parity Regime (Even $n_x, n_y$, Odd $n_z$)}
The odd parity of $n_z$ removes the sign constraint $\sigma_z$ via coordinate inversion. Evaluating the canonical positive branch, the transverse mirror symmetry allows folding the 2D integration into the first quadrant. Using the same weighted polar coordinates and substituting $q = \sqrt{|E'|}x$, the DOS resolves into:
\begin{equation}
g_A^{1-odd}(E) = \frac{\Gamma\left(\frac{1}{n_x}\right) \Gamma\left(\frac{1}{n_y}\right) |E - \varepsilon_0|^{\nu - 1}}{\pi^3 n_x n_y n_z C_{xyz} \Gamma\left(\frac{1}{n_x} + \frac{1}{n_y}\right)}
\begin{cases} 
\mathcal{I}_{A-}^{1-odd} & (E \le \varepsilon_0) \\
\mathcal{I}_{A+}^{1-odd} & (E > \varepsilon_0) 
\end{cases} 
\end{equation}
where $\mathcal{I}_{A-}^{1-odd} = \int_0^\infty x^{\frac{2}{n_x} + \frac{2}{n_y} - 1} ( 1 + x^2 )^{\frac{1}{n_z}-1} dx$ and $\mathcal{I}_{A+}^{1-odd} = \int_0^\infty x^{\frac{2}{n_x} + \frac{2}{n_y} - 1} | 1 - x^2 |^{\frac{1}{n_z}-1} dx$. For hyperbolic topologies ($\sigma_y = -1$), the coupled integrations resolve into $g_B^{1-odd}(E) = \mathcal{C}_{A/B}^{1-odd} |E - \varepsilon_0|^{\nu - 1}$.

\subsubsection{Double-Odd ($n_x$ even) and Fully Non-TRIM (All $n_i$ odd) Regimes}
When two or more exponents are odd, topological distinctions collapse into a unified canonical dispersion. In the **Double-Odd** case, the energy factors out as $|E'|^{\nu - 1}$, yielding a piecewise DOS $g^{2-odd}(E) = \mathcal{C}^{2-odd} |E - \varepsilon_0|^{\nu - 1}$ with dimensionless weights $\mathcal{J}_{\pm}^{2-odd}$.

In the **Fully Non-TRIM Regime**, all axes lack inversion symmetry. Decomposing the integral into four quadrants and applying $k_i' = |E'|^{1/n_i} x_i$, we define the weights $\mathcal{K}_{\pm}^{3-odd} = \sum_{\sigma_{x,y}} \int_0^\infty \int_0^\infty | 1 \mp \sigma_x x^{n_x} \mp \sigma_y y^{n_y} |^{\frac{1}{n_z}-1} dx dy$. By index re-mapping $\sigma_i \to -\sigma_i$, we prove $\mathcal{K}_-^{3-odd} \equiv \mathcal{K}_+^{3-odd}$, confirming a perfectly symmetric energy-sign reversal:
\begin{equation}
g^{3-odd}(E) = \mathcal{C}^{3-odd} |E - \varepsilon_0|^{\frac{1}{n_x} + \frac{1}{n_y} + \frac{1}{n_z} - 1}
\end{equation}

\subsubsection{Universal Hallmarks of $T_3$-type HOVHS: The Ultimate Flat Band}
Synthesizing all parity configurations, the $T_3$-type HOVHS represents the limit of band flattening, governed by two mathematical absolutes:

\begin{itemize}
    \item \textbf{Power-Law Inevitability}: For any system with at least one even-parity exponent ($n_{even} \ge 4, n_{odd} \ge 3$), the exponent index is bounded by $\nu_{max} = 11/12$. This mandates $\alpha = \nu-1 \le -1/12 < 0$, guaranteeing an unconditional power-law divergence ($g(E) \to \infty$) manifesting as an extreme macroscopic cusp.
    
    \item \textbf{The Singular $(3,3,3)$ Logarithmic Anomaly}: A logarithmic DOS ($\alpha = 0$) strictly requires the Diophantine solution $1/n_x + 1/n_y + 1/n_z = 1$. Under the constraint $n_i \ge 3$, the unique solution is $n_x = n_y = n_z = 3$. This anomaly is permissible only in the \textbf{fully non-TRIM} regime. Any deviation (e.g., $n_z > 3$) instantly ruptures the logarithmic scaling back into a power-law singularity.
\end{itemize}
Ultimately, the $T_3$ HOVHS quenches the 3D phase-space volume expansion via complete elimination of quadratic energy, providing an optimal frontier for strong electronic correlations.

\section{DOS near the Noncritical Ordinary Van Hove Singularity}
\label{sec:iv}

As established in the main text, noncritical ordinary VHSs ($N$-type) occur at momentum points where exactly one direction is linear ($n_\alpha = 1$) while the remaining two are quadratic ($n_{\beta,\gamma} = 2$). The indices $\alpha, \beta, \gamma$ denote a generic permutation of the Cartesian axes $x, y, z$, which physically depends on the local symmetry of the specific momentum point. The topological subtypes ($N_0, N_1, N_2$) are determined by the signs of the reduced $2 \times 2$ Hessian in the stationary plane (see Fig.~2, Fig.~\ref{figs6} and Table.~\ref{table_Ordinary_Noncritical_DOS}).

To derive the analytical DOS without loss of generality, we align the non-stationary direction ($\alpha$) with the longitudinal $k_z$ axis and the stationary subspace ($\beta, \gamma$) with the transverse $k_x\text{-}k_y$ plane. The canonical energy dispersion is formulated as:
\begin{equation}
\varepsilon(\mathbf{k}) = \varepsilon_0 + \sigma_x c_x k_x^2 + \sigma_y c_y k_y^2 + c_z k_z
\end{equation}
where $c_i > 0$ and $\sigma_{x,y} \in \{+1, -1\}$. Note that the sign of the linear term is fixed as $+c_z k_z$ without loss of generality, as its monotonicity ensures the DOS is invariant under $\sigma_z \to -\sigma_z$.

The total 3D DOS $g(E)$ is evaluated via the dimensionality reduction method (slice integration). By defining the effective transverse energy as $\Lambda = E' - c_z k_z$ (where $E' = E - \varepsilon_0$), the 3D DOS is obtained by integrating the 2D subsystem DOS $g_{2D}(\Lambda)$ over the longitudinal Brillouin zone $k_z \in [-R, R]$:
\begin{equation}
g(E) = \frac{1}{2\pi} \int_{-R}^{R} g_{2D}(E' - c_z k_z) \, dk_z
\label{eq:slice_integration_N}
\end{equation}
In the following subsections, we apply this framework to each subtype, revealing the characteristic linear or quadratic energy corrections superimposed on a constant background $g_0 \propto R$.

\subsection{$N_0$-type}
\label{sec:iv.1}
For the $N_0$-type noncritical VHS, the energy dispersion is $\varepsilon(\mathbf{k}) = \varepsilon_0 + c_x k_x^2 + c_y k_y^2 + c_z k_z$. For a fixed longitudinal momentum $k_z$, the transverse dispersion defines an elliptic area $S(\Lambda) = \pi \Lambda / \sqrt{c_x c_y}$ in momentum space, where $\Lambda = E' - c_z k_z$ is the effective transverse energy. This yields a step-like 2D DOS:
\begin{equation}
g_{2D}(\Lambda) = \frac{1}{4\pi^2} \frac{dS}{d\Lambda} = \frac{1}{4\pi \sqrt{c_x c_y}} \Theta(\Lambda)
\end{equation}
where the Heaviside function $\Theta(\Lambda)$ enforces the physical requirement of non-negative transverse kinetic energy.

Substituting $g_{2D}$ into the 3D slice integral (Eq.~\ref{eq:slice_integration_N}), the constraint $\Lambda \ge 0$ dynamically dictates the upper integration limit as $k_z \le E'/c_z$. In the low-energy regime ($|E'| \ll c_z R$), this limit falls within the Brillouin zone boundaries $[-R, R]$. The total DOS is thus:
\begin{equation}
g(E) = \frac{1}{8\pi^2 \sqrt{c_x c_y}} \int_{-R}^{E'/c_z} dk_z = \underbrace{\frac{R}{8\pi^2 \sqrt{c_x c_y}}}_{g_0} + \underbrace{\frac{E - \varepsilon_0}{8\pi^2 c_z \sqrt{c_x c_y}}}_{\delta g(E)}
\end{equation}
This confirms that the $N_0$ singularity is characterized by a finite constant background $g_0$ superimposed with a positive linear energy correction $\delta g(E) \propto (E - \varepsilon_0)$, matching the results in Table. ~\ref{table_Ordinary_Noncritical_DOS}.

\subsection{$N_1$-type}
\label{sec:iv.2}
For the $N_1$-type noncritical VHS, the dispersion is $\varepsilon(\mathbf{k}) = \varepsilon_0 + c_x k_x^2 - c_y k_y^2 + c_z k_z$. The transverse subsystem exhibits a hyperbolic saddle geometry $c_x k_x^2 - c_y k_y^2 = \Lambda$, which inherently generates a logarithmic divergence regulated by an in-plane bandwidth cutoff $W$:
\begin{equation}
g_{2D}(\Lambda) \approx \frac{1}{2\pi^2 \sqrt{c_x c_y}} \ln\left( \frac{W}{|\Lambda|} \right)
\end{equation}

Substituting $g_{2D}$ into the slice integral and defining $A = (4\pi^3 \sqrt{c_x c_y})^{-1}$, the total DOS decomposes into a constant term and an energy-dependent integral $I_2$:
\begin{equation}
g(E) = 2RA \ln W - A \underbrace{\int_{-R}^{R} \ln|E' - c_z k_z| \, d k_z}_{I_2}
\end{equation}
Using the substitution $u = E' - c_z k_z$ and the longitudinal half-bandwidth $\Delta_z = c_z R$, the core integral evaluates to:
\begin{equation}
I_2 = \frac{1}{c_z} \Big[ (\Delta_z + E') \ln(\Delta_z + E') + (\Delta_z - E') \ln(\Delta_z - E') - 2\Delta_z \Big]
\end{equation}
Crucially, $I_2$ is an even function of $E'$, meaning first-order linear corrections identically vanish. Performing a second-order Taylor expansion in the limit $|E'| \ll \Delta_z$ yields the final analytical form:
\begin{equation}
g(E) =  \underbrace{\frac{R}{2\pi^3 \sqrt{c_x c_y}} \left[ 1 + \ln\left( \frac{W}{c_z R} \right) \right]}_{g_0} -  \underbrace{\frac{(E - \varepsilon_0)^2}{4\pi^3 c_z^2 R \sqrt{c_x c_y}}}_{\delta g(E)}
\end{equation}
This proves that the longitudinal linear dispersion quenches the transverse logarithmic singularity into a finite background $g_0$, accompanied by a symmetric quadratic decay $\delta g(E) \propto -(E - \varepsilon_0)^2$, matching the results in Table.~\ref{table_Ordinary_Noncritical_DOS}. This mathematically establishes a broad, plateau-like quadratic maximum. Physically, this topological shape ensures that the integrated effective DOS remains remarkably robust against experimental convolutions (e.g., impurity disorder broadening $\Gamma$ or doping fluctuations $\Delta E_F$), fundamentally distinguishing it from the fragile derivative-divergent kinks of ordinary 3D $M$-type critical points.

\subsection{$N_2$-type}
\label{sec:iv.3}
For the $N_2$-type noncritical VHS, the energy dispersion is $\varepsilon(\mathbf{k}) = \varepsilon_0 - c_x k_x^2 - c_y k_y^2 + c_z k_z$. The transverse subsystem follows a downward-opening elliptic paraboloid, requiring the effective transverse energy $\Lambda = E' - c_z k_z$ to be non-positive ($\Lambda \le 0$) for physical states. This leads to the step-like 2D DOS:
\begin{equation}
g_{2D}(\Lambda) = \frac{1}{4\pi \sqrt{c_x c_y}} \Theta(-\Lambda)
\end{equation}
where the Heaviside function $\Theta(-\Lambda)$ ensures states exist only when the longitudinal energy exceeds the total energy offset.

Substituting $g_{2D}$ into the slice integral, the constraint $\Lambda \le 0$ dynamically truncates the lower integration limit as $k_z \ge E'/c_z$. For low energies where this bound resides within $[-R, R]$, the integration interval becomes $[E'/c_z, R]$, yielding:
\begin{equation}
g(E) = \frac{1}{8\pi^2 \sqrt{c_x c_y}} \int_{E'/c_z}^{R} dk_z = \underbrace{\frac{R}{8\pi^2 \sqrt{c_x c_y}}}_{g_0} - \underbrace{\frac{E - \varepsilon_0}{8\pi^2 c_z \sqrt{c_x c_y}}}_{\delta g(E)}
\end{equation}
Unlike the $N_0$ case, the $N_2$ singularity exhibits a negative linear energy correction. This reflects the requisite geometric increase in the longitudinal momentum threshold $k_z$ to compensate for the inverted transverse energy contribution as $E$ increases, consistent with the results in Table.~\ref{table_Ordinary_Noncritical_DOS}.

\section{DOS near the Noncritical Higher-Order Van Hove Singularity}
\label{sec:v}

Extending the noncritical framework to higher orders, $S$-type singularities arise when the reduced $2 \times 2$ Hessian matrix in the stationary plane becomes singular. This topological transition occurs when at least one transverse exponent $n_i$ exceeds 2, reflecting a local flattening of the energy surface along specific directions (see Fig.~\ref{figs6}, Tables. ~\ref{table_S1_DOS} and ~\ref{table_S2_DOS}).

Following the coordinate alignment and slice integration method established in Sec.~\ref{sec:iv}, we formulate the generalized dispersion for $S$-type HOVHS as:
\begin{equation}
\varepsilon(\mathbf{k}) = \varepsilon_0 + \sigma_x c_x k_x^{n_x} + \sigma_y c_y k_y^{n_y} + c_z k_z
\end{equation}
where $n_x, n_y \ge 2$, and at least one $n_i \ge 3$. The specific universality class is dictated by the number of zero eigenvalues of the reduced Hessian, which directly maps to the transverse exponents:
\begin{itemize}
    \item \textbf{$S_1$-type HOVHS ($n_x = 2, n_y \ge 3$):} One transverse direction remains quadratic while the other exhibits higher-order scaling, yielding \textit{one zero eigenvalue}.
    \item \textbf{$S_2$-type HOVHS ($n_x \ge 3, n_y \ge 3$):} Both transverse directions exhibit higher-order scaling, rendering the reduced Hessian a null matrix with \textit{two zero eigenvalues}.
\end{itemize}

The total 3D DOS is evaluated by integrating the higher-order 2D subsystem DOS $g_{2D}(\Lambda)$ over the bounded longitudinal Brillouin zone $k_z \in [-R, R]$:
\begin{equation}
g(E) = \frac{1}{2\pi} \int_{-R}^{R} g_{2D}(E' - c_z k_z) \, dk_z
\end{equation}
In the following derivations, we demonstrate that unlike the $N$-type case, the higher-order nature of the transverse plane leads to non-trivial power-law corrections that depend on the specific parity and scaling exponents $n_x$ and $n_y$.

\subsection{$S_1$-type HOVHS}
\label{sec:v.1}
For the $S_1$-type HOVHS, one transverse direction is strictly quadratic while the other is higher-order ($n_y \ge 3$). The canonical dispersion is formulated as:
\begin{equation}
\varepsilon(\mathbf{k}) = \varepsilon_0 + c_x k_x^2 + \sigma_y c_y k_y^{n_y} + c_z k_z
\end{equation}
Applying dimension-specific scaling $k_x' = \sqrt{c_x} k_x$ and $k_y' = c_y^{1/n_y} k_y$ yields the Jacobian $J = (c_x c_y^{2/n_y})^{-1/2}$. To derive a unified framework, we consider the scaling behavior of the 2D transverse DOS $g_{2D}(\Lambda)$ under the effective energy $\Lambda = E' - c_z k_z$. 

Due to the scaling invariance of the enclosed momentum area ($S \propto \Lambda^{1/2 + 1/n_y}$), the 2D DOS universally satisfies $g_{2D}(\lambda \Lambda) = \lambda^{\mu-1} g_{2D}(\Lambda)$, where $\mu = \frac{1}{2} + \frac{1}{n_y}$. This allows for a generalized representation:
\begin{equation}
g_{2D}(\Lambda) = \frac{1}{\sqrt{c_x} c_y^{1/n_y}} |\Lambda|^{\mu - 1} \Big[ C_{+} \Theta(\Lambda) + C_{-} \Theta(-\Lambda) \Big]
\label{eq:S1_2D_generalized}
\end{equation}
where $C_\pm$ are strictly dimensionless constants defining the phase-space weights for positive and negative effective energies:
\begin{equation}
C_{\pm} = \frac{1}{(2\pi)^2} \iint_{-\infty}^{\infty} \delta\left(\pm 1 - (k_x')^2 - \sigma_y(k_y')^{n_y}\right) dk_x' dk_y'
\end{equation}

The total 3D DOS is obtained by substituting Eq.~(\ref{eq:S1_2D_generalized}) into the slice integral. In the low-energy limit ($|E'| \ll \Delta_z$, where $\Delta_z = c_z R$), we perform a Taylor expansion $(\Delta_z \pm E')^\mu \approx \Delta_z^\mu \pm \mu \Delta_z^{\mu - 1} E'$, yielding:
\begin{equation}
g(E) \approx \underbrace{\frac{(C_+ + C_-) \Delta_z^\mu}{2\pi c_z \mu \sqrt{c_x} c_y^{1/n_y}}}_{ g_0} + \underbrace{\frac{(C_+ - C_-) \Delta_z^{\mu - 1}}{2\pi c_z \sqrt{c_x} c_y^{1/n_y}} (E - \varepsilon_0)}_{ \delta g(E)}
\end{equation}
This establishes that the survival of a first-order linear energy correction $\delta g(E)$ is strictly dictated by the phase-space asymmetry factor $(C_+ - C_-)$. We evaluate this factor across different regimes below.

\subsubsection{TRIM Regime (Even $n_y$)}
In the centrosymmetric regime, $n_y$ is an even integer. The weights $C_\pm$ are determined by the topological category:

\noindent \textbf{Category A: Elliptic Transverse Dispersions ($\sigma_y = +1$)}:
The condition $-1 - (k_x')^2 - (k_y')^{n_y} = 0$ has no real solutions, rendering $C_- = 0$. The resulting DOS exhibits a strictly positive linear correction:
\begin{equation}
g_{elliptic}(E) \approx g_0 + \frac{C_+ \Delta_z^{\mu - 1}}{2\pi c_z \sqrt{c_x} c_y^{1/n_y}} (E - \varepsilon_0)
\end{equation}

\noindent \textbf{Category B: Hyperbolic Transverse Dispersions ($\sigma_y = -1$)}:
States exist for both $\Lambda > 0$ and $\Lambda < 0$. The weights resolve into distinct one-dimensional integrals $C_+ \propto B(1/n_y, 1/2 - 1/n_y)$ and $C_- \propto B(1/2, 1/2 - 1/n_y)$. Since $n_y \ge 3$, these Beta functions are unequal ($C_+ \neq C_-$), ensuring the survival of the linear slope:
\begin{equation}
g_{hyperbolic}(E) \approx g_0 + \frac{(C_+ - C_-) \Delta_z^{\mu - 1}}{2\pi c_z \sqrt{c_x} c_y^{1/n_y}} (E - \varepsilon_0)
\end{equation}

\subsubsection{Non-TRIM Regime (Odd $n_y$)}
When $n_y$ is an odd integer, the transverse dispersion lacks inversion symmetry. The $C_+$ integral bounds a closed, asymmetric elliptic region, whereas $C_-$ evaluates an open saddle region. Due to this inherent geometric inequality ($C_+ \neq C_-$), the total DOS robustly preserves the linear term:
\begin{equation}
g_{odd}(E) \approx g_0 + \frac{(C_+ - C_-) \Delta_z^{\mu - 1}}{2\pi c_z \sqrt{c_x} c_y^{1/n_y}} (E - \varepsilon_0) 
\end{equation}

In summary, the integration over the linear longitudinal dimension quenches the transverse fractional singularity into a stable background $g_0$, while the intrinsic asymmetry ($C_+ \neq C_-$) ensures a linear scaling $\delta g \propto E$ (see Table.~\ref{table_S1_DOS}). This distinguishes $S_1$-type HOVHS from the perfectly symmetric quadratic corrections of $S_2$-type geometries.

\subsection{$S_2$-type HOVHS}
\label{sec:V.2}
For the $S_2$-type singularity, both transverse principal directions exhibit higher-order scaling ($n_x, n_y \ge 3$), rendering the reduced $2 \times 2$ Hessian a null matrix. The generalized dispersion is:
\begin{equation}
\varepsilon(\mathbf{k}) = \varepsilon_0 + \sigma_x c_x k_x^{n_x} + \sigma_y c_y k_y^{n_y} + c_z k_z
\end{equation}

Applying the dimension-specific scaling $k_i' = c_i^{1/n_i} k_i$, we define the transverse scaling exponent $\nu = 1/n_x + 1/n_y$. For all $n_i \ge 3$, the condition $\nu \le 2/3$ ensures that the 2D subsystem DOS is inherently \textbf{UV convergent}. Decoupled from transverse boundary effects, the 2D DOS manifests as a pure fractional power law:
\begin{equation}
g_{2D}(\Lambda) = \frac{1}{c_x^{1/n_x} c_y^{1/n_y}} |\Lambda|^{\nu - 1} \Big[ C_{+} \Theta(\Lambda) + C_{-} \Theta(-\Lambda) \Big]
\label{eq:S2_2D_DOS}
\end{equation}
where $C_\pm$ are dimensionless weights determined by the transverse topology. Substituting this into the 3D slice integral and performing a second-order Taylor expansion in the low-energy limit ($|E'| \ll \Delta_z$), we establish the following general DOS framework:
\begin{equation}
g(E) \approx \underbrace{\frac{(C_+ + C_-) \Delta_z^\nu}{2\pi c_z \nu c_x^{1/n_x} c_y^{1/n_y}}}_{g_0} + \underbrace{\frac{(C_+ - C_-) \Delta_z^{\nu - 1}}{2\pi c_z c_x^{1/n_x} c_y^{1/n_y}} E'}_{\delta g^{(1)}(E)} + \underbrace{\frac{(\nu - 1)(C_+ + C_-) \Delta_z^{\nu - 2}}{4\pi c_z c_x^{1/n_x} c_y^{1/n_y}} (E')^2}_{\delta g^{(2)}(E)}
\label{eq:S2_Taylor_Unified}
\end{equation}
where $E' = E - \varepsilon_0$. We now evaluate specific parity and topological regimes.

\subsubsection{TRIM Regime (Even $n_x, n_y$)}

\noindent \textbf{Category A: Elliptic Transverse Dispersions ($\sigma_x = \sigma_y = +1$)}:
The phase-space is asymmetric with $C_- = 0$. The DOS follows:
\begin{equation}
g_{elliptic}(E) \approx g_0 + \frac{C_+ \Delta_z^{\nu - 1}}{2\pi c_z c_x^{1/n_x} c_y^{1/n_y}} (E - \varepsilon_0) + \frac{(\nu - 1) C_+ \Delta_z^{\nu - 2}}{4\pi c_z c_x^{1/n_x} c_y^{1/n_y}} (E - \varepsilon_0)^2
\end{equation}

\noindent \textbf{Category B: Hyperbolic Transverse Dispersions ($\sigma_x \sigma_y = -1$)}:
\begin{itemize}
    \item For \textbf{anisotropic saddle points} ($n_x \neq n_y$), $C_+ \neq C_-$, leading to:
    \begin{equation}
    g_{hyperbolic}(E) \approx g_0 + \frac{(C_+ - C_-) \Delta_z^{\nu - 1}}{2\pi c_z c_x^{1/n_x} c_y^{1/n_y}} (E - \varepsilon_0) + \delta g^{(2)}(E)
    \end{equation}
    \item For \textbf{isotropic saddle points} ($n_x = n_y = n$), discrete exchange symmetry enforces $C_+ = C_- \equiv C_{iso}$. The linear term identically cancels, yielding a \textbf{pure quadratic correction}:
    \begin{equation}
    g_{isotropic}(E) \approx g_0 + \frac{(\nu - 1) C_{iso} \Delta_z^{\nu - 2}}{2\pi c_z c_x^{1/n} c_y^{1/n}} (E - \varepsilon_0)^2
    \end{equation}
\end{itemize}

\subsubsection{Fully Non-TRIM Regime (Odd $n_x, n_y$)}
As specified in the previous sections, coordinate inversion for odd exponents establishes that $g_{2D}(\Lambda)$ is an even function, which mandates $C_+ = C_- \equiv C_{odd}$. Substituting this symmetry into Eq.~(\ref{eq:S2_Taylor_Unified}), the first-order linear term vanishes, and the DOS universally manifests a strictly quadratic decay:
\begin{equation}
g_{odd}(E) \approx g_0 + \frac{(\nu - 1) C_{odd} \Delta_z^{\nu - 2}}{2\pi c_z c_x^{1/n_x} c_y^{1/n_y}} (E - \varepsilon_0)^2
\end{equation}

\subsubsection{Mixed-Parity Regime (Even $n_x$, Odd $n_y$)}
The lack of inversion symmetry results in topologically distinct contours ($C_+$ closed, $C_-$ open), ensuring $C_+ \neq C_-$. The leading-order linear correction is preserved, accompanied by a second-order parabolic bending:
\begin{equation}
g_{mixed}(E) \approx g_0 + \frac{(C_+ - C_-) \Delta_z^{\nu - 1}}{2\pi c_z c_x^{1/n_x} c_y^{1/n_y}} (E - \varepsilon_0) + \frac{(\nu - 1)(C_+ + C_-) \Delta_z^{\nu - 2}}{4\pi c_z c_x^{1/n_x} c_y^{1/n_y}} (E - \varepsilon_0)^2
\end{equation}
Since $\nu < 1$, the curvature coefficient is negative, manifesting as an asymmetric downward bending of the DOS relative to the linear slope as energy departs from $\varepsilon_0$.

\vspace{8pt}
In summary, the $S_2$-type HOVHS represents a regime where the dominant energy correction (linear vs quadratic) is dictated by the symmetry-protected weights of the transverse phase space, as summarized in Table.~\ref{table_S2_DOS}.

\section{Tight-binding models of pyrochlore lattice}
As explicitly introduced via the compact Pauli-matrix formalism in the main text, this section provides the fully expanded massive matrix forms and the detailed eigenvalue derivations for both the non-SOC and SOC-inclusive models.

\subsection{s-orbital TB model with nearest-neighbor hoppings without SOC }
For TB model based on \textit{s} orbitals, the Hamiltonian matrix in momentum space can be written as
\begin{equation}
H = \begin{pmatrix}
H_{11} & H_{12} & H_{13} & H_{14} \\
H_{12}^\dagger & H_{22} & H_{23} & H_{24} \\
H_{13}^\dagger & H_{23}^\dagger & H_{33} & H_{34} \\
H_{14}^\dagger & H_{24}^\dagger & H_{34}^\dagger & H_{44}
\end{pmatrix}
\label{TBH0}
\end{equation}
in which the matrix elements of Eq.~\ref{TBH0} would be 
\begin{equation}
\begin{aligned}
& H_{11} = H_{22} = H_{33} = H_{44} = e_{1} , 
H_{12} = e^{-\frac{i k_x}{2}}t_{1} + e^{\frac{i k_x}{2}}t_{1}, 
H_{13} = e^{-\frac{i k_y}{2}}t_{1} + e^{\frac{i k_y}{2}}t_{1}, \\
& H_{14} = e^{-\frac{i k_z}{2}}t_{1} + e^{\frac{i k_z}{2}}t_{1}, 
H_{23} = e^{i(\frac{k_x}{2}-\frac{k_y}{2})}t_{1} + e^{i(-\frac{k_x}{2}+\frac{k_y}{2})}t_{1}, \\
& H_{24} = e^{i(\frac{k_x}{2}-\frac{k_z}{2})}t_{1} + e^{i(-\frac{k_x}{2}+\frac{k_z}{2})}t_{1}, 
H_{34} = e^{i(\frac{k_y}{2}-\frac{k_z}{2})}t_{1} + e^{i(-\frac{k_y}{2}+\frac{k_z}{2})}t_{1}. \\
\end{aligned}
\end{equation}
The four eigenvalues $\varepsilon_{n}(\textbf{k})$ ($n = 1,2,3,4$) are then obtained as
\begin{equation}
    \begin{aligned}
        & \varepsilon_{1}(\textbf{k}) = e_{1} +2t_{1} - \vert t_1 \rvert \sqrt{4 + 2\cos(k_x) + 2\cos(k_y) + 2\cos(k_z) 
        + 2\cos(k_{x}-k_{y}) +  2\cos(k_{x}-k_{z}) + 2\cos(k_{y}-k_{z})}, \\
        & \varepsilon_{2}(\textbf{k}) = e_{1} +2t_{1} + \vert t_1 \rvert \sqrt{4 + 2\cos(k_x) + 2\cos(k_y) + 2\cos(k_z) 
        + 2\cos(k_{x}-k_{y}) +  2\cos(k_{x}-k_{z}) + 2\cos(k_{y}-k_{z})}, \\
        & \varepsilon_{3}(\textbf{k}) = \varepsilon_{4}(\textbf{k}) =e_{1} - 2t_{1}, \\
    \end{aligned}
    \label{eigen4s0}
\end{equation}
where $e_{1}$ is the on-site energy and $t_1$ denotes the conventional spin-conserving nearest-neighbor (NN) hopping parameter \cite{S_zhang2022magnetictb} in the non-SOC limit. The resulting band structure features a twofold-degenerate flat band, corresponding to $\varepsilon_{3}(\textbf{k})$ and $\varepsilon_{4}(\textbf{k})$, which remains dispersionless throughout the entire Brillouin zone. The remaining two singlet bands, $\varepsilon_{1}(\textbf{k})$ and $\varepsilon_{2}(\textbf{k})$, are symmetric with respect to the energy $\varepsilon(\textbf{k}) = e_{1} + 2t_{1}$. Notably, the energies at the high-symmetry points $\mathrm{K}$ (3/8, 3/8, 3/4) and $\mathrm{U}$ (5/8, 1/4, 5/8) are equal within each band branch: $\varepsilon_{1}(\textbf{K}) = \varepsilon_{1}(\textbf{U}) = e_{1} + 2t_{1} - \sqrt{(6-4\sqrt{2})t_{1}^{2}}$, $\varepsilon_{2}(\textbf{K}) = \varepsilon_{2}(\textbf{U}) = e_{1} + 2t_{1} + \sqrt{(6-4\sqrt{2})t_{1}^{2}}$, $\varepsilon_{3}(\textbf{K}) = \varepsilon_{3}(\textbf{U}) = e_{1} - 2t_{1}$, and $\varepsilon_{4}(\textbf{K}) = \varepsilon_{4}(\textbf{U}) = e_{1} - 2t_{1}$. For simplicity, $e_{1}$ and $t_{1}$ were set to $0$ and $-1$, respectively. Under this parametrization, Eq.~\eqref{eigen4s0} reduces to a simplified form. The resulting band dispersions and related properties are presented in Figs.~\ref{figs2} and \ref{figs3}.

\subsection{s-orbital TB model with SOC and nearest-neighbor hoppings}
When spin-orbit coupling is introduced, the basis expands, and the nearest-neighbor interactions are decoupled into two distinct physical channels. Here, $e_1$ represents the on-site energy, $t_2$ governs the spin-conserving hopping, and $t_1$ dictates the spin-flip hopping processes induced by SOC. For the TB model based on \textit{s} orbitals with SOC parameterized by $t_1$ and $t_2$, each matrix element in Eq.~\ref{TBH0} becomes a $2\times2$ matrix
\begin{equation}
\begin{aligned}
&H_{11} = H_{22} = H_{33} = H_{44} = \begin{pmatrix}
e_1 & 0  \\
0 & e_1  \\
\end{pmatrix},\\
&H_{12} = \begin{pmatrix}
e^{-\frac{i k_{x}}{2}} (- i t_1 + t_2 ) + e^{\frac{i k_{x}}{2}} (-i t_1 + t_2) & e^{-\frac{i k_{x}}{2}} t_1 + e^{\frac{i k_{x}}{2}} t_1  \\
-e^{-\frac{i k_{x}}{2}} t_1 - e^{\frac{i k_{x}}{2}} t_1 & e^{-\frac{i k_{x}}{2}} (i t_1 + t_2 ) + e^{\frac{i k_{x}}{2}} (i t_1 + t_2)  \\
\end{pmatrix},
\\
&H_{13} = \begin{pmatrix}
e^{-\frac{i k_{y}}{2}} (i t_1 + t_2) + e^{\frac{i k_{y}}{2}} (i t_1 + t_2) & -i e^{-\frac{i k_{y}}{2}} t_1 -i e^{\frac{i k_{y}}{2}} t_1  \\
-i e^{-\frac{i k_{y}}{2}} t_1 -i e^{\frac{i k_{y}}{2}} t_1 & e^{-\frac{i k_{y}}{2}} (-i t_1 + t_2) + e^{\frac{i k_{y}}{2}} (-i t_1 + t_2)  \\
\end{pmatrix},
\\
&H_{14} = \begin{pmatrix}
e^{-\frac{i k_{z}}{2}} t_2 + e^{\frac{i k_{z}}{2}} t_2 & (-1+i)e^{-\frac{i k_{z}}{2}} t_1 - (1-i)e^{\frac{i k_{z}}{2}} t_1  \\
(1+i) e^{-\frac{i k_{z}}{2}} t_1 + (1+i)e^{\frac{i k_{z}}{2}} t_1 & e^{-\frac{i k_{z}}{2}} t_2 + e^{\frac{i k_{z}}{2}} t_2 \\
\end{pmatrix},
\\
&H_{23} = \begin{pmatrix}
t_2 e^{i \left( \frac{k_x}{2} - \frac{k_y}{2} \right)} + t_2 e^{i \left( \frac{k_y}{2} - \frac{k_x}{2} \right)} 
& (1+i)t_1 e^{i \left( \frac{k_x}{2} - \frac{k_y}{2} \right)} + (1+i)t_1 e^{i \left( \frac{k_y}{2} - \frac{k_x}{2} \right)}  \\

(-1+i)t_1 e^{i \left( \frac{k_x}{2} - \frac{k_y}{2} \right)} - (1-i)t_1 e^{i \left( \frac{k_y}{2} - \frac{k_x}{2} \right)}
& t_2 e^{i \left( \frac{k_x}{2} - \frac{k_y}{2} \right)} + t_2 e^{i \left( \frac{k_y}{2} - \frac{k_x}{2} \right)}  \\
\end{pmatrix},
\\
&H_{24} = \begin{pmatrix}
(-i t_1 + t_2) e^{i \left( \frac{k_x}{2} - \frac{k_z}{2} \right)} + (-i t_1 + t_2) e^{i \left( \frac{k_z}{2} - \frac{k_x}{2} \right)} 
& - i t_{1} e^{i \left( \frac{k_x}{2} - \frac{k_z}{2} \right)}  - i t_{1} e^{i \left( \frac{k_z}{2} - \frac{k_x}{2} \right)} \\

- i t_{1} e^{i \left( \frac{k_x}{2} - \frac{k_z}{2} \right)}  - i t_{1} e^{i \left( \frac{k_z}{2} - \frac{k_x}{2} \right)} 
& (i t_1 + t_2) e^{i \left( \frac{k_x}{2} - \frac{k_z}{2} \right)} + (i t_1 + t_2) e^{i \left( \frac{k_z}{2} - \frac{k_x}{2} \right)} \\
\end{pmatrix},
\\
&H_{34} = \begin{pmatrix}
(i t_1 + t_2) e^{i \left( \frac{k_y}{2} - \frac{k_z}{2} \right)} + (i t_1 + t_2) e^{i \left( \frac{k_z}{2} - \frac{k_y}{2} \right)} 
&  e^{i \left( \frac{k_y}{2} - \frac{k_z}{2} \right)} t_{1}  + e^{i \left( \frac{k_z}{2} - \frac{k_y}{2} \right)} t_{1} \\

-e^{i \left( \frac{k_y}{2} - \frac{k_z}{2} \right)} t_{1}  - e^{i \left( \frac{k_z}{2} - \frac{k_y}{2} \right)} t_{1}
& (-i t_1 + t_2) e^{i \left( \frac{k_y}{2} - \frac{k_z}{2} \right)} + (-i t_1 + t_2) e^{i \left( \frac{k_z}{2} - \frac{k_y}{2} \right)} \\
\end{pmatrix}.
\end{aligned}
\label{ham4ssoc}
\end{equation}

For simplicity, the on-site energy $e_1$ and the spin-flip hopping parameter $t_1$ are set to $-1.0$, respectively, to serve as the baseline energy scale. The spin-conserving hopping parameter $t_2$ takes values of $0$, $\pm0.5$, and $\pm1$. As an example, for $t_2 = -1$, the eight eigenvalues $\varepsilon_{n}(\mathbf{k})$ ($n = 1, \ldots, 8$) are given by
\begin{equation}
    \begin{aligned}
        &  \varepsilon_{1}(\textbf{k}) = \varepsilon_{2}(\textbf{k}) = - \sqrt{3\alpha +  3\sqrt{\alpha^{2}-\beta}}, \quad
        \varepsilon_{3}(\textbf{k}) = \varepsilon_{4}(\textbf{k}) = - \sqrt{3\alpha -  3\sqrt{\alpha^{2}-\beta}}, \\
        &   \varepsilon_{5}(\textbf{k}) = \varepsilon_{6}(\textbf{k}) = \sqrt{3\alpha -  3\sqrt{\alpha^{2}-\beta}}, \quad
        \varepsilon_{7}(\textbf{k}) = \varepsilon_{8}(\textbf{k}) =   \sqrt{3\alpha +  3\sqrt{\alpha^{2}-\beta}}, \\ 
        with \quad 
        & \alpha= 6 + \cos(k_x) + \cos(k_x - k_y) + \cos(k_y) + \cos(k_x - k_z) + \cos(k_y - k_z) + \cos(k_z)\\
         &\beta =  8\bigl[3 + 2\cos(k_x)+ 2\cos(k_x - k_y)+ 2\cos(k_y) + 2\cos(k_x - k_z) + \cos(k_x - k_y - k_z)  \\
           &\quad + 2\cos(k_y - k_z) + \cos(k_x + k_y - k_z) + 2\cos(k_z)+ \cos(k_x - k_y + k_z)\bigr]
    \end{aligned}
    \label{eigen4ssocType1}
\end{equation}

Similarly, the eigenvalues for $t_2 = -0.5$, $0$, $0.5$, and $1$ can be obtained, and the corresponding band dispersions are presented in Figs.~\ref{figs4} and \ref{figs5}.

\subsection{Topological Equivalence of Noncritical Ordinary Van Hove Singularities at K and U}

To elucidate the topological nature of the electronic structure, we perform a rigorous classification of the singularity located at the high-symmetry points $\mathrm{K}$ and $\mathrm{U}$. For simplicity, we focus on the dispersive bands derived from the $s$-orbital tight-binding model without SOC (Fig.~\ref{figs2}), described by $\varepsilon_{n}(\mathbf{k}) = e_{1} + 2t_{1} \mp |t_1| \sqrt{\Omega(\mathbf{k})}$, where $n=1,2$ denotes the band index and $\Omega(\mathbf{k})$ encapsulates the momentum dependence. Given that $\varepsilon_{n}(\mathrm{K}) = \varepsilon_{n}(\mathrm{U})$, we demonstrate here that for any non-zero hopping $t_1$, the singularity at $\mathrm{K}$ and $\mathrm{U}$ are not only iso-energetic but also belong to the identical topological class of noncritical ordinary Van Hove singularities (n-o-VHSs).

The classification protocol first requires establishing that these points are noncritical in the three-dimensional space, characterized by a non-vanishing gradient along a specific momentum direction. At the $\mathrm{K}$ point ($3\pi/4, 3\pi/4, 3\pi/2$), an analytical evaluation of the auxiliary function $\Omega(\mathbf{k})$ reveals that the transverse derivatives vanish ($\partial_{k_x} \Omega = \partial_{k_y} \Omega = 0$), whereas the longitudinal derivative is non-zero ($\partial_{k_z} \Omega = 2(1+\sqrt{2}) \neq 0$). This explicitly categorizes $\mathrm{K}$ as a noncritical ordinary point, necessitating a dimensionality reduction onto the stationary $k_x$-$k_y$ subspace (corresponding to the explicit mapping $\alpha \to z$, and $\beta, \gamma \to x, y$). Similarly, at the $\mathrm{U}$ point ($5\pi/4, \pi/2, 5\pi/4$), symmetry dictates a non-vanishing gradient solely along the $k_y$ direction ($\partial_{k_y} \Omega = -2(1+\sqrt{2}) \neq 0$), confining the critical behavior to the reduced $k_x$-$k_z$ plane (corresponding to the explicit mapping $\alpha \to y$, and $\beta, \gamma \to x, z$).

We proceed by constructing the $2 \times 2$ Hessian matrices for the reduced two-dimensional energy dispersion $E_{\text{2D}}$ within these stationary subspaces. The matrix elements are proportional to the Hessian of $\Omega(\mathbf{k})$ via the chain rule: $H_{ij}^{E} \propto \mp H_{ij}^{\Omega}$. 
At the $\mathrm{K}$ point, projecting onto the $k_x$-$k_y$ plane, the off-diagonal term is governed by $\cos(k_x-k_y)$. Substituting the coordinates, we find $\cos(3\pi/4 - 3\pi/4) = 1$. The diagonal terms are determined by $\cos k_x = -1/\sqrt{2}$ and the summation of difference terms. Explicit calculation yields the reduced Hessian:
\begin{equation}
    H_{\mathrm{K}}^{\Omega} = 
    \begin{pmatrix}
    \partial^2_{xx}\Omega & \partial^2_{xy}\Omega \\
    \partial^2_{yx}\Omega & \partial^2_{yy}\Omega
    \end{pmatrix}_{\mathrm{K}}
    =
    \begin{pmatrix}
    2\sqrt{2}-2 & 2 \\
    2 & 2\sqrt{2}-2
    \end{pmatrix}.
\end{equation}
For the $\mathrm{U}$ point, the stationary subspace is spanned by $k_x$ and $k_z$. Crucially, the local topology mirrors that of $\mathrm{K}$ due to the permutation of the $y$ and $z$ axes. Specifically, the off-diagonal coupling at $\mathrm{U}$ is $\cos(k_x-k_z) = \cos(5\pi/4 - 5\pi/4) = 1$, which is numerically identical to the $x$-$y$ coupling at $\mathrm{K}$. Similarly, the diagonal elements at $\mathrm{U}$ involve terms such as $\cos k_x = -1/\sqrt{2}$ and $\cos(k_x-k_y) = -1/\sqrt{2}$, reproducing the exact same values found at $\mathrm{K}$. Thus, we arrive at the identity:
\begin{equation}
    H_{\mathrm{U}}^{\Omega} = 
    \begin{pmatrix}
    \partial^2_{xx}\Omega & \partial^2_{xz}\Omega \\
    \partial^2_{zx}\Omega & \partial^2_{zz}\Omega
    \end{pmatrix}_{\mathrm{U}}
    =
    \begin{pmatrix}
    2\sqrt{2}-2 & 2 \\
    2 & 2\sqrt{2}-2
    \end{pmatrix}.
\end{equation}
The eigenvalues of this characteristic matrix are $\lambda_{1} = 2\sqrt{2}$ and $\lambda_{2} = 2\sqrt{2}-4$. This spectrum contains exactly one positive and one negative eigenvalue ($\lambda_1 > 0, \lambda_2 < 0$). Since the Hessian of the actual energy band $H^{E}$ is a scalar multiple of $H^{\Omega}$, this mixed signature is preserved (or globally inverted) for both bands $n=1$ and $n=2$. Consequently, the reduced energy surfaces at both $\mathrm{K}$ and $\mathrm{U}$ exhibit indefinite curvature. We thus conclude that $\mathrm{K}$ and $\mathrm{U}$ host topologically equivalent n-o-VHS of type $N_1$ (saddle points), underpinning the symmetric features observed in the DOS.

\section{DOS for the HOVHS of pyrochlore lattice}
\subsection{$T_1$-type HOVHS}
For TB bands constructed by $s$ orbitals with SOC, when $e_1=0$, $t_1=-1$, and $t_2=-1$, one can find a $T_1$-type HOVHS at the $\mathrm{L}$ point (see Fig.3 of the main text). Setting the energy at the $\mathrm{L}$ point of band 2 as the reference zero ($\varepsilon_0 = -6$ eV), the Taylor expansion at the $\mathrm{L}$ point under the canonical transformation coordinate is:
\begin{equation}
\begin{aligned}
\varepsilon(\mathbf{k}) & =  \frac{1}{2}(1.0 k_x^{2}+1.0 k_y^{2}+1.6757\times 10^{-7} k_z^{2})\\
&  + \frac{1}{4} (-0.502709 k_x^{2}k_y^{2}+0.00532907 k_x^{2}k_z^{2}-0.00177636 k_y^{2}k_z^{2})\\
& + \frac{1}{6}(-0.00444089 k_x^{3}k_y-0.00577316 k_x k_y^{3}+0.579981 k_x^{3}k_z-0.403677 k_y^{3}k_z \\ 
& \quad +0.00266454 k_x k_z^{3}+0.000888178 k_y k_z^{3}) \\
& + \frac{1}{24}(-1.49569 k_x^{4}-1.5099 k_y^{4}+0.00355271 k_z^{4})
\end{aligned}
\label{T1-HVHS-case1}
\end{equation}

We first perform physical simplification on the energy dispersion expressions to extract the dominant behavior, following the principles below:

1. Neglect higher-order small quantities, except for specific dominant higher-order terms (e.g., when there are no low-order terms in the \(k_z\) direction, higher-order terms need to be retained).
2. Symmetry: Terms involving odd functions cancel out due to symmetry when integrated over the entire space.
3. Anisotropic averaging: The \(k_x\) and \(k_y\) directions are approximately isotropic, and cross terms can be incorporated into the dominant terms through angular averaging.

Applying these simplification principles, we obtain the symmetric energy dispersion expression up to the fourth order:
\begin{equation}
\begin{aligned}
\varepsilon(\mathbf{k}) & \approx \frac{1}{2}(1.0 k_x^{2}+1.0 k_y^{2}) + \frac{1}{4}(-0.502709 k_x^{2}k_y^{2}+0.00532907 k_x^{2}k_z^{2}-0.00177636 k_y^{2}k_z^{2})\\
&+\frac{1}{24}(-1.49569 k_x^{4}-1.5099 k_y^{4}+0.00355271 k_z^{4})
\end{aligned}
\label{Eq_simplified_full}
\end{equation}

To rigorously justify the truncation of the effective Hamiltonian to a decoupled form, we perform an asymptotic scaling analysis in the low-energy limit ($E \to 0$). The scaling behavior of each wave vector component $\mathbf{k} = (k_x, k_y, k_z)$ is strictly dictated by the leading non-vanishing terms in their respective directions:
Since the transverse directions ($k_x, k_y$) possess non-vanishing quadratic coefficients, their momenta scale as $k_x, k_y \sim E^{1/2}$. Conversely, the longitudinal direction ($k_z$) has a vanishingly small quadratic coefficient ($\sim 10^{-7}$), making its leading order the quartic term $k_z^4$, which dictates $k_z \sim E^{1/4}$. 

Substituting these scaling relations into the higher-order terms of Eq.~(\ref{Eq_simplified_full}), we evaluate the asymptotic order of magnitude for each term with respect to the energy scale $E$:
\begin{itemize}
    \item Pure transverse quartic terms: $k_x^4, k_y^4 \sim (E^{1/2})^4 = \mathcal{O}(E^2)$
    \item In-plane cross term: $k_x^2 k_y^2 \sim (E^{1/2})^2 (E^{1/2})^2 = \mathcal{O}(E^2)$
    \item Out-of-plane cross terms: $k_x^2 k_z^2, k_y^2 k_z^2 \sim (E^{1/2})^2 (E^{1/4})^2 = E^1 \cdot E^{1/2} = \mathcal{O}(E^{3/2})$
    \item Pure longitudinal quartic term: $k_z^4 \sim (E^{1/4})^4 = \mathcal{O}(E^1)$
\end{itemize}

For low-energy excitations ($E \ll 1$), higher-power terms vanish significantly faster than lower-power terms. Because $\mathcal{O}(E^2) \ll \mathcal{O}(E^{3/2}) \ll \mathcal{O}(E^1)$, the pure transverse quartic terms and all mixed cross terms ($k_x^2k_y^2, k_x^2k_z^2, k_y^2k_z^2$) are strictly higher-order infinitesimals compared to the leading $\mathcal{O}(E^1)$ terms ($k_x^2, k_y^2, k_z^4$). It is important to emphasize that this scaling truncation fundamentally assumes the Taylor expansion coefficients are of natural order. Barring rare extreme events with anomalously large cross-term coefficients (e.g., near critical band inversions), these cross terms only provide negligible subleading corrections and can be safely truncated without losing the leading topological signature. 

Consequently, the energy dispersion relation perfectly decouples and is approximated as:
\begin{equation}
\varepsilon(\mathbf{k}) \approx a(k_x^2 + k_y^2) + b k_z^4
\end{equation}
where the coefficients are directly extracted from the dominant $\mathcal{O}(E^1)$ terms of the original expansion: $a = 0.5$ and $b = 0.00355271 / 24 \approx 0.0001480$. Note that since all dominant coefficients are positive, the $\mathrm{L}$ point acts as a local generalized minimum, meaning physical states only exist for $E \ge 0$ (assuming the reference energy is $\varepsilon_0 = 0$).

The DOS is given by:
\begin{equation}
g(E) = \frac{1}{(2\pi)^3} \iiint \delta( E -  \varepsilon(\mathbf{k})) \, dk_x  dk_y  dk_z
\end{equation}
Transforming the in-plane coordinates to polar form ($q^2 = k_x^2 + k_y^2$, $dk_x dk_y = 2\pi q \, dq$) and integrating over $q$, we get:
\begin{equation}
\begin{aligned}
g(E) &= \frac{1}{(2\pi)^2} \int dk_z \int_{0}^{R} q \, dq \, \delta( E - a q^2 - b k_z^4) \\
&= \frac{1}{4\pi^2 a} \int dk_z \, \Theta( E - b k_z^4 ) \, \Theta( aR^2 - E + b k_z^4 )
\end{aligned}
\end{equation}
The first Heaviside step function requires $|k_z| \le (E/b)^{1/4}$, directly confirming $E \ge 0$. The second step function places a lower bound on $|k_z|$ if $E > aR^2$. Evaluating the $k_z$ integral with symmetric bounds yields:
\begin{equation}
g(E) = 
\begin{cases} 
0 & (E < 0) \\
\frac{1}{2\pi^2 a b^{1/4}} E^{1/4} & (0 \le E \le aR^2) \\
\frac{1}{2\pi^2 a b^{1/4}} \left[ E^{1/4} - (E - aR^2)^{1/4} \right] & (E > aR^2)
\end{cases}
\end{equation}
Substituting the numerical values of $a=0.5$ and $b \approx 0.0001480$, the prefactor is calculated as $\frac{1}{2\pi^2 (0.5) (0.0001480)^{1/4}} \approx 0.918$. The analytical DOS for the relevant low-energy scale is thus strictly expressed as:
\begin{equation}
g(E) \approx 
\begin{cases} 
0 & (E < 0) \\
0.918 E^{1/4} & (0 \le E \le 0.5 R^2) \\
0.918 \left[ E^{1/4} - (E - 0.5 R^2)^{1/4} \right] & (E > 0.5 R^2)
\end{cases}
\end{equation}

\subsection{$T_3$-type HOVHS}
For TB bands constructed by $s$ orbitals with SOC, when $e_1=0$, $t_1=-1$, and $t_2=0$, one can find a $T_3$-type HOVHS at the $\mathrm{\Gamma}$ point (see Fig.~\ref{figs4}). The Taylor expansion at the $\mathrm{\Gamma}$ point is:
\begin{equation}
\begin{aligned}
\varepsilon(\mathbf{k}) & = 4.03347 \times 10^{-16} + \frac{1}{2} \left(-4.40672 \times 10^{-9} \, k_x^2 + 1.19541 \times 10^{-9} \, k_y^2 - 1.1954 \times 10^{-9} \, k_z^2\right) \\
&\quad + \frac{1}{4} \left(-0.0692704 \, k_x^2 k_y^2 - 0.0528748 \, k_x^2 k_z^2 - 0.131415 \, k_y^2 k_z^2\right) \\
&\quad + \frac{1}{6} \left(0.180624 \, k_x^3 k_y - 0.21603 \, k_x k_y^3 - 0.136391 \, k_x^3 k_z + 0.124405 \, k_y^3 k_z - 0.0633544 \, k_x k_z^3 - 0.269827 \, k_y k_z^3\right) \\
&\quad + \frac{1}{24} \left(-0.104757 \, k_x^4 - 0.162414 \, k_y^4 - 0.106302 \, k_z^4\right)
\end{aligned}
\label{HVHS-case3}
\end{equation}

Following the asymptotic scaling logic established in the $T_1$ section, we first perform a rigorous simplification. The zero-point offset and the quadratic coefficients ($\sim 10^{-9}$) are artifacts of computational precision; analytically, the Hessian matrix is identically a null matrix for a $T_3$ singularity, meaning all quadratic terms are strictly zero. Furthermore, all cubic terms in Eq.~(\ref{HVHS-case3}) consist exclusively of odd-parity wavevector combinations (e.g., $k_x^3 k_y$, $k_x k_y^3$). Due to inversion symmetry, their total contribution to the DOS strictly vanishes upon full solid-angle integration. 

Because the quadratic terms are completely quenched, all three momentum components share identical low-energy scaling behavior: $k_x, k_y, k_z \sim |E|^{1/4}$. Consequently, \textit{unlike the $T_1$ case}, both pure quartic terms ($k_i^4$) and mixed cross-quartic terms ($k_i^2 k_j^2$) strictly belong to the same leading-order $\mathcal{O}(E^1)$. Thus, all fourth-order terms must be completely retained, yielding the symmetric dominant effective dispersion:
\begin{equation}
\begin{aligned}
\varepsilon(\mathbf{k}) &\approx\frac{1}{4} \left(-0.0692704 \, k_x^2 k_y^2 - 0.0528748 \, k_x^2 k_z^2 - 0.131415 \, k_y^2 k_z^2\right)\\
&\quad +\frac{1}{24} \left(-0.104757 \, k_x^4 - 0.162414 \, k_y^4 - 0.106302 \, k_z^4\right)
\end{aligned}
\end{equation}

Since the variables cannot be decoupled dimensionally, we introduce a spherical coordinate parameterization to handle the fully coupled geometry. Let $k_x = k\sin\theta\cos\varphi$, $k_y = k\sin\theta\sin\varphi$, $k_z = k\cos\theta$, where $\theta\in[0,\pi]$ and $\varphi\in[0,2\pi]$ are the polar and azimuthal angles, respectively, and $k = |\mathbf{k}|$ is the magnitude of the wave vector. Substituting these parameterizations into the simplified energy dispersion formula, the energy $\varepsilon(\mathbf{k})$ can be rewritten as:
\begin{equation}
\varepsilon(\mathbf{k}) = k^4 F(\theta,\varphi)
\end{equation}
where the angular function $F(\theta,\varphi)$ is structurally negative across the entire solid angle:
\begin{equation}
\begin{aligned}
F(\theta,\varphi) &= -0.0173176\sin^4\theta\cos^2\varphi\sin^2\varphi - 0.0132187\sin^2\theta\cos^2\theta\cos^2\varphi \\
&\quad - 0.0328538\sin^2\theta\cos^2\theta\sin^2\varphi - 0.00436488\sin^4\theta\cos^4\varphi \\
&\quad - 0.00676725\sin^4\theta\sin^4\varphi - 0.00442925\cos^4\theta
\end{aligned}
\end{equation}

The DOS is formally calculated as:
\begin{equation}
g(E) = \frac{1}{(2\pi)^3}\iiint \delta(E - \varepsilon(\mathbf{k})) \, dk_x dk_y dk_z
\end{equation}
Substituting $\varepsilon(\mathbf{k}) = k^4 F(\theta,\varphi)$ into the DOS formula and converting to spherical coordinates ($dk_x dk_y dk_z = k^2\sin\theta \, dk d\theta d\varphi$), we obtain:
\begin{equation}
g(E) = \frac{1}{(2\pi)^3}\int_{0}^{\infty} \int_0^\pi\int_0^{2\pi} \delta(E - k^4 F(\theta,\varphi)) \, k^2 \sin\theta \, d\varphi d\theta dk
\end{equation}
Taking advantage of the properties of the Dirac delta function, and noting that both $E$ and $F(\theta, \varphi)$ are strictly non-positive (which implies physical states only exist for $E \le 0$), we can rewrite the delta function argument using absolute values as $\delta(-|E| + k^4 |F|)$. Applying the composite derivative rule $\delta(f(k)) = \sum \delta(k - k_0)/|f'(k_0)|$, we first integrate over the radial momentum $k$:
\begin{equation}
\delta(-|E| + k^4|F|) = \frac{1}{4k^3|F|} \delta\left[k-\left(\frac{|E|}{|F|}\right)^{1/4}\right]
\end{equation}
Substituting this directly into the $k$-integral yields:
\begin{equation}
\int_0^\infty \delta(-|E| + k^4|F|) \, k^2 dk = \left. \frac{k^2}{4k^3|F|} \right|_{k=\left(\frac{|E|}{|F|}\right)^{1/4}} = \frac{1}{4|E|^{1/4} |F|^{3/4}}
\end{equation}
After integrating over $k$, we are left with a purely geometric integral over the angular variables $\theta$ and $\varphi$:
\begin{equation}
g(E) = \frac{1}{4(2\pi)^3}|E|^{-1/4}\int_0^\pi\int_0^{2\pi} |F(\theta,\varphi)|^{-3/4}\sin\theta \, d\varphi d\theta
\end{equation}
Define the dimensionless angular integral constant as $I = \int_0^\pi\int_0^{2\pi}|F(\theta,\varphi)|^{-3/4}\sin\theta \, d\varphi d\theta$. At this point, the analytical expression for the DOS manifests the characteristic $T_3$-type fractional power-law scaling:
\begin{equation}
g(E) = \frac{I}{4(2\pi)^3}|E|^{-1/4}
\end{equation}
Evaluating the angular integral numerically, the final DOS $g(E)$ behavior near the critical point is accurately captured as:
\begin{equation}
\begin{aligned}
g(E) &\approx 
\begin{cases} 
0.222 (-E)^{-1/4} & (E \leq 0) \\
0 & (E > 0) 
\end{cases}
\end{aligned}
\end{equation}

\section{DOS near the noncritical ordinary VHS of pyrochlore lattice}
\subsection{$N_{0}$-type}
Here, we take the tight-binding (TB) bands constructed by $s$ orbitals with spin-orbit coupling (SOC) as an example to evaluate the DOS near the $N_0$-type noncritical ordinary VHS (n-o-VHS) of the pyrochlore lattice. When $e_1=0, t_{1}=-1$ and $t_{2}=-\frac{1}{2}$ (see Fig.~\ref{figs4}), the relevant band $\varepsilon_{4}(\mathbf{k})$ is identified as the second root (energy-sorted from lowest to highest) of the characteristic polynomial $f(\varepsilon) = 0$:
\begin{equation}
\begin{aligned}      
f(\varepsilon) &= 2\varepsilon^4 + \left( -54 - 9\cos(k_x) - 9\cos(k_x - k_y) - 9\cos(k_y) - 9\cos(k_x - k_z) - 9\cos(k_y - k_z) - 9\cos(k_z) \right)\varepsilon^2 \\
&\quad + \left( 20 + 10\cos(k_x) + 10\cos(k_x - k_y) + 10\cos(k_y) + 10\cos(k_x - k_z) + 10\cos(k_y - k_z) + 10\cos(k_z) \right)\varepsilon \\
&\quad + 192 + 111\cos(k_x) + 111\cos(k_x - k_y) + 111\cos(k_y) + 111\cos(k_x - k_z) + 64\cos(k_x - k_y - k_z) \\
&\quad + 111\cos(k_y - k_z) + 64\cos(k_x + k_y - k_z) + 111\cos(k_z) + 64\cos(k_x - k_y + k_z)
\end{aligned}
\label{eig_s-SOC_e4}
\end{equation}

For the $\mathrm{K}$ point located at $\mathbf{k} = (\frac{3\pi}{4}, \frac{3\pi}{4}, \frac{3\pi}{2})$, the first-order derivatives are $\partial_{k_x} = \partial_{k_y} = 0$ and $\partial_{k_z} = -1.17817$. Because the gradient along $k_z$ is non-vanishing ($v_z \neq 0$), the topological classification of this noncritical critical point is strictly determined by the transverse 2D Hessian matrix evaluated at the stationary plane $k_{z} = \frac{3\pi}{2}$:
\begin{equation}
H_{2D} = \begin{pmatrix}
0.363361 & -0.275082  \\
-0.275082 & 0.363361  \\
\end{pmatrix},
\label{Hessian_s-SOC_K4_2D}
\end{equation}
of which the two eigenvalues are both positive: $\lambda_{1} \approx 0.6384$ and $\lambda_{2} \approx 0.0883$. This upward-opening transverse parabolic dispersion alongside a longitudinal linear gradient unambiguously confirms an $N_0$-type n-o-VHS at the $\mathrm{K}$ point.

To construct the effective low-energy Hamiltonian, we evaluate the explicit Taylor expansion at the $\mathrm{K}$ point in terms of the momentum displacements $\Delta k_x = k_x - \frac{3\pi}{4}$, $\Delta k_y = k_y - \frac{3\pi}{4}$, and $\Delta k_z = k_z - \frac{3\pi}{2}$:
\begin{equation}
\begin{aligned}
\varepsilon_{4}(\mathbf{k}) & = -1.2137 - 1.1782 \Delta k_z \\
& + \frac{1}{2} \bigl( 0.3634 \Delta k_x^2 - 0.5502 \Delta k_x \Delta k_y + 0.3634 \Delta k_y^2 - 0.0883 \Delta k_x \Delta k_z - 0.0883 \Delta k_y \Delta k_z - 0.4200 \Delta k_z^2 \bigr) \\
& + \frac{1}{6} \bigl( 0.0001 \Delta k_x^3 - 0.0002 \Delta k_x^2 \Delta k_y + 0.0002 \Delta k_x \Delta k_y^2 - 0.0001 \Delta k_y^3 + 2.7042 \Delta k_x^2 \Delta k_z \\
& - 6.2050 \Delta k_x \Delta k_y \Delta k_z + \dots \bigr) + \mathcal{O}(\Delta \mathbf{k}^4)
\end{aligned}
\label{HVHS-case1}
\end{equation}

To rigorously extract the dominant terms and justify the truncation of the mixed cross terms, we apply asymptotic scaling analysis in the low-energy limit ($\Delta E \to 0$). The scaling behavior of the momenta is dictated by the leading non-vanishing terms in their respective directions: 
Because the longitudinal direction has a non-zero linear term, its momentum scales identically with energy: $\Delta k_z \sim \mathcal{O}(E^1)$. Conversely, the transverse directions are governed by quadratic terms, mandating $\Delta k_x, \Delta k_y \sim \mathcal{O}(E^{1/2})$. 

Substituting these scaling relations into Eq.~(\ref{HVHS-case1}), we evaluate the asymptotic order of magnitude for each dispersion component:
\begin{itemize}
    \item Leading longitudinal linear term: $\Delta k_z \sim \mathcal{O}(E^1)$
    \item Pure and mixed transverse quadratic terms: $\Delta k_x^2, \Delta k_x \Delta k_y, \Delta k_y^2 \sim (E^{1/2})^2 = \mathcal{O}(E^1)$
    \item Transverse-longitudinal mixed cross terms: $\Delta k_{x,y} \Delta k_z \sim E^{1/2} \cdot E^1 = \mathcal{O}(E^{1.5})$
    \item Pure longitudinal quadratic term: $\Delta k_z^2 \sim (E^1)^2 = \mathcal{O}(E^2)$
    \item All cubic and quartic terms: $\ge \mathcal{O}(E^{1.5})$
\end{itemize}

Since $\mathcal{O}(E^1) \gg \mathcal{O}(E^{1.5}) \gg \mathcal{O}(E^2)$ for $E \ll 1$, the mixed terms ($\Delta k_x \Delta k_z$, $\Delta k_y \Delta k_z$) and the longitudinal quadratic term ($\Delta k_z^2$) are strictly higher-order infinitesimals compared to the $\mathcal{O}(E^1)$ principal terms. Thus, they provide merely subleading corrections and must be systematically truncated.

By discarding the $\mathcal{O}(E^{1.5})$ and higher terms, and diagonalizing the $\mathcal{O}(E^1)$ transverse quadratic form using the eigenvectors of the 2D Hessian (Eq.~\ref{Hessian_s-SOC_K4_2D}), we transform the system into generalized principal coordinates $q_1$ and $q_2$. Setting $q_z = \Delta k_z$, the effective Hamiltonian maps perfectly to the canonical decoupled $N_0$-type form:
\begin{equation}
\varepsilon(\mathbf{q}) \approx \varepsilon_0 + v_z q_z + \frac{1}{2}\lambda_1 q_1^2 + \frac{1}{2}\lambda_2 q_2^2
\end{equation}
Here, the reference energy is $\varepsilon_0 = -1.2137$ eV, the group velocity is $v_z = -1.1782$, and the curvature parameters match the transverse eigenvalues: $\lambda_1 \approx 0.6384$ and $\lambda_2 \approx 0.0883$. 

Based on the universal DOS integration derived in the preceding section, substituting the effective dispersion parameters ($c_x = \lambda_1/2$, $c_y = \lambda_2/2$, and $c_z = |v_z|$) reveals a rigorous linear relationship:
\begin{equation}
g(E) \simeq g_0 + \frac{1}{4\pi^2 |v_z| \sqrt{\lambda_1 \lambda_2}} (E - \varepsilon_0)
\label{eq:N0_DOS_analytical}
\end{equation}
Substituting the derived numerical coefficients directly yields the slope $\alpha$:
\begin{equation}
\alpha = \frac{1}{4\pi^2 |-1.1782| \sqrt{0.6384 \times 0.0883}} \approx 0.091 \, 
\end{equation}
This positive linear scaling behavior ($\delta g(E) \propto E$) superimposed on a finite background represents the defining signature of the $N_0$-type singularity originating from the bulk bands.

\subsection{$N_1$-type} 
Here, we take the TB bands constructed by $s$ orbitals with SOC as an example to evaluate the DOS near the $N_1$-type n-o-VHS of the pyrochlore lattice. With parameters set to $e_1=0$, $t_{1}=-1$, and $t_{2}=-1$ (see Fig.~3 of the main text), the dispersion relation for the fourth band $\varepsilon_{4}(\mathbf{k})$ is given by:
\begin{equation}
\begin{aligned}
\varepsilon_{4}(\mathbf{k}) &= -2 \Bigg\{ \frac{9}{2} + \frac{3}{4}\cos k_x + \frac{3}{4}\cos(k_x - k_y) + \frac{3}{4}\cos k_y + \frac{3}{4}\cos(k_x - k_z) + \frac{3}{4}\cos(k_y - k_z)\\
&\quad  + \frac{3}{4}\cos k_z - \frac{1}{4} \Bigg[ \Big( -18 - 3\cos k_x - 3\cos(k_x - k_y) - 3\cos k_y - 3\cos(k_x - k_z)\\
&\quad  - 3\cos(k_y - k_z) - 3\cos k_z \Big)^2 - 8 \Big( 27 + 18\cos k_x + 18\cos(k_x - k_y) + 18\cos k_y\\
&\quad  + 18\cos(k_x - k_z) + 9\cos(k_x - k_y - k_z) + 18\cos(k_y - k_z) \\
&\quad  + 9\cos(k_x + k_y - k_z) + 18\cos k_z + 9\cos(k_x - k_y + k_z) \Big) \Bigg]^{1/2} \Bigg\}^{1/2} 
\end{aligned}
\label{N1-eig_s_soc}
\end{equation}

For the $\mathrm{K}$ point located at fractional coordinates $(3/8, 3/8, 3/4)$ in reciprocal space, the first-order derivatives are $\partial_{k_x} = \partial_{k_y} = 0$ and $\partial_{k_z} \approx -1.8646$. Because the longitudinal gradient $v_z$ is non-vanishing, the critical topology is determined by the 2D transverse Hessian matrix evaluated at the symmetry plane $k_z = 3\pi/2$:
\begin{equation}
H_{\mathrm{2D}} = \begin{pmatrix}
0.3199 & -1.4121 \\
-1.4121 & 0.3199 \\
\end{pmatrix},
\label{Hessian_s_K4_2D_soc}
\end{equation}
with eigenvalues $\lambda_{1} \approx 1.7321$ and $\lambda_{2} \approx -1.0922$. The opposite signs of the transverse eigenvalues indicate a downward- and upward-opening saddle point in the $k_x\text{-}k_y$ plane, unambiguously identifying it as an $N_1$-type n-o-VHS.

The explicit Taylor expansion at the $\mathrm{K}$ point, in terms of momentum displacements $\Delta k_x = k_x - \frac{3\pi}{4}$, $\Delta k_y = k_y - \frac{3\pi}{4}$, and $\Delta k_z = k_z - \frac{3\pi}{2}$, is expressed up to the second order as:
\begin{equation}
\begin{aligned}
\varepsilon_{4}(\mathbf{k}) \approx & -1.0146 - 1.8646 \Delta k_z \\
& + \frac{1}{2} \bigl( 0.3199 \Delta k_x^2 + 0.3199 \Delta k_y^2 - 0.2281 \Delta k_z^2 - 2.8243 \Delta k_x \Delta k_y+ 1.0922 \Delta k_x \Delta k_z + 1.0922 \Delta k_y \Delta k_z \bigr) \\
& + \dots + \mathcal{O}(\Delta \mathbf{k}^3)
\end{aligned}
\label{HVHS-case2_soc}
\end{equation}

Following the asymptotic scaling analysis established in the $N_0$-type section, the substantial group velocity along the $k_z$ direction mandates the energy scaling behavior $\Delta k_z \sim \mathcal{O}(E^1)$, whereas the transverse quadratic terms dictate $\Delta k_{x,y} \sim \mathcal{O}(E^{1/2})$. Consequently, the pure transverse quadratic terms ($\Delta k_x^2, \Delta k_x \Delta k_y, \Delta k_y^2$) scale as $\mathcal{O}(E^1)$ and act as the principal components. By contrast, the longitudinal quadratic term ($\Delta k_z^2 \sim \mathcal{O}(E^2)$) and the transverse-longitudinal mixed terms ($\Delta k_{x,y}\Delta k_z \sim \mathcal{O}(E^{1.5})$) represent strictly higher-order infinitesimals in the low-energy limit. 

By systematically truncating these $\ge \mathcal{O}(E^{1.5})$ terms and diagonalizing the $\mathcal{O}(E^1)$ transverse quadratic form via the eigenvectors of Eq.~\ref{Hessian_s_K4_2D_soc}, we introduce the generalized principal coordinates $q_1$ and $q_2$. Setting $q_z = \Delta k_z$, the effective Hamiltonian maps perfectly to the canonical hybrid saddle-linear form:
\begin{equation}
\varepsilon(\mathbf{q}) \approx \varepsilon_0 + v_z q_z + \frac{1}{2}\lambda_1 q_1^2 - \frac{1}{2}|\lambda_2| q_2^2
\end{equation}
Here, the reference energy is $\varepsilon_0 = -1.0146$ eV, the group velocity is $v_z = -1.8646$, and the planar curvatures are $\lambda_1 \approx 1.7321$ (convex) and $\lambda_2 \approx -1.0922$ (concave), confirming the $N_1$ saddle-point geometry. 

The DOS is obtained by integrating over the Brillouin zone with a longitudinal cutoff scale $R$. As rigorously derived analytically for the $N_1$-type singularity, the symmetric longitudinal integration over the transverse saddle-point logarithmic singularity strictly cancels out the first-order linear correction. The result follows a purely quadratic energy decay superimposed on a logarithmically enhanced constant background:
\begin{equation}
g(E) \simeq g_0 - \frac{1}{4\pi^3 v_z^2 R \sqrt{c_x c_y}} (E - \varepsilon_0)^2
\label{eq:N1_DOS_analytical}
\end{equation}
Mapping the parameters to the canonical form $c_x = \lambda_1/2$ and $c_y = |\lambda_2|/2$, the coefficient of the quadratic term simplifies to:
\begin{equation}
\frac{1}{2\pi^3 v_z^2 R \sqrt{\lambda_1 |\lambda_2|}}
\end{equation}
Using the numerical values extracted from the effective Hamiltonian ($|v_z| \approx 1.8646$ and $\sqrt{\lambda_1 |\lambda_2|} \approx \sqrt{1.7321 \times 1.0922} \approx 1.3754$), we explicitly evaluate the prefactor $\beta$ governing the quadratic decay curvature (where $\delta g(E) = -\frac{\beta}{R} (E - \varepsilon_0)^2$):
\begin{equation}
\beta = \frac{1}{2\pi^3 (1.8646)^2 (1.3754)} \approx 0.00337 \, 
\end{equation}
Thus, the DOS exhibits a symmetric quadratic decay modified by an inverse cutoff dependence ($1/R$). This distinct quadratic signature, devoid of any linear term, clearly distinguishes the $N_1$-type saddle-linear hybrid topology from the purely linear behavior of the $N_0$-type noncritical point.

\subsection{$N_2$-type}
Here, we take the tight-binding (TB) bands constructed by $s$ orbitals with spin-orbit coupling (SOC) as an example to evaluate the DOS near the $N_2$-type n-o-VHS of the pyrochlore lattice. When $e_1=0$, $t_{1}=-1$, and $t_{2}=0$ (see Fig.~\ref{figs4}), the relevant band $\varepsilon_{6}(\mathbf{k})$ is identified as twice the third root (energy-sorted from lowest to highest) of the characteristic polynomial $f(\varepsilon) = 0$.

The polynomial $f(\varepsilon)$ is explicitly defined as:
\begin{equation}
\begin{aligned}      
f(\varepsilon) &= 2\varepsilon^4 + \left( -12 - 2\cos(k_x) - 2\cos(k_x - k_y) - 2\cos(k_y) - 2\cos(k_x - k_z) - 2\cos(k_y - k_z) - 2\cos(k_z) \right)\varepsilon^2 \\
&\quad + \left( 8 + 4\cos(k_x) + 4\cos(k_x - k_y) + 4\cos(k_y) + 4\cos(k_x - k_z) + 4\cos(k_y - k_z) + 4\cos(k_z) \right)\varepsilon \\
&\quad + 3 - \cos(k_x) - \cos(k_x - k_y) - \cos(k_y) - \cos(k_x - k_z) + \cos(k_x - k_y - k_z) \\
&\quad - \cos(k_y - k_z) + \cos(k_x + k_y - k_z) - \cos(k_z) + \cos(k_x - k_y + k_z)
\end{aligned}
\label{eig_s-SOC_e6}
\end{equation}
Thus, the energy dispersion is formally given by $\varepsilon_{6}(\mathbf{k}) = 2 \cdot \operatorname{Root}\left[ f(\varepsilon), 3 \right]$.

For the $\mathrm{K}$ point located at $\mathbf{k} = (\frac{3\pi}{4}, \frac{3\pi}{4}, \frac{3\pi}{2})$, the first-order derivatives are $\partial_{k_x} = \partial_{k_y} = 0$ and $\partial_{k_z} \approx 0.4051$. Because the longitudinal gradient $v_z$ is non-vanishing, the critical topology is strictly determined by the 2D transverse Hessian matrix evaluated at the symmetry plane $k_z = 3\pi/2$:
\begin{equation}
H_{\mathrm{2D}} = \begin{pmatrix}
-0.149723 & -0.000240  \\
-0.000240 & -0.149723  \\
\end{pmatrix},
\label{Hessian_s-SOC_U6_2D}
\end{equation}
of which the two eigenvalues are $\lambda_{1} \approx -0.1499$ and $\lambda_{2} \approx -0.1495$. The negative signs of both planar eigenvalues unambiguously indicate a local maximum (downward-opening paraboloid) in the $k_x\text{-}k_y$ plane, signifying an $N_2$-type n-o-VHS at the $\mathrm{K}$ point.

The explicit Taylor expansion at the $\mathrm{K}$ point, in terms of the momentum displacements $\Delta k_x = k_x - \frac{3\pi}{4}$, $\Delta k_y = k_y - \frac{3\pi}{4}$, and $\Delta k_z = k_z - \frac{3\pi}{2}$, is expressed up to the second order as:
\begin{equation}
\begin{aligned}
\varepsilon_{6}(\mathbf{k})  & \approx 1.7330 + 0.4051 \Delta k_z \\
& + \frac{1}{2} \bigl( -0.1497 \Delta k_x^2 - 0.0005 \Delta k_x \Delta k_y - 0.1497 \Delta k_y^2 + 0.1500 \Delta k_x \Delta k_z \\
& + 0.1500 \Delta k_y \Delta k_z + 0.1969 \Delta k_z^2 \bigr) + \dots+ \mathcal{O}(\Delta \mathbf{k}^3).
\end{aligned}
\label{N2-HVHS-case3}
\end{equation}

Consistent with the asymptotic scaling analysis established in the preceding sections, the non-zero longitudinal group velocity ($v_z \approx 0.4051$) dictates the linear energy scaling $\Delta k_z \sim \mathcal{O}(E^1)$, while the transverse quadratic terms enforce $\Delta k_{x,y} \sim \mathcal{O}(E^{1/2})$. As a result, the pure transverse quadratic terms ($\Delta k_x^2, \Delta k_x \Delta k_y, \Delta k_y^2$) scale as $\mathcal{O}(E^1)$ and dominate the transverse dispersion. The longitudinal quadratic term ($\Delta k_z^2 \sim \mathcal{O}(E^2)$) and the transverse-longitudinal mixed terms ($\Delta k_{x,y} \Delta k_z \sim \mathcal{O}(E^{1.5})$) vanish much faster in the low-energy limit ($\Delta E \to 0$) and must be strictly truncated as subleading higher-order infinitesimals.

By discarding these $\ge \mathcal{O}(E^{1.5})$ terms and diagonalizing the $\mathcal{O}(E^1)$ transverse quadratic form using the eigenvectors of the 2D Hessian (Eq.~\ref{Hessian_s-SOC_U6_2D}), we map the effective Hamiltonian perfectly to the canonical decoupled $N_2$-type form in the generalized principal coordinates ($q_1, q_2, q_z = \Delta k_z$):
\begin{equation}
\varepsilon(\mathbf{q}) \approx \varepsilon_0 + v_z q_z - \frac{1}{2}|\lambda_1| q_1^2 - \frac{1}{2}|\lambda_2| q_2^2
\end{equation}
Here, the reference energy is $\varepsilon_0 = 1.7330$ eV, the longitudinal group velocity is $v_z = 0.4051$, and the planar curvatures correspond to the negative 2D Hessian eigenvalues: $\lambda_1 \approx -0.1500$ and $\lambda_2 \approx -0.1495$. 

Based on the universal DOS integration established in Table.~\ref{table_Ordinary_Noncritical_DOS} for the $N_2$-type topology ($c_x = |\lambda_1|/2$, $c_y = |\lambda_2|/2$, and $c_z = |v_z|$), the analytical DOS resolves to a linear step-like decay:
\begin{equation}
g(E) \simeq g_0 - \frac{1}{4\pi^2 |v_z| \sqrt{|\lambda_1 \lambda_2|}} (E - \varepsilon_0)
\label{eq:N2_DOS_analytical}
\end{equation}
Substituting the derived effective parameters ($|v_z| \approx 0.4051$ and $\sqrt{|\lambda_1 \lambda_2|} \approx 0.1497$), we calculate the decay slope $\gamma$ associated with the linear correction ($\delta g(E) = -\gamma(E - \varepsilon_0)$):
\begin{equation}
\gamma = \frac{1}{4\pi^2 (0.4051) (0.1497)} \approx 0.418 \,
\end{equation}
This confirms that the $N_2$-type singularity at the $\mathrm{K}$ point manifests as a negative linear decay ($\delta g(E) \propto -E$) superimposed on a constant background, accurately reflecting the phase-space geometry of a downward-opening transverse paraboloid hybridized with a longitudinal linear dispersion.

\section{DOS near the noncritical higher-order VHS of pyrochlore lattice}
\subsection{$S_1$-type}
Here, we classify the noncritical higher-order Van Hove singularity (n-HOVHS) into the $S_1$- and $S_2$-type, corresponding to one and two zero eigenvalues of the $2\times 2$ transverse Hessian matrix, respectively. For tight-binding (TB) bands of the pyrochlore lattice constructed by $s$ orbitals with spin-orbit coupling (SOC), when $e_1=0$, $t_1=-1$, and $t_2=-1$ (see Fig.~3 of the main text), the relevant band $\varepsilon_{2}(\mathbf{k})$ is given by:
\begin{gather*}
\varepsilon_{2}(\mathbf{k}) = -\sqrt{3} \sqrt{\alpha + \sqrt{\beta}}, \\
\text{with} \quad \alpha = 6 + \cos(k_x) + \cos(k_x - k_y) + \cos(k_y) + \cos(k_x - k_z) + \cos(k_y - k_z) + \cos(k_z), \\
\beta = \alpha^2 - 16\alpha + 72  - 8\bigl[ \cos(k_x - k_y - k_z) + \cos(k_x + k_y - k_z) + \cos(k_x - k_y + k_z) \bigr].
\label{eig_s_e4}
\end{gather*}

For the $\mathrm{K}$ point located at $\mathbf{k} = (\frac{3\pi}{4}, \frac{3\pi}{4}, \frac{3\pi}{2})$, the first-order derivatives are $\partial_{k_x}=\partial_{k_y}= 0$ and $\partial_{k_z} \approx 0.6398$. To demonstrate the intrinsic higher-order flatness of this critical point, we first evaluate the full 3D Hessian matrix via a Taylor series expansion:
\begin{equation}
H = \begin{pmatrix}
-0.3199080  & -0.3199080 & 0.3199080 \\
-0.3199080 & -0.3199080 & 0.3199080 \\
 0.3199080 &  0.3199080 & -0.0255714 \\
\end{pmatrix},
\label{Hessian_s-SOC_K2_3D}
\end{equation}
of which the three eigenvalues are $\lambda_{1} \approx -0.8795$, $\lambda_{2} \approx 0.2141$, and critically, $\lambda_{3} = 0$. 

Because the longitudinal gradient is non-vanishing ($v_z \approx 0.6398$), the topological classification relies on the projected 2D Hessian matrix evaluated at the stationary plane $k_z = \frac{3\pi}{2}$:
\begin{equation}
H_{\mathrm{2D}} = \begin{pmatrix}
-0.319908 & -0.319908  \\
-0.319908 & -0.319908  \\
\end{pmatrix},
\label{Hessian_s-SOC_K2_2D}
\end{equation}
of which the two planar eigenvalues are $\lambda_{1} \approx -0.6398$ and $\lambda_{2} = 0$. Unlike ordinary $N$-type singularities where both planar eigenvalues are non-zero, the presence of exactly one strictly zero eigenvalue alongside a non-zero eigenvalue clearly signifies an $S_1$-type noncritical higher-order Van Hove singularity.

To rigorously extract the effective Hamiltonian and properly handle the higher-order terms, we present the explicit Taylor expansion at the $\mathrm{K}$ point in terms of momentum displacements ($\Delta k_x, \Delta k_y, \Delta k_z$):
\begin{equation}
\begin{aligned}
\varepsilon_{2}(\mathbf{k})  & \approx -4.8990 + 0.6398 \Delta k_z \\
& + \frac{1}{2} \bigl( -0.3199 \Delta k_x^2 - 0.6398 \Delta k_x \Delta k_y - 0.3199 \Delta k_y^2 + 0.6398 \Delta k_x \Delta k_z \\
&\quad + 0.6398 \Delta k_y \Delta k_z - 0.0256 \Delta k_z^2 \bigr) \\
& + \frac{1}{6} \bigl( -0.7524 \Delta k_x^2 \Delta k_z - 0.7524 \Delta k_y^2 \Delta k_z + 0.6252 \Delta k_x \Delta k_y \Delta k_z + \dots \bigr) \\
& + \frac{1}{24} \bigl( 0.4368 \Delta k_x^4 + 0.4368 \Delta k_y^4 + 0.2123 \Delta k_z^4 \bigr) + \mathcal{O}(\Delta \mathbf{k}^5).
\end{aligned}
\label{HVHS-case4}
\end{equation}

We apply a higher-order asymptotic scaling analysis in the low-energy limit ($\Delta E \to 0$) to justify the truncation of the mixed terms. The substantial group velocity along the $z$-direction enforces $\Delta k_z \sim \mathcal{O}(E^1)$. In the transverse plane, by rotating to the principal eigenbasis $(q_1, q_2)$ of Eq.~\ref{Hessian_s-SOC_K2_2D}, the non-degenerate direction scales as $q_1 \sim \mathcal{O}(E^{1/2})$ due to $\lambda_1 \neq 0$. Crucially, because $\lambda_2 = 0$, the $q_2$ direction is flat at the quadratic level and is exclusively governed by higher-order terms (e.g., $q_2^{n_y}$ with $n_y \ge 3$), dictating a fractional scaling $q_2 \sim \mathcal{O}(E^{1/n_y})$.

Evaluating the cross terms under this dimensional hierarchy: the longitudinal-transverse couplings scale as $q_1 \Delta k_z \sim \mathcal{O}(E^{1.5})$ and $q_2 \Delta k_z \sim \mathcal{O}(E^{1 + 1/n_y})$. Both are strictly higher-order infinitesimals compared to the $\mathcal{O}(E^1)$ principal terms ($\Delta k_z$ and $q_1^2$). Similarly, the longitudinal quadratic term $\Delta k_z^2 \sim \mathcal{O}(E^2)$ is mathematically negligible. 

This rigorous scaling ensures that the longitudinal linear dispersion completely and safely decouples from the transverse plane. The effective dispersion naturally takes the form of a quasi-1D parabolic cylinder combined with a longitudinal linear term. While a strictly topological evaluation would retain the $\mathcal{O}(E^1)$ higher-order term for the flat $q_2$ direction, its exceptionally non-dispersive nature allows us to approximate it as a strictly flat band bounded by a characteristic Brillouin zone cutoff length $L_\perp$:
\begin{equation}
\varepsilon(\mathbf{q}) \approx \varepsilon_0 + v_z q_z - \frac{1}{2}|\lambda_1| q_1^2.
\end{equation}
Here, the reference energy is $\varepsilon_0 = -4.8990$ eV, $v_z = 0.6398$ is the group velocity along $k_z$, and $|\lambda_1| \approx 0.6398$ dictates the curvature of the downward-opening parabolic cylinder. 

The DOS is calculated by integrating over the Brillouin zone. The integration along the flat direction $q_2$ introduces the transverse cutoff $L_\perp$. Because the transverse parabolic dispersion is downward-opening (negative curvature), the resulting DOS exhibits a negative linear energy dependence superimposed on a constant background, entirely consistent with our theoretical derivation:
\begin{equation}
g(E) \simeq g_0 - \frac{L_\perp}{8\pi^3 |v_z| \sqrt{|v_z| R \cdot \frac{|\lambda_1|}{2}}} (E - \varepsilon_0),
\label{eq:S1_DOS_analytical}
\end{equation}
where $R$ is the cutoff wavevector along the $k_z$ direction. Substituting the numerical values ($|v_z| \approx 0.6398, |\lambda_1| \approx 0.6398$), we obtain the magnitude of the slope $\delta$ (where $\delta g(E) = -\delta (E - \varepsilon_0)$):
\begin{equation}
\delta = \frac{L_\perp}{8\pi^3 (0.6398) \sqrt{(0.6398)(0.3199) R}} \approx \frac{0.014 L_\perp}{\sqrt{R}}.
\end{equation}
Unlike the isotropic $N_0$- or $N_2$-type singularities, the slope $\delta$ for the $S_1$-type heavily depends on the geometric anisotropy ratio of the Brillouin zone cutoffs ($L_\perp/\sqrt{R}$), effectively reflecting the robust directional criticality of the flat band manifold.

\section{Topological Tunability in High-Dimensional Parameter Space}
\label{sec:viii_high_dimensional}
To explicitly demonstrate the robust tunability of the singularity types without requiring drastic changes to the primary hopping backbone ($t_2/t_1$), we expand the $s$-orbital tight-binding model with SOC on the pyrochlore lattice into a high-dimensional parameter space. In realistic physical systems, the electronic structure within the $Fd\bar{3}m$ (SG-227) space group is governed not only by nearest-neighbor hoppings ($t_1$ and $t_2$) but also by extended interactions, including second-neighbor spin-orbit couplings ($r_1, r_2, r_3$) and third-neighbor scalar hoppings ($s_1, s_2$). 

\subsection{Explicit Hamiltonian for the Pyrochlore Lattice with SOC and Extended Interactions}
Following the exact same block-matrix structure defined in Eq.~(\ref{TBH0}) for the pyrochlore framework, where the total Hamiltonian is partitioned into $4 \times 4$ sublattice blocks and each element $H_{ij}$ is a $2 \times 2$ sub-matrix spanning the spin space, we formulate the extended interactions. The diagonal elements $H_{ii}$ are proportional to the identity matrix $\sigma_0$. They cleanly capture the extended scalar third-neighbor interactions ($s_1, s_2$) within the tetrahedral network alongside the on-site energy $e_1$:
\begin{equation}
\begin{aligned}
H_{11} &= \Big\{ e_1 + 2s_1[\cos(k_x-k_y)+\cos(k_x-k_z)+\cos(k_y-k_z)] + 2s_2[\cos k_x + \cos k_y + \cos k_z] \Big\} \sigma_0 \\
H_{22} &= \Big\{ e_1 + 2s_1[\cos k_y + \cos k_z + \cos(k_y-k_z)] + 2s_2[\cos k_x + \cos(k_x-k_y) + \cos(k_x-k_z)] \Big\} \sigma_0 \\
H_{33} &= \Big\{ e_1 + 2s_1[\cos k_x + \cos k_z + \cos(k_x-k_z)] + 2s_2[\cos k_y + \cos(k_x-k_y) + \cos(k_y-k_z)] \Big\} \sigma_0 \\
H_{44} &= \Big\{ e_1 + 2s_1[\cos k_x + \cos k_y + \cos(k_x-k_y)] + 2s_2[\cos k_z + \cos(k_x-k_z) + \cos(k_y-k_z)] \Big\} \sigma_0
\end{aligned}
\end{equation}

The off-diagonal blocks $H_{ij}$ ($i \neq j$) incorporate the complex SOC and inter-sublattice hopping channels characteristic of the pyrochlore symmetry. Governed by time-reversal symmetry, they universally adopt the generic spin-coupled structure $H_{ij} = \begin{pmatrix} A_{ij} & B_{ij} \\ -B_{ij}^* & A_{ij}^* \end{pmatrix}$. Through rigorous analytic reduction of the spatial lattice vectors, the exact matrix elements are elegantly decoupled into trigonometric form as:
\begin{equation}
\begin{aligned}
A_{12} &= 2(-ir_2+r_3)[\cos(\tfrac{k_x}{2}-k_y)+\cos(\tfrac{k_x}{2}-k_z)] + 2(-it_1+t_2)\cos(\tfrac{k_x}{2}) \\
B_{12} &= 2(ir_1+r_2)\cos(\tfrac{k_x}{2}-k_y) + 2(-ir_1+r_2)\cos(\tfrac{k_x}{2}-k_z) + 2t_1\cos(\tfrac{k_x}{2}) \\[6pt]
A_{13} &= 2(ir_2+r_3)[\cos(k_x-\tfrac{k_y}{2}) + \cos(\tfrac{k_y}{2}-k_z)] + 2(it_1+t_2)\cos(\tfrac{k_y}{2}) \\
B_{13} &= 2(-r_1-ir_2)\cos(k_x-\tfrac{k_y}{2}) + 2(r_1-ir_2)\cos(\tfrac{k_y}{2}-k_z) - 2it_1\cos(\tfrac{k_y}{2}) \\[6pt]
A_{14} &= 2(-ir_1+r_3)\cos(k_y-\tfrac{k_z}{2}) + 2(ir_1+r_3)\cos(k_x-\tfrac{k_z}{2}) + 2t_2\cos(\tfrac{k_z}{2}) \\
B_{14} &= 2(-1+i)r_2\cos(k_x-\tfrac{k_z}{2}) - 2(1-i)r_2\cos(k_y-\tfrac{k_z}{2}) - 2(1-i)t_1\cos(\tfrac{k_z}{2}) \\[6pt]
A_{23} &= 2(-ir_1+r_3)\cos(\tfrac{k_x}{2}+\tfrac{k_y}{2}) + 2(ir_1+r_3)\cos(\tfrac{k_x}{2}+\tfrac{k_y}{2}-k_z) + 2t_2\cos(\tfrac{k_x}{2}-\tfrac{k_y}{2}) \\
B_{23} &= 2(1+i)r_2[\cos(\tfrac{k_x}{2}+\tfrac{k_y}{2}) + \cos(\tfrac{k_x}{2}+\tfrac{k_y}{2}-k_z)] + 2(1+i)t_1\cos(\tfrac{k_x}{2}-\tfrac{k_y}{2}) \\[6pt]
A_{24} &= 2(-ir_2+r_3)[\cos(\tfrac{k_x}{2}+\tfrac{k_z}{2}) + \cos(\tfrac{k_x}{2}-k_y+\tfrac{k_z}{2})] + 2(-it_1+t_2)\cos(\tfrac{k_x}{2}-\tfrac{k_z}{2}) \\
B_{24} &= 2(-r_1-ir_2)\cos(\tfrac{k_x}{2}-k_y+\tfrac{k_z}{2}) + 2(r_1-ir_2)\cos(\tfrac{k_x}{2}+\tfrac{k_z}{2}) - 2it_1\cos(\tfrac{k_x}{2}-\tfrac{k_z}{2}) \\[6pt]
A_{34} &= 2(ir_2+r_3)[\cos(\tfrac{k_y}{2}+\tfrac{k_z}{2}) + \cos(k_x-\tfrac{k_y}{2}-\tfrac{k_z}{2})] + 2(it_1+t_2)\cos(\tfrac{k_y}{2}-\tfrac{k_z}{2}) \\
B_{34} &= 2(-ir_1+r_2)\cos(\tfrac{k_y}{2}+\tfrac{k_z}{2}) + 2(ir_1+r_2)\cos(k_x-\tfrac{k_y}{2}-\tfrac{k_z}{2}) + 2t_1\cos(\tfrac{k_y}{2}-\tfrac{k_z}{2})
\end{aligned}
\end{equation}

\subsection{Specific Band Realization and the $S_2$-type Singularity}
To intuitively illustrate the profound impact of these extended interactions on the electronic structure, we evaluate the band dispersion and DOS for a specific, physically representative parameter set. Rather than selecting arbitrary values, we intentionally pinpoint a precise coordinate in the high-dimensional parameter space that lies exactly on an $S_2$-type topological hub (which will be systematically mapped in the subsequent phase diagrams). By setting the on-site energy $e_1 = 0$ and the spin-flip hopping $t_1 = -1.0 $ as the baseline, this specific critical point is reached at $t_2 = 0.4274$, $r_1 = 0.4078$, $r_2 = 0.3187$, $r_3 = -0.0825$, $s_1 = -0.4177$, and $s_2 = 0.5776$. 

\begin{figure}[htbp]
    \centering
    \includegraphics[width=0.75\linewidth]{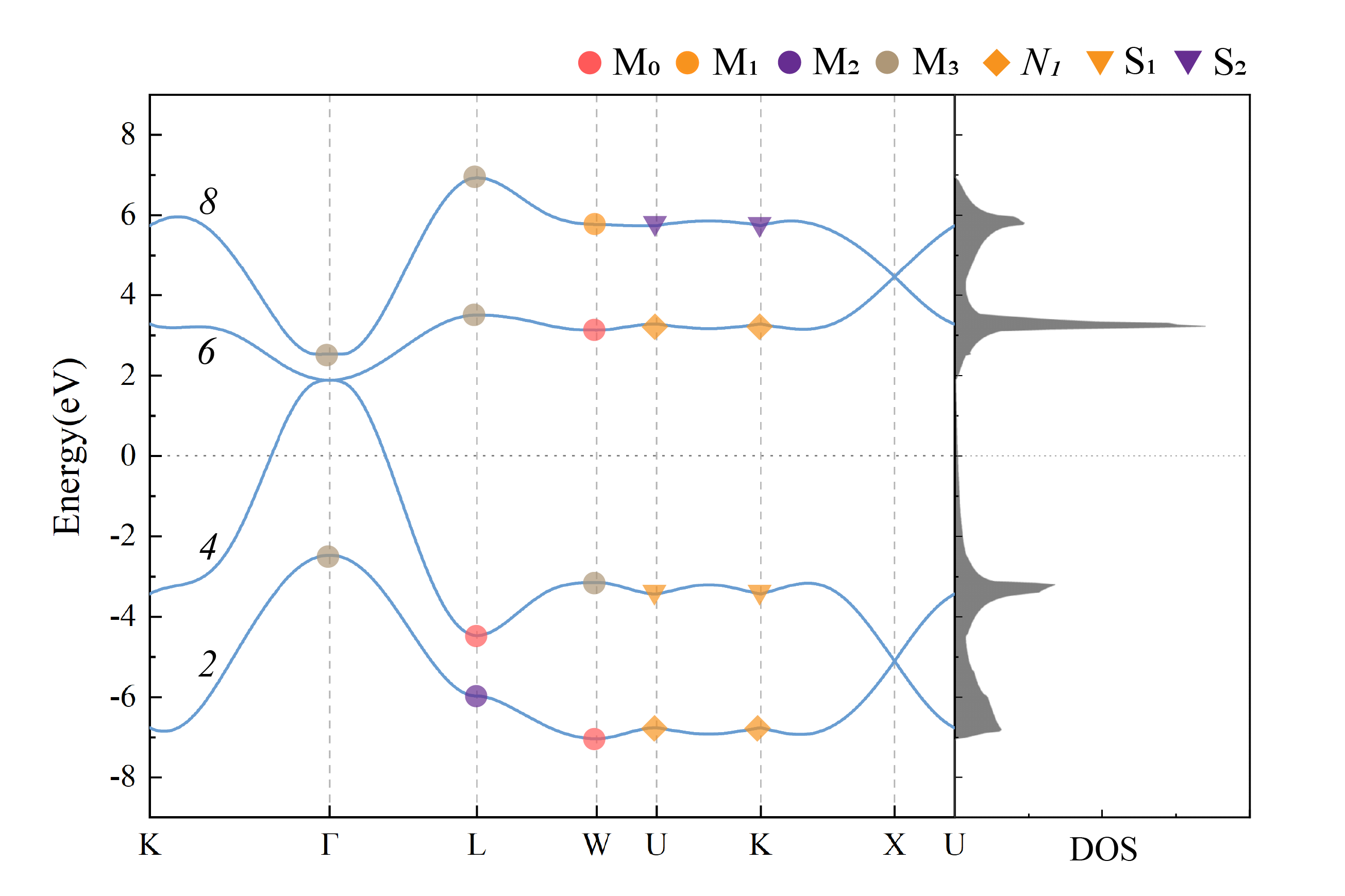} 
    \caption{\textbf{Band structure and corresponding density of states (DOS) for the pyrochlore lattice with extended interactions.} The specific tight-binding parameters are detailed in the text above. The blue curves represent the TB band structure along the high-symmetry path; numbers (2, 4, 6, 8) label distinct bands, and colored symbols (legend-coded) mark the corresponding Van Hove singularity types. The right panel displays the calculated DOS.}
    \label{figs8}
\end{figure}

By diagonalizing the $8 \times 8$ Hamiltonian, we obtain the electronic band structure and the corresponding DOS, as presented in Fig.~\ref{figs8}. The inclusion of extended interactions introduces rich topological features across multiple bands. Strikingly, on the uppermost bands 7 and 8, $S_2$-type noncritical higher-order Van Hove singularities (purple downward-pointing triangles) are perfectly realized at both the high-symmetry $\mathrm{K}$ and $\mathrm{U}$ points. Because both transverse planar curvatures mathematically vanish simultaneously ($\lambda_1 = \lambda_2 = 0$) while the longitudinal gradient remains non-zero, this $S_2$-type singularity natively generates an extended quasi-2D transverse flat manifold. As shown in the right panel of Fig.~\ref{figs8}, this topological feature directly translates into a pronounced DOS peak characterized by sub-leading quadratic corrections, perfectly corroborating our analytical derivations.

Furthermore, the intermediate bands 3 and 4 also host noncritical higher-order features, manifesting as $S_1$-type singularities (orange downward-pointing triangles) at both $\mathrm{K}$ and $\mathrm{U}$ points. The simultaneous emergence of these singularities at both $\mathrm{K}$ and $\mathrm{U}$ across different bands consistently reflects their nature as symmetry-enforced topological partners within the $Fd\bar{3}m$ space group.

\subsection{Phase Diagrams and Topological Boundaries}
\begin{figure}[htbp]
    \centering
    \includegraphics[width=0.98\linewidth]{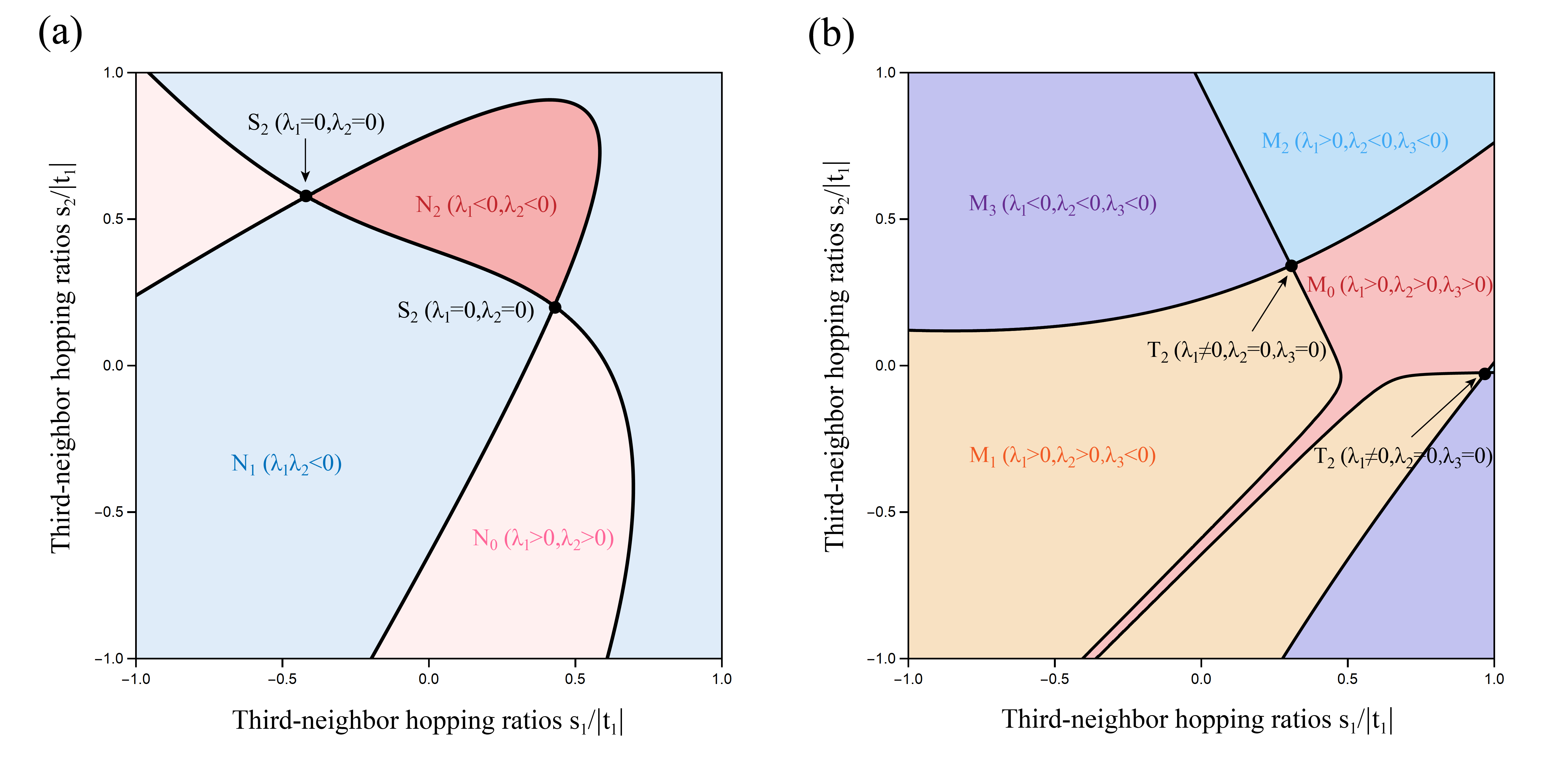} 
    \caption{\textbf{Cross-sectional topological phase diagrams of Van Hove singularities for the SG-227 pyrochlore lattice (bands 7 and 8) in the high-dimensional parameter space.} To demonstrate the tunability of singularity types via small variations, the phase space is mapped by the dimensionless third-neighbor hoppings $s_1/|t_1|$ and $s_2/|t_1|$. The background parameters are fixed identically to those in Fig.~\ref{figs8}: $t_2/|t_1| = 0.427$, $r_1/|t_1| = 0.408$, $r_2/|t_1| = 0.319$, and $r_3/|t_1| = -0.083$, ensuring that the nearest-neighbor scale $|t_1|$ strictly dominates. (a) Noncritical VHSs at the $K$ point. Permitted by local symmetries, a non-vanishing longitudinal band gradient remains natively present. Colored regions denote distinct $N$-type noncritical phases governed by the signs of the reduced $2\times2$ Hessian eigenvalues ($\lambda_1, \lambda_2$ evaluated in the stationary transverse plane): $N_0$ (pink), $N_1$ (blue), and $N_2$ (red). Solid black lines represent the $S_1$-type higher-order noncritical boundaries ($\lambda_1\lambda_2 = 0$), whose intersections pinpoint the $S_2$-type transverse flat-band hubs ($\lambda_1 = \lambda_2 = 0$). (b) Critical VHSs at the $L$ point. At this Time-Reversal Invariant Momentum, the band gradient strictly vanishes. The colored regions denote distinct ordinary $M$-type critical phases determined by the Morse index (the number of negative eigenvalues among the three 3D Hessian principal curvatures $\lambda_1, \lambda_2, \lambda_3$): $M_0$ (red), $M_1$ (orange), $M_2$ (light blue), and $M_3$ (dark purple). The solid black lines represent the $T_1$-type higher-order critical boundaries ($\operatorname{Det}(H) = 0$, where exactly one principal curvature vanishes, i.e., $\lambda_i = 0$), whose intersections pinpoint the $T_2$-type critical hubs featuring the simultaneous vanishing of two principal curvatures ($\lambda_i = \lambda_j = 0$).}
    \label{figs9}
\end{figure}

Recognizing that the specific $S_2$ realization in Fig.~\ref{figs8} is merely one singular point within a vast topological landscape, we investigate the continuous evolution of these singularities. When these realistic extended interactions are considered, the isolated topological critical points expand into continuous, densely packed topological boundaries (manifolds). We extract two representative two-dimensional cross-sectional phase diagrams (slices) from this high-dimensional space, as summarized in Fig.~\ref{figs9}. Both slices share the exact same physically optimized background parameters used in Fig.~\ref{figs8}---where the primary nearest-neighbor hopping $|t_1|$ strictly dominates ($t_2/|t_1| = 0.427$, $r_1/|t_1| = 0.408$, $r_2/|t_1| = 0.319$, and $r_3/|t_1| = -0.083$)---and are mapped as continuous functions of the dimensionless third-neighbor hoppings $s_1/|t_1|$ and $s_2/|t_1|$.

For the noncritical singularities at the $\mathrm{K}$ point (Fig.~\ref{figs9}(a)), where a non-zero longitudinal gradient is natively permitted by local point-group symmetries, our 2D slice for the highest valence bands reveals that noncritical phases (e.g., $N_1, N_0$) are densely packed. They are separated by continuous higher-order boundaries ($S_1$-type, where exactly one planar curvature vanishes). Crucially, the intersection hubs of these $S_1$ boundaries precisely pinpoint the $S_2$-type transverse flat-band hubs (where both planar curvatures vanish simultaneously), with the specific parameter set from Fig.~\ref{figs8} serving as one explicit realization of such a hub.

Conversely, at the highly symmetric $\mathrm{L}$ point (Fig.~\ref{figs9}(b)), where the band gradient strictly vanishes in all directions, the boundaries separating ordinary critical VHSs (e.g., $M_1$ and $M_2$) naturally manifest as continuous manifolds of the higher-order critical $T_1$-type singularities. The intersections of these $T_1$ manifolds precisely pinpoint the $T_2$-type critical hubs. It is worth noting that the ultimate $T_3$-type singularity does not appear in this specific 2D slice. Mathematically, simultaneously zeroing three independent principal curvatures is a co-dimension 3 topological condition. One would not expect this co-dimension 3 singularity to accidentally fall within a generic 2D cross-section where the primary backbone is intentionally locked away from the strong geometric frustration limit. This multidimensional phase space definitively proves that the pyrochlore structure natively supports the continuous evolution and realization of the complete higher-order singularity hierarchy, including arbitrary higher-order vanishing derivatives.

This high-dimensional phase space mapping conclusively demonstrates that if a synthesized compound naturally lies in the vicinity of these dense topological boundaries, experimentally accessible small physical perturbations (such as hydrostatic pressure or epitaxial strain) that subtly modulate the extended hopping integrals ($s_1, s_2$) are completely sufficient to drive the system across the phase boundaries.

\makeatletter
\setcounter{NAT@ctr}{0}
\renewcommand{\bibnumfmt}[1]{[1]}
\renewcommand{\citenumfont}[1]{1}
\makeatother


\begin{thebibliography}{55}%
\makeatletter
\providecommand \@ifxundefined [1]{%
 \@ifx{#1\undefined}
}%
\providecommand \@ifnum [1]{%
 \ifnum #1\expandafter \@firstoftwo
 \else \expandafter \@secondoftwo
 \fi
}%
\providecommand \@ifx [1]{%
 \ifx #1\expandafter \@firstoftwo
 \else \expandafter \@secondoftwo
 \fi
}%
\providecommand \natexlab [1]{#1}%
\providecommand \enquote  [1]{``#1''}%
\providecommand \bibnamefont  [1]{#1}%
\providecommand \bibfnamefont [1]{#1}%
\providecommand \citenamefont [1]{#1}%
\providecommand \href@noop [0]{\@secondoftwo}%
\providecommand \href [0]{\begingroup \@sanitize@url \@href}%
\providecommand \@href[1]{\@@startlink{#1}\@@href}%
\providecommand \@@href[1]{\endgroup#1\@@endlink}%
\providecommand \@sanitize@url [0]{\catcode `\\12\catcode `\$12\catcode `\&12\catcode `\#12\catcode `\^12\catcode `\_12\catcode `\%12\relax}%
\providecommand \@@startlink[1]{}%
\providecommand \@@endlink[0]{}%
\providecommand \url  [0]{\begingroup\@sanitize@url \@url }%
\providecommand \@url [1]{\endgroup\@href {#1}{\urlprefix }}%
\providecommand \urlprefix  [0]{URL }%
\providecommand \Eprint [0]{\href }%
\providecommand \doibase [0]{https://doi.org/}%
\providecommand \selectlanguage [0]{\@gobble}%
\providecommand \bibinfo  [0]{\@secondoftwo}%
\providecommand \bibfield  [0]{\@secondoftwo}%
\providecommand \translation [1]{[#1]}%
\providecommand \BibitemOpen [0]{}%
\providecommand \bibitemStop [0]{}%
\providecommand \bibitemNoStop [0]{.\EOS\space}%
\providecommand \EOS [0]{\spacefactor3000\relax}%
\providecommand \BibitemShut  [1]{\csname bibitem#1\endcsname}%
\let\auto@bib@innerbib\@empty
\bibitem [{\citenamefont {Van~Hove}(1953)}]{van1953occurrence}%
  \BibitemOpen
  \bibfield  {author} {\bibinfo {author} {\bibfnamefont {L.}~\bibnamefont {Van~Hove}},\ }\bibfield  {title} {\bibinfo {title} {The occurrence of singularities in the elastic frequency distribution of a crystal},\ }\href {https://doi.org/10.1103/PhysRev.89.1189} {\bibfield  {journal} {\bibinfo  {journal} {Phys. Rev}\ }\textbf {\bibinfo {volume} {89}},\ \bibinfo {pages} {1189} (\bibinfo {year} {1953})}\BibitemShut {NoStop}%
\bibitem [{\citenamefont {Yuan}\ and\ \citenamefont {Fu}(2020)}]{yuan2020classification}%
  \BibitemOpen
  \bibfield  {author} {\bibinfo {author} {\bibfnamefont {N.~F.}\ \bibnamefont {Yuan}}\ and\ \bibinfo {author} {\bibfnamefont {L.}~\bibnamefont {Fu}},\ }\bibfield  {title} {\bibinfo {title} {Classification of critical points in energy bands based on topology, scaling, and symmetry},\ }\href {https://doi.org/10.1103/PhysRevB.101.125120} {\bibfield  {journal} {\bibinfo  {journal} {Phys. Rev. B}\ }\textbf {\bibinfo {volume} {101}},\ \bibinfo {pages} {125120} (\bibinfo {year} {2020})}\BibitemShut {NoStop}%
\bibitem [{\citenamefont {Patra}\ \emph {et~al.}(2025)\citenamefont {Patra}, \citenamefont {Mukherjee},\ and\ \citenamefont {Singh}}]{patra2025high}%
  \BibitemOpen
  \bibfield  {author} {\bibinfo {author} {\bibfnamefont {B.}~\bibnamefont {Patra}}, \bibinfo {author} {\bibfnamefont {A.}~\bibnamefont {Mukherjee}},\ and\ \bibinfo {author} {\bibfnamefont {B.}~\bibnamefont {Singh}},\ }\bibfield  {title} {\bibinfo {title} {High-order van hove singularities and nematic instability in the kagome superconductor {$CsTi_3Bi_5$}},\ }\href {https://doi.org/10.1103/PhysRevB.111.045135} {\bibfield  {journal} {\bibinfo  {journal} {Phys. Rev. B}\ }\textbf {\bibinfo {volume} {111}},\ \bibinfo {pages} {045135} (\bibinfo {year} {2025})}\BibitemShut {NoStop}%
\bibitem [{\citenamefont {Efremov}\ \emph {et~al.}(2019)\citenamefont {Efremov}, \citenamefont {Shtyk}, \citenamefont {Rost}, \citenamefont {Chamon}, \citenamefont {Mackenzie},\ and\ \citenamefont {Betouras}}]{efremov2019multicritical}%
  \BibitemOpen
  \bibfield  {author} {\bibinfo {author} {\bibfnamefont {D.~V.}\ \bibnamefont {Efremov}}, \bibinfo {author} {\bibfnamefont {A.}~\bibnamefont {Shtyk}}, \bibinfo {author} {\bibfnamefont {A.~W.}\ \bibnamefont {Rost}}, \bibinfo {author} {\bibfnamefont {C.}~\bibnamefont {Chamon}}, \bibinfo {author} {\bibfnamefont {A.~P.}\ \bibnamefont {Mackenzie}},\ and\ \bibinfo {author} {\bibfnamefont {J.~J.}\ \bibnamefont {Betouras}},\ }\bibfield  {title} {\bibinfo {title} {Multicritical fermi surface topological transitions},\ }\href {https://doi.org/10.1103/PhysRevLett.123.207202} {\bibfield  {journal} {\bibinfo  {journal} {Phys. Rev. Lett}\ }\textbf {\bibinfo {volume} {123}},\ \bibinfo {pages} {207202} (\bibinfo {year} {2019})}\BibitemShut {NoStop}%
\bibitem [{\citenamefont {Yuan}\ \emph {et~al.}(2019)\citenamefont {Yuan}, \citenamefont {Isobe},\ and\ \citenamefont {Fu}}]{yuan2019magic}%
  \BibitemOpen
  \bibfield  {author} {\bibinfo {author} {\bibfnamefont {N.~F.}\ \bibnamefont {Yuan}}, \bibinfo {author} {\bibfnamefont {H.}~\bibnamefont {Isobe}},\ and\ \bibinfo {author} {\bibfnamefont {L.}~\bibnamefont {Fu}},\ }\bibfield  {title} {\bibinfo {title} {Magic of high-order van hove singularity},\ }\href {https://doi.org/10.1038/s41467-019-13670-9} {\bibfield  {journal} {\bibinfo  {journal} {Nat. Commun}\ }\textbf {\bibinfo {volume} {10}},\ \bibinfo {pages} {5769} (\bibinfo {year} {2019})}\BibitemShut {NoStop}%
\bibitem [{\citenamefont {Classen}\ and\ \citenamefont {Betouras}(2025)}]{classen2025high}%
  \BibitemOpen
  \bibfield  {author} {\bibinfo {author} {\bibfnamefont {L.}~\bibnamefont {Classen}}\ and\ \bibinfo {author} {\bibfnamefont {J.~J.}\ \bibnamefont {Betouras}},\ }\bibfield  {title} {\bibinfo {title} {High-order van hove singularities and their connection to flat bands},\ }\href {https://doi.org/10.1146/annurev-conmatphys-042924-015000} {\bibfield  {journal} {\bibinfo  {journal} {Annu Rev Condens Matter Phys}\ }\textbf {\bibinfo {volume} {16}},\ \bibinfo {pages} {229} (\bibinfo {year} {2025})}\BibitemShut {NoStop}%
\bibitem [{\citenamefont {Hausoel}\ \emph {et~al.}(2017)\citenamefont {Hausoel}, \citenamefont {Karolak}, \citenamefont {{\c{S}}a{\c{s}}$\iota$o{\u{g}}lu}, \citenamefont {Lichtenstein}, \citenamefont {Held}, \citenamefont {Katanin}, \citenamefont {Toschi},\ and\ \citenamefont {Sangiovanni}}]{hausoel2017local}%
  \BibitemOpen
  \bibfield  {author} {\bibinfo {author} {\bibfnamefont {A.}~\bibnamefont {Hausoel}}, \bibinfo {author} {\bibfnamefont {M.}~\bibnamefont {Karolak}}, \bibinfo {author} {\bibfnamefont {E.}~\bibnamefont {{\c{S}}a{\c{s}}$\iota$o{\u{g}}lu}}, \bibinfo {author} {\bibfnamefont {A.}~\bibnamefont {Lichtenstein}}, \bibinfo {author} {\bibfnamefont {K.}~\bibnamefont {Held}}, \bibinfo {author} {\bibfnamefont {A.}~\bibnamefont {Katanin}}, \bibinfo {author} {\bibfnamefont {A.}~\bibnamefont {Toschi}},\ and\ \bibinfo {author} {\bibfnamefont {G.}~\bibnamefont {Sangiovanni}},\ }\bibfield  {title} {\bibinfo {title} {Local magnetic moments in iron and nickel at ambient and earth’s core conditions},\ }\href {https://doi.org/10.1038/ncomms16062} {\bibfield  {journal} {\bibinfo  {journal} {Nat. Commun}\ }\textbf {\bibinfo {volume} {8}},\ \bibinfo {pages} {16062} (\bibinfo {year} {2017})}\BibitemShut {NoStop}%
\bibitem [{\citenamefont {Sala}\ \emph {et~al.}(2021)\citenamefont {Sala}, \citenamefont {Stone}, \citenamefont {Rai}, \citenamefont {May}, \citenamefont {Laurell}, \citenamefont {Garlea}, \citenamefont {Butch}, \citenamefont {Lumsden}, \citenamefont {Ehlers}, \citenamefont {Pokharel} \emph {et~al.}}]{sala2021van}%
  \BibitemOpen
  \bibfield  {author} {\bibinfo {author} {\bibfnamefont {G.}~\bibnamefont {Sala}}, \bibinfo {author} {\bibfnamefont {M.~B.}\ \bibnamefont {Stone}}, \bibinfo {author} {\bibfnamefont {B.~K.}\ \bibnamefont {Rai}}, \bibinfo {author} {\bibfnamefont {A.~F.}\ \bibnamefont {May}}, \bibinfo {author} {\bibfnamefont {P.}~\bibnamefont {Laurell}}, \bibinfo {author} {\bibfnamefont {V.~O.}\ \bibnamefont {Garlea}}, \bibinfo {author} {\bibfnamefont {N.~P.}\ \bibnamefont {Butch}}, \bibinfo {author} {\bibfnamefont {M.~D.}\ \bibnamefont {Lumsden}}, \bibinfo {author} {\bibfnamefont {G.}~\bibnamefont {Ehlers}}, \bibinfo {author} {\bibfnamefont {G.}~\bibnamefont {Pokharel}}, \emph {et~al.},\ }\bibfield  {title} {\bibinfo {title} {Van hove singularity in the magnon spectrum of the antiferromagnetic quantum honeycomb lattice},\ }\href {https://doi.org/10.1038/s41467-020-20335-5} {\bibfield  {journal} {\bibinfo  {journal} {Nat. Commun}\ }\textbf {\bibinfo {volume} {12}},\ \bibinfo {pages} {171} (\bibinfo {year} {2021})}\BibitemShut
  {NoStop}%
\bibitem [{\citenamefont {Makogon}\ \emph {et~al.}(2011)\citenamefont {Makogon}, \citenamefont {Van~Gelderen}, \citenamefont {Rold{\'a}n},\ and\ \citenamefont {Smith}}]{makogon2011spin}%
  \BibitemOpen
  \bibfield  {author} {\bibinfo {author} {\bibfnamefont {D.}~\bibnamefont {Makogon}}, \bibinfo {author} {\bibfnamefont {R.}~\bibnamefont {Van~Gelderen}}, \bibinfo {author} {\bibfnamefont {R.}~\bibnamefont {Rold{\'a}n}},\ and\ \bibinfo {author} {\bibfnamefont {C.~M.}\ \bibnamefont {Smith}},\ }\bibfield  {title} {\bibinfo {title} {Spin-density-wave instability in graphene doped near the van hove singularity},\ }\href {https://doi.org/10.1103/PhysRevB.84.125404} {\bibfield  {journal} {\bibinfo  {journal} {Phys. Rev. B}\ }\textbf {\bibinfo {volume} {84}},\ \bibinfo {pages} {125404} (\bibinfo {year} {2011})}\BibitemShut {NoStop}%
\bibitem [{\citenamefont {Liu}\ \emph {et~al.}(2018)\citenamefont {Liu}, \citenamefont {Zhang}, \citenamefont {Chen},\ and\ \citenamefont {Yang}}]{liu2018chiral}%
  \BibitemOpen
  \bibfield  {author} {\bibinfo {author} {\bibfnamefont {C.-C.}\ \bibnamefont {Liu}}, \bibinfo {author} {\bibfnamefont {L.-D.}\ \bibnamefont {Zhang}}, \bibinfo {author} {\bibfnamefont {W.-Q.}\ \bibnamefont {Chen}},\ and\ \bibinfo {author} {\bibfnamefont {F.}~\bibnamefont {Yang}},\ }\bibfield  {title} {\bibinfo {title} {Chiral spin density wave and d+ id superconductivity in the magic-angle-twisted bilayer graphene},\ }\href {https://doi.org/10.1103/PhysRevLett.121.217001} {\bibfield  {journal} {\bibinfo  {journal} {Phys. Rev. Lett}\ }\textbf {\bibinfo {volume} {121}},\ \bibinfo {pages} {217001} (\bibinfo {year} {2018})}\BibitemShut {NoStop}%
\bibitem [{\citenamefont {Isobe}\ \emph {et~al.}(2018)\citenamefont {Isobe}, \citenamefont {Yuan},\ and\ \citenamefont {Fu}}]{isobe2018unconventional}%
  \BibitemOpen
  \bibfield  {author} {\bibinfo {author} {\bibfnamefont {H.}~\bibnamefont {Isobe}}, \bibinfo {author} {\bibfnamefont {N.~F.}\ \bibnamefont {Yuan}},\ and\ \bibinfo {author} {\bibfnamefont {L.}~\bibnamefont {Fu}},\ }\bibfield  {title} {\bibinfo {title} {Unconventional superconductivity and density waves in twisted bilayer graphene},\ }\href {https://doi.org/10.1103/PhysRevX.8.041041} {\bibfield  {journal} {\bibinfo  {journal} {Phys. Rev. X}\ }\textbf {\bibinfo {volume} {8}},\ \bibinfo {pages} {041041} (\bibinfo {year} {2018})}\BibitemShut {NoStop}%
\bibitem [{\citenamefont {Wilson}\ and\ \citenamefont {Ortiz}(2024)}]{wilson2024v3sb5}%
  \BibitemOpen
  \bibfield  {author} {\bibinfo {author} {\bibfnamefont {S.~D.}\ \bibnamefont {Wilson}}\ and\ \bibinfo {author} {\bibfnamefont {B.~R.}\ \bibnamefont {Ortiz}},\ }\bibfield  {title} {\bibinfo {title} {{$AV_3Sb_5$} kagome superconductors},\ }\href {https://doi.org/10.1038/s41578-024-00677-y} {\bibfield  {journal} {\bibinfo  {journal} {Nat. Rev. Mater}\ }\textbf {\bibinfo {volume} {9}},\ \bibinfo {pages} {420} (\bibinfo {year} {2024})}\BibitemShut {NoStop}%
\bibitem [{\citenamefont {Yin}\ \emph {et~al.}(2022)\citenamefont {Yin}, \citenamefont {Lian},\ and\ \citenamefont {Hasan}}]{yin2022topological}%
  \BibitemOpen
  \bibfield  {author} {\bibinfo {author} {\bibfnamefont {J.-X.}\ \bibnamefont {Yin}}, \bibinfo {author} {\bibfnamefont {B.}~\bibnamefont {Lian}},\ and\ \bibinfo {author} {\bibfnamefont {M.~Z.}\ \bibnamefont {Hasan}},\ }\bibfield  {title} {\bibinfo {title} {Topological kagome magnets and superconductors},\ }\href {https://doi.org/10.1038/s41586-022-05516-0} {\bibfield  {journal} {\bibinfo  {journal} {Nat}\ }\textbf {\bibinfo {volume} {612}},\ \bibinfo {pages} {647} (\bibinfo {year} {2022})}\BibitemShut {NoStop}%
\bibitem [{\citenamefont {Teng}\ \emph {et~al.}(2022)\citenamefont {Teng}, \citenamefont {Chen}, \citenamefont {Ye}, \citenamefont {Rosenberg}, \citenamefont {Liu}, \citenamefont {Yin}, \citenamefont {Jiang}, \citenamefont {Oh}, \citenamefont {Hasan}, \citenamefont {Neubauer} \emph {et~al.}}]{teng2022discovery}%
  \BibitemOpen
  \bibfield  {author} {\bibinfo {author} {\bibfnamefont {X.}~\bibnamefont {Teng}}, \bibinfo {author} {\bibfnamefont {L.}~\bibnamefont {Chen}}, \bibinfo {author} {\bibfnamefont {F.}~\bibnamefont {Ye}}, \bibinfo {author} {\bibfnamefont {E.}~\bibnamefont {Rosenberg}}, \bibinfo {author} {\bibfnamefont {Z.}~\bibnamefont {Liu}}, \bibinfo {author} {\bibfnamefont {J.-X.}\ \bibnamefont {Yin}}, \bibinfo {author} {\bibfnamefont {Y.-X.}\ \bibnamefont {Jiang}}, \bibinfo {author} {\bibfnamefont {J.~S.}\ \bibnamefont {Oh}}, \bibinfo {author} {\bibfnamefont {M.~Z.}\ \bibnamefont {Hasan}}, \bibinfo {author} {\bibfnamefont {K.~J.}\ \bibnamefont {Neubauer}}, \emph {et~al.},\ }\bibfield  {title} {\bibinfo {title} {Discovery of charge density wave in a kagome lattice antiferromagnet},\ }\href {https://doi.org/10.1038/s41586-022-05034-z} {\bibfield  {journal} {\bibinfo  {journal} {Nat}\ }\textbf {\bibinfo {volume} {609}},\ \bibinfo {pages} {490} (\bibinfo {year} {2022})}\BibitemShut {NoStop}%
\bibitem [{\citenamefont {Teng}\ \emph {et~al.}(2023)\citenamefont {Teng}, \citenamefont {Oh}, \citenamefont {Tan}, \citenamefont {Chen}, \citenamefont {Huang}, \citenamefont {Gao}, \citenamefont {Yin}, \citenamefont {Chu}, \citenamefont {Hashimoto}, \citenamefont {Lu} \emph {et~al.}}]{teng2023magnetism}%
  \BibitemOpen
  \bibfield  {author} {\bibinfo {author} {\bibfnamefont {X.}~\bibnamefont {Teng}}, \bibinfo {author} {\bibfnamefont {J.~S.}\ \bibnamefont {Oh}}, \bibinfo {author} {\bibfnamefont {H.}~\bibnamefont {Tan}}, \bibinfo {author} {\bibfnamefont {L.}~\bibnamefont {Chen}}, \bibinfo {author} {\bibfnamefont {J.}~\bibnamefont {Huang}}, \bibinfo {author} {\bibfnamefont {B.}~\bibnamefont {Gao}}, \bibinfo {author} {\bibfnamefont {J.-X.}\ \bibnamefont {Yin}}, \bibinfo {author} {\bibfnamefont {J.-H.}\ \bibnamefont {Chu}}, \bibinfo {author} {\bibfnamefont {M.}~\bibnamefont {Hashimoto}}, \bibinfo {author} {\bibfnamefont {D.}~\bibnamefont {Lu}}, \emph {et~al.},\ }\bibfield  {title} {\bibinfo {title} {Magnetism and charge density wave order in kagome {$FeGe$}},\ }\href {https://doi.org/10.1038/s41567-023-01985-w} {\bibfield  {journal} {\bibinfo  {journal} {Nat. Phys}\ }\textbf {\bibinfo {volume} {19}},\ \bibinfo {pages} {814} (\bibinfo {year} {2023})}\BibitemShut {NoStop}%
\bibitem [{\citenamefont {Hu}\ \emph {et~al.}(2024)\citenamefont {Hu}, \citenamefont {Ma}, \citenamefont {Li}, \citenamefont {Jiang}, \citenamefont {Gawryluk}, \citenamefont {Hu}, \citenamefont {Teyssier}, \citenamefont {Multian}, \citenamefont {Yin}, \citenamefont {Xu} \emph {et~al.}}]{hu2024phonon}%
  \BibitemOpen
  \bibfield  {author} {\bibinfo {author} {\bibfnamefont {Y.}~\bibnamefont {Hu}}, \bibinfo {author} {\bibfnamefont {J.}~\bibnamefont {Ma}}, \bibinfo {author} {\bibfnamefont {Y.}~\bibnamefont {Li}}, \bibinfo {author} {\bibfnamefont {Y.}~\bibnamefont {Jiang}}, \bibinfo {author} {\bibfnamefont {D.~J.}\ \bibnamefont {Gawryluk}}, \bibinfo {author} {\bibfnamefont {T.}~\bibnamefont {Hu}}, \bibinfo {author} {\bibfnamefont {J.}~\bibnamefont {Teyssier}}, \bibinfo {author} {\bibfnamefont {V.}~\bibnamefont {Multian}}, \bibinfo {author} {\bibfnamefont {Z.}~\bibnamefont {Yin}}, \bibinfo {author} {\bibfnamefont {S.}~\bibnamefont {Xu}}, \emph {et~al.},\ }\bibfield  {title} {\bibinfo {title} {Phonon promoted charge density wave in topological kagome metal {$ScV_6Sn_6$}},\ }\href {https://doi.org/10.1038/s41467-024-45859-y} {\bibfield  {journal} {\bibinfo  {journal} {Nat. Commun}\ }\textbf {\bibinfo {volume} {15}},\ \bibinfo {pages} {1658} (\bibinfo {year} {2024})}\BibitemShut {NoStop}%
\bibitem [{\citenamefont {Tan}\ \emph {et~al.}(2021)\citenamefont {Tan}, \citenamefont {Liu}, \citenamefont {Wang},\ and\ \citenamefont {Yan}}]{tan2021charge}%
  \BibitemOpen
  \bibfield  {author} {\bibinfo {author} {\bibfnamefont {H.}~\bibnamefont {Tan}}, \bibinfo {author} {\bibfnamefont {Y.}~\bibnamefont {Liu}}, \bibinfo {author} {\bibfnamefont {Z.}~\bibnamefont {Wang}},\ and\ \bibinfo {author} {\bibfnamefont {B.}~\bibnamefont {Yan}},\ }\bibfield  {title} {\bibinfo {title} {Charge density waves and electronic properties of superconducting kagome metals},\ }\href {https://doi.org/10.1103/PhysRevLett.127.046401} {\bibfield  {journal} {\bibinfo  {journal} {Phys. Rev. Lett}\ }\textbf {\bibinfo {volume} {127}},\ \bibinfo {pages} {046401} (\bibinfo {year} {2021})}\BibitemShut {NoStop}%
\bibitem [{\citenamefont {Gonzalez}\ and\ \citenamefont {Stauber}(2019)}]{gonzalez2019kohn}%
  \BibitemOpen
  \bibfield  {author} {\bibinfo {author} {\bibfnamefont {J.}~\bibnamefont {Gonzalez}}\ and\ \bibinfo {author} {\bibfnamefont {T.}~\bibnamefont {Stauber}},\ }\bibfield  {title} {\bibinfo {title} {Kohn-luttinger superconductivity in twisted bilayer graphene},\ }\href {https://doi.org/10.1103/PhysRevLett.122.026801} {\bibfield  {journal} {\bibinfo  {journal} {Phys. Rev. Lett}\ }\textbf {\bibinfo {volume} {122}},\ \bibinfo {pages} {026801} (\bibinfo {year} {2019})}\BibitemShut {NoStop}%
\bibitem [{\citenamefont {Hao}\ \emph {et~al.}(2021)\citenamefont {Hao}, \citenamefont {Zimmerman}, \citenamefont {Ledwith}, \citenamefont {Khalaf}, \citenamefont {Najafabadi}, \citenamefont {Watanabe}, \citenamefont {Taniguchi}, \citenamefont {Vishwanath},\ and\ \citenamefont {Kim}}]{hao2021electric}%
  \BibitemOpen
  \bibfield  {author} {\bibinfo {author} {\bibfnamefont {Z.}~\bibnamefont {Hao}}, \bibinfo {author} {\bibfnamefont {A.}~\bibnamefont {Zimmerman}}, \bibinfo {author} {\bibfnamefont {P.}~\bibnamefont {Ledwith}}, \bibinfo {author} {\bibfnamefont {E.}~\bibnamefont {Khalaf}}, \bibinfo {author} {\bibfnamefont {D.~H.}\ \bibnamefont {Najafabadi}}, \bibinfo {author} {\bibfnamefont {K.}~\bibnamefont {Watanabe}}, \bibinfo {author} {\bibfnamefont {T.}~\bibnamefont {Taniguchi}}, \bibinfo {author} {\bibfnamefont {A.}~\bibnamefont {Vishwanath}},\ and\ \bibinfo {author} {\bibfnamefont {P.}~\bibnamefont {Kim}},\ }\bibfield  {title} {\bibinfo {title} {Electric field--tunable superconductivity in alternating-twist magic-angle trilayer graphene},\ }\href {https://doi.org/10.1126/science.abg0399} {\bibfield  {journal} {\bibinfo  {journal} {Science}\ }\textbf {\bibinfo {volume} {371}},\ \bibinfo {pages} {1133} (\bibinfo {year} {2021})}\BibitemShut {NoStop}%
\bibitem [{\citenamefont {Wu}\ \emph {et~al.}(2021)\citenamefont {Wu}, \citenamefont {Schwemmer}, \citenamefont {M{\"u}ller}, \citenamefont {Consiglio}, \citenamefont {Sangiovanni}, \citenamefont {Di~Sante}, \citenamefont {Iqbal}, \citenamefont {Hanke}, \citenamefont {Schnyder}, \citenamefont {Denner} \emph {et~al.}}]{wu2021nature}%
  \BibitemOpen
  \bibfield  {author} {\bibinfo {author} {\bibfnamefont {X.}~\bibnamefont {Wu}}, \bibinfo {author} {\bibfnamefont {T.}~\bibnamefont {Schwemmer}}, \bibinfo {author} {\bibfnamefont {T.}~\bibnamefont {M{\"u}ller}}, \bibinfo {author} {\bibfnamefont {A.}~\bibnamefont {Consiglio}}, \bibinfo {author} {\bibfnamefont {G.}~\bibnamefont {Sangiovanni}}, \bibinfo {author} {\bibfnamefont {D.}~\bibnamefont {Di~Sante}}, \bibinfo {author} {\bibfnamefont {Y.}~\bibnamefont {Iqbal}}, \bibinfo {author} {\bibfnamefont {W.}~\bibnamefont {Hanke}}, \bibinfo {author} {\bibfnamefont {A.~P.}\ \bibnamefont {Schnyder}}, \bibinfo {author} {\bibfnamefont {M.~M.}\ \bibnamefont {Denner}}, \emph {et~al.},\ }\bibfield  {title} {\bibinfo {title} {Nature of unconventional pairing in the kagome superconductors {$AV_3Sb_5$ (A= K, Rb, Cs)}},\ }\href {https://doi.org/10.1103/PhysRevLett.127.177001} {\bibfield  {journal} {\bibinfo  {journal} {Phys. Rev. Lett}\ }\textbf {\bibinfo {volume} {127}},\ \bibinfo {pages} {177001} (\bibinfo {year}
  {2021})}\BibitemShut {NoStop}%
\bibitem [{\citenamefont {Xu}\ \emph {et~al.}(2021)\citenamefont {Xu}, \citenamefont {Al~Ezzi}, \citenamefont {Balakrishnan}, \citenamefont {Garcia-Ruiz}, \citenamefont {Tsim}, \citenamefont {Mullan}, \citenamefont {Barrier}, \citenamefont {Xin}, \citenamefont {Piot}, \citenamefont {Taniguchi} \emph {et~al.}}]{xu2021tunable}%
  \BibitemOpen
  \bibfield  {author} {\bibinfo {author} {\bibfnamefont {S.}~\bibnamefont {Xu}}, \bibinfo {author} {\bibfnamefont {M.~M.}\ \bibnamefont {Al~Ezzi}}, \bibinfo {author} {\bibfnamefont {N.}~\bibnamefont {Balakrishnan}}, \bibinfo {author} {\bibfnamefont {A.}~\bibnamefont {Garcia-Ruiz}}, \bibinfo {author} {\bibfnamefont {B.}~\bibnamefont {Tsim}}, \bibinfo {author} {\bibfnamefont {C.}~\bibnamefont {Mullan}}, \bibinfo {author} {\bibfnamefont {J.}~\bibnamefont {Barrier}}, \bibinfo {author} {\bibfnamefont {N.}~\bibnamefont {Xin}}, \bibinfo {author} {\bibfnamefont {B.~A.}\ \bibnamefont {Piot}}, \bibinfo {author} {\bibfnamefont {T.}~\bibnamefont {Taniguchi}}, \emph {et~al.},\ }\bibfield  {title} {\bibinfo {title} {Tunable van hove singularities and correlated states in twisted monolayer--bilayer graphene},\ }\href {https://doi.org/10.1038/s41567-021-01172-9} {\bibfield  {journal} {\bibinfo  {journal} {Nat. Phys}\ }\textbf {\bibinfo {volume} {17}},\ \bibinfo {pages} {619} (\bibinfo {year} {2021})}\BibitemShut {NoStop}%
\bibitem [{\citenamefont {Steinke}\ \emph {et~al.}(2017)\citenamefont {Steinke}, \citenamefont {Zhao}, \citenamefont {Barber}, \citenamefont {Scaffidi}, \citenamefont {Jerzembek}, \citenamefont {Rosner}, \citenamefont {Gibbs}, \citenamefont {Maeno}, \citenamefont {Simon}, \citenamefont {Mackenzie} \emph {et~al.}}]{stepke2017strong}%
  \BibitemOpen
  \bibfield  {author} {\bibinfo {author} {\bibfnamefont {A.}~\bibnamefont {Steinke}}, \bibinfo {author} {\bibfnamefont {L.}~\bibnamefont {Zhao}}, \bibinfo {author} {\bibfnamefont {M.}~\bibnamefont {Barber}}, \bibinfo {author} {\bibfnamefont {T.}~\bibnamefont {Scaffidi}}, \bibinfo {author} {\bibfnamefont {F.}~\bibnamefont {Jerzembek}}, \bibinfo {author} {\bibfnamefont {H.}~\bibnamefont {Rosner}}, \bibinfo {author} {\bibfnamefont {A.}~\bibnamefont {Gibbs}}, \bibinfo {author} {\bibfnamefont {Y.}~\bibnamefont {Maeno}}, \bibinfo {author} {\bibfnamefont {S.}~\bibnamefont {Simon}}, \bibinfo {author} {\bibfnamefont {A.~P.}\ \bibnamefont {Mackenzie}}, \emph {et~al.},\ }\bibfield  {title} {\bibinfo {title} {Strong peak in {$T_c$} of {Sr$_2$RuO$_4$} under uniaxial pressure},\ }\href {https://doi.org/10.1126/science.aaf9398} {\bibfield  {journal} {\bibinfo  {journal} {Science}\ }\textbf {\bibinfo {volume} {355}},\ \bibinfo {pages} {eaaf9398} (\bibinfo {year} {2017})}\BibitemShut {NoStop}%
\bibitem [{\citenamefont {Chichinadze}\ \emph {et~al.}(2022)\citenamefont {Chichinadze}, \citenamefont {Classen}, \citenamefont {Wang},\ and\ \citenamefont {Chubukov}}]{chichinadze20224}%
  \BibitemOpen
  \bibfield  {author} {\bibinfo {author} {\bibfnamefont {D.~V.}\ \bibnamefont {Chichinadze}}, \bibinfo {author} {\bibfnamefont {L.}~\bibnamefont {Classen}}, \bibinfo {author} {\bibfnamefont {Y.}~\bibnamefont {Wang}},\ and\ \bibinfo {author} {\bibfnamefont {A.~V.}\ \bibnamefont {Chubukov}},\ }\bibfield  {title} {\bibinfo {title} {Su (4) symmetry in twisted bilayer graphene: An itinerant perspective},\ }\href {https://doi.org/10.1103/PhysRevLett.128.227601} {\bibfield  {journal} {\bibinfo  {journal} {Phys. Rev. Lett}\ }\textbf {\bibinfo {volume} {128}},\ \bibinfo {pages} {227601} (\bibinfo {year} {2022})}\BibitemShut {NoStop}%
\bibitem [{\citenamefont {Shtyk}\ \emph {et~al.}(2017)\citenamefont {Shtyk}, \citenamefont {Goldstein},\ and\ \citenamefont {Chamon}}]{shtyk2017electrons}%
  \BibitemOpen
  \bibfield  {author} {\bibinfo {author} {\bibfnamefont {A.}~\bibnamefont {Shtyk}}, \bibinfo {author} {\bibfnamefont {G.}~\bibnamefont {Goldstein}},\ and\ \bibinfo {author} {\bibfnamefont {C.}~\bibnamefont {Chamon}},\ }\bibfield  {title} {\bibinfo {title} {Electrons at the monkey saddle: A multicritical lifshitz point},\ }\href {https://doi.org/10.1103/PhysRevB.95.035137} {\bibfield  {journal} {\bibinfo  {journal} {Phys. Rev. B}\ }\textbf {\bibinfo {volume} {95}},\ \bibinfo {pages} {035137} (\bibinfo {year} {2017})}\BibitemShut {NoStop}%
\bibitem [{\citenamefont {Park}\ \emph {et~al.}(2024)\citenamefont {Park}, \citenamefont {Ghioldi}, \citenamefont {May}, \citenamefont {Kolopus}, \citenamefont {Podlesnyak}, \citenamefont {Calder}, \citenamefont {Paddison}, \citenamefont {Trumper}, \citenamefont {Manuel}, \citenamefont {Batista} \emph {et~al.}}]{park2024anomalous}%
  \BibitemOpen
  \bibfield  {author} {\bibinfo {author} {\bibfnamefont {P.}~\bibnamefont {Park}}, \bibinfo {author} {\bibfnamefont {E.~A.}\ \bibnamefont {Ghioldi}}, \bibinfo {author} {\bibfnamefont {A.~F.}\ \bibnamefont {May}}, \bibinfo {author} {\bibfnamefont {J.~A.}\ \bibnamefont {Kolopus}}, \bibinfo {author} {\bibfnamefont {A.~A.}\ \bibnamefont {Podlesnyak}}, \bibinfo {author} {\bibfnamefont {S.}~\bibnamefont {Calder}}, \bibinfo {author} {\bibfnamefont {J.~A.}\ \bibnamefont {Paddison}}, \bibinfo {author} {\bibfnamefont {A.~E.}\ \bibnamefont {Trumper}}, \bibinfo {author} {\bibfnamefont {L.~O.}\ \bibnamefont {Manuel}}, \bibinfo {author} {\bibfnamefont {C.~D.}\ \bibnamefont {Batista}}, \emph {et~al.},\ }\bibfield  {title} {\bibinfo {title} {Anomalous continuum scattering and higher-order van hove singularity in the strongly anisotropic {S}= 1/2 triangular lattice antiferromagnet},\ }\href {https://doi.org/10.1038/s41467-024-51618-w} {\bibfield  {journal} {\bibinfo  {journal} {Nat. Commun}\ }\textbf {\bibinfo {volume}
  {15}},\ \bibinfo {pages} {7264} (\bibinfo {year} {2024})}\BibitemShut {NoStop}%
\bibitem [{\citenamefont {Wu}\ \emph {et~al.}(2024)\citenamefont {Wu}, \citenamefont {Shi}, \citenamefont {Ozerov}, \citenamefont {Du}, \citenamefont {Wang}, \citenamefont {Ni}, \citenamefont {Meng}, \citenamefont {Jiang}, \citenamefont {Wang}, \citenamefont {Hao} \emph {et~al.}}]{wu2024discovery}%
  \BibitemOpen
  \bibfield  {author} {\bibinfo {author} {\bibfnamefont {W.}~\bibnamefont {Wu}}, \bibinfo {author} {\bibfnamefont {Z.}~\bibnamefont {Shi}}, \bibinfo {author} {\bibfnamefont {M.}~\bibnamefont {Ozerov}}, \bibinfo {author} {\bibfnamefont {Y.}~\bibnamefont {Du}}, \bibinfo {author} {\bibfnamefont {Y.}~\bibnamefont {Wang}}, \bibinfo {author} {\bibfnamefont {X.-S.}\ \bibnamefont {Ni}}, \bibinfo {author} {\bibfnamefont {X.}~\bibnamefont {Meng}}, \bibinfo {author} {\bibfnamefont {X.}~\bibnamefont {Jiang}}, \bibinfo {author} {\bibfnamefont {G.}~\bibnamefont {Wang}}, \bibinfo {author} {\bibfnamefont {C.}~\bibnamefont {Hao}}, \emph {et~al.},\ }\bibfield  {title} {\bibinfo {title} {The discovery of three-dimensional van hove singularity},\ }\href {https://doi.org/10.1038/s41467-024-46626-9} {\bibfield  {journal} {\bibinfo  {journal} {Nat. Commun}\ }\textbf {\bibinfo {volume} {15}},\ \bibinfo {pages} {2313} (\bibinfo {year} {2024})}\BibitemShut {NoStop}%
\bibitem [{\citenamefont {Tan}\ \emph {et~al.}(2024)\citenamefont {Tan}, \citenamefont {Jiang}, \citenamefont {McCandless}, \citenamefont {Chan},\ and\ \citenamefont {Yan}}]{tan2024three}%
  \BibitemOpen
  \bibfield  {author} {\bibinfo {author} {\bibfnamefont {H.}~\bibnamefont {Tan}}, \bibinfo {author} {\bibfnamefont {Y.}~\bibnamefont {Jiang}}, \bibinfo {author} {\bibfnamefont {G.~T.}\ \bibnamefont {McCandless}}, \bibinfo {author} {\bibfnamefont {J.~Y.}\ \bibnamefont {Chan}},\ and\ \bibinfo {author} {\bibfnamefont {B.}~\bibnamefont {Yan}},\ }\bibfield  {title} {\bibinfo {title} {Three-dimensional higher-order saddle-point-induced flat bands in {Co}-based kagome metals},\ }\href {https://doi.org/10.1103/PhysRevResearch.6.043132} {\bibfield  {journal} {\bibinfo  {journal} {Phys Rev Res}\ }\textbf {\bibinfo {volume} {6}},\ \bibinfo {pages} {043132} (\bibinfo {year} {2024})}\BibitemShut {NoStop}%
\bibitem [{\citenamefont {B{\"o}hm}(1991)}]{bohm1991material}%
  \BibitemOpen
  \bibfield  {author} {\bibinfo {author} {\bibfnamefont {M.~C.}\ \bibnamefont {B{\"o}hm}},\ }\bibfield  {title} {\bibinfo {title} {Material properties of low-dimensional charge-transfer salts. {II}. mode-softening, peierls transitions and van hove singularities},\ }\href {https://doi.org/10.1016/0301-0104(91)87005-G} {\bibfield  {journal} {\bibinfo  {journal} {Chem. Phys}\ }\textbf {\bibinfo {volume} {155}},\ \bibinfo {pages} {49} (\bibinfo {year} {1991})}\BibitemShut {NoStop}%
\bibitem [{\citenamefont {Zeng}\ \emph {et~al.}(2008)\citenamefont {Zeng}, \citenamefont {Kent}, \citenamefont {Kim}, \citenamefont {Li},\ and\ \citenamefont {Weitering}}]{zeng2008charge}%
  \BibitemOpen
  \bibfield  {author} {\bibinfo {author} {\bibfnamefont {C.}~\bibnamefont {Zeng}}, \bibinfo {author} {\bibfnamefont {P.}~\bibnamefont {Kent}}, \bibinfo {author} {\bibfnamefont {T.-H.}\ \bibnamefont {Kim}}, \bibinfo {author} {\bibfnamefont {A.-P.}\ \bibnamefont {Li}},\ and\ \bibinfo {author} {\bibfnamefont {H.~H.}\ \bibnamefont {Weitering}},\ }\bibfield  {title} {\bibinfo {title} {Charge-order fluctuations in one-dimensional silicides},\ }\href {https://doi.org/10.1038/nmat2209} {\bibfield  {journal} {\bibinfo  {journal} {Nat. Mater}\ }\textbf {\bibinfo {volume} {7}},\ \bibinfo {pages} {539} (\bibinfo {year} {2008})}\BibitemShut {NoStop}%
\bibitem [{\citenamefont {Li}\ \emph {et~al.}(2010)\citenamefont {Li}, \citenamefont {Luican}, \citenamefont {Lopes~dos Santos}, \citenamefont {Castro~Neto}, \citenamefont {Reina}, \citenamefont {Kong},\ and\ \citenamefont {Andrei}}]{li2010observation}%
  \BibitemOpen
  \bibfield  {author} {\bibinfo {author} {\bibfnamefont {G.}~\bibnamefont {Li}}, \bibinfo {author} {\bibfnamefont {A.}~\bibnamefont {Luican}}, \bibinfo {author} {\bibfnamefont {J.}~\bibnamefont {Lopes~dos Santos}}, \bibinfo {author} {\bibfnamefont {A.}~\bibnamefont {Castro~Neto}}, \bibinfo {author} {\bibfnamefont {A.}~\bibnamefont {Reina}}, \bibinfo {author} {\bibfnamefont {J.}~\bibnamefont {Kong}},\ and\ \bibinfo {author} {\bibfnamefont {E.}~\bibnamefont {Andrei}},\ }\bibfield  {title} {\bibinfo {title} {Observation of van hove singularities in twisted graphene layers},\ }\href {https://doi.org/10.1038/nphys1463} {\bibfield  {journal} {\bibinfo  {journal} {Nat. Phys}\ }\textbf {\bibinfo {volume} {6}},\ \bibinfo {pages} {109} (\bibinfo {year} {2010})}\BibitemShut {NoStop}%
\bibitem [{\citenamefont {Seiler}\ \emph {et~al.}(2022)\citenamefont {Seiler}, \citenamefont {Geisenhof}, \citenamefont {Winterer}, \citenamefont {Watanabe}, \citenamefont {Taniguchi}, \citenamefont {Xu}, \citenamefont {Zhang},\ and\ \citenamefont {Weitz}}]{seiler2022quantum}%
  \BibitemOpen
  \bibfield  {author} {\bibinfo {author} {\bibfnamefont {A.~M.}\ \bibnamefont {Seiler}}, \bibinfo {author} {\bibfnamefont {F.~R.}\ \bibnamefont {Geisenhof}}, \bibinfo {author} {\bibfnamefont {F.}~\bibnamefont {Winterer}}, \bibinfo {author} {\bibfnamefont {K.}~\bibnamefont {Watanabe}}, \bibinfo {author} {\bibfnamefont {T.}~\bibnamefont {Taniguchi}}, \bibinfo {author} {\bibfnamefont {T.}~\bibnamefont {Xu}}, \bibinfo {author} {\bibfnamefont {F.}~\bibnamefont {Zhang}},\ and\ \bibinfo {author} {\bibfnamefont {R.~T.}\ \bibnamefont {Weitz}},\ }\bibfield  {title} {\bibinfo {title} {Quantum cascade of correlated phases in trigonally warped bilayer graphene},\ }\href {https://doi.org/10.1038/s41586-022-04937-1} {\bibfield  {journal} {\bibinfo  {journal} {Nat}\ }\textbf {\bibinfo {volume} {608}},\ \bibinfo {pages} {298} (\bibinfo {year} {2022})}\BibitemShut {NoStop}%
\bibitem [{\citenamefont {Tamai}\ \emph {et~al.}(2008)\citenamefont {Tamai}, \citenamefont {Allan}, \citenamefont {Mercure}, \citenamefont {Meevasana}, \citenamefont {Dunkel}, \citenamefont {Lu}, \citenamefont {Perry}, \citenamefont {Mackenzie}, \citenamefont {Singh}, \citenamefont {Shen} \emph {et~al.}}]{tamai2008fermi}%
  \BibitemOpen
  \bibfield  {author} {\bibinfo {author} {\bibfnamefont {A.}~\bibnamefont {Tamai}}, \bibinfo {author} {\bibfnamefont {M.~P.}\ \bibnamefont {Allan}}, \bibinfo {author} {\bibfnamefont {J.-F.}\ \bibnamefont {Mercure}}, \bibinfo {author} {\bibfnamefont {W.}~\bibnamefont {Meevasana}}, \bibinfo {author} {\bibfnamefont {R.}~\bibnamefont {Dunkel}}, \bibinfo {author} {\bibfnamefont {D.}~\bibnamefont {Lu}}, \bibinfo {author} {\bibfnamefont {R.~S.}\ \bibnamefont {Perry}}, \bibinfo {author} {\bibfnamefont {A.}~\bibnamefont {Mackenzie}}, \bibinfo {author} {\bibfnamefont {D.~J.}\ \bibnamefont {Singh}}, \bibinfo {author} {\bibfnamefont {Z.-X.}\ \bibnamefont {Shen}}, \emph {et~al.},\ }\bibfield  {title} {\bibinfo {title} {Fermi surface and van hove singularities in the itinerant metamagnet {$Sr_3Ru_2O_7$}},\ }\href {https://doi.org/10.1103/PhysRevLett.101.026407} {\bibfield  {journal} {\bibinfo  {journal} {Phys. Rev. Lett}\ }\textbf {\bibinfo {volume} {101}},\ \bibinfo {pages} {026407} (\bibinfo {year} {2008})}\BibitemShut
  {NoStop}%
\bibitem [{\citenamefont {Volovik}(2003)}]{volovik2003universe}%
  \BibitemOpen
  \bibfield  {author} {\bibinfo {author} {\bibfnamefont {G.~E.}\ \bibnamefont {Volovik}},\ }\href@noop {} {\emph {\bibinfo {title} {{The Universe in a Helium Droplet}}}}\ (\bibinfo  {publisher} {Oxford University Press},\ \bibinfo {address} {Oxford},\ \bibinfo {year} {2003})\BibitemShut {NoStop}%
\bibitem [{\citenamefont {Milnor}(1963)}]{milnor1963morse}%
  \BibitemOpen
  \bibfield  {author} {\bibinfo {author} {\bibfnamefont {J.}~\bibnamefont {Milnor}},\ }\href@noop {} {\emph {\bibinfo {title} {{Morse Theory}}}}\ (\bibinfo  {publisher} {Princeton University Press},\ \bibinfo {address} {Princeton, NJ},\ \bibinfo {year} {1963})\BibitemShut {NoStop}%
\bibitem [{\citenamefont {Zhang}\ \emph {et~al.}(2022)\citenamefont {Zhang}, \citenamefont {Yu}, \citenamefont {Liu},\ and\ \citenamefont {Yao}}]{zhang2022magnetictb}%
  \BibitemOpen
  \bibfield  {author} {\bibinfo {author} {\bibfnamefont {Z.}~\bibnamefont {Zhang}}, \bibinfo {author} {\bibfnamefont {Z.-M.}\ \bibnamefont {Yu}}, \bibinfo {author} {\bibfnamefont {G.-B.}\ \bibnamefont {Liu}},\ and\ \bibinfo {author} {\bibfnamefont {Y.}~\bibnamefont {Yao}},\ }\bibfield  {title} {\bibinfo {title} {Magnetic{TB}: A package for tight-binding model of magnetic and non-magnetic materials},\ }\href {https://doi.org/10.1016/j.cpc.2021.108153} {\bibfield  {journal} {\bibinfo  {journal} {Comput. Phys. Commun}\ }\textbf {\bibinfo {volume} {270}},\ \bibinfo {pages} {108153} (\bibinfo {year} {2022})}\BibitemShut {NoStop}%
\bibitem [{\citenamefont {Hicks}\ \emph {et~al.}(2014)\citenamefont {Hicks}, \citenamefont {Brodsky}, \citenamefont {Yelland}, \citenamefont {Gibbs}, \citenamefont {Bruin}, \citenamefont {Barber}, \citenamefont {Edkins}, \citenamefont {Nishimura}, \citenamefont {Yonezawa}, \citenamefont {Maeno},\ and\ \citenamefont {Mackenzie}}]{hicks2014strong}%
  \BibitemOpen
  \bibfield  {author} {\bibinfo {author} {\bibfnamefont {C.~W.}\ \bibnamefont {Hicks}}, \bibinfo {author} {\bibfnamefont {D.~O.}\ \bibnamefont {Brodsky}}, \bibinfo {author} {\bibfnamefont {E.~A.}\ \bibnamefont {Yelland}}, \bibinfo {author} {\bibfnamefont {A.~S.}\ \bibnamefont {Gibbs}}, \bibinfo {author} {\bibfnamefont {J.~A.~N.}\ \bibnamefont {Bruin}}, \bibinfo {author} {\bibfnamefont {M.~E.}\ \bibnamefont {Barber}}, \bibinfo {author} {\bibfnamefont {S.~D.}\ \bibnamefont {Edkins}}, \bibinfo {author} {\bibfnamefont {K.}~\bibnamefont {Nishimura}}, \bibinfo {author} {\bibfnamefont {S.}~\bibnamefont {Yonezawa}}, \bibinfo {author} {\bibfnamefont {Y.}~\bibnamefont {Maeno}},\ and\ \bibinfo {author} {\bibfnamefont {A.~P.}\ \bibnamefont {Mackenzie}},\ }\bibfield  {title} {\bibinfo {title} {{Strong Increase of {T$_c$} of {Sr$_2$RuO$_4$} Under Both Tensile and Compressive Strain}},\ }\href {https://doi.org/10.1126/science.1248292} {\bibfield  {journal} {\bibinfo  {journal} {Science}\ }\textbf {\bibinfo {volume} {344}},\
  \bibinfo {pages} {283} (\bibinfo {year} {2014})}\BibitemShut {NoStop}%
\bibitem [{\citenamefont {Lin}\ \emph {et~al.}(2024)\citenamefont {Lin}, \citenamefont {Consiglio}, \citenamefont {Forslund}, \citenamefont {K{\"u}spert}, \citenamefont {Denner}, \citenamefont {Lei}, \citenamefont {Louat}, \citenamefont {Watson}, \citenamefont {Kim}, \citenamefont {Cacho} \emph {et~al.}}]{lin2024uniaxial}%
  \BibitemOpen
  \bibfield  {author} {\bibinfo {author} {\bibfnamefont {C.}~\bibnamefont {Lin}}, \bibinfo {author} {\bibfnamefont {A.}~\bibnamefont {Consiglio}}, \bibinfo {author} {\bibfnamefont {O.~K.}\ \bibnamefont {Forslund}}, \bibinfo {author} {\bibfnamefont {J.}~\bibnamefont {K{\"u}spert}}, \bibinfo {author} {\bibfnamefont {M.~M.}\ \bibnamefont {Denner}}, \bibinfo {author} {\bibfnamefont {H.}~\bibnamefont {Lei}}, \bibinfo {author} {\bibfnamefont {A.}~\bibnamefont {Louat}}, \bibinfo {author} {\bibfnamefont {M.~D.}\ \bibnamefont {Watson}}, \bibinfo {author} {\bibfnamefont {T.~K.}\ \bibnamefont {Kim}}, \bibinfo {author} {\bibfnamefont {C.}~\bibnamefont {Cacho}}, \emph {et~al.},\ }\bibfield  {title} {\bibinfo {title} {{Uniaxial strain tuning of charge modulation and singularity in a kagome superconductor}},\ }\href {https://doi.org/10.1038/s41467-024-53737-w} {\bibfield  {journal} {\bibinfo  {journal} {Nat. Commun}\ }\textbf {\bibinfo {volume} {15}},\ \bibinfo {pages} {10466} (\bibinfo {year} {2024})}\BibitemShut {NoStop}%
\bibitem [{\citenamefont {Park}\ \emph {et~al.}(2021)\citenamefont {Park}, \citenamefont {Cao}, \citenamefont {Watanabe}, \citenamefont {Taniguchi},\ and\ \citenamefont {Jarillo-Herrero}}]{park2021tunable}%
  \BibitemOpen
  \bibfield  {author} {\bibinfo {author} {\bibfnamefont {J.~M.}\ \bibnamefont {Park}}, \bibinfo {author} {\bibfnamefont {Y.}~\bibnamefont {Cao}}, \bibinfo {author} {\bibfnamefont {K.}~\bibnamefont {Watanabe}}, \bibinfo {author} {\bibfnamefont {T.}~\bibnamefont {Taniguchi}},\ and\ \bibinfo {author} {\bibfnamefont {P.}~\bibnamefont {Jarillo-Herrero}},\ }\bibfield  {title} {\bibinfo {title} {Tunable strongly coupled superconductivity in magic-angle twisted trilayer graphene},\ }\href {https://doi.org/10.1038/s41586-021-03192-0} {\bibfield  {journal} {\bibinfo  {journal} {Nat}\ }\textbf {\bibinfo {volume} {590}},\ \bibinfo {pages} {249} (\bibinfo {year} {2021})}\BibitemShut {NoStop}%
\bibitem [{\citenamefont {Kiesel}\ \emph {et~al.}(2013)\citenamefont {Kiesel}, \citenamefont {Platt},\ and\ \citenamefont {Thomale}}]{kiesel2013unconventional}%
  \BibitemOpen
  \bibfield  {author} {\bibinfo {author} {\bibfnamefont {M.~L.}\ \bibnamefont {Kiesel}}, \bibinfo {author} {\bibfnamefont {C.}~\bibnamefont {Platt}},\ and\ \bibinfo {author} {\bibfnamefont {R.}~\bibnamefont {Thomale}},\ }\bibfield  {title} {\bibinfo {title} {{Unconventional Fermi surface instabilities in the kagome Hubbard model}},\ }\href {https://doi.org/10.1103/PhysRevLett.110.126405} {\bibfield  {journal} {\bibinfo  {journal} {Phys. Rev. Lett}\ }\textbf {\bibinfo {volume} {110}},\ \bibinfo {pages} {126405} (\bibinfo {year} {2013})}\BibitemShut {NoStop}%
\bibitem [{\citenamefont {Classen}\ \emph {et~al.}(2020)\citenamefont {Classen}, \citenamefont {Chubukov}, \citenamefont {Honerkamp},\ and\ \citenamefont {Scherer}}]{classen2020competing}%
  \BibitemOpen
  \bibfield  {author} {\bibinfo {author} {\bibfnamefont {L.}~\bibnamefont {Classen}}, \bibinfo {author} {\bibfnamefont {A.}~\bibnamefont {Chubukov}}, \bibinfo {author} {\bibfnamefont {C.}~\bibnamefont {Honerkamp}},\ and\ \bibinfo {author} {\bibfnamefont {M.}~\bibnamefont {Scherer}},\ }\bibfield  {title} {\bibinfo {title} {{Competing orders at higher-order Van Hove points}},\ }\href {https://doi.org/10.1103/PhysRevB.102.125141} {\bibfield  {journal} {\bibinfo  {journal} {Phys. Rev. B}\ }\textbf {\bibinfo {volume} {102}},\ \bibinfo {pages} {125141} (\bibinfo {year} {2020})}\BibitemShut {NoStop}%
\bibitem [{\citenamefont {Newns}\ \emph {et~al.}(1992)\citenamefont {Newns}, \citenamefont {Krishnamurthy}, \citenamefont {Pattnaik}, \citenamefont {Tsuei},\ and\ \citenamefont {Kane}}]{newns1992saddle}%
  \BibitemOpen
  \bibfield  {author} {\bibinfo {author} {\bibfnamefont {D.~M.}\ \bibnamefont {Newns}}, \bibinfo {author} {\bibfnamefont {H.~R.}\ \bibnamefont {Krishnamurthy}}, \bibinfo {author} {\bibfnamefont {P.~C.}\ \bibnamefont {Pattnaik}}, \bibinfo {author} {\bibfnamefont {C.~C.}\ \bibnamefont {Tsuei}},\ and\ \bibinfo {author} {\bibfnamefont {C.~L.}\ \bibnamefont {Kane}},\ }\bibfield  {title} {\bibinfo {title} {{Saddle-point pairing: An electronic mechanism for superconductivity}},\ }\href {https://doi.org/10.1103/PhysRevLett.69.1264} {\bibfield  {journal} {\bibinfo  {journal} {Phys. Rev. Lett}\ }\textbf {\bibinfo {volume} {69}},\ \bibinfo {pages} {1264} (\bibinfo {year} {1992})}\BibitemShut {NoStop}%
\bibitem [{\citenamefont {Markiewicz}(1997)}]{markiewicz1997survey}%
  \BibitemOpen
  \bibfield  {author} {\bibinfo {author} {\bibfnamefont {R.}~\bibnamefont {Markiewicz}},\ }\bibfield  {title} {\bibinfo {title} {{A survey of the Van Hove scenario for high-{$t_c$} superconductivity with special emphasis on pseudogaps and striped phases}},\ }\href {https://doi.org/10.1016/S0022-3697(97)00025-5} {\bibfield  {journal} {\bibinfo  {journal} {J. Phys. Chem. Solids}\ }\textbf {\bibinfo {volume} {58}},\ \bibinfo {pages} {1179} (\bibinfo {year} {1997})}\BibitemShut {NoStop}%
\bibitem [{\citenamefont {Nandkishore}\ \emph {et~al.}(2012)\citenamefont {Nandkishore}, \citenamefont {Levitov},\ and\ \citenamefont {Chubukov}}]{nandkishore2012chiral}%
  \BibitemOpen
  \bibfield  {author} {\bibinfo {author} {\bibfnamefont {R.}~\bibnamefont {Nandkishore}}, \bibinfo {author} {\bibfnamefont {L.~S.}\ \bibnamefont {Levitov}},\ and\ \bibinfo {author} {\bibfnamefont {A.~V.}\ \bibnamefont {Chubukov}},\ }\bibfield  {title} {\bibinfo {title} {{Chiral superconductivity from repulsive interactions in doped graphene}},\ }\href {https://doi.org/10.1038/nphys2208} {\bibfield  {journal} {\bibinfo  {journal} {Nat. Phys}\ }\textbf {\bibinfo {volume} {8}},\ \bibinfo {pages} {158} (\bibinfo {year} {2012})}\BibitemShut {NoStop}%
\bibitem [{\citenamefont {Bruus}\ and\ \citenamefont {Flensberg}(2004)}]{bruus2004many}%
  \BibitemOpen
  \bibfield  {author} {\bibinfo {author} {\bibfnamefont {H.}~\bibnamefont {Bruus}}\ and\ \bibinfo {author} {\bibfnamefont {K.}~\bibnamefont {Flensberg}},\ }\href@noop {} {\emph {\bibinfo {title} {{Many-body quantum theory in condensed matter physics: an introduction}}}}\ (\bibinfo  {publisher} {Oxford university press},\ \bibinfo {address} {Oxford},\ \bibinfo {year} {2004})\BibitemShut {NoStop}%
\bibitem [{\citenamefont {Pullasseri}\ and\ \citenamefont {Santos}(2024)}]{pullasseri2024chern}%
  \BibitemOpen
  \bibfield  {author} {\bibinfo {author} {\bibfnamefont {L.}~\bibnamefont {Pullasseri}}\ and\ \bibinfo {author} {\bibfnamefont {L.~H.}\ \bibnamefont {Santos}},\ }\bibfield  {title} {\bibinfo {title} {Chern bands with higher-order van hove singularities on topological moir{\'e} surface states},\ }\href {https://doi.org/10.1103/PhysRevB.110.115125} {\bibfield  {journal} {\bibinfo  {journal} {Phys. Rev. B}\ }\textbf {\bibinfo {volume} {110}},\ \bibinfo {pages} {115125} (\bibinfo {year} {2024})}\BibitemShut {NoStop}%
\bibitem [{\citenamefont {Hsu}\ \emph {et~al.}(2021)\citenamefont {Hsu}, \citenamefont {Wu},\ and\ \citenamefont {Das~Sarma}}]{hsu2021spin}%
  \BibitemOpen
  \bibfield  {author} {\bibinfo {author} {\bibfnamefont {Y.-T.}\ \bibnamefont {Hsu}}, \bibinfo {author} {\bibfnamefont {F.}~\bibnamefont {Wu}},\ and\ \bibinfo {author} {\bibfnamefont {S.}~\bibnamefont {Das~Sarma}},\ }\bibfield  {title} {\bibinfo {title} {Spin-valley locked instabilities in moir{\'e} transition metal dichalcogenides with conventional and higher-order van hove singularities},\ }\href {https://doi.org/10.1103/PhysRevB.104.195134} {\bibfield  {journal} {\bibinfo  {journal} {Phys. Rev. B}\ }\textbf {\bibinfo {volume} {104}},\ \bibinfo {pages} {195134} (\bibinfo {year} {2021})}\BibitemShut {NoStop}%
\bibitem [{\citenamefont {Wang}\ \emph {et~al.}(2021)\citenamefont {Wang}, \citenamefont {Yuan},\ and\ \citenamefont {Fu}}]{wang2021moire}%
  \BibitemOpen
  \bibfield  {author} {\bibinfo {author} {\bibfnamefont {T.}~\bibnamefont {Wang}}, \bibinfo {author} {\bibfnamefont {N.~F.}\ \bibnamefont {Yuan}},\ and\ \bibinfo {author} {\bibfnamefont {L.}~\bibnamefont {Fu}},\ }\bibfield  {title} {\bibinfo {title} {Moir{\'e} surface states and enhanced superconductivity in topological insulators},\ }\href {https://doi.org/10.1103/PhysRevX.11.021024} {\bibfield  {journal} {\bibinfo  {journal} {Phys. Rev. X}\ }\textbf {\bibinfo {volume} {11}},\ \bibinfo {pages} {021024} (\bibinfo {year} {2021})}\BibitemShut {NoStop}%
\bibitem [{\citenamefont {Wu}\ \emph {et~al.}(2023)\citenamefont {Wu}, \citenamefont {Wu},\ and\ \citenamefont {Wu}}]{wu2023pair}%
  \BibitemOpen
  \bibfield  {author} {\bibinfo {author} {\bibfnamefont {Z.}~\bibnamefont {Wu}}, \bibinfo {author} {\bibfnamefont {Y.-M.}\ \bibnamefont {Wu}},\ and\ \bibinfo {author} {\bibfnamefont {F.}~\bibnamefont {Wu}},\ }\bibfield  {title} {\bibinfo {title} {Pair density wave and loop current promoted by van hove singularities in moir{\'e} systems},\ }\href {https://doi.org/10.1103/PhysRevB.107.045122} {\bibfield  {journal} {\bibinfo  {journal} {Phys. Rev. B}\ }\textbf {\bibinfo {volume} {107}},\ \bibinfo {pages} {045122} (\bibinfo {year} {2023})}\BibitemShut {NoStop}%
\bibitem [{\citenamefont {Lu}\ and\ \citenamefont {Santos}(2024)}]{lu2024fractional}%
  \BibitemOpen
  \bibfield  {author} {\bibinfo {author} {\bibfnamefont {T.}~\bibnamefont {Lu}}\ and\ \bibinfo {author} {\bibfnamefont {L.~H.}\ \bibnamefont {Santos}},\ }\bibfield  {title} {\bibinfo {title} {Fractional chern insulators in twisted bilayer {$MoTe_2$}: A composite fermion perspective},\ }\href {https://doi.org/10.1103/PhysRevLett.133.186602} {\bibfield  {journal} {\bibinfo  {journal} {Phys. Rev. Lett}\ }\textbf {\bibinfo {volume} {133}},\ \bibinfo {pages} {186602} (\bibinfo {year} {2024})}\BibitemShut {NoStop}%
\bibitem [{\citenamefont {Hu}\ \emph {et~al.}(2022)\citenamefont {Hu}, \citenamefont {Wu}, \citenamefont {Ortiz}, \citenamefont {Ju}, \citenamefont {Han}, \citenamefont {Ma}, \citenamefont {Plumb}, \citenamefont {Radovic}, \citenamefont {Thomale}, \citenamefont {Wilson} \emph {et~al.}}]{hu2022rich}%
  \BibitemOpen
  \bibfield  {author} {\bibinfo {author} {\bibfnamefont {Y.}~\bibnamefont {Hu}}, \bibinfo {author} {\bibfnamefont {X.}~\bibnamefont {Wu}}, \bibinfo {author} {\bibfnamefont {B.~R.}\ \bibnamefont {Ortiz}}, \bibinfo {author} {\bibfnamefont {S.}~\bibnamefont {Ju}}, \bibinfo {author} {\bibfnamefont {X.}~\bibnamefont {Han}}, \bibinfo {author} {\bibfnamefont {J.}~\bibnamefont {Ma}}, \bibinfo {author} {\bibfnamefont {N.~C.}\ \bibnamefont {Plumb}}, \bibinfo {author} {\bibfnamefont {M.}~\bibnamefont {Radovic}}, \bibinfo {author} {\bibfnamefont {R.}~\bibnamefont {Thomale}}, \bibinfo {author} {\bibfnamefont {S.~D.}\ \bibnamefont {Wilson}}, \emph {et~al.},\ }\bibfield  {title} {\bibinfo {title} {Rich nature of van hove singularities in kagome superconductor {$CsV_3Sb_5$}},\ }\href {https://doi.org/10.1038/s41467-022-29828-x} {\bibfield  {journal} {\bibinfo  {journal} {Nat. Commun}\ }\textbf {\bibinfo {volume} {13}},\ \bibinfo {pages} {2220} (\bibinfo {year} {2022})}\BibitemShut {NoStop}%
\bibitem [{\citenamefont {Kang}\ \emph {et~al.}(2022)\citenamefont {Kang}, \citenamefont {Fang}, \citenamefont {Kim}, \citenamefont {Ortiz}, \citenamefont {Ryu}, \citenamefont {Kim}, \citenamefont {Yoo}, \citenamefont {Sangiovanni}, \citenamefont {Di~Sante}, \citenamefont {Park} \emph {et~al.}}]{kang2022twofold}%
  \BibitemOpen
  \bibfield  {author} {\bibinfo {author} {\bibfnamefont {M.}~\bibnamefont {Kang}}, \bibinfo {author} {\bibfnamefont {S.}~\bibnamefont {Fang}}, \bibinfo {author} {\bibfnamefont {J.-K.}\ \bibnamefont {Kim}}, \bibinfo {author} {\bibfnamefont {B.~R.}\ \bibnamefont {Ortiz}}, \bibinfo {author} {\bibfnamefont {S.~H.}\ \bibnamefont {Ryu}}, \bibinfo {author} {\bibfnamefont {J.}~\bibnamefont {Kim}}, \bibinfo {author} {\bibfnamefont {J.}~\bibnamefont {Yoo}}, \bibinfo {author} {\bibfnamefont {G.}~\bibnamefont {Sangiovanni}}, \bibinfo {author} {\bibfnamefont {D.}~\bibnamefont {Di~Sante}}, \bibinfo {author} {\bibfnamefont {B.-G.}\ \bibnamefont {Park}}, \emph {et~al.},\ }\bibfield  {title} {\bibinfo {title} {Twofold van hove singularity and origin of charge order in topological kagome superconductor {$CsV_3Sb_5$}},\ }\href {https://doi.org/10.1038/s41567-021-01451-5} {\bibfield  {journal} {\bibinfo  {journal} {Nat. Phys.}\ }\textbf {\bibinfo {volume} {18}},\ \bibinfo {pages} {301} (\bibinfo {year} {2022})}\BibitemShut
  {NoStop}%
\bibitem [{\citenamefont {Wang}\ \emph {et~al.}(2025)\citenamefont {Wang}, \citenamefont {Pullasseri},\ and\ \citenamefont {Santos}}]{wang2025higher}%
  \BibitemOpen
  \bibfield  {author} {\bibinfo {author} {\bibfnamefont {E.}~\bibnamefont {Wang}}, \bibinfo {author} {\bibfnamefont {L.}~\bibnamefont {Pullasseri}},\ and\ \bibinfo {author} {\bibfnamefont {L.~H.}\ \bibnamefont {Santos}},\ }\bibfield  {title} {\bibinfo {title} {Higher-order van hove singularities in kagome topological bands},\ }\href {https://doi.org/10.1103/PhysRevB.111.075114} {\bibfield  {journal} {\bibinfo  {journal} {Phys. Rev. B}\ }\textbf {\bibinfo {volume} {111}},\ \bibinfo {pages} {075114} (\bibinfo {year} {2025})}\BibitemShut {NoStop}%
\bibitem [{\citenamefont {Yao}\ and\ \citenamefont {Yang}(2015)}]{yao2015topological}%
  \BibitemOpen
  \bibfield  {author} {\bibinfo {author} {\bibfnamefont {H.}~\bibnamefont {Yao}}\ and\ \bibinfo {author} {\bibfnamefont {F.}~\bibnamefont {Yang}},\ }\bibfield  {title} {\bibinfo {title} {Topological odd-parity superconductivity at type-{II} two-dimensional van hove singularities},\ }\href {https://doi.org/10.1103/PhysRevB.92.035132} {\bibfield  {journal} {\bibinfo  {journal} {Phys. Rev. B}\ }\textbf {\bibinfo {volume} {92}},\ \bibinfo {pages} {035132} (\bibinfo {year} {2015})}\BibitemShut {NoStop}%
\bibitem [{\citenamefont {Meng}\ \emph {et~al.}(2015)\citenamefont {Meng}, \citenamefont {Yang}, \citenamefont {Chen}, \citenamefont {Yao},\ and\ \citenamefont {Kee}}]{meng2015evidence}%
  \BibitemOpen
  \bibfield  {author} {\bibinfo {author} {\bibfnamefont {Z.~Y.}\ \bibnamefont {Meng}}, \bibinfo {author} {\bibfnamefont {F.}~\bibnamefont {Yang}}, \bibinfo {author} {\bibfnamefont {K.-S.}\ \bibnamefont {Chen}}, \bibinfo {author} {\bibfnamefont {H.}~\bibnamefont {Yao}},\ and\ \bibinfo {author} {\bibfnamefont {H.-Y.}\ \bibnamefont {Kee}},\ }\bibfield  {title} {\bibinfo {title} {Evidence for spin-triplet odd-parity superconductivity close to type-{II} van hove singularities},\ }\href {https://doi.org/10.1103/PhysRevB.91.184509} {\bibfield  {journal} {\bibinfo  {journal} {Phys. Rev. B}\ }\textbf {\bibinfo {volume} {91}},\ \bibinfo {pages} {184509} (\bibinfo {year} {2015})}\BibitemShut {NoStop}%
\bibitem [{\citenamefont {Morita}\ \emph {et~al.}(2026)\citenamefont {Morita}, \citenamefont {Li}, \citenamefont {Wei}, \citenamefont {Nakayama}, \citenamefont {Wang}, \citenamefont {Li}, \citenamefont {Kato}, \citenamefont {Souma}, \citenamefont {Tanaka}, \citenamefont {Ozawa}, \citenamefont {Yin}, \citenamefont {Takahashi}, \citenamefont {Kuang}, \citenamefont {Yao},\ and\ \citenamefont {Sato}}]{morita2026saddlepoints}%
  \BibitemOpen
  \bibfield  {author} {\bibinfo {author} {\bibfnamefont {Y.}~\bibnamefont {Morita}}, \bibinfo {author} {\bibfnamefont {Y.}~\bibnamefont {Li}}, \bibinfo {author} {\bibfnamefont {Y.-H.}\ \bibnamefont {Wei}}, \bibinfo {author} {\bibfnamefont {K.}~\bibnamefont {Nakayama}}, \bibinfo {author} {\bibfnamefont {Z.}~\bibnamefont {Wang}}, \bibinfo {author} {\bibfnamefont {H.-Y.}\ \bibnamefont {Li}}, \bibinfo {author} {\bibfnamefont {T.}~\bibnamefont {Kato}}, \bibinfo {author} {\bibfnamefont {S.}~\bibnamefont {Souma}}, \bibinfo {author} {\bibfnamefont {K.}~\bibnamefont {Tanaka}}, \bibinfo {author} {\bibfnamefont {K.}~\bibnamefont {Ozawa}}, \bibinfo {author} {\bibfnamefont {J.-X.}\ \bibnamefont {Yin}}, \bibinfo {author} {\bibfnamefont {T.}~\bibnamefont {Takahashi}}, \bibinfo {author} {\bibfnamefont {M.-Q.}\ \bibnamefont {Kuang}}, \bibinfo {author} {\bibfnamefont {Y.}~\bibnamefont {Yao}},\ and\ \bibinfo {author} {\bibfnamefont {T.}~\bibnamefont {Sato}},\ }\bibfield  {title} {\bibinfo {title} {Type-{I} and type-{II} saddle
  points and a topological flat band in a {Bi}-pyrochlore superconductor {CsBi$_2$}},\ }\bibfield  {journal} {\bibinfo  {journal} {arXiv:2604.07805}\ }\href {https://doi.org/10.48550/arXiv.2604.07805} {10.48550/arXiv.2604.07805} (\bibinfo {year} {2026})\BibitemShut {NoStop}%
\end{thebibliography}

\begin{thebibliography}{1}%
\makeatletter
\providecommand \@ifxundefined [1]{%
 \@ifx{#1\undefined}
}%
\providecommand \@ifnum [1]{%
 \ifnum #1\expandafter \@firstoftwo
 \else \expandafter \@secondoftwo
 \fi
}%
\providecommand \@ifx [1]{%
 \ifx #1\expandafter \@firstoftwo
 \else \expandafter \@secondoftwo
 \fi
}%
\providecommand \natexlab [1]{#1}%
\providecommand \enquote  [1]{``#1''}%
\providecommand \bibnamefont  [1]{#1}%
\providecommand \bibfnamefont [1]{#1}%
\providecommand \citenamefont [1]{#1}%
\providecommand \href@noop [0]{\@secondoftwo}%
\providecommand \href [0]{\begingroup \@sanitize@url \@href}%
\providecommand \@href[1]{\@@startlink{#1}\@@href}%
\providecommand \@@href[1]{\endgroup#1\@@endlink}%
\providecommand \@sanitize@url [0]{\catcode `\\12\catcode `\$12\catcode `\&12\catcode `\#12\catcode `\^12\catcode `\_12\catcode `\%12\relax}%
\providecommand \@@startlink[1]{}%
\providecommand \@@endlink[0]{}%
\providecommand \url  [0]{\begingroup\@sanitize@url \@url }%
\providecommand \@url [1]{\endgroup\@href {#1}{\urlprefix }}%
\providecommand \urlprefix  [0]{URL }%
\providecommand \Eprint [0]{\href }%
\providecommand \doibase [0]{https://doi.org/}%
\providecommand \selectlanguage [0]{\@gobble}%
\providecommand \bibinfo  [0]{\@secondoftwo}%
\providecommand \bibfield  [0]{\@secondoftwo}%
\providecommand \translation [1]{[#1]}%
\providecommand \BibitemOpen [0]{}%
\providecommand \bibitemStop [0]{}%
\providecommand \bibitemNoStop [0]{.\EOS\space}%
\providecommand \EOS [0]{\spacefactor3000\relax}%
\providecommand \BibitemShut  [1]{\csname bibitem#1\endcsname}%
\let\auto@bib@innerbib\@empty

\bibitem [{\citenamefont {Zhang}\ \emph {et~al.}(2022)\citenamefont {Zhang}, \citenamefont {Yu}, \citenamefont {Liu},\ and\ \citenamefont {Yao}}]{S_zhang2022magnetictb}%
  \BibitemOpen
  \bibfield  {author} {\bibinfo {author} {\bibfnamefont {Z.}~\bibnamefont {Zhang}}, \bibinfo {author} {\bibfnamefont {Z.-M.}\ \bibnamefont {Yu}}, \bibinfo {author} {\bibfnamefont {G.-B.}\ \bibnamefont {Liu}},\ and\ \bibinfo {author} {\bibfnamefont {Y.}~\bibnamefont {Yao}},\ }\bibfield  {title} {\bibinfo {title} {Magnetic{TB}: A package for tight-binding model of magnetic and non-magnetic materials},\ }\href {https://doi.org/10.1016/j.cpc.2021.108153} {\bibfield  {journal} {\bibinfo  {journal} {Comput. Phys. Commun.}\ }\textbf {\bibinfo {volume} {270}},\ \bibinfo {pages} {108153} (\bibinfo {year} {2022})}\BibitemShut {NoStop}%
\end{thebibliography}
\end{document}